\documentclass[longauth]{aa}
\usepackage{graphicx}
\usepackage{multirow}
\usepackage[colorlinks=true, linkcolor=blue, citecolor=blue, draft=False]{hyperref}
\usepackage{gensymb}
\usepackage{color}

\usepackage{txfonts}
\usepackage{natbib}
\usepackage{ulem}
\begin{document}

\title{GOODS-ALMA: 1.1\,mm galaxy survey - I. Source catalogue and optically dark galaxies}

\author{M.~Franco\inst{\ref{inst1}}\thanks{E-mail: maximilien.franco@cea.fr}
\and D.~Elbaz\inst{\ref{inst1}} 
\and M.~B\'ethermin\inst{\ref{inst2}}
\and B.~Magnelli\inst{\ref{inst3}}
\and C.~Schreiber\inst{\ref{inst4}}
\and L.~Ciesla\inst{\ref{inst1},\ref{inst2}}
\and M.~Dickinson\inst{\ref{inst5}}
\and N.~Nagar\inst{\ref{inst6}}
\and J.~Silverman\inst{\ref{inst7}}
\and E.~Daddi\inst{\ref{inst1}}
\and D.~M.~Alexander\inst{\ref{inst8}}
\and T.~Wang\inst{\ref{inst1},\ref{inst9}}
\and M.~Pannella\inst{\ref{inst10}}
\and E.~Le Floc'h\inst{\ref{inst1}}
\and A.~Pope\inst{\ref{inst11}}
\and M.~Giavalisco\inst{\ref{inst11}}
\and A.~J.~Maury\inst{\ref{inst1},\ref{inst12}}
\and F.~Bournaud\inst{\ref{inst1}}
\and R.~Chary\inst{\ref{inst13}}
\and R.~Demarco\inst{\ref{inst6}}
\and H.~Ferguson\inst{\ref{inst14}}
\and S.~L.~Finkelstein\inst{\ref{inst15}}
\and H.~Inami\inst{\ref{inst16}}
\and D.~Iono\inst{\ref{inst17},\ref{inst18}}
\and S.~Juneau\inst{\ref{inst1},\ref{inst5}}
\and G.~Lagache\inst{\ref{inst2}}
\and R.~Leiton\inst{\ref{inst19},\ref{inst6},\ref{inst1}}
\and L.~Lin\inst{\ref{inst20}}
\and G.~Magdis\inst{\ref{inst21},\ref{inst22}}
\and H.~Messias\inst{\ref{inst23},\ref{inst24}}
\and K.~Motohara\inst{\ref{inst25}}
\and J.~Mullaney\inst{\ref{inst26}}
\and K.~Okumura\inst{\ref{inst1}}
\and C.~Papovich\inst{\ref{inst27},\ref{inst28}}
\and J.~Pforr\inst{\ref{inst2},\ref{inst29}}
\and W.~Rujopakarn\inst{\ref{inst30},\ref{inst31},\ref{inst32}}
\and M.~Sargent\inst{\ref{inst33}}
\and X.~Shu\inst{\ref{inst34}}
\and L.~Zhou\inst{\ref{inst1},\ref{inst35}}}

\institute{AIM, CEA, CNRS, Universit\'{e} Paris-Saclay, Universit\'{e} Paris Diderot, Sorbonne Paris Cit\'{e}, F-91191 Gif-sur-Yvette, France\label{inst1}
\and Aix Marseille Univ, CNRS, LAM, Laboratoire d'Astrophysique de Marseille, Marseille, France\label{inst2}
\and Argelander-Institut f\"{u}r Astronomie, Universit\"{a}t Bonn, Auf dem H\"{u}gel 71, D-53121 Bonn, Germany\label{inst3}
\and Leiden Observatory, Leiden University, NL-2300 RA Leiden, The Netherlands\label{inst4}
\and National Optical Astronomy Observatory, 950 North Cherry Avenue, Tucson, AZ 85719, USA \label{inst5}
\and Department of Astronomy, Universidad de Concepci\'{o}n, Casilla 160-C Concepci\'{o}n, Chile \label{inst6}
\and Kavli Institute for the Physics and Mathematics of the Universe (WPI), The University of Tokyo Institutes for Advanced Study, The University of Tokyo, Kashiwa, Chiba 277-8583, Japan\label{inst7}
\and Centre for Extragalactic Astronomy, Department of Physics, Durham University, Durham DH1 3LE, UK\label{inst8}
\and Institute of Astronomy, University of Tokyo, 2-21-1 Osawa, Mitaka, Tokyo 181-0015, Japan\label{inst9}
\and Fakult\"{a}t f\"{u}r Physik der Ludwig-Maximilians-Universit\"{a}t, D-81679 M\"{u}nchen, Germany\label{inst10}
\and Astronomy Department, University of Massachusetts, Amherst, MA 01003, USA\label{inst11}
\and Harvard-Smithsonian Center for Astrophysics, Cambridge, MA 02138, USA\label{inst12}
\and Infrared Processing and Analysis Center, MS314-6, California Institute of Technology, Pasadena, CA 91125, USA\label{inst13}
\and Space Telescope Science Institute, 3700 San Martin Drive, Baltimore, MD 21218, USA\label{inst14}
\and Department of Astronomy, The University of Texas at Austin, Austin, TX 78712, USA\label{inst15}
\and Univ Lyon, Univ Lyon1, ENS de Lyon, CNRS, Centre de Recherche Astrophysique de Lyon (CRAL) UMR5574, F-69230, Saint-Genis-Laval, France\label{inst16}
\and National Astronomical Observatory of Japan, National Institutes of Natural Sciences, 2-21-1 Osawa, Mitaka, Tokyo 181-8588, Japan\label{inst17}
\and SOKENDAI (The Graduate University for Advanced Studies), 2-21-1 Osawa, Mitaka, Tokyo 181-8588, Japan\label{inst18}
\and Instituto de F\'{i}sica y Astronom\'{i}a, Universidad de Valpara\'{i}so, Avda. Gran Breta\~{n}a 1111, Valparaiso, Chile\label{inst19}
\and Institute of Astronomy \& Astrophysics, Academia Sinica, Taipei 10617, Taiwan\label{inst20}
\and Dark Cosmology Centre, Niels Bohr Institute, University of Copenhagen, Juliane Mariesvej 30, 2100, Copenhagen, Denmark\label{inst21}
\and Institute for Astronomy, Astrophysics, Space Applications and Remote Sensing, National Observatory of Athens, GR-15236 Athens, Greece\label{inst22}
\and Joint ALMA Observatory, Alonso de C\'{o}rdova 3107, Vitacura 763-0355, Santiago, Chile\label{inst23}
\and European Southern Observatory, Alonso de C\'{o}rdova 3107, Vitacura, Casilla 19001, 19 Santiago, Chile\label{inst24}
\and Institute of Astronomy, Graduate School of Science, The University of Tokyo, 2-21-1 Osawa, Mitaka, Tokyo 181-0015, Japan\label{inst25}
\and Department of Physics and Astronomy, The University of Sheffield, Hounsfield Road, Sheffield S3 7RH, UK\label{inst26}
\and Department of Physics and Astronomy, Texas A\&M University, College Station, TX, 77843-4242, USA\label{inst27}
\and George P. and Cynthia Woods Mitchell Institute for Fundamental Physics and Astronomy, Texas A\&M University, College Station, TX, 77843-4242, USA\label{inst28}
\and Scientific Support Office, ESA/ESTEC, Noordwijk, The Netherlands\label{inst29}
\and Department of Physics, Faculty of Science, Chulalongkorn University, 254 Phayathai Road, Pathumwan, Bangkok 10330, Thailand\label{inst30}
\and National Astronomical Research Institute of Thailand (Public Organization), Donkaew, Maerim, Chiangmai 50180, Thailand\label{inst31}
\and Kavli Institute for the Physics and Mathematics of the Universe (WPI), The University of Tokyo Institutes for Advanced Study, The University of Tokyo, Kashiwa, Chiba 277-8583, Japan \label{inst32}
\and Astronomy Centre, Department of Physics and Astronomy, University of Sussex, Brighton, BN1 9QH, UK\label{inst33}
\and Department of Physics, Anhui Normal University, Wuhu, Anhui, 241000, China\label{inst34}
\and School of Astronomy and Space Science, Nanjing University, Nanjing 210093, China\label{inst35}
}
\date{Received 2018 March 1; accepted 2018 July 30}

\abstract{}
{We present a 69 arcmin$^2$ ALMA survey at 1.1mm, GOODS--ALMA, matching the deepest HST--WFC3 $H$-band part of the GOODS--\textit{South} field.}
{We taper the 0\arcsec24 original image with a homogeneous and circular synthesized beam of 0\arcsec60 to reduce the number of independent beams -- thus reducing the number of purely statistical spurious detections -- and optimize the sensitivity to point sources. We extract a catalogue of galaxies purely selected by ALMA and identify sources with and without HST counterparts down to a 5$\sigma$ limiting depth of H=28.2 AB (HST/WFC3 F160W).}
{ALMA detects 20 sources brighter than 0.7\,mJy at 1.1mm in the 0\arcsec60 tapered mosaic (rms sensitivity $\sigma$ $\simeq$ 0.18\,mJy.beam$^{-1}$) with a purity greater than 80\%. Among these detections, we identify three sources with no HST nor \textit{Spitzer}-IRAC counterpart, consistent with the expected number of spurious galaxies from the analysis of the inverted image; their definitive status will require additional investigation. An additional three sources with HST counterparts are detected either at high significance in the higher resolution map, or with different detection-algorithm parameters ensuring a purity greater than 80\%. Hence we identify in total 20 robust detections.}
{Our wide contiguous survey allows us to push further in redshift the blind detection of massive galaxies with ALMA with a median redshift of $z$\,=\,2.92 and a median stellar mass of $\text{M}_{\star}$\,=\,1.1\,$\times$\,10$^{11}$M$_\sun$. Our sample includes 20\% HST--dark galaxies (4 out of 20), all detected in the mid-infrared with \textit{Spitzer}--IRAC. The near-infrared based photometric redshifts of two of them ($z$$\sim$4.3 and 4.8) suggest that these sources have redshifts $z$\,$>$\, 4. At least 40\% of the ALMA sources host an X-ray AGN, compared to $\sim$14\% for other galaxies of similar mass and redshift. The wide area of our ALMA survey provides lower values at the bright end of number counts than single-dish telescopes affected by confusion.}


    \maketitle

\section{Introduction}
In the late 1990s a population of galaxies was discovered at submillimetre wavelengths using the Submillimetre Common-User Bolometer Array (SCUBA; \citealt{Holland1999}) on the James Clerk Maxwell Telescope (see e.g. \citealt{Smail1997,Hughes1998,Barger1998,Blain2002}). These "submillimetre galaxies" or SMGs are highly obscured by dust, typically located around $z$ $\sim$2--2.5 \citep[e.g.][]{Chapman2003,Wardlow2011,Yun2012}, massive \citep[M$_\star$\,$>$\,7\,$\times$\,10$^{10} $M$_\sun$ ; e.g.][]{Chapman2005,Hainline2011,Simpson2014}, gas-rich \citep[f$_{gas}$\,$>$\,50\%; e.g.][]{Daddi2010}, with huge star formation rates (SFR) - often greater than 100 M$_\sun$year$^{-1}$ \citep[e.g.][]{Magnelli2012, Swinbank2014} - making them significant contributors to the cosmic star formation \citep[e.g.][]{Casey2013}, often driven by mergers \citep[e.g.][]{Tacconi2008, Narayanan2010} and often host an active galactic nucleus (AGN; e.g. \citealt{Alexander2008, Pope2008,Wang2013}. These SMGs are plausible progenitors of present-day massive early-type galaxies \citep[e.g.][]{Cimatti2008, Michaowski2010}. 

Recently, thanks to the advent of the Atacama Large Millimetre/submillimetre Array (ALMA) and its capabilities to perform both high-resolution and high-sensitivity observations, our view of SMGs has become increasingly refined. The high angular resolution compared to single-dish observations reduces drastically the uncertainties of source confusion and blending, and affords new opportunities for robust galaxy identification and flux measurement. The ALMA sensitivity allows for the detection of sources down to 0.1\,mJy \citep[e.g.][]{Carniani2015}, the analysis of populations of dust-poor high-$z$ galaxies \citep{Fujimoto2016} or Main Sequence (MS; \citealt{Noeske2007,Rodighiero2011, Elbaz2011}) galaxies \cite[e.g.][]{Papovich2016,Dunlop2017,Schreiber2017b}, and also demonstrates that the Extragalactic Background Light (EBL) can be resolved partially or totally by faint galaxies (S\,$<$\,1\,mJy; e.g. \citealt{Hatsukade2013, Ono2014, Carniani2015, Fujimoto2016}). Thanks to this new domain of sensitivity, ALMA is able to unveil less extreme objects, bridging the gap between massive starbursts and more normal galaxies: SMGs no longer stand apart from the general galaxy population.

However, many previous ALMA studies have been based on biased samples, with prior selection (pointing) or a posteriori selection (e.g. based on HST detections) of galaxies, or in a relatively limited region. In this study we present an unbiased view of a large (69 arcmin$^2$) region of the sky, without prior or a posteriori selection based on already known galaxies, in order to improve our understanding of dust-obscured star formation and investigate the main properties of these objects. We take advantage of one of the most uncertain and potentially transformational outputs of ALMA - its ability to reveal a new class of galaxies through serendipitous detections. This is one of the main reasons for performing blind extragalactic surveys.

Thanks to the availability of very deep, panchromatic photometry at rest-frame UV, optical and NIR in legacy fields such as Great Observatories Origins Deep Survey--\textit{South} (GOODS--\textit{South}), which also includes among the deepest available X-ray and radio maps, precise multi-wavelength analysis that include the crucial FIR region is now possible with ALMA. In particular, a population of high redshift (2 $<z<$ 4) galaxies, too faint to be detected in the deepest HST-WFC3 images of the GOODS--\textit{South} field has been revealed, thanks to the thermal dust emission seen by ALMA. Sources without an HST counterpart in the $H$-band, the reddest available (so-called HST-dark) have been previously found by colour selection \cite[e.g.][]{Huang2012, Caputi2012, Caputi2015, Wang2016}, by serendipitous detection of line emitters \cite[e.g.][]{Ono2014} or in the continuum \cite[e.g.][]{Fujimoto2016}. We will show that $\sim$20\% of the sources detected in the survey described in this paper are HST-dark, and strong evidence suggests that they are not spurious detections.

The aim of the work presented in this paper is to exploit a 69 arcmin$^2$ ALMA image reaching a sensitivity of 0.18\,mJy at a resolution of 0\arcsec60.  We use the leverage of the excellent multi-wavelength supporting data in the GOODS--\textit{South} field: the Cosmic Assembly Near-infrared Deep Extragalactic Legacy Survey \mbox{\citep{Koekemoer2011,Grogin2011}}, the \textit{Spitzer} Extended Deep Survey \citep{Ashby2013}, the GOODS--\textit{Herschel} Survey \citep{Elbaz2011}, the \textit{Chandra} Deep Field-\textit{South} \citep{Luo2017} and ultra-deep radio imaging with the VLA \citep{Rujopakarn2016}, to construct a robust catalogue and derive physical properties of ALMA-detected galaxies. 
The region covered by ALMA in this survey corresponds to the region with the deepest HST-WFC3 coverage, and has also been chosen for a guaranteed time observation (GTO) program with the James Webb Space Telescope (JWST).

This paper is organized as follows: in $\S$\ref{sec:Survey_data} we describe our ALMA survey, the data reduction, and the multi-wavelength ancillary data which support our studies. In $\S$\ref{sec:Source_Detection}, we present the methodology and criteria used to detect sources, we also present the procedures used to compute the completeness and the fidelity of our flux measurements. In $\S$\ref{sec:catalogue} we detail the different steps we conducted to construct a catalogue  of our detections.  In $\S$\ref{sec:Number_counts} we estimate the differential and cumulative number counts from our detections. We compare these counts with other (sub)millimetre studies. In  $\S$\ref{sec:Properties} we investigate some properties of our galaxies such as redshift and mass distributions. Other properties will be analysed in Franco et al. (in prep) and finally in $\S$\ref{sec:Conclusion}, we summarize the main results of this study.  Throughout this paper, we adopt a spatially flat $\Lambda$CDM cosmological model with H$_0$\,=\,70 kms$^{-1}$Mpc$^{-1}$, $\Omega_m$\,=\,0.7 and $\Omega_{\Lambda}$\,=\,0.3. We assume a Salpeter \citep{Salpeter1955} Initial Mass Function (IMF). We use the conversion factor of M$_\star$ (\citealt{Salpeter1955} IMF)\,=\,1.7\,$\times$\,M$_\star$ (\citealt{Chabrier2003} IMF). All magnitudes are quoted in the AB system \citep{Oke1983}.

\section{ALMA GOODS--\textit{South} Survey Data}\label{sec:Survey_data}
\subsection{Survey description}
\begin{figure*}
   \centering
   \includegraphics[width=0.9\hsize]{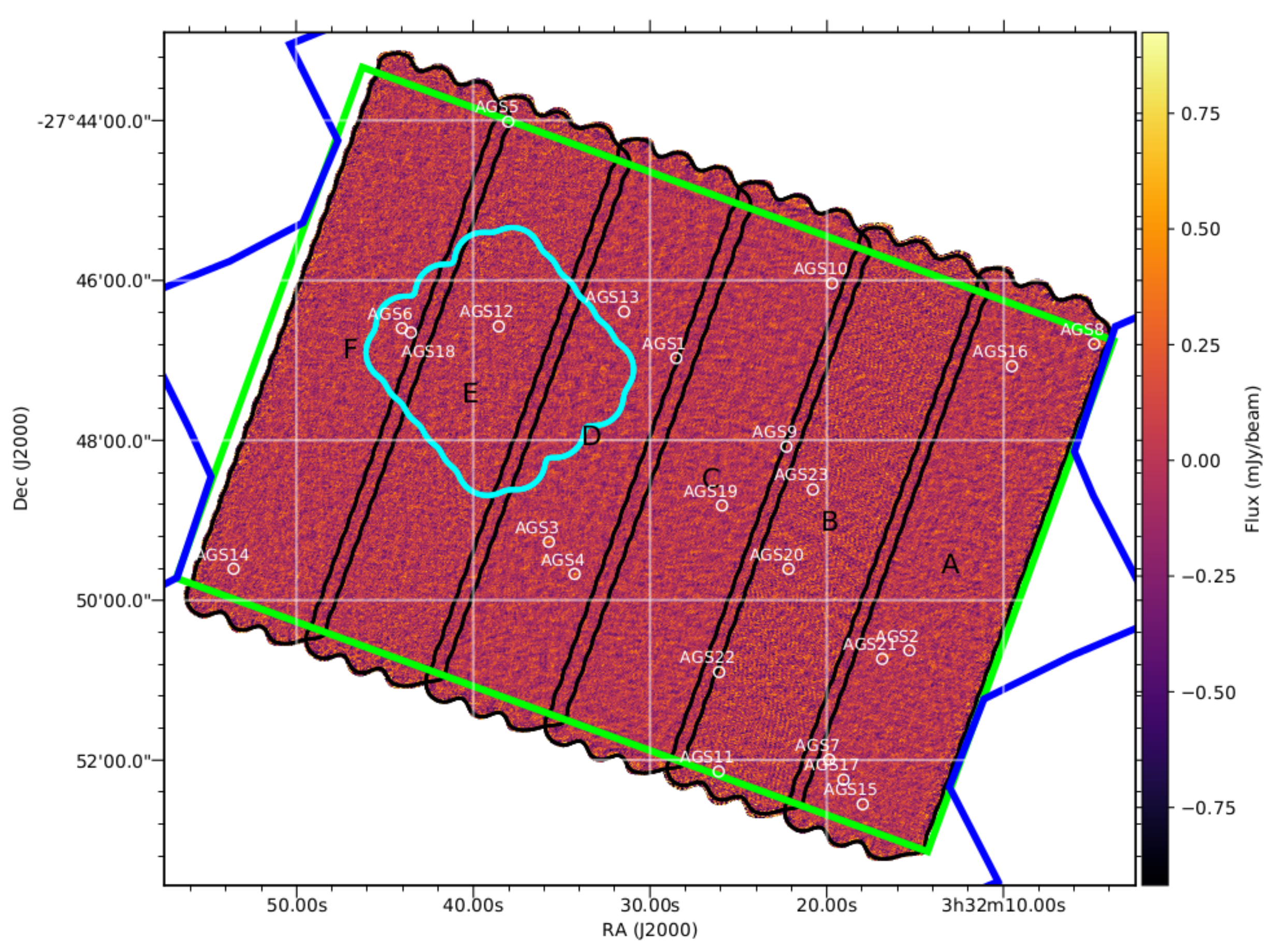}
      \caption{ALMA 1.1 mm image tapered at 0\arcsec60. The white circles have a diameter of 4 arcseconds and indicate the positions of the galaxies listed in Table~\ref{catalogue}. Black contours show the different slices (labelled A to F) used to compose the homogeneous 1.1 mm coverage, with a median RMS-noise of 0.18\,mJy per beam.  Blue lines show the limits of the HST/ACS field and green lines indicate the HST-WFC3 deep region. The cyan contour represents the limit of the \cite{Dunlop2017} survey covering all the Hubble Ultra Deep Field region. All of the ALMA-survey field is encompassed by the \textit{Chandra} Deep Field-\textit{South}.}
         \label{map_detection}
   \end{figure*}

Our ALMA coverage extends over an effective area of 69 arcmin$^2$ within the GOODS--\textit{South} field (Fig.~\ref{map_detection}), centred at $\alpha$\,=\,3$^{\rm h}$ 32$^{\rm m}$ 30.0$^{\rm s}$ , $\delta$\,=\,-27$\degree$ 48$\arcmin$ 00$\arcsec$ (J2000; 2015.1.00543.S; PI: D. Elbaz).
To cover this $\sim$10\arcmin\,$\times$\,7\arcmin  region (comoving scale of 15.1 Mpc\,$\times$\,10.5 Mpc at $z$\,=\,2),  we designed a 846-pointing mosaic, each pointing being separated by $0.8$ times the antenna Half Power Beam Width (i.e. HPBW$\,\sim\,$23\arcsec3).

To accommodate such a large number of pointings within the ALMA Cycle 3 observing mode restrictions, we divided this mosaic into six parallel, slightly overlapping, sub-mosaics of 141 pointing each. To get a homogeneous pattern over the 846 pointings, we computed the offsets between the sub-mosaics so that they connect with each other without breaking the hexagonal pattern of the ALMA mosaics.

Each sub-mosaic (or slice) has a length of 6.8 arcmin, a width of 1.5 arcmin and an inclination (PA) of 70 deg (see Fig.~\ref{map_detection}). This required three execution blocks (EBs), yielding a total on-source integration time of $\sim\,$60 seconds per pointing (Table~\ref{Planning_observation}).
We determined that the highest frequencies of the band 6 is the optimal setup for a continuum survey and we thus set the ALMA correlator to Time Division Multiplexing (TDM) mode and optimised the setup for continuum detection at 264.9\,GHz ($\lambda$\,=\,$1.13$\,mm) using four 1875\,MHz-wide spectral windows centered at 255.9\,GHz, 257.9\,GHz, 271.9\,GHz and 273.9\,GHz, covering a total bandwidth of $7.5\,$GHz.
The TDM mode has 128 channels per spectral window, providing us with $\sim$37 km/s velocity channels.

Observations were taken between the 1$^{\rm{st}}$ of August and the $2^{\rm{nd}}$ of September 2016, using $\sim$40 antennae (see Table~\ref{Planning_observation}) in configuration C40-5 with a maximum baseline of $\sim\,$1500\,m.
J0334-4008 and J0348-2749 (VLBA calibrator and hence has a highly precise position) were systematically used as flux and phase calibrators, respectively.
In 14 EBs, J0522-3627 was used as bandpass calibrator, while in the remaining 4 EBs J0238+1636 was used.
Observations were taken under nominal weather conditions with a typical precipitable water vapour of $\sim\,$1\,mm.

\subsection{Data reduction}\label{sec:data_reduction}

All EBs were calibrated with CASA \citep{McMullin2007} using the scripts provided by the ALMA project.
Calibrated visibilities were systematically inspected and few additional flaggings were added to the original calibration scripts.
Flux calibrations were validated by verifying the accuracy of our phase and bandpass calibrator flux density estimations. 
Finally, to reduce computational time for the forthcoming continuum imaging, we time- and frequency-averaged our calibrated EBs over 120\,seconds and 8 channels, respectively.

Imaging was done in CASA using the multi-frequency synthesis algorithm implemented within the task \texttt{CLEAN}.
Sub-mosaics were produced separately and combined subsequently using a weighted mean based on their noise maps.
As each sub-mosaic was observed at different epochs and under different weather conditions, they exhibit different synthesized beams and sensitivities (Table~\ref{Planning_observation}). Sub-mosaics were produced and primary beam corrected separately, to finally be combined using a weighted mean based on their noise maps.
To obtain a relatively homogeneous and circular synthesized beam across our final mosaic, we applied different $u$, $v$ tapers to each sub-mosaic.
The best balance between spatial resolution and sensitivity was found with a homogeneous and circular synthesized beam of $0\arcsec29$ Full Width Half Maximum (FWHM; hereafter 0\arcsec29-mosaic; Table~\ref{Planning_observation}). This resolution corresponds to the highest resolution for which a circular beam can be synthesized for the full mosaic.
We also applied this tapering method to create a second mosaic with an homogeneous and circular synthesized beam of $0\arcsec60$ FWHM (hereafter 0\arcsec60-mosaic; Table~\ref{Planning_observation}), i.e., optimised for the detection of extended sources.
Mosaics with even coarser spatial resolution could not be created because of drastic sensitivity and synthesized beam shape degradations.

\begin{figure}
   \centering
   \includegraphics[width=0.9\hsize]{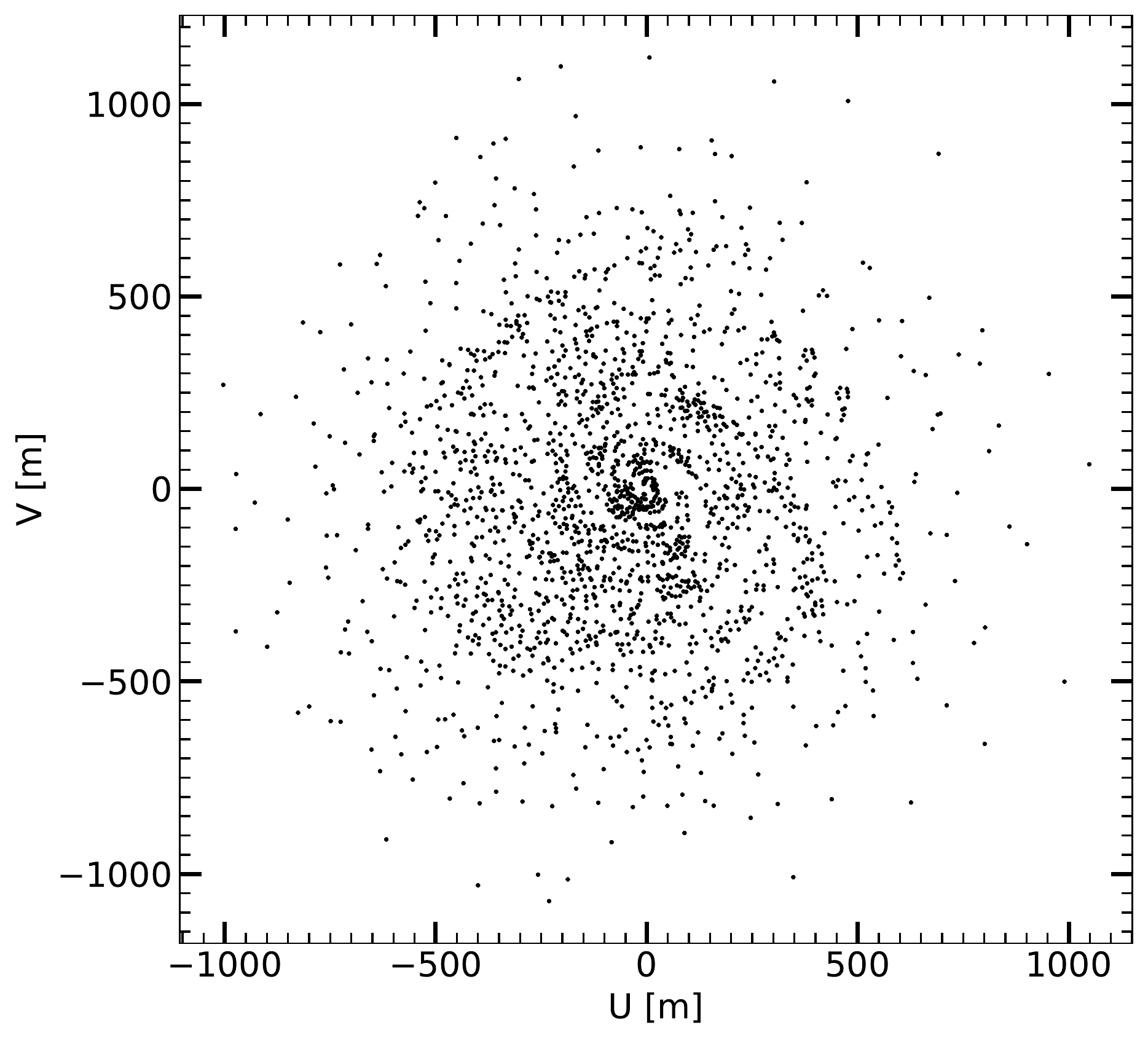}
      \caption{\textit{uv}-coverage of one of the 846 ALMA pointings constituting this survey. This \textit{uv}-coverage allows us to perform the source detection in the dirty map.}
         \label{UV_Coverage}
   \end{figure}

Due to the good coverage in the \textit{uv}-plane (see Fig.~\ref{UV_Coverage}) and the absence of very bright sources (the sources present in our image do not cover a large dynamic range in flux densities; see Sect.~\ref{sec:catalogue}), we decided to work with the dirty map. This prevents introducing potential biases during the \texttt{CLEAN} process and we noticed that the noise in the clean map is not significantly different (<\,1\%).

\begin{table*}\footnotesize    
\centering          
\begin{tabular}{l l c c c c c c c c c c c c c}   
\hline     
\rule{0pt}{2.5ex}&   &  &  &  & & \multicolumn{2}{c}{Original Mosaic} & &\multicolumn{2}{c}{0\arcsec29-Mosaic} &  &\multicolumn{2}{c}{0\arcsec60-Mosaic}\\
\cline{7-8} \cline{10-11} \cline{13-14}
\rule{0pt}{2.5ex}Slice & Date  & \# & t on target & total t & & Beam & $\sigma$ & & Beam  & $\sigma$ & & Beam  & $\sigma$ \\
&&&{\tiny min} & {\tiny min} & & {\tiny mas$\,\times\,$mas} & {\tiny\,$\mu$Jy.beam$^{-1}$} & &{\tiny mas$\,\times\,$mas} & {\tiny\,$\mu$Jy.beam$^{-1}$} & & {\tiny mas$\,\times\,$mas} & {\tiny\,$\mu$Jy.beam$^{-1}$}\\
\hline
\hline
\multirow{3}{*}{A}  &August 17  &42 & 46.52& 72.12 & & \multirow{3}{*}{240$\,\times\,$200} & \multirow{3}{*}{98} & & \multirow{3}{*}{297$\,\times\,$281} & \multirow{3}{*}{108}&  & \multirow{3}{*}{618$\,\times\,$583} & \multirow{3}{*}{171}\\
&August 31  &39 & 50.36& 86.76 \\ 
& August 31  &39 & 46.61& 72.54 \\ 

\hline
\multirow{3}{*}{B}&September 1  &38 & 46.87& 72.08 & & \multirow{3}{*}{206$\,\times\,$184} & \multirow{3}{*}{113} & & \multirow{3}{*}{296$\,\times\,$285} & \multirow{3}{*}{134}&  & \multirow{3}{*}{614$\,\times\,$591} & \multirow{3}{*}{224}\\
&September 1  &38 & 48.16& 72.48 \\ 
& September 2  &39 & 46.66& 75.06 \\ 
\hline
\multirow{3}{*}{C}&August 16  &37 & 46.54& 73.94 & & \multirow{3}{*}{243$\,\times\,$231} & \multirow{3}{*}{102} & & \multirow{3}{*}{295$\,\times\,$288} & \multirow{3}{*}{107}&  & \multirow{3}{*}{608$\,\times\,$593} & \multirow{3}{*}{166}\\
&August 16 &37 & 46.54& 71.58 \\ 
& August 27  &42 & 46.52& 74.19 \\ 
\hline
\multirow{3}{*}{D}
&August 16  &37 & 46.54& 71.69  & & \multirow{3}{*}{257$\,\times\,$231} & \multirow{3}{*}{107} & & \multirow{3}{*}{292$\,\times\,$289} & \multirow{3}{*}{111}&  & \multirow{3}{*}{612$\,\times\,$582} & \multirow{3}{*}{164}\\
&August 27  &44 & 46.52& 72.00 \\ 
& August 27  &44 & 46.52& 72.08 \\ 
\hline
\multirow{3}{*}{E}
&August 01  &39 & 46.54& 71.84 & & \multirow{3}{*}{285$\,\times\,$259} & \multirow{3}{*}{123} & & \multirow{3}{*}{292$\,\times\,$286} & \multirow{3}{*}{124}&  & \multirow{3}{*}{619$\,\times\,$588} & \multirow{3}{*}{186}\\
&August 01  &39 & 46.53& 72.20 \\ 
& August 02  &40 & 46.53& 74.46 \\ 
\hline
\multirow{3}{*}{F}
&August 02  &40 & 46.53& 72.04 & & \multirow{3}{*}{293$\,\times\,$256} & \multirow{3}{*}{118} & & \multirow{3}{*}{292$\,\times\,$284} & \multirow{3}{*}{120}&  & \multirow{3}{*}{613$\,\times\,$582} & \multirow{3}{*}{178}\\
&August 02  &41 & 46.53& 71.61 \\ 
& August 02  &39 & 46.53& 71.55 \\ 
\hline
Mean & & 40 & 46.86&73.35& & 254$\,\times\,$227&110& & 294$\,\times\,$286&117 & &614$\,\times\,$587&182\\
\hline
Total & & & 843.55 & 1320.22 & &&&&& \\
\hline
\hline                  
\end{tabular}
\caption{Summary of the observations. The slice ID, the date, the number of antennae, the time on target, the total time (time on target + calibration time), the resolution and the 1-$\sigma$ noise of the slice are given.}             
\label{Planning_observation}      
\end{table*}

\subsection{Building of the noise map}
We build the RMS-map of the ALMA survey by a k-$\sigma$ clipping method. In steps of 4 pixels on the image map, the standard deviation was computed in a square of 100\,$\times$\,100 pixels around the central pixel. The pixels, inside this box, with values greater than 3 times the standard deviation ($\sigma$) from the median value were masked. This procedure was repeated 3 times. Finally, we assign the value of the standard deviation of the non-masked pixels to the central pixel. This box size corresponds to the smallest size for which the value of the median pixel of the rms map converges to the typical value of the noise in the ALMA map while taking into account the local variation of noise. The step of 4 pixels corresponds to a sub-sampling of the beam so, the noise should not vary significantly on this scale. The median value of the standard deviation is 0.176\,mJy.beam$^{-1}$. In comparison, the Gaussian fit of the unclipped map gives a standard deviation of 0.182\,mJy.beam$^{-1}$. We adopt a general value of rms sensitivity $\sigma$\,=\,0.18\,mJy.beam$^{-1}$. The average values for the 0\arcsec29-mosaic and the untapered mosaic are given in Table~\ref{Planning_observation}.

\subsection{Ancillary data}\label{sec:Ancillary_data}
The area covered by this survey is ideally located, in that it profits from ancillary data from some of the deepest sky surveys at infrared (IR), optical and X-ray wavelengths. In this section, we describe all of the data that were used in the analysis of the ALMA detected sources in this paper.
\subsubsection{Optical/near-infrared imaging}\label{sec:Optical/near-infrared_imaging}
We have supporting data from the Cosmic Assembly Near-IR Deep Extragalactic Legacy Survey (CANDELS; \citealt{Grogin2011}) with images obtained with the Wide Field Camera 3 / infrared channel (WFC3/IR) and UVIS channel, along with the Advanced Camera for Surveys (ACS; \citealt{Koekemoer2011}. The area covered by this survey lies in the deep region of the CANDELS program (central one-third of the field). The 5-$\sigma$ detection depth for a point-source reaches a magnitude of 28.16 for the $H_{160}$ filter (measured within a fixed aperture of 0.17$\arcsec$ \citealt{Guo2013}). The CANDELS/Deep program also provides images in 7 other bands: the $Y_{125}$, $J_{125}$, $B_{435}$, $V_{606}$, $i_{775}$, $i_{814}$ and $z_{850}$ filters, reaching 5-$\sigma$ detection depths of 28.45, 28.35, 28.95, 29.35, 28.55, 28.84, and 28.77 mag respectively.

The \cite{Guo2013} catalogue  also includes galaxy magnitudes from the VLT, taken in the $U$-band with VIMOS \mbox{\citep{Nonino2009}}, and in the $K_s$-band with ISAAC \citep{Retzlaff2010} and HAWK-I \citep{Fontana2014}.

In addition, we use data coming from the \texttt{FourStar} Galaxy Evolution Survey (ZFOURGE, PI: I. Labb\'{e}) on the 6.5 m Magellan Baade Telescope. The \texttt{FourStar} instrument \mbox{\citep{Persson2013}} observed the CDFS (encompassing the GOODS--\textit{South} Field) through 5 near-IR medium-bandwidth filters ($J_1$, $J_2$, $J_3$, $H_s$, $H_l$) as well as broad-band $K_s$. By combination of the \text{FourStar} observations in the $K_s$-band and previous deep and ultra-deep surveys in the $K$-band, VLT/ISAAC/$K$ (v2.0) from GOODS \citep{Retzlaff2010}, VLT/HAWK-I/$K$ from HUGS \citep{Fontana2014}, CFHST/WIRCAM/$K$ from TENIS \citep{Hsieh2012} and Magellan/PANIC/$K$ in HUDF (PI: I. Labb\'e), a super-deep detection image has been produced. The ZFOURGE catalogue reaches a completeness greater than 80\% to $K_s$\,$<$\, 25.3 - 25.9 \citep{Straatman2016}.

We use the stellar masses and redshifts from the ZFOURGE catalogue, except when spectroscopic redshifts are available. Stellar masses have been derived from \cite{Bruzual2003} models \citep{Straatman2016} assuming exponentially declining star formation histories and a dust attenuation law as described by \cite{Calzetti2000}.

\subsubsection{Mid/far-infrared imaging}

Data in the mid and far-IR are provided by the Infrared Array Camera (\text{IRAC}; \citealt{Fazio2004}) at 3.6, 4.5, 5.8, and 8\,$\mu$m, \textit{Spitzer} Multiband Imaging Photometer (MIPS; \citealt{Rieke2004}) at 24\,$\mu$m, \textit{Herschel} Photodetector Array Camera and Spectrometer (PACS, \citealt{Poglitsch2010}) at 70, 100 and 160\,$\mu$m, and \textit{Herschel} Spectral and Photometric Imaging REceiver (SPIRE, \citealt{Griffin2010}) at 250, 350, and 500 \,$\mu$m.

The IRAC observations in the GOODS--\textit{South} field were taken in February 2004 and August 2004 by the GOODS \textit{Spitzer} Legacy project (PI: M. Dickinson). These data have been supplemented by the \textit{Spitzer} Extended Deep Survey (SEDS;  PI: G. Fazio) at 3.6 and 4.5\,$\mu$m \citep{Ashby2013} as well as the \textit{Spitzer}-Cosmic Assembly Near-Infrared Deep Extragalactic Survey (S-CANDELS; \citealt{Ashby2015}) and recently by the ultradeep \text{IRAC} imaging  at 3.6 and 4.5\,$\mu$m \citep{Labbe2015}.

The flux extraction and deblending in 24 $\mu$m imaging have been provided by \cite{Magnelli2009} to reach a depth of $S_{24}$ $\sim$30\,$\mu$Jy.
\textit{Herschel} images come from a 206.3 h GOODS--\textit{South} observational program \citep{Elbaz2011} and combined by \cite{Magnelli2013} with the PACS Evolutionary Probe (PEP) observations \citep{Lutz2011}. Because the SPIRE confusion limit is very high, we use the catalogue of T. Wang et al. (in prep), which is built with a state-of-the art de-blending method using optimal prior sources positions from 24\,$\mu$m and \textit{Herschel} PACS detections.

\subsubsection{Complementary ALMA data}
As the GOODS--\textit{South} Field encompasses the Hubble Ultra Deep Field (HUDF), we take advantage of deep 1.3-mm ALMA data of the HUDF. The ALMA image of the full HUDF reaches a $\sigma_{1.3mm}$\,=\,35\,$\mu$Jy \citep{Dunlop2017}, over an area of 4.5 arcmin$^2$ that was observed using a 45-pointing mosaic at a tapered resolution of 0.7\arcsec. These observations were taken in two separate periods from July to September 2014. In this region, 16 galaxies were detected by \cite{Dunlop2017}, 3 of them with a high SNR (SNR\,$>$\, 14), the other 13 with lower SNRs (3.51\,$<$\, SNR\,$<$\, 6.63).

\subsubsection{Radio imaging}
We also use radio imaging at 5\,cm from the Karl G. Jansky Very Large Array (VLA). These data were observed during 2014 March - 2015 September for a total of 177\,hours in the A, B, and C configurations (PI: W. Rujopakarn). The images have a 0\arcsec31\,$\times$\,0\arcsec61 synthesized beam and an rms noise at the pointing centre of 0.32\,$\mu$Jy.beam$^{-1}$\citep{Rujopakarn2016}. Here, 179 galaxies were detected with a significance greater than 3\,$\sigma$ over an area of 61\,arcmin$^2$ around the HUDF field, with a rms sensitivity better than 1\,$\mu$Jy.beam$^{-1}$. However, this radio survey does not cover the entire ALMA area presented in this paper.

\subsubsection{X-ray}
The \textit{Chandra} Deep Field-\textit{South} (CDF-S) was observed for 7 Msec between 2014 June and 2016 March. These observations cover a total area of 484.2\,arcmin$^2$, offset by just 32\arcsec from the centre of our survey, in three X-ray bands: 0.5-7.0\,keV, 0.5-2.0\,keV, and 2-7\,keV \citep{Luo2017}. The average flux limits over the central region are 1.9\,$\times$\,10$^{-17}$, 6.4\,$\times$\,10$^{-18}$, and 2.7\,$\times$\,10$^{-17}$\,erg cm$^{-2}$ s$^{-1}$ respectively. This survey enhances the previous X-ray catalogues in this field, the 4 Msec \textit{Chandra} exposure \citep{Xue2011} and the 3 Msec XMM-Newton exposure \citep{Ranalli2013}. We will use this X-ray catalogue to identify candidate X-ray active galactic nuclei (AGN) among our ALMA detections.

\section{Source Detection}\label{sec:Source_Detection}
The search for faint sources in high-resolution images with moderate source densities faces a major limitation. At the native resolution (0\arcsec25\,$\times$\,0\arcsec23), the untapered ALMA mosaic encompasses almost 4 million independent beams, where the beam area is $A_{\rm beam}$\,=\,$\pi$\,$\times$\,FWHM$^2/(4ln(2))$. It results that a search for sources above a detection threshold of 4-$\sigma$ would include as many as 130 spurious sources assuming a Gaussian statistics.
Identifying the real sources from such catalogue is not possible. In order to increase the detection quality to a level that ensures a purity greater than 80\% -- i.e., the excess of sources in the original mosaic needs to be 5 times greater than the number of detections in the mosaic multiplied by (-1) -- we have decided to use a tapered image and adapt the detection threshold accordingly.

By reducing the weight of the signal originating from the most peripheral ALMA antennae, the tapering reduces the angular resolution hence the number of independent beams at the expense of collected light. The lower angular resolution presents the advantage of optimizing the sensitivity to point sources -- we recall that 0\arcsec24 corresponds to a proper size of only 2 kpc at $z$$\sim$1--3 --  and therefore will result in an enhancement of the signal-to-noise ratio for the sources larger than the resolution.

We chose to taper the image with an homogeneous and circular synthesized beam of $0\arcsec60$ FWHM -- corresponding to a proper size of 5 kpc at $z$$\sim$1--3 -- after having tested various kernels and found that this beam was optimised for our mosaic, avoiding both a beam degradation and a too heavy loss of sensitivity. This tapering reduces by nearly an order of magnitude the number of spurious sources expected at a 4-$\sigma$ level down to about 19 out of 600 000 independent beams. However, we will check in a second step whether we may have missed in the process some compact sources by analysing as well the 0\arcsec29 tapered map. 

We also excluded the edges of the mosaic, where the standard deviation is larger than 0.30\,mJy.beam$^{-1}$ in the 0\arcsec60-mosaic. The effective area is thus reduced by 4.9\% as compared to the full mosaic (69.46 arcmin$^2$ out of 72.83 arcmin$^2$). 

To identify the galaxies present on the image, we use \textsc{Blobcat} \citep{Hales2012}. \textsc{Blobcat} is a source extraction software using a "flood fill" algorithm to detect and catalogue  blobs (see \citealt{Hales2012}). A blob is defined by two criteria:
\begin{itemize}
\item at least one pixel has to be above a threshold ($\sigma_p$)
\item all the adjacent surrounding pixels must be above a floodclip threshold ($\sigma_f$)
\end{itemize}
where $\sigma_p$ and $\sigma_f$ are defined in number of $\sigma$, the local RMS of the mosaic.

\begin{figure}
\centering
\resizebox{\hsize}{!} {
\includegraphics[width=5cm,clip]{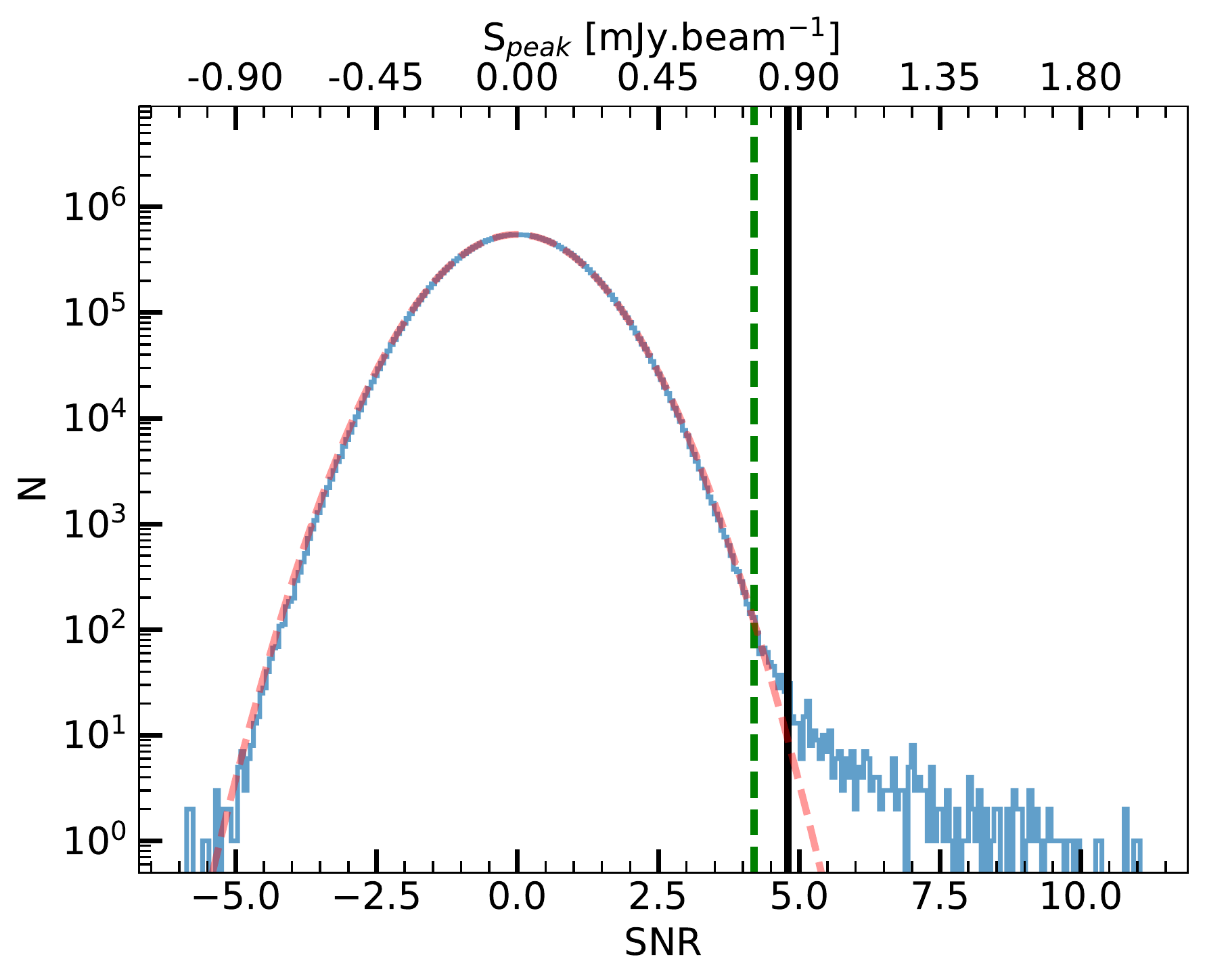}
}
\caption{Histogram of pixels of the signal to noise map, where pixels with noise\,$>$\, 0.3\,mJy.beam$^{-1}$ have been removed. The red dashed line is the best Gaussian fit. The green dashed line is indicative and shows where the pixel brightness distribution moves away from the Gaussian fit. This is also the 4.2\,$\sigma$ level corresponding to a peak flux of 0.76\,mJy for a typical noise per beam of 0.18\,mJy. The solid black line corresponds to our peak threshold of 4.8\,$\sigma$ (0.86\,mJy).}
\label{histogramme_SN}
\end{figure}

A first guess to determine the detection threshold $\sigma_p$ is provided by the examination of the pixel distribution of the signal to noise map (SNR-map). The SNR-map has been created by dividing the 0\arcsec60 tapered map by the noise map. Fig.~\ref{histogramme_SN} shows that the SNR-map follows an almost perfect Gaussian below SNR\,=\,4.2. Above this threshold, a significant difference can be observed that is characteristic of the excess of positive signal expected in the presence of real sources in the image. However, this histogram alone cannot be used to estimate a number of sources because the pixels inside one beam are not independent from one another. Hence although the non Gaussian behaviour appears around SNR\,=\,4.2 we perform simulations to determine the optimal values of $\sigma_p$ and $\sigma_f$.

We first conduct positive and negative -- on the continuum map multiplied by (-1) -- detection analysis for a range of $\sigma_p$ and $\sigma_f$ values ranging from $\sigma_p$\,=\,4 to 6 and $\sigma_f$\,=\,2.5 to 4
with intervals of 0.05 and imposing each time $\sigma_p$ $\ge$ $\sigma_f$. The difference between positive and negative detections for each pair of ($\sigma_p$, $\sigma_f$) values provides the expected number of real sources. 

\begin{figure*}
\centering
\begin{minipage}[t]{1.0\textwidth}
\resizebox{\hsize}{!} {
\includegraphics[width=5.17cm,clip]{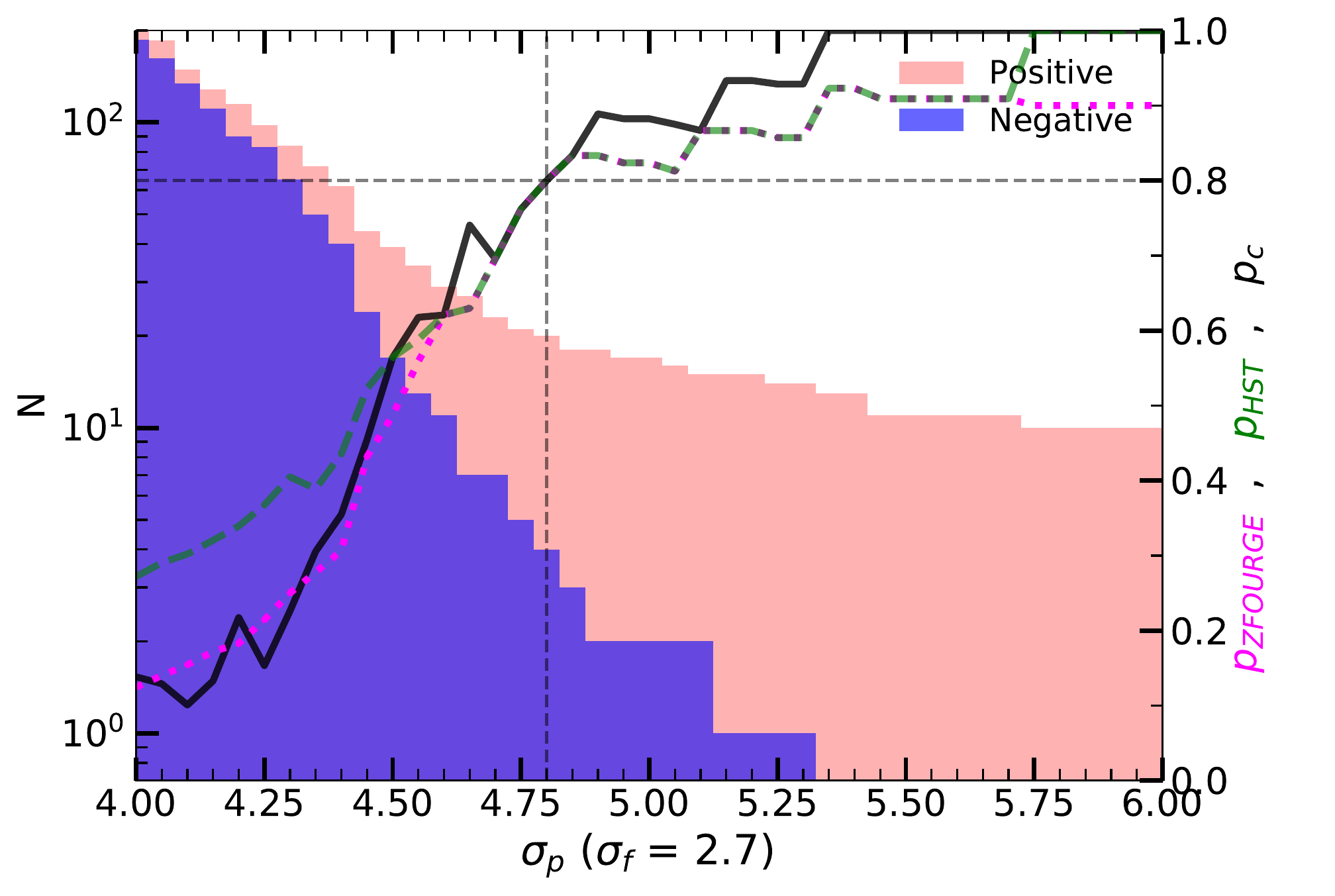}
\includegraphics[width=5cm,clip]{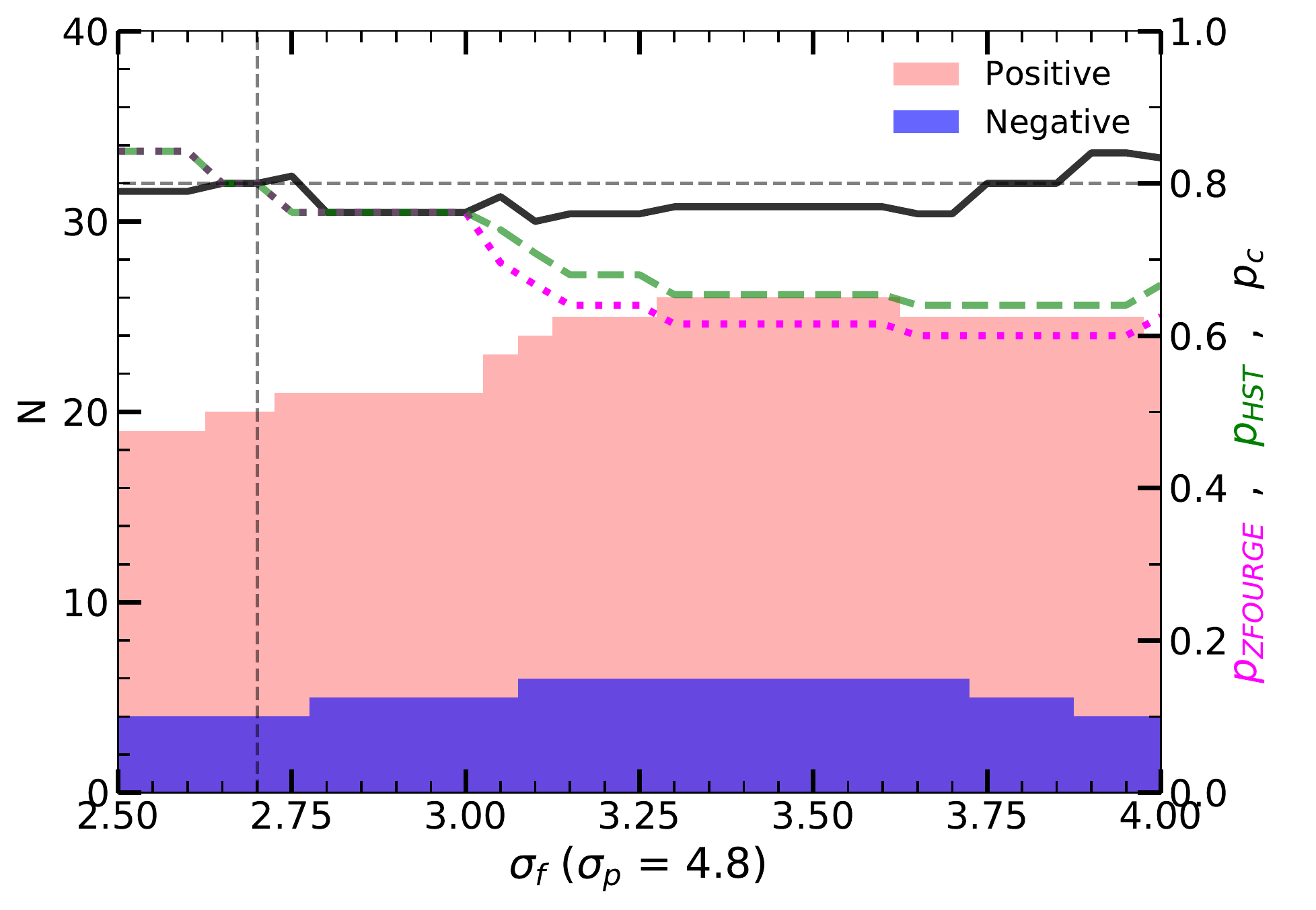}
}
\end{minipage}
\caption{Cumulative number of positive (red histogram) and negative (blue histogram) detections as a function of the $\sigma_p$ (at a fixed $\sigma_f$, left panel) and $\sigma_f$ (at a fixed $\sigma_p$, right panel) in units of $\sigma$. Solid black line represents the purity criterion  $p_{c}$ define by Eq.~\ref{quality_criteria}, green dashed-line represents the percentage of positive detection with HST-WFC3 counterpart  $p_{HST}$ and magenta dashed-line represents the percentage of positive detection with ZFOURGE counterpart  $p_{ZFOURGE}$. Grey dashed-lines show the thresholds $\sigma_p$\,=\,4.8\,$\sigma$ and $\sigma_f$\,=\,2.7\,$\sigma$  and the 80\% purity limit.}
\label{positive_negative}
\end{figure*}

Then we search for the pair of threshold parameters to find the best compromise between \textit{(i)} providing the maximum number of detections, and \textit{(ii)} minimising the number spurious sources.
The later purity criterion, $p_c$, is defined as:
\begin{equation}
p_c\,=\,\frac{N_p -N_n}{N_p}
\label{quality_criteria}
\end{equation}
\noindent where $N_p$ and $N_n$ are the numbers of positive and negative detections respectively. To ensure a purity of 80\% as discussed above, we enforce $p_c$ $\geq$ 0.8. This leads to $\sigma_p$\,=\,4.8\,$\sigma$ when fixing the value of $\sigma_f$\,=\,2.7\,$\sigma$ (see Fig.~\ref{positive_negative}-left). Below $\sigma_p$\,=\,4.8\,$\sigma$, the purity criterion rapidly drops below 80\% whereas above this value it only mildly rises. 
Fixing $\sigma_p$\,=\,4.8\,$\sigma$, the purity remains roughly constant at $\sim$80$\pm$5\% when varying $\sigma_f$.
We do see an increase in the difference between the number of positive and negative detections with increasing $\sigma_f$.
However, the size of the sources above $\sigma_f$\,=\,2.7\,$\sigma$ drops below the 0\arcsec60 FWHM and tends to become pixel-like, hence non physical. This is due to the fact that an increase of $\sigma_f$ results in a reduction of the number of pixels above the floodclip threshold ($\sigma_f$) that will be associated to a given source. This parameter can be seen as a percolation criterion that sets the size of the sources in number of pixels. Reversely reducing $\sigma_f$ below 2.7\,$\sigma$ results in adding more noise than signal and in reducing the number of detections. We therefore decided to set $\sigma_f$ to 2.7\,$\sigma$.

While we do not wish to impose a criterion on the existence of optical counterparts to define our ALMA catalogue, we do find that high values of $\sigma_f$ not only generates the problem discussed above, but also generates a rapid drop of the fraction of ALMA detections with an HST counterpart in the \cite{Guo2013} catalogue, $p_{HST}$\,=\,$N_{HST}/N_p$. $N_{HST}$ is the number of ALMA sources with an HST counterpart within 0\arcsec60 (corresponding to the size of the beam). The fraction falls rapidly from around $\sim$80\% to $\sim$60\%, which we interpret as being due to a rise of the proportion of spurious sources since the faintest optical sources, e.g., detected by HST-WFC3, are not necessarily associated with the faintest ALMA sources due to the negative K-correction at 1.1mm.
This rapid drop can be seen in the dashed green and dotted pink lines of Fig.~\ref{positive_negative}-right. This confirms that the sources that are added to our catalogue with a floodclip threshold greater than 2.7\,$\sigma$ are most probably spurious. Similarly, we can see in Fig.~\ref{positive_negative}-left that increasing the number of ALMA detections to fainter flux densities by reducing $\sigma_p$ below 4.8\,$\sigma$ leads to a rapid drop of the fraction of ALMA detections with an HST counterpart. Again there is no well-established physical reason to expect the number of ALMA detections with an optical counterpart to decrease with decreasing S/N ratio in the ALMA catalogue.

Hence we decided to set $\sigma_p$\,=\,4.8\,$\sigma$ and $\sigma_f$\,=\,2.7\,$\sigma$ to produce our catalogue of ALMA detections. We note that we only discussed the existence of HST counterparts as a complementary test on the definition of the detection thresholds but our approach is not set to limit in any way our ALMA detections to galaxies with HST counterparts.

Indeed, evidence for the existence of ALMA detections with no HST-WFC3 counterparts already exist in the literature. \cite{Wang2016} identified $H$-dropouts galaxies, i.e. galaxies detected above the $H$-band with \textit{Spitzer}-IRAC at 4.5\,$\mu$m but undetected in the $H$-band and in the optical. The median flux density of these galaxies is F$_{870\mu m}\simeq\,$1.6\,mJy (T. Wang et al., in prep.). By scaling this median value to our wavelength of 1.1 mm (the details of this computation are given in Sect.~\ref{sec:Cumulative_number_counts}), we obtain a flux density of 0.9\,mJy, close to the typical flux of our detections (median flux $\sim$1\,mJy, see Table~\ref{catalogue}).

\section{Catalogue}\label{sec:catalogue}
\subsection{Creation of the catalogue}\label{sec:Main_catalogue}
Using the optimal parameters of $\sigma_p$\,=\,4.8\,$\sigma$ and $\sigma_f$\,=\,2.7\,$\sigma$ described in Sect.~\ref{sec:Source_Detection}, we obtain a total of 20 detections down to a flux density limit of $S_{1.1mm}$\,$\approx$\,880\,$\mu$Jy that constitute our main catalogue. These detections can be seen ranked by their SNR in Fig.~\ref{map_detection}. The comparison of negative and positive detections suggests the presence of 4$\pm$2 (assuming a Poissonian uncertainty on the difference between the number of positive and negative detections) spurious sources in this sample. 

In the following, we assume that the galaxies detected in the 0\arcsec60-mosaic are point-like. This hypothesis will later on be discussed and justified in Sect.~\ref{sec:galaxy_sizes}. In order to check the robustness of our flux density measurements, we compared different flux extraction methods and softwares: \texttt{PyBDSM} \mbox{\citep{Mohan2015}}; \texttt{Galfit} \citep{Peng2010}; \texttt{Blobcat} \citep{Hales2012}. The peak flux value determined by \texttt{Blobcat} refers to the peak of the surface brightness corrected for peak bias (see \citealt{Hales2012}). The different results are consistent, with a median ratio of F$_{peak}^{\texttt{Blobcat}}$/F$_{peak}^{\texttt{PyBDSM}}$\,=\,1.04$\pm$0.20 and F$_{peak}^{\texttt{Blobcat}}$/F$_{PSF}^{\texttt{Galfit}}$\,=\,0.93$\pm$0.20. The flux measured using psf-fitting (\texttt{Galfit}) and peak flux measurement (\texttt{Blobcat}) for each galaxy are listed in Table~\ref{catalogue}. We also ran CASA \textit{fitsky} and a simple aperture photometry corrected for the ALMA PSF and also found consistent results. The psf-fitting with \texttt{Galfit} was performed inside a box of 5$\times$5\arcsec centred on the source.

The main characteristics of these detections (redshift, flux, SNR, stellar mass, counterpart) are given in Table~\ref{catalogue}. We use redshifts and stellar masses from the ZFOURGE catalogue (see Sect.~\ref{sec:Optical/near-infrared_imaging}).

We compare the presence of galaxies between the 0\arcsec60-mosaic and the 0\arcsec29-mosaic. Of the 20 detections found in the 0\arcsec60 map, 14 of them are also detected in the 0\arcsec29 map. The presence of a detection in both maps reinforces the plausibility of a detection. However, a detection in only one of these two maps may be a consequence of the intrinsic source size. An extended source is more likely to be detected with a larger beam, whereas a more compact source is more likely to be missed in the maps with larger tapered sizes and reduced point source sensitivity.

A first method to identify potential false detections is to compare our results with a deeper survey overlapping with our area of the sky. We compare the positions of our catalogue sources with the positions of sources found by \cite{Dunlop2017} in the HUDF. This 1.3-mm image is deeper than our survey and reaches a $\sigma \simeq$ 35\,$\mu$Jy (corresponding to $\sigma$\,=\,52\,$\mu$Jy at 1.1mm) but overlaps with only $\sim$6.5\% of our survey area. The final sample of \cite{Dunlop2017} was compiled by selecting sources with $S_{1.3}\,>\,$ 120\,$\mu$Jy to avoid including spurious sources due to the large number of beams in the mosaics and due to their choice of including only ALMA detections with optical counterpart seen with HST.

With our flux density limit of $S_{1.1mm}$\,$\approx$\,880\,$\mu$Jy any non-spurious detection should be associated to a source seen at 1.3mm in the HUDF 1.3 mm survey, the impact of the wavelength difference being much smaller than this ratio. We detect 3 galaxies that were also detected by \cite{Dunlop2017}, UDF1, UDF2 and UDF3, all of which having $S_{1.3mm}$\,$>$\, 0.8\,mJy. The other galaxies detected by \cite{Dunlop2017} have a flux density at 1.3 mm lower than 320\,$\mu$Jy, which makes them undetectable with our sensitivity. 

We note however that we did not impose as a strict criterion the existence of an optical counterpart to our detections whereas \cite{Dunlop2017} did. Hence if we had detected a source with no optical counterpart within the HUDF, this source may not be included in the \cite{Dunlop2017} catalogue. However, as we will see, the projected density of such sources is small and none of our candidate optically dark sources falls within the limited area of the HUDF. We also note that the presence of an HST-WFC3 source within a radius of 0\arcsec6 does not necessarily imply that is the correct counterpart. As we will discuss in detail in Sect.~\ref{sec:galaxies_identification}, due to the depth of the HST-WFC3 observation and the large number of galaxies listed in the CANDELS catalogue, a match between the HST and ALMA positions may be possible by chance alignment alone (see Sect.~\ref{sec:galaxies_identification}). 

\begin{table*}\footnotesize     
\centering          
\begin{tabular}{r r r c c c c c c c c c}     
\hline    
ID&ID$_{CLS}$&ID$_{ZF}$&RA$_{\rm ALMA}$ &Dec$_{\rm ALMA}$ &RA$_{HST}$ &Dec$_{HST}$ & $\Delta _{HST_1} $ &$\Delta _{HST_2} $  &$(\Delta\alpha)_{HST}$ &$(\Delta\delta)_{HST}$ & $\Delta_\text{IRAC}$\\
&& &$\deg$  &$\deg$&$\deg$ &$\deg$ &arcsec&arcsec&arcsec&arcsec&arcsec\\
(1)&(2)&(3)&(4)&(5)&(6)&(7)&(8)&(9)&(10)&(11)&(12)\\
\hline    
\hline                  
 AGS1  	&  14876  	&  17856 	&  53.118815 & -27.782889 &  53.118790 & -27.782818 &    0.27 &    0.03 &    0.091 &   -0.278 & 0.16 \\
 AGS2  	&   7139  	&  10316 	&  53.063867 & -27.843792 &  53.063831 & -27.843655 &    0.51 &    0.23 &    0.163 &   -0.269 & 0.04\\
 AGS3  	&   9834  	&  13086  	&  53.148839 & -27.821192 &  53.148827 & -27.821121 &    0.26 &    0.06 &    0.099 &   -0.262 & 0.10\\
 AGS4 	&   8923b  &  12333 	&  53.142778 & -27.827888 &  53.142844 & -27.827890 &    0.21 &    0.40 &    0.087 &   -0.264 & 0.09\\
 AGS5 	&  20765  	&  23898  &  53.158392 & -27.733607 &  53.158345 & -27.733485 &    0.46 &    0.13 &    0.087 &   -0.329 & 0.26\\
 AGS6 	&  15669  	&  -   	&  53.183458 & -27.776654 &  53.183449 & -27.776584 &    0.26 &    0.03 &    0.054 &   -0.267 & 0.40\\
 AGS7 	&   4854  	&   7867	&  53.082738 & -27.866577 &  53.082705 & -27.866567 &    0.11 &    0.19 &    0.124 &   -0.225 & 0.03\\
 AGS8 	&  15261  &  18282 	&  53.020356 & -27.779905 &  53.020297 & -27.779829 &    0.33 &    0.03 &    0.159 &   -0.275 & 0.20\\
 AGS9  	&  12016  &  15639	&  53.092844 & -27.801330 &  53.092807 & -27.801208 &    0.45 &    0.16 &    0.100 &   -0.276 & 0.18\\
AGS10 	&  16972  &  19833 	&  53.082118 & -27.767299 &  53.081957 & -27.767202 &    0.62 &    0.39 &    0.128 &   -0.300 & 0.40\\
AGS11 	&      -  	&   7589 	&  53.108818 & -27.869055 &     -      	&     -      	      &    -      &    -      &     -         &     -      & 0.12\\
AGS12 	&  15876  &  18701	&  53.160634 & -27.776273 &  53.160594 & -27.776129 &    0.53 &    0.28 &    0.076 &   -0.242 & 0.51\\
AGS13 	&  16274  &  19033 	&  53.131122 & -27.773194 &  53.131080 & -27.773108 &    0.34 &    0.05 &    0.087 &   -0.291 & 0.14\\
AGS14 	&      -  	&      - 	&  53.223156 & -27.826771 &     -      	&     -      	      &    -       &    -      &     -        &     -      & -\\
AGS15 	&   3818b 	&   6755	&  53.074847 & -27.875880 &  53.074755 & -27.875976 &    0.45 &    0.57 &    0.125 &   -0.195 & 0.12$^1$\\
AGS16 	&      -  	&      - 	&  53.039724 & -27.784557 &     -      	&     -              &    -       &    -      &     -      	&     -      & -\\
AGS17 	&   4414b 	&   6964 	&  53.079374 & -27.870770 &  53.079327 & -27.870781 &    0.16 &    0.27 &    0.122 &   -0.231 & 0.06\\
AGS18 	&  15639  &  18645	&  53.181355 & -27.777544 &  53.181364 & -27.777501 &    0.16 &    0.12 &    0.043 &   -0.256 & 0.10\\
AGS19 	&      -  	&      - 	&  53.108041 & -27.813610 &     -      	&     -              &    -       &    -      &     -      	&     -      & -\\
AGS20 	&   9089  	&  12416	&  53.092365 & -27.826829 &  53.092381 & -27.826828 &    0.05 &    0.29 &    0.116 &   -0.247 & 0.18\\
\hline
AGS21 	&   6905  	&  10152 	&  53.070274 & -27.845586 &  53.070230 & -27.845533 &    0.24 &    0.06 &    0.143 &   -0.249 & 0.07\\
AGS22 	&  28952  &      - 	&  53.108695 & -27.848332 &  53.108576 & -27.848242 &    0.50 &    0.29 &    0.106 &   -0.226 & -\\
AGS23  	&  10954  &  14543	&  53.086623 & -27.810272 &  53.086532 & -27.810217 &    0.35 &    0.19 &    0.111 &   -0.263 & 0.16\\
\hline                  
\end{tabular}
\caption{Details of the positional differences between ALMA and HST-WFC3 for our catalogue of galaxies identified in the 1.1mm-continuum map. Columns: (1) Source ID; (2),(3) IDs of the HST-WFC3 (from the CANDELS catalogue) and ZFOURGE counterparts of these detections (the cross correlations between ALMA and HST-WFC3 and between ALMA and ZFOURGE are discussed in Sect.~\ref{sec:galaxies_identification}). b indicates HST-WFC3 galaxies located in a radius of 0\arcsec6 around the ALMA detection, although strong evidence presented in Sect.~\ref{sec:HST-dark} suggests they are not the optical counterparts of our detections; (4), (5) RA and Dec of the sources in the ALMA image (J2000);  (6), (7) Positions of HST-WFC3 $H$-band counterparts when applicable from \cite{Guo2013}, (8), (9) Distances between the ALMA and HST source positions \textit{before} ($\Delta _{HST_1}$) and \textit{after} ($\Delta _{HST_2}$) applying the offset correction derived from the comparison with Pan-STARRS and GAIA;  (10), (11) Offset to be applied to the HST source positions, which includes both the global systematic offset and the local offset; (12) Distance from the closest IRAC galaxy. $^1$ For AGS15 we use the distance given in the ZFOURGE catalogue (see Sect.~\ref{sec:HST-dark}).}
\label{table_offset}  
\end{table*}

\subsection{Supplementary catalogue}\label{sec:Add-on_catalogue}

After the completion of the main catalogue, three sources that did not satisfy the criteria of the main catalogue presented strong evidences of being robust detections. We therefore enlarged our catalogue, in order to incorporate these sources into a supplementary catalogue.

These three sources are each detected using a combination of $\sigma_p$ and $\sigma_f$ giving a purity factor greater than 80\%, whilst also ensuring the existence of an HST counterpart.

The galaxy AGS21 has an SNR\,=\,5.83 in the 0\arcsec29 tapered map, but is not detected in the 0\arcsec60 tapered map. The non-detection of this source is most likely caused by its size. Due to its dilution in the 0\arcsec60-mosaic, a very compact galaxy detected at 5\,$\sigma$ in the 0\arcsec29-mosaic map could be below the detection limit in the 0\arcsec60-mosaic. The ratio of the mean RMS of the two tapered maps is 1.56, meaning that for a point source of certain flux, a 5.83\,$\sigma$ measurement in the 0\arcsec29-mosaic becomes 3.74$\,\sigma$ in the 0\arcsec60-map.

The galaxy AGS22 has been detected with an SNR\,=\,4.9 in the 0\arcsec60 tapered map ($\sigma_p$\,=\,4.9 and $\sigma_f$\,=\,3.1). With $\sigma_p$ and $\sigma_f$ values more stringent than the thresholds chosen for the main catalogue, it may seem paradoxical that this source does not appear in the main catalogue. With a floodclip criterion of 2.7\,$\sigma$, this source would have an SNR just below 4.8 excluding it from the main catalogue. This source is associated with a faint galaxy that has been detected by HST-WFC3 (ID$_{CANDELS}$\,=\,28952) at 1.6\,$\mu$m (6.6\,$\sigma$) at a position close to the ALMA detection (0\arcsec28). Significant flux has also been measured at 1.25\,$\mu$m (3.6\,$\sigma$) for this galaxy. In all of the other filters, the flux measurement is not significant ($<$\,3\,$\sigma$). Due to this lack of information, it has not been possible to compute its redshift. AGS22 is not detected in the 0\arcsec29-mosaic map with $p_c$\,$>$\,0.8. The optical counterpart of this source has a low $H$-band magnitude (26.8$\pm$0.2 AB), which corresponds to a range for which the \cite{Guo2013} catalogue is no longer complete. This is the only galaxy (except the three galaxies most likely to be spurious: AGS14, AGS16 and AGS19) that has not been detected by IRAC (which could possibly be explained by a low stellar mass). The probability of the ALMA detection being spurious, within the association radius 0\arcsec6 of a $H$-band source of this magnitude or brighter, is 5.5\%. For these reasons we do not consider it as spurious.

The galaxy AGS23 was detected in the 0\arcsec60 map just below our threshold at 4.8\,$\sigma$, with a combination $\sigma_p$\,=\,4.6 and $\sigma_f$\,=\,2.9 giving a purity criterion greater than 0.9. This detection is associated with an HST-WFC3 counterpart. It is for these two reasons that we include this galaxy in the supplementary catalogue. The photometric redshift ($z$\,=\,2.36) and stellar mass (10$^{11.26}$ M$_\sun$) both reinforce the plausibility of this detection.

\begin{table*}\footnotesize 
\centering          
\begin{tabular}{l  c c c c c c c c c c c c c c }   
\hline    
ID & z & SNR&S$_{peak}^{\texttt{Blobcat}}$& f$_\text{deboost}$& S$_{PSF}^{\texttt{Galfit}}$&  log$_{10}$M$_\star$&0\arcsec60&0\arcsec29&S$_\text{6GHz}$ &$L_X$/10$^{42}$& ID$_{\text{sub(mm)}}$\\
&  & &mJy&&mJy &M$_\sun$&&&$\mu$Jy& erg.s$^{-1}$&\\
(1)&(2)&(3)&(4)&(5)&(6)&(7)&(8)&(9)&(10)&(11)&(12)\\
\hline                  
\hline    
AGS1 	&  2.309 	&11.26 	& 1.90$\pm$ 0.20 		&1.03& 1.99$\pm$ 0.15 & 11.05 	&  1 & 1 & 18.38$\pm$0.71 	& 1.93 	&GS6, ASA1	\\
 AGS2 	&  2.918 	&10.47 	& 1.99$\pm$ 0.22 		&1.03& 2.13$\pm$ 0.15 & 10.90 	&  1 & 1 &  - 				& 51.31 	&			\\
 AGS3	&  2.582 	&9.68 	& 1.84$\pm$ 0.21 		&1.03& 2.19$\pm$ 0.15 & 11.33 	&  1 & 1 & 19.84$\pm$0.93 	& 34.54 	&GS5, ASA2  	\\
 AGS4   	&  4.32 	& 9.66 	& 1.72$\pm$ 0.20 		&1.03&1.69$\pm$ 0.18 & 11.45 	&  1 & 1 & 8.64$\pm$0.77 	& 10.39 	&			\\
 AGS5   	&  3.46 	& 8.95 	& 1.56$\pm$ 0.19 		&1.03& 1.40$\pm$ 0.18 & 11.13 	&  1 & 1 & 14.32$\pm$1.05 	& 37.40 	&         	        \\
 AGS6    	&  3.00 	&7.63 	& 1.27$\pm$ 0.18 		&1.05& 1.26$\pm$ 0.16 & 10.93 	&  1 & 1 & 9.02$\pm$0.57 	& 83.30  	& UDF1 , ASA3	\\
 AGS7   	&  3.29 	&7.26 	& 1.15$\pm$ 0.17 		&1.05& 1.20$\pm$ 0.13 & 11.43 	&  1 & 1 & - 				& 24.00 	&			\\
 AGS8   	&  1.95 	&7.10 	& 1.43$\pm$ 0.22 		&1.05& 1.98$\pm$ 0.20 & 11.53 	&  1 & 1 & -  				& 3.46 	&   LESS18	\\
 AGS9  	&  3.847 	&6.19 	& 1.25$\pm$ 0.21 		&1.05& 1.39$\pm$ 0.17 & 10.70 	&  1 & 1 & 14.65$\pm$1.12	& - 		&			\\
AGS10  	&  2.41 	&6.10 	& 0.88$\pm$ 0.15 		&1.06& 1.04$\pm$ 0.13 & 11.32 	&  1 & 1 & -  				& 2.80 	&			\\
AGS11  	&  4.82 	&5.71 	& 1.34$\pm$ 0.25 		&1.08& 1.58$\pm$ 0.22 & 10.55 	&  1 & 1 & -  				& - 		&			\\
AGS12  	&  2.543 	&5.42 	& 0.93$\pm$ 0.18 		&1.10& 1.13$\pm$ 0.15 & 10.72 	&  1 & 1 & 12.65$\pm$0.55 	& 4.53 	& UDF3, C1, ASA8       \\
AGS13   	&  2.225 	&5.41 	& 0.78$\pm$ 0.15 		&1.10& 0.47$\pm$ 0.14 & 11.40 	&  1 & 0 & 22.52$\pm$0.81 	& 13.88 	&     ASA12	\\
AGS14* 	&  -   	&5.30 	& 0.86$\pm$ 0.17 		&1.10& 1.17$\pm$ 0.15 &  -   	 	&  1 & 0 & -  				& - 		&			\\
AGS15   	&  -  		&5.22 	& 0.80$\pm$ 0.16 		&1.11& 0.64$\pm$ 0.15 &  - 	 	&  1 & 1 & -  				& - 		&LESS34		\\
AGS16*  	&  -   	&5.05 	& 0.82$\pm$ 0.17 		&1.12& 0.99$\pm$ 0.17 &  -   		&  1 & 0 & -  				& - 		&			\\
AGS17  	&  -  		&5.01 	& 0.93$\pm$ 0.19\dag 	& 1.14&1.37$\pm$ 0.18 &  - 		&  1 & 0 & -  				& - 		&   LESS10	\\
AGS18   	&  2.794 	&4.93 	& 0.85$\pm$ 0.18\dag 	&1.15& 0.79$\pm$ 0.15 & 11.01 	&  1 & 0 & 6.21$\pm$0.57 	& - 		& UDF2 , ASA6 \\
AGS19*  	&  -  		&4.83 	& 0.69$\pm$ 0.15 		&1.16& 0.72$\pm$ 0.13 &  - 		&  1 & 0 & -  				& - 		&			\\
AGS20   	&  2.73 	&4.81 	& 1.11$\pm$ 0.24 		&1.16& 1.18$\pm$ 0.23 & 10.76 	&  1 & 1 & 12.79$\pm$1.40 	& 4.02 	&			\\
\hline      
AGS21  	&  3.76 	&5.83 	& 0.64$\pm$ 0.11 		&1.07& 0.88$\pm$ 0.19 & 10.63 	&  0 & 1 & -  				& 19.68 	&			\\
AGS22  	&   -   	&4.90 	& 1.05$\pm$ 0.22 		&1.15& 1.26$\pm$ 0.22 &   -   		&  1 & 0 & - 				& - 		&			\\
AGS23  	&  2.36 	&4.68 	& 0.98$\pm$ 0.21 		&1.19& 1.05$\pm$ 0.20 & 11.26 	&  1 & 0 & -  				& - 		&			\\
\hline                       
\end{tabular}
\caption{Details of the final sample of sources detected in the ALMA GOODS--\textit{South} continuum map, from the primary catalogue in the main part of the table and from the supplementary catalogue below the solid line (see Sect.~\ref{sec:Main_catalogue} and Sect.~\ref{sec:Add-on_catalogue}). Columns: (1) IDs of the sources as shown in Fig.~\ref{map_detection}. The sources are sorted by SNR. * indicates galaxies that are most likely spurious, i.e., not detected at any other wavelength; (2) Redshifts from the ZFOURGE catalogue. Spectroscopic redshifts are shown with three decimal places. As AGS6 is not listed in the ZFOURGE catalogue, we use the redshift computed by \cite{Dunlop2017}; (3) Signal to noise ratio of the detections in the 0\arcsec60 mosaic (except for AGS21). This SNR is computed using the flux from \texttt{Blobcat} and is corrected for peak bias; (4) Peak fluxes measured using \texttt{Blobcat} in the 0\arcsec60-mosaic image before de-boosting correction; (5) Deboosting factor; (6) Fluxes measured by PSF-fitting with \texttt{Galfit} in the 0\arcsec60-mosaic image before de-boosting correction; (7) Stellar masses from the ZFOURGE catalogue; (8), (9) Flags for detection by \texttt{Blobcat} in the 0\arcsec60-mosaic and 0\arcsec29-mosaic images, where at least one combination of $\sigma_p$ and $\sigma_f$ gives a purity factor (Eq.~\ref{quality_criteria}) greater than 80\%; (10) Flux for detection greater than 3\,$\sigma$ by VLA (5 cm). Some of these sources are visible in the VLA image but not detected with a threshold\,$>$\, 3\,$\sigma$. AGS8 and AGS16 are not in the field of the VLA survey; (11) Absorption-corrected intrinsic 0.5-7.0 keV luminosities. The X-ray luminosities have been corrected to account for the redshift difference between the redshifts provided in the catalogue of \cite{Luo2017} and those used in the present table, when necessary. For this correction we used Eq.~1 from \cite{Alexander2003}, and assuming a photon index of $\Gamma$\,=\,2; (12) Corresponding IDs for detections of the sources in previous (sub)millimetre ancillary data. UDF is for Hubble Ultra Deep Field survey \citep{Dunlop2017} at 1.3 mm, C indicates the ALMA Spectroscopic Survey in the Hubble Ultra Deep Field (ASPECS) at 1.2 mm \citep{Aravena2016}, LESS indicates data at 870\,$\mu$m presented in \cite{Hodge2013},  GS indicates data at 870\,$\mu$m presented in \cite{Elbaz2017},  ASA indicates the ALMA 26 arcmin$^2$ Survey of GOODS-S at One-millimeter (ASAGAO).  We also note the pointed observations of AGS1 presented in \cite{Barro2017}, and those of AGS13 by \cite{Talia2018}. For the two sources marked by a \dag, the hypothesis a of a point-like source is no longer valid. We therefore apply correction factors of 2.3 and 1.7 to the peak flux values of AGS17 and AGS18 respectively, to take into account the extended flux emission of these sources.}
\label{catalogue}  
\end{table*}

\subsection{Astrometric correction}\label{sec:Astrometric_correction}

The comparison of our ALMA detections with HST (Sect.~\ref{sec:Main_catalogue}) in the previous section was carried out after correcting for an astrometric offset, which we outline here. In order to perform the most rigorous counterpart identification and take advantage of the accuracy of ALMA, we carefully investigated the astrometry of our images. Before correction, the galaxy positions viewed by HST are systematically offset from the ALMA positions. This offset has already been identified in previous studies \citep[e.g.][]{Maiolino2015, Rujopakarn2016, Dunlop2017}. 

In order to quantify this effect, we compared the HST source positions with detections from the Panoramic Survey Telescope and Rapid Response System (Pan-STARRS). This survey has the double advantage to cover a large portion of the sky, notably the GOODS--\textit{South} field, and to observe the sky at a wavelength similar to HST-WFC3. We use the Pan-STARRS DR1 catalogue provided by \cite{Flewelling2016} and also include the corresponding regions issued from the \textit{GAIA} DR1 \citep{Gaia_Collaboration2016}.

Cross-matching was done within a radius of 0\arcsec5. In order to minimize the number of false identifications, we subtracted the median offset between the two catalogues from the \cite{Guo2013} catalogue positions, after the first round of matching. We iterated this process three times. In this way, 3\,587 pairs were found over the GOODS--\textit{South} field.  

\begin{figure}
   \centering
   \includegraphics[width=\hsize]{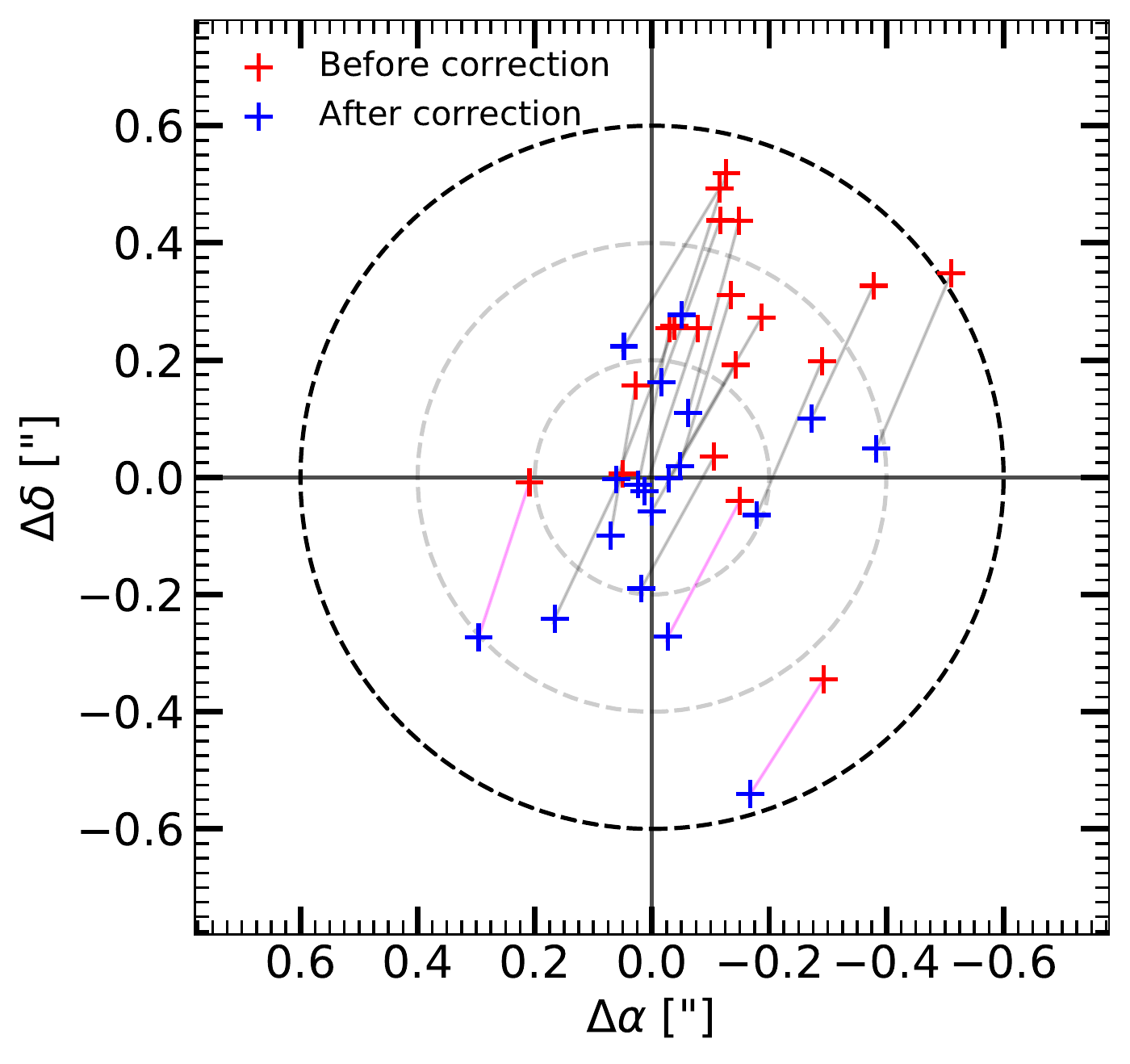}
      \caption{Positional offset (RA$_{HST}$ - RA$_\text{ALMA}$, DEC$_{HST}$ - DEC$_\text{ALMA}$) between HST and ALMA before (red crosses) and after (blue crosses) the correction of both a global systematic offset and a local offset. The black dashed circle corresponds to the cross-matching limit radius of 0\arcsec6. The grey dashed circles show a positional offset of 0\arcsec2 and 0\arcsec4 respectively. The magenta lines indicate the HST galaxies previously falsely associated with ALMA detections.}
         \label{offset_plot}
\end{figure}

To correct for the median offset between the HST and ALMA images, the HST image coordinates must be corrected by $-$94$\pm$42 mas in right ascension, $\alpha$, and 262$\pm$50 mas in declination, $\delta$, where the uncertainties correspond to the standard deviation of the 3\,587 offset measurements. This offset is consistent with that found by \cite{Rujopakarn2016} of $\Delta \alpha$\,=\,$-$80$\pm$110 mas and  $\Delta \delta$\,=\,260$\pm$130 mas. The latter offsets were calculated by comparing the HST source positions with 2MASS and VLA positions. In all cases, it is the HST image that presents an offset, whereas ALMA, Pan-STARRS, \textit{GAIA}, 2MASS and VLA are all in agreement.  We therefore deduce that it is the astrometric solution used to build the HST mosaic that introduced this offset. As discussed in Dickinson et al. (in prep.), the process of building the HST mosaic also introduced less significant local offsets, that can be considered equivalent to a distortion of the HST image. These local offsets are larger in the periphery of GOODS--\textit{South} than in the centre, and close to zero in the HUDF field. The local offsets can be considered as a distortion effect. The offsets listed in Table~\ref{table_offset} include both effects, i.e., the global and local offsets. The separation between HST and ALMA detections before and after offset correction, and the individual offsets applied for each of the galaxies are indicated in Table~\ref{table_offset} and can be visualized in Fig.~\ref{offset_plot}. We applied the same offset corrections to the galaxies listed in the ZFOURGE catalogue.

\begin{figure*}
\centering

\begin{minipage}[t]{0.9\textwidth}
\resizebox{\hsize}{!} {
\includegraphics[width=3.21cm,clip]{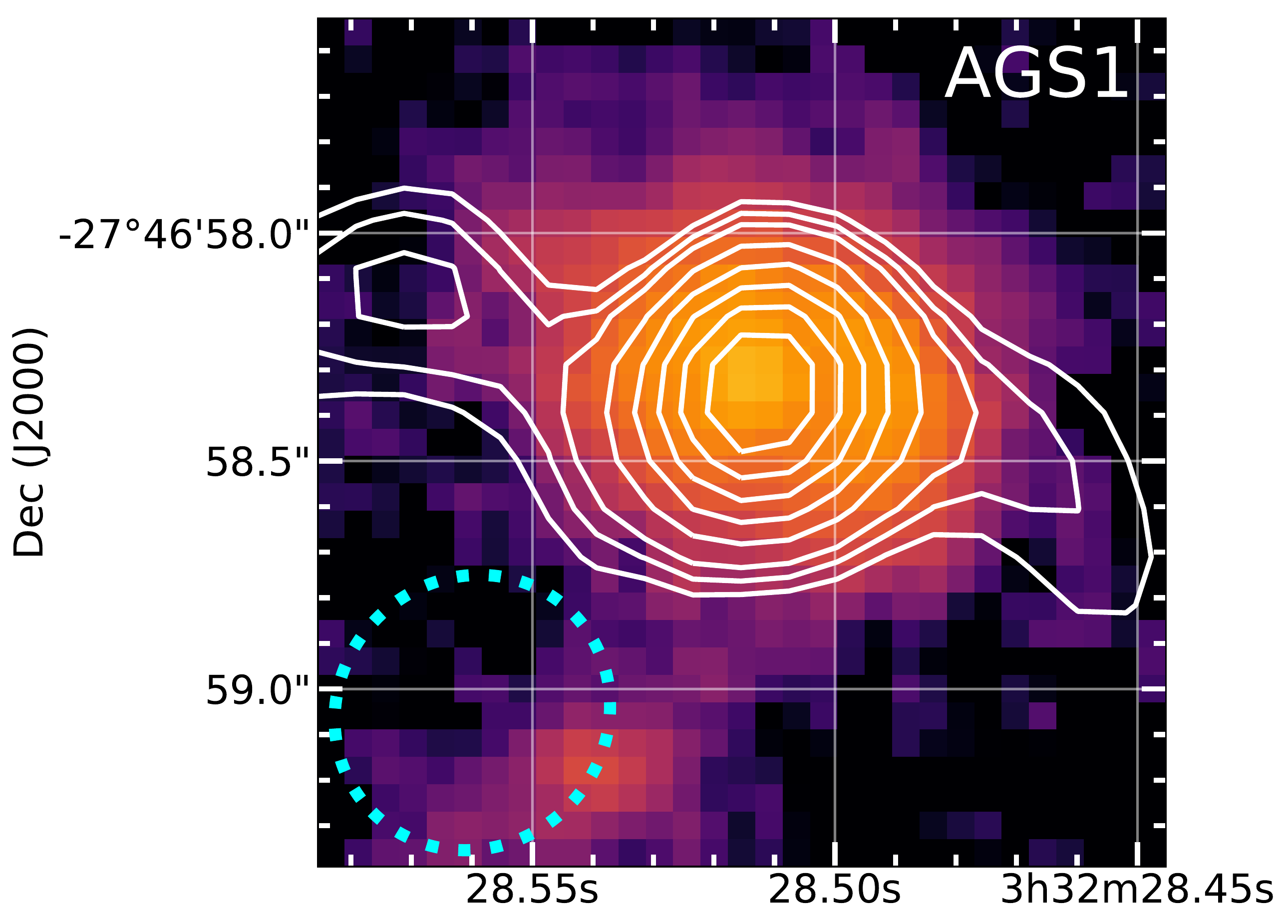}
\includegraphics[width=3cm,clip]{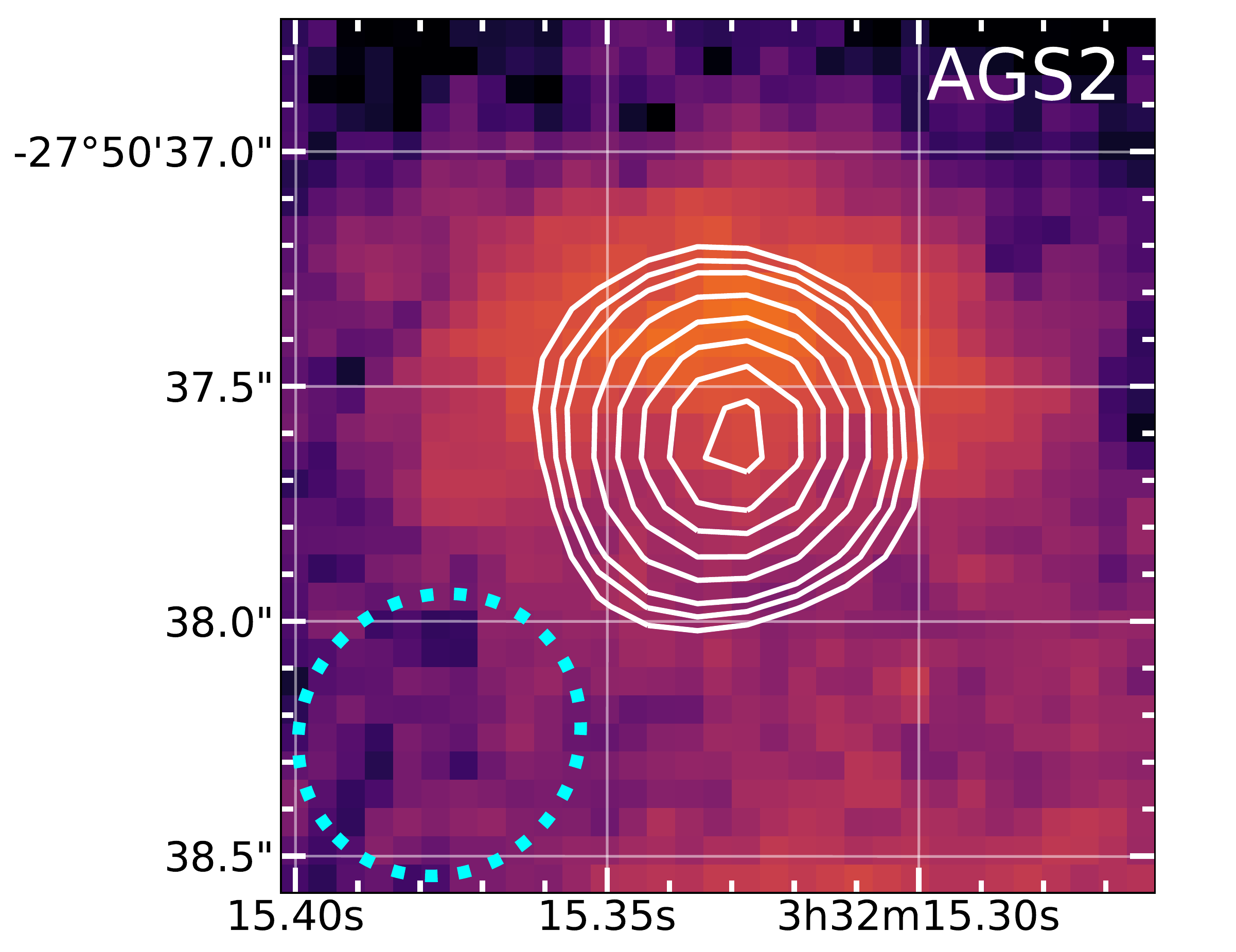}
\includegraphics[width=3cm,clip]{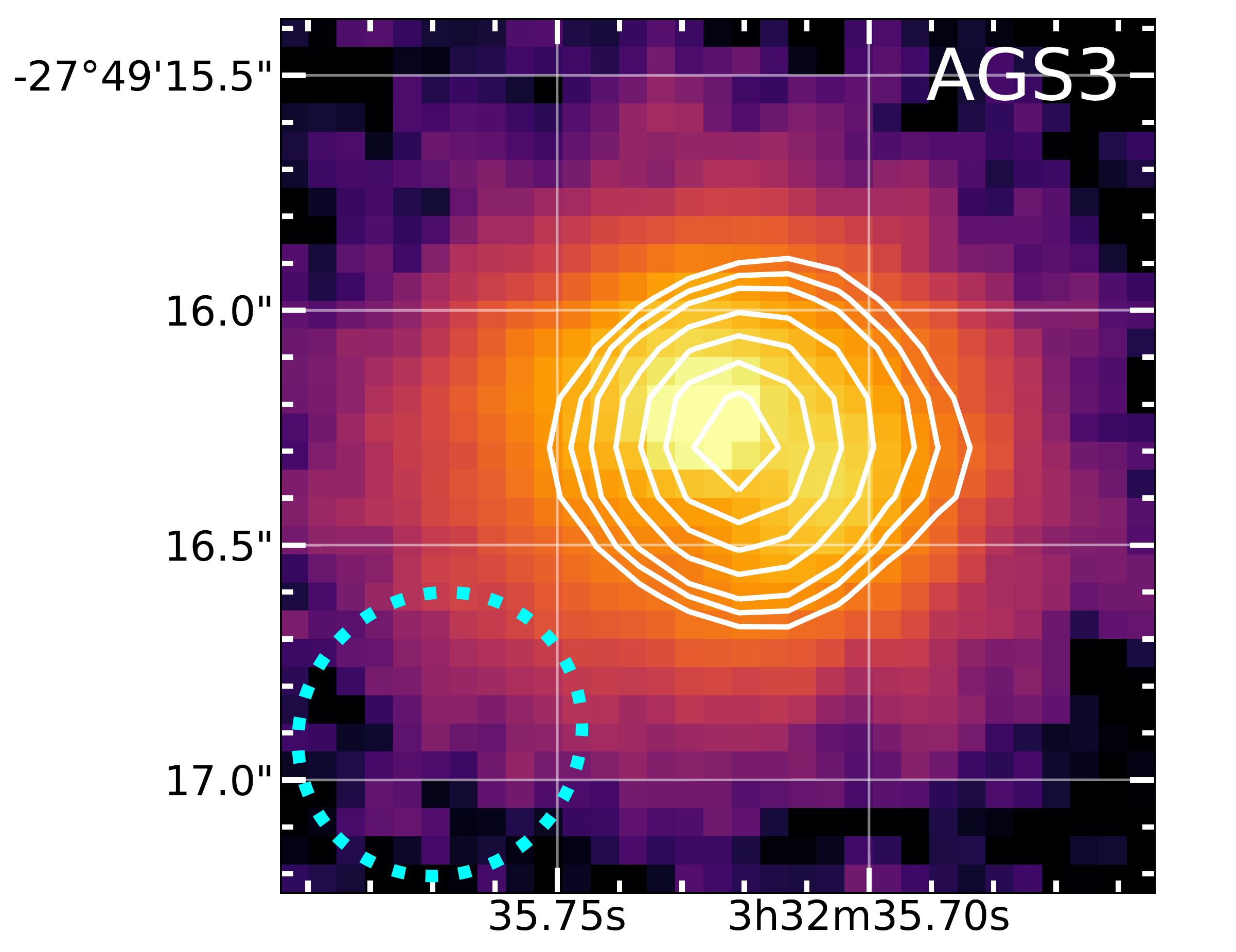}
\includegraphics[width=3cm,clip]{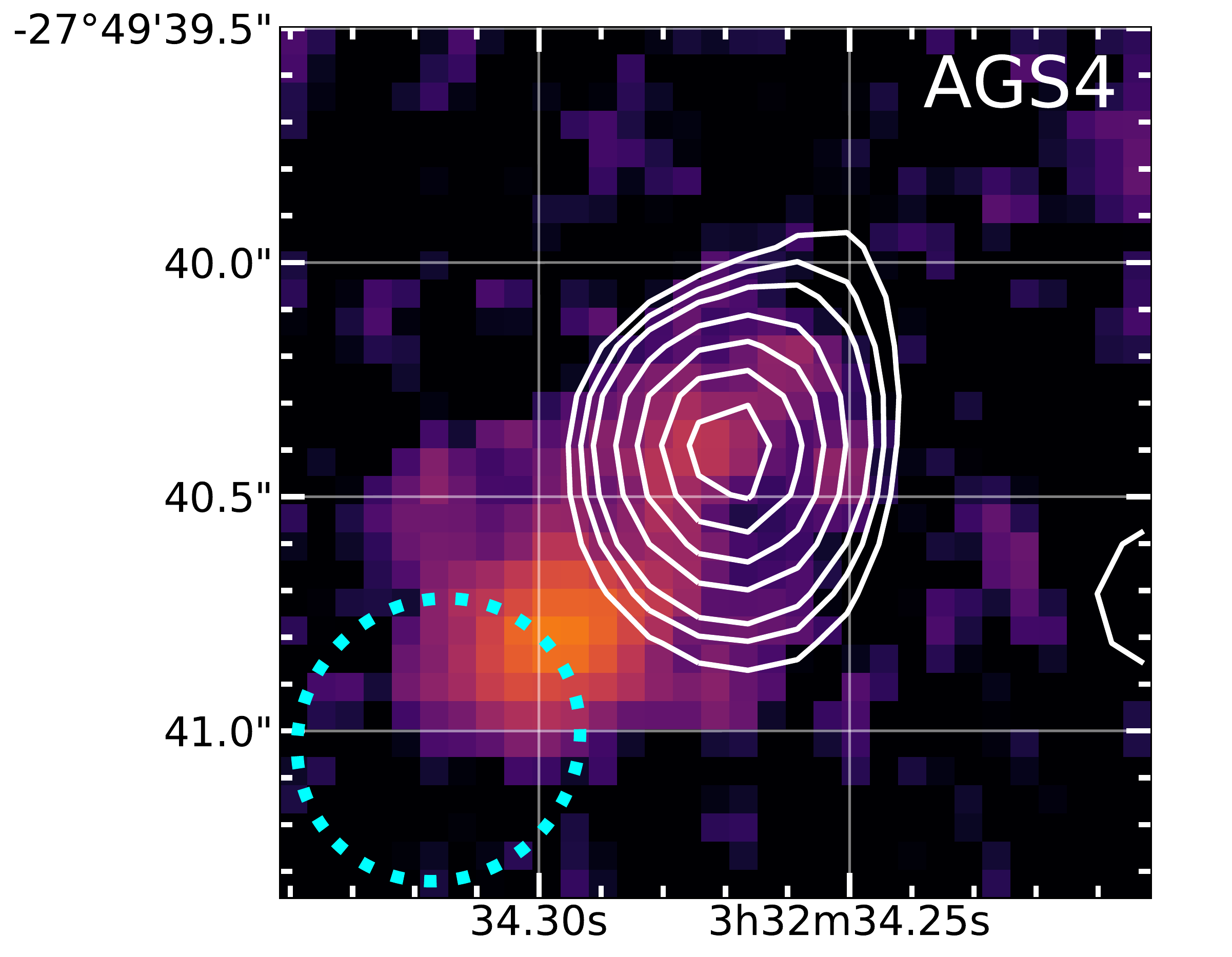}
}
\end{minipage}
\begin{minipage}[t]{0.9\textwidth}
\resizebox{\hsize}{!} {
\includegraphics[width=3.23cm,clip]{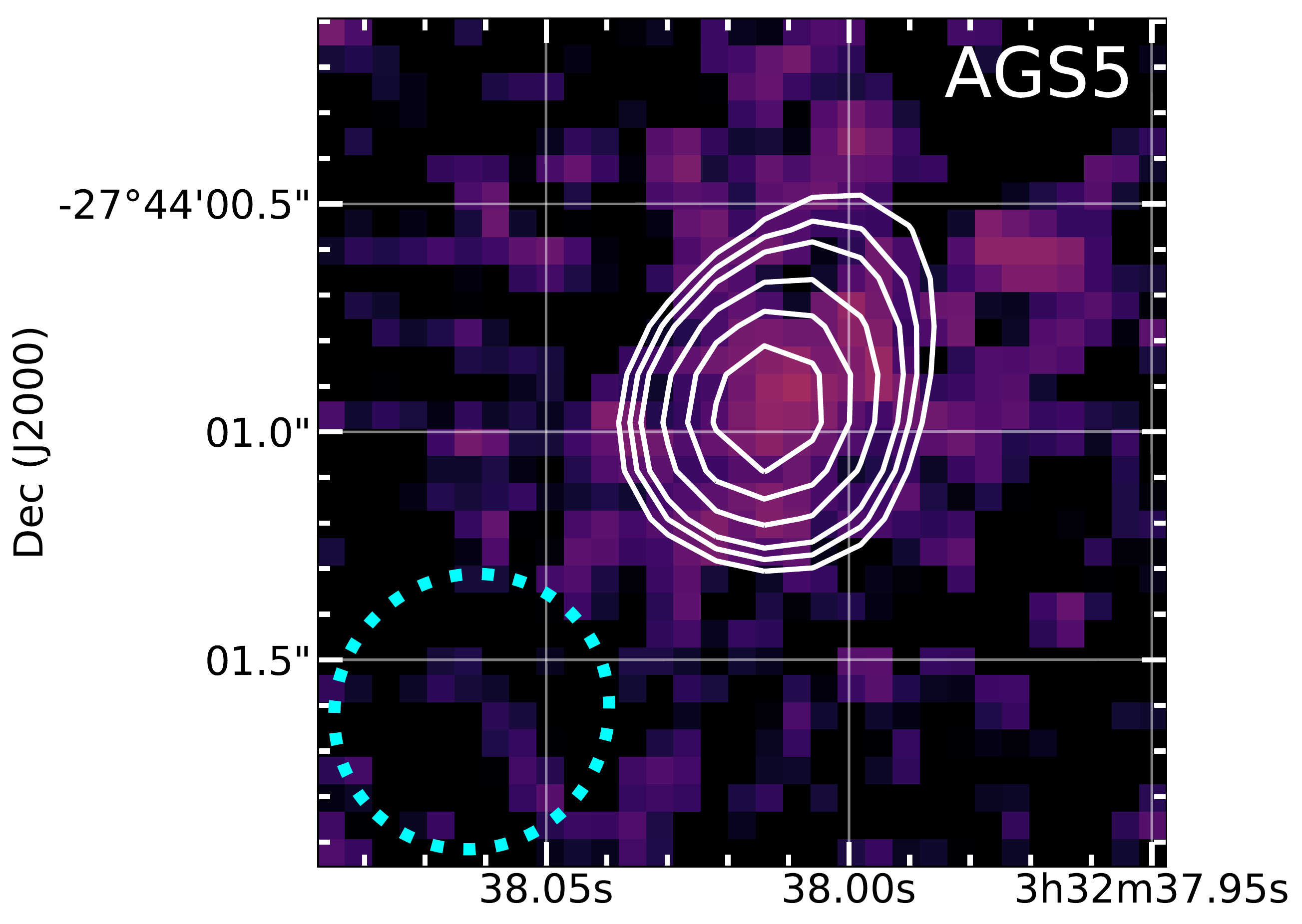}
\includegraphics[width=3cm,clip]{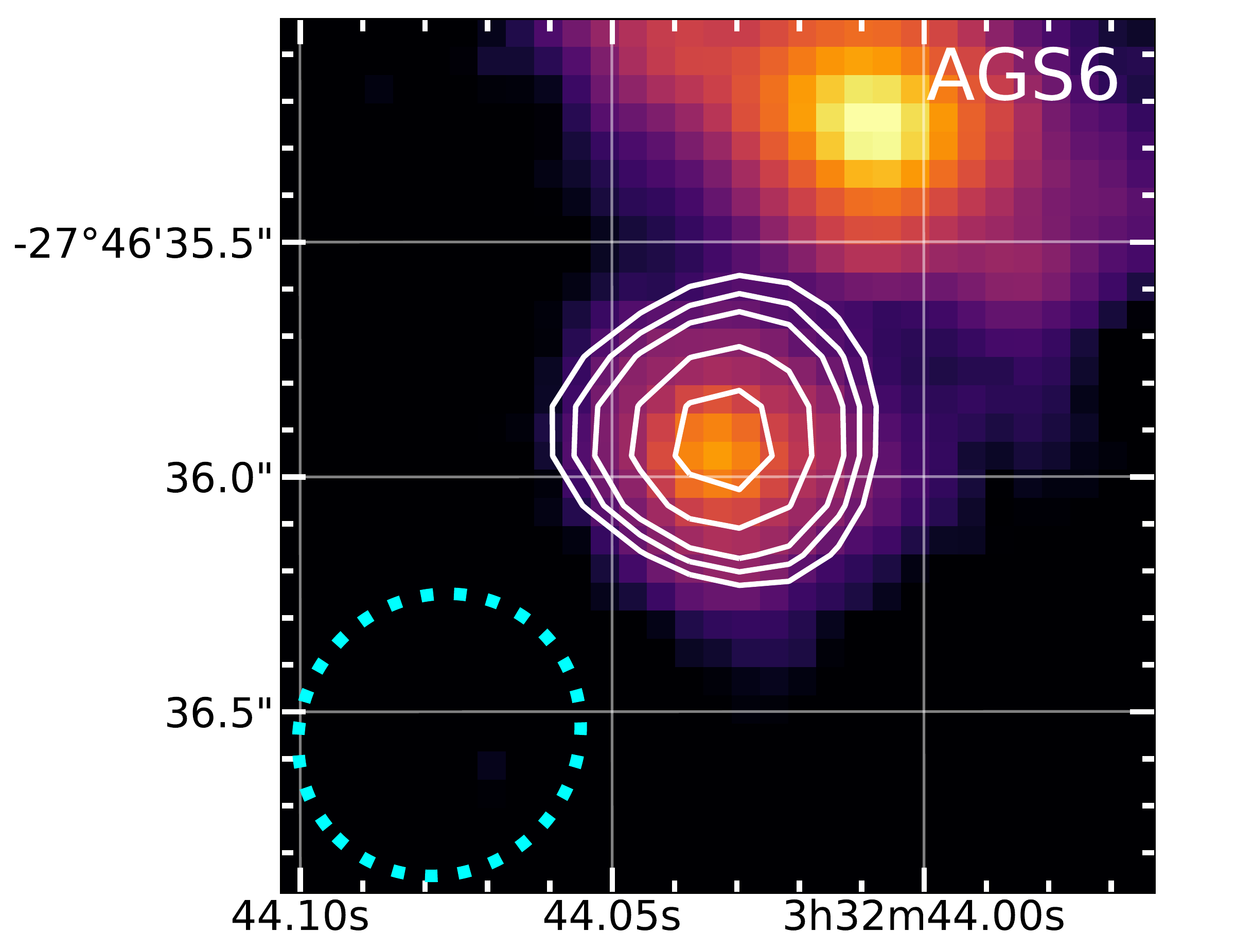}
\includegraphics[width=3.05cm,clip]{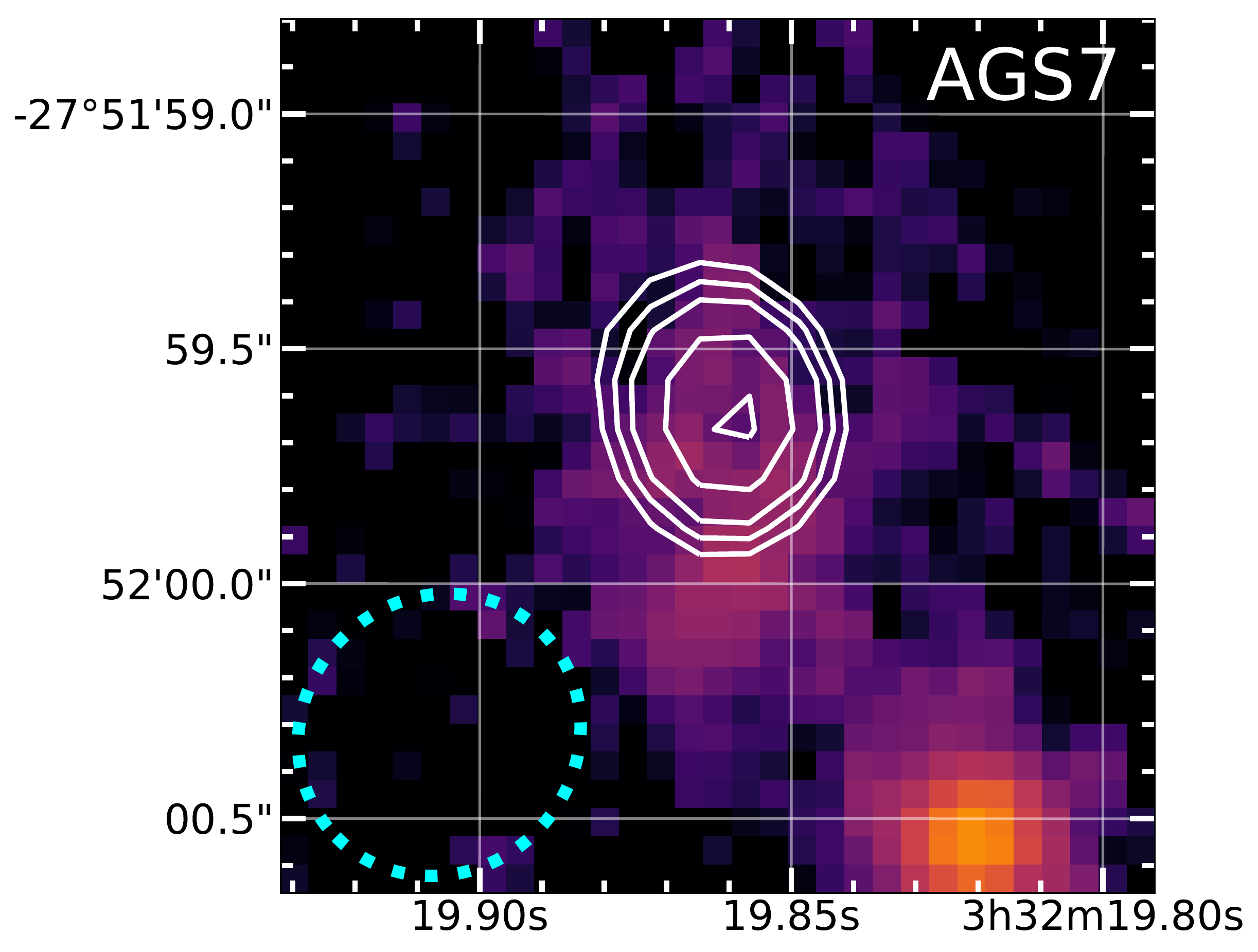}
\includegraphics[width=3cm,clip]{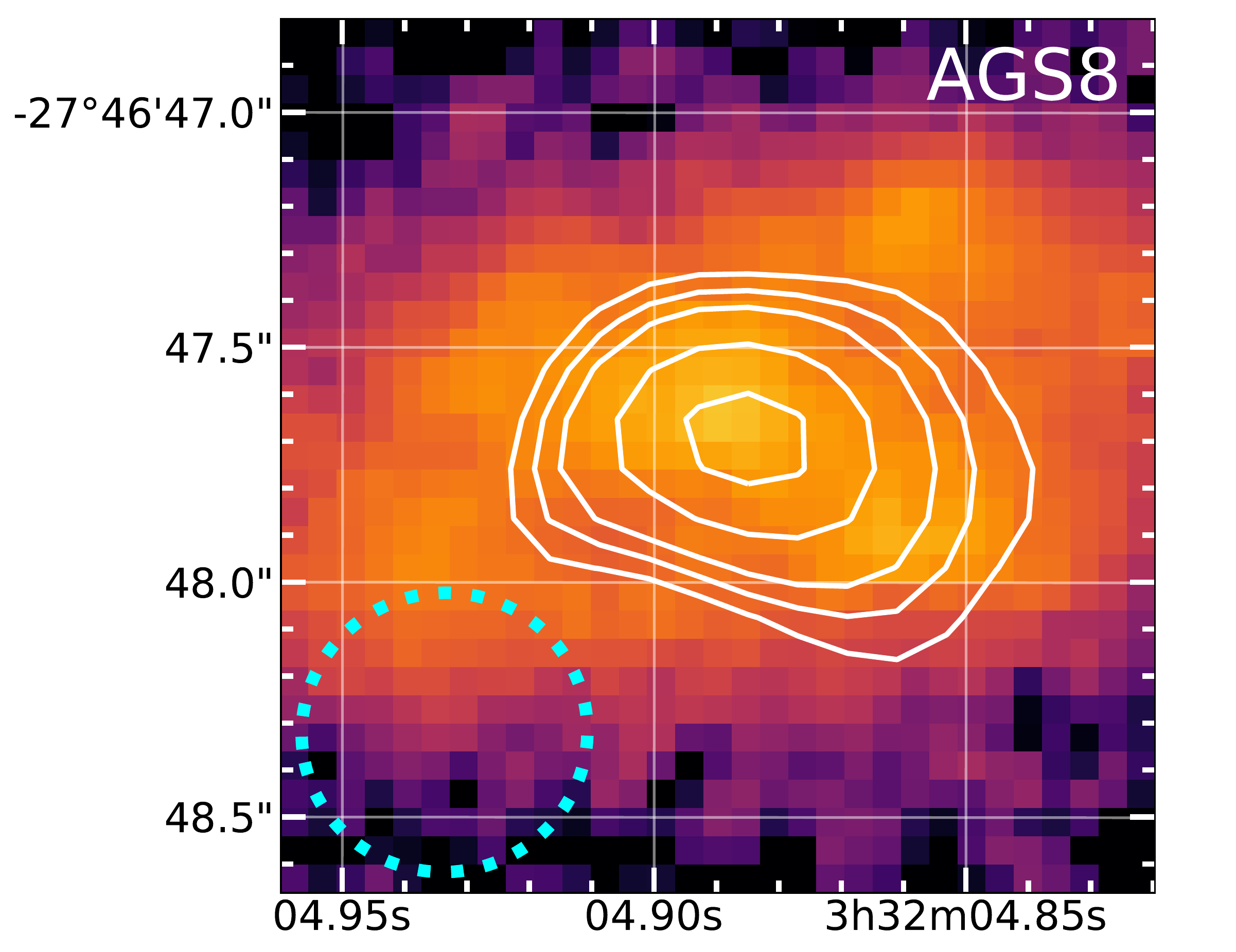}
}
\end{minipage}

\begin{minipage}[t]{0.9\textwidth}
\resizebox{\hsize}{!} {
\includegraphics[width=3.23cm,clip]{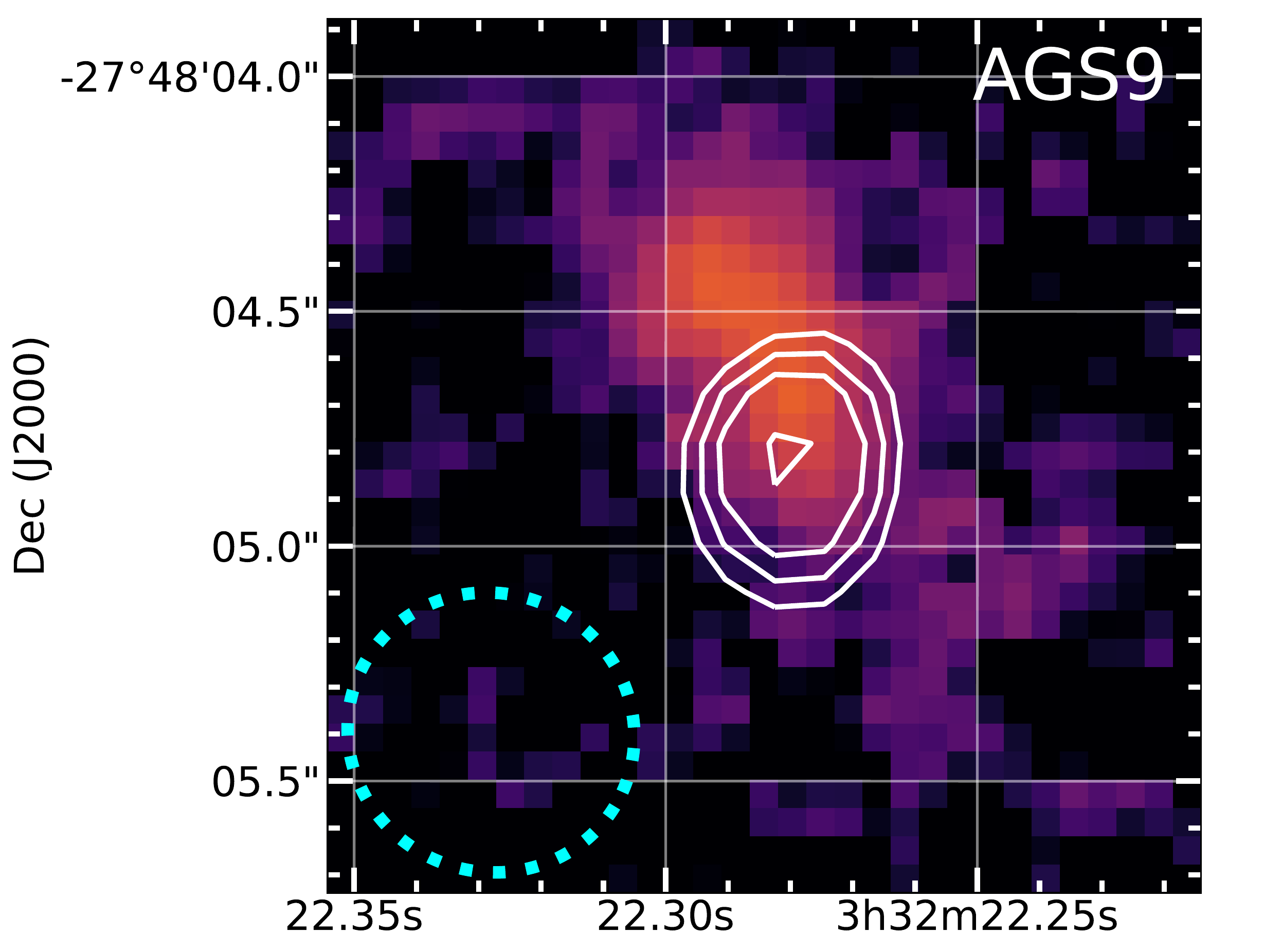}
\includegraphics[width=3.1cm,clip]{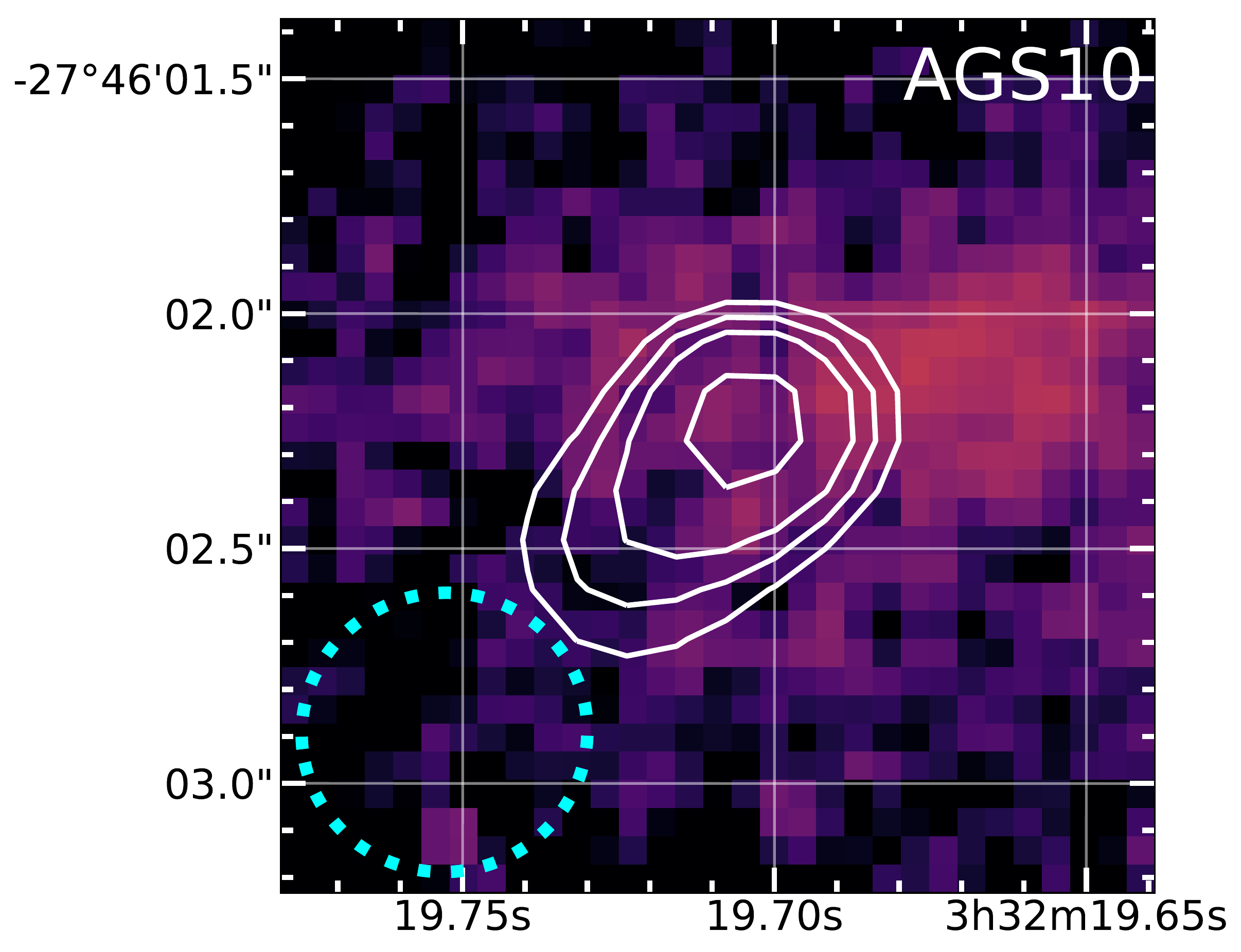}
\includegraphics[width=3.23cm,clip]{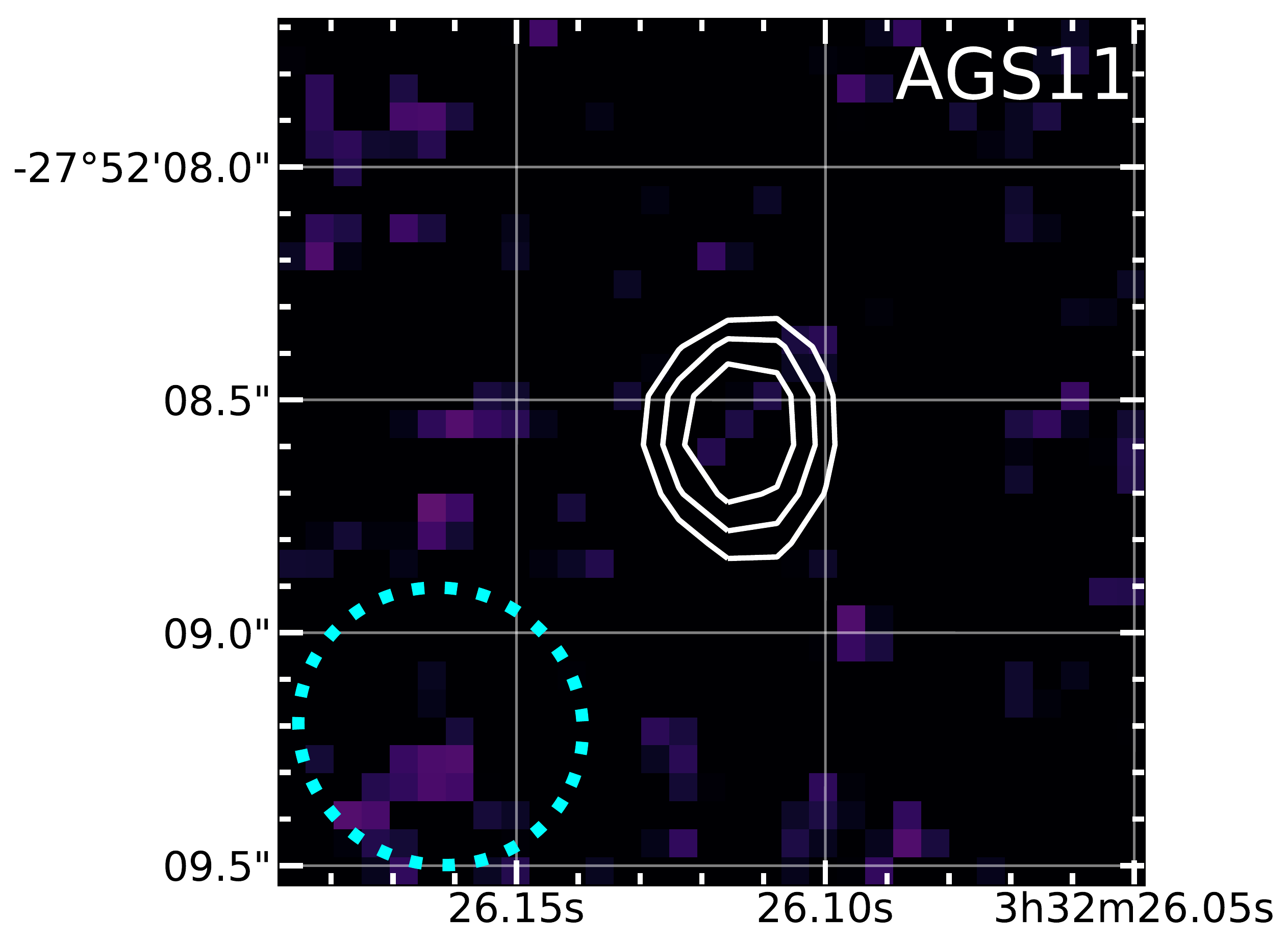}
\includegraphics[width=3.1cm,clip]{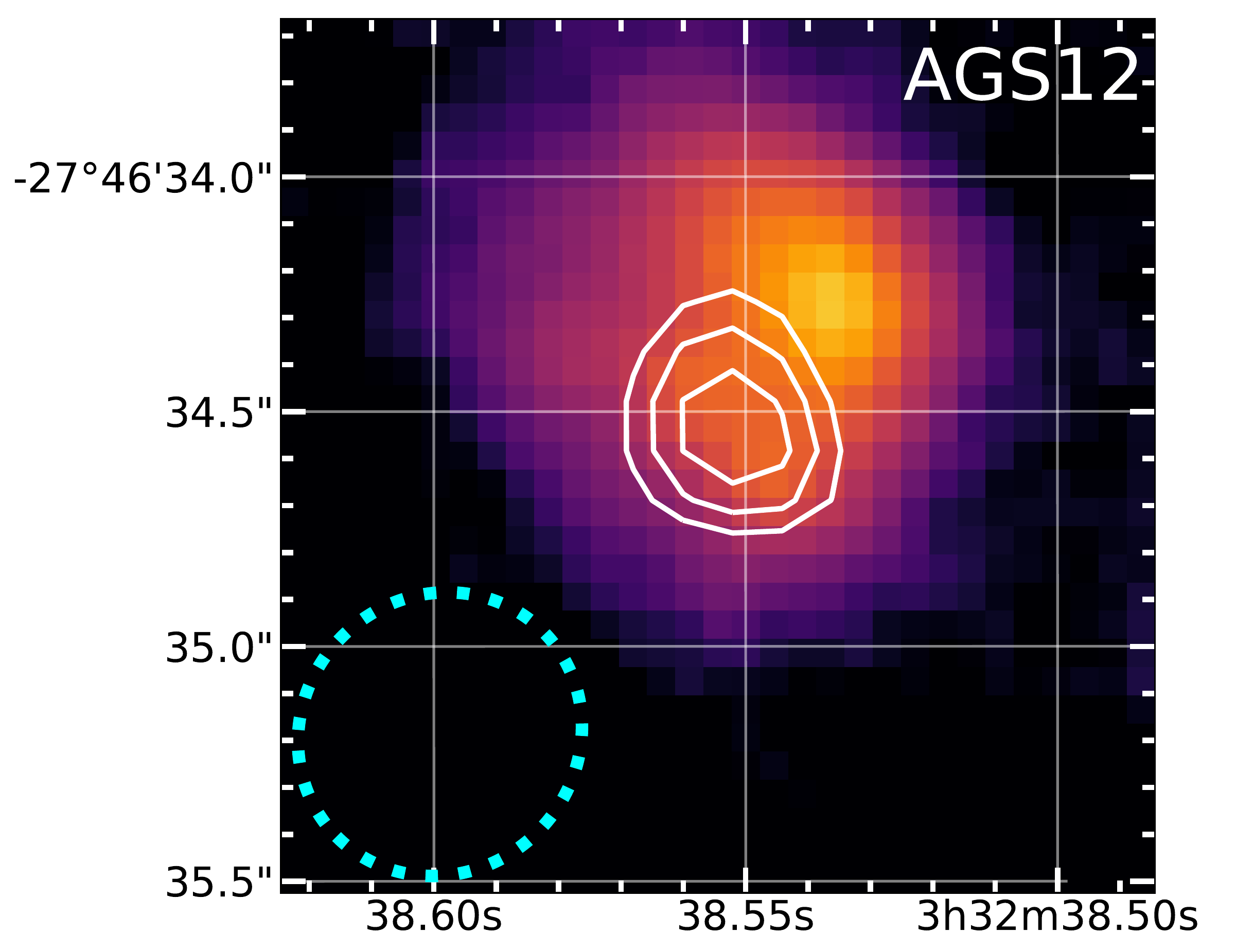}
}
\end{minipage}
\begin{minipage}[t]{0.9\textwidth}
\resizebox{\hsize}{!} {
\includegraphics[width=3.23cm,clip]{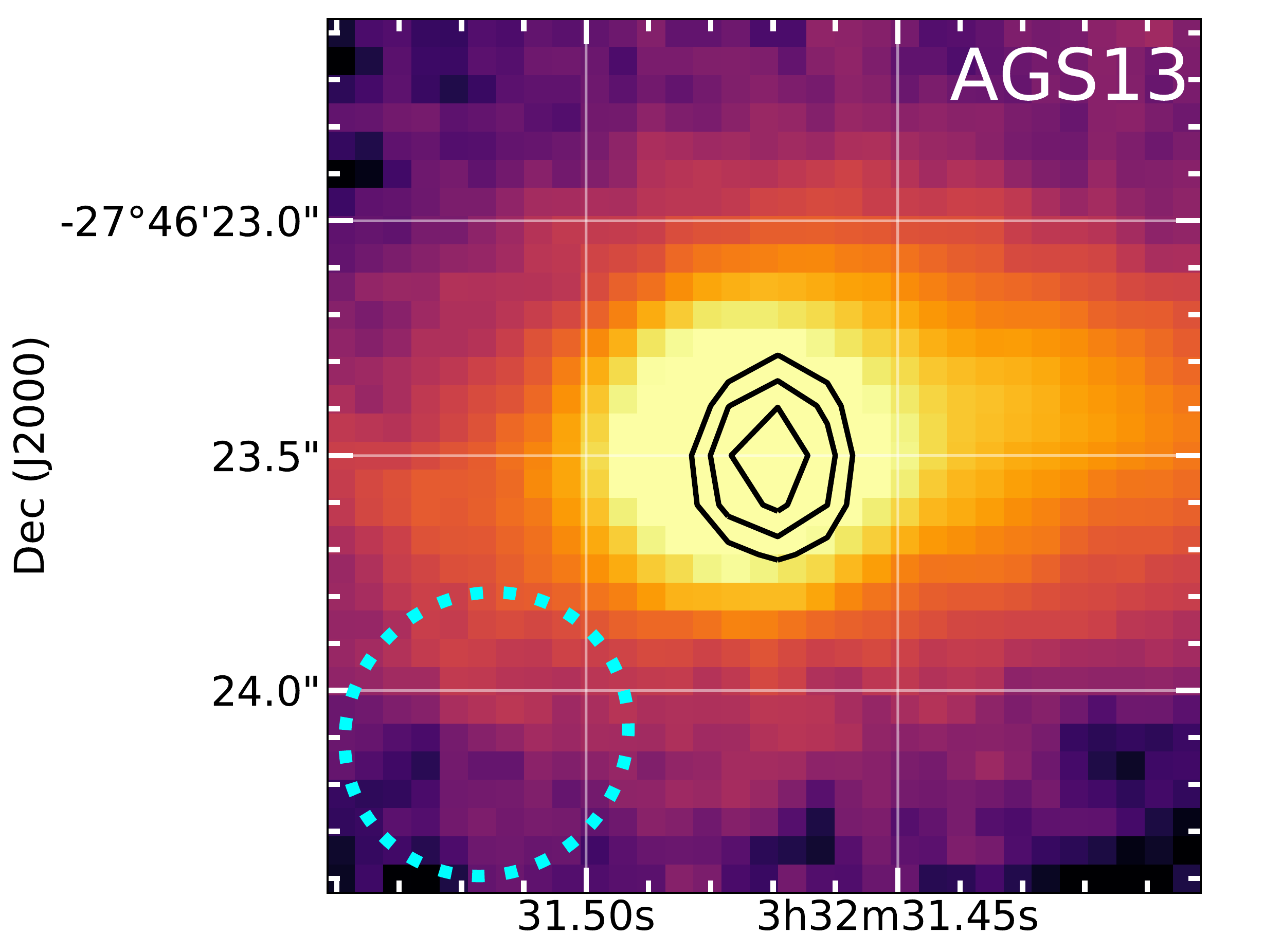}
\includegraphics[width=3.13cm,clip]{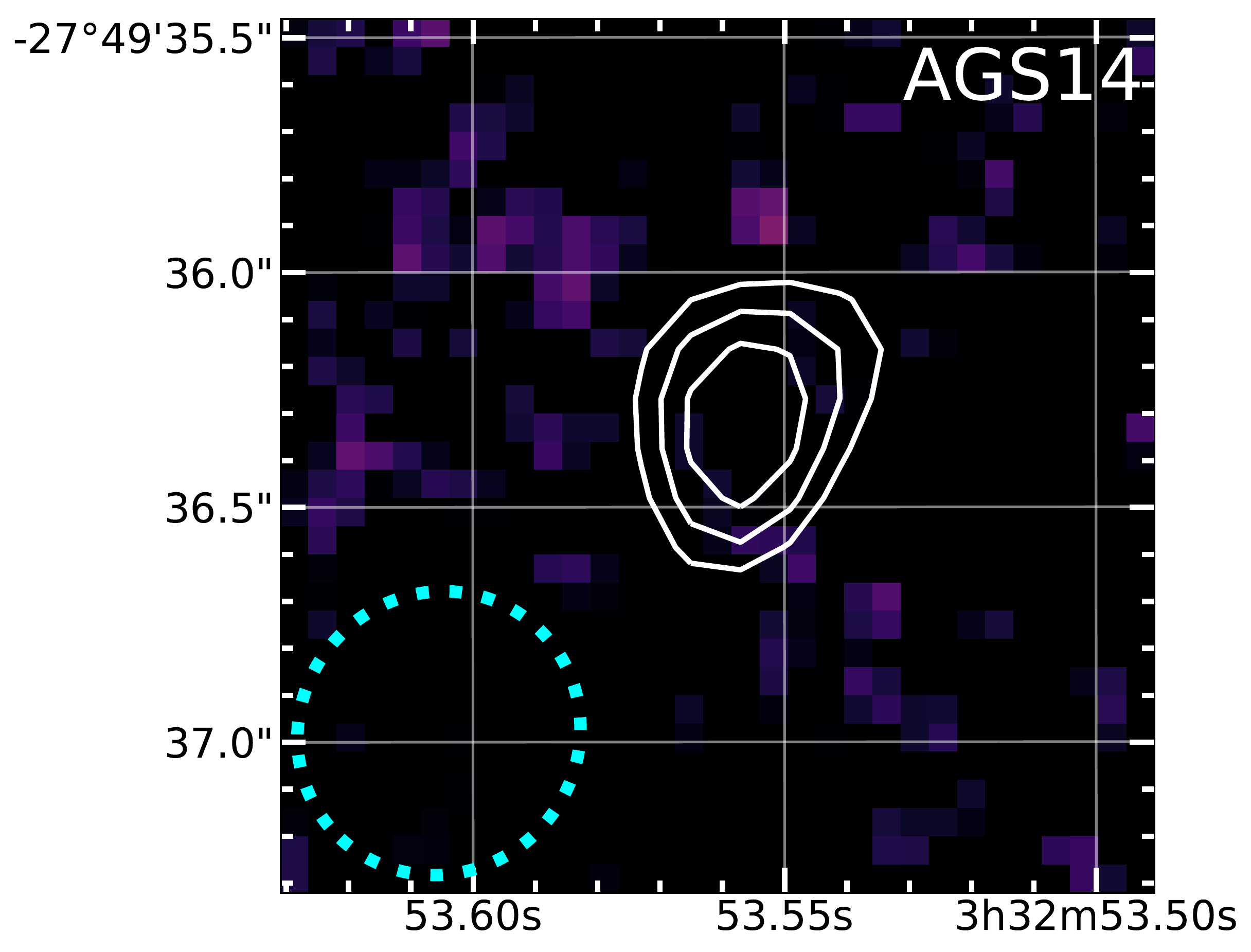}
\includegraphics[width=3.26cm,clip]{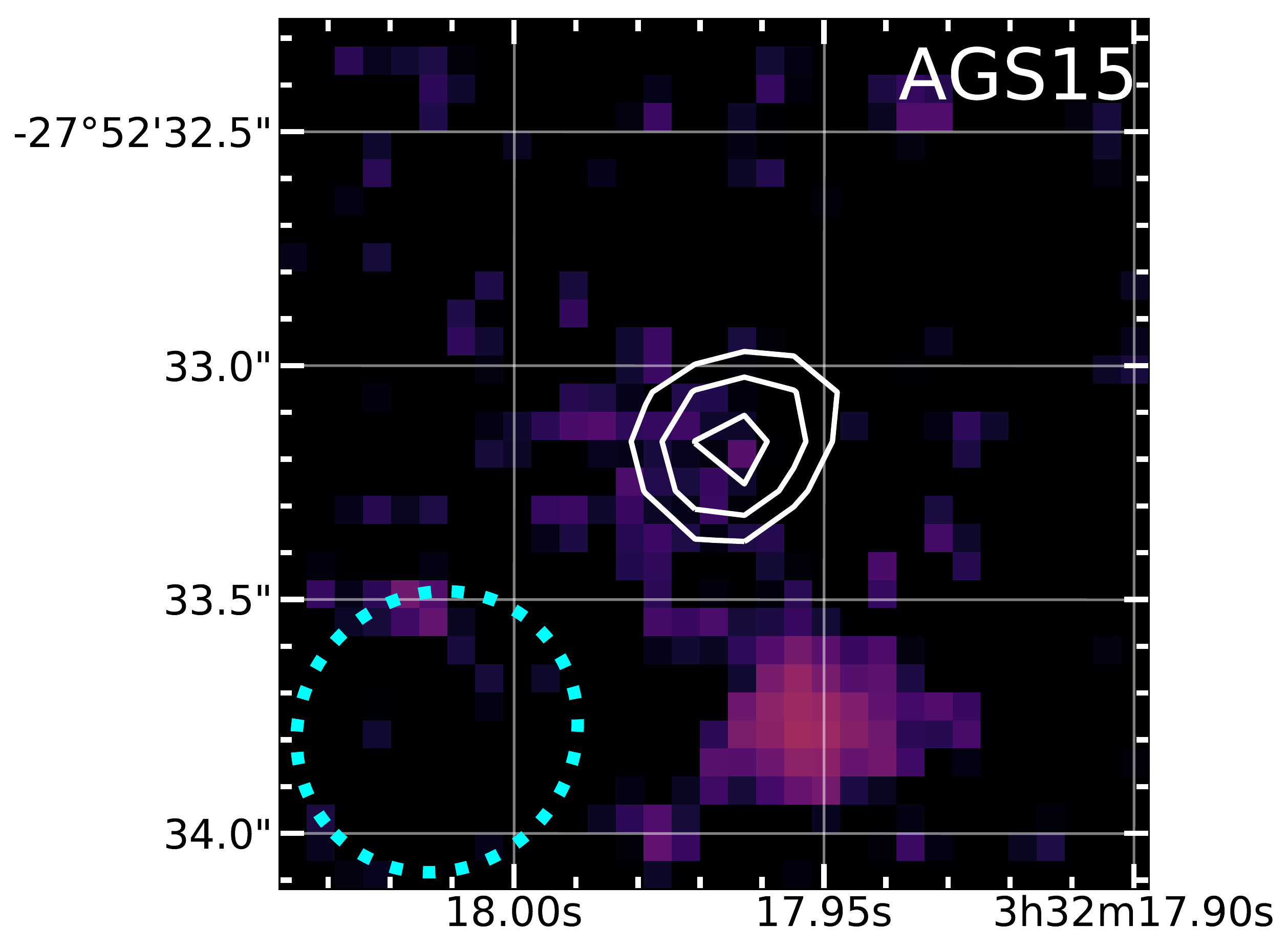}
\includegraphics[width=3.13cm,clip]{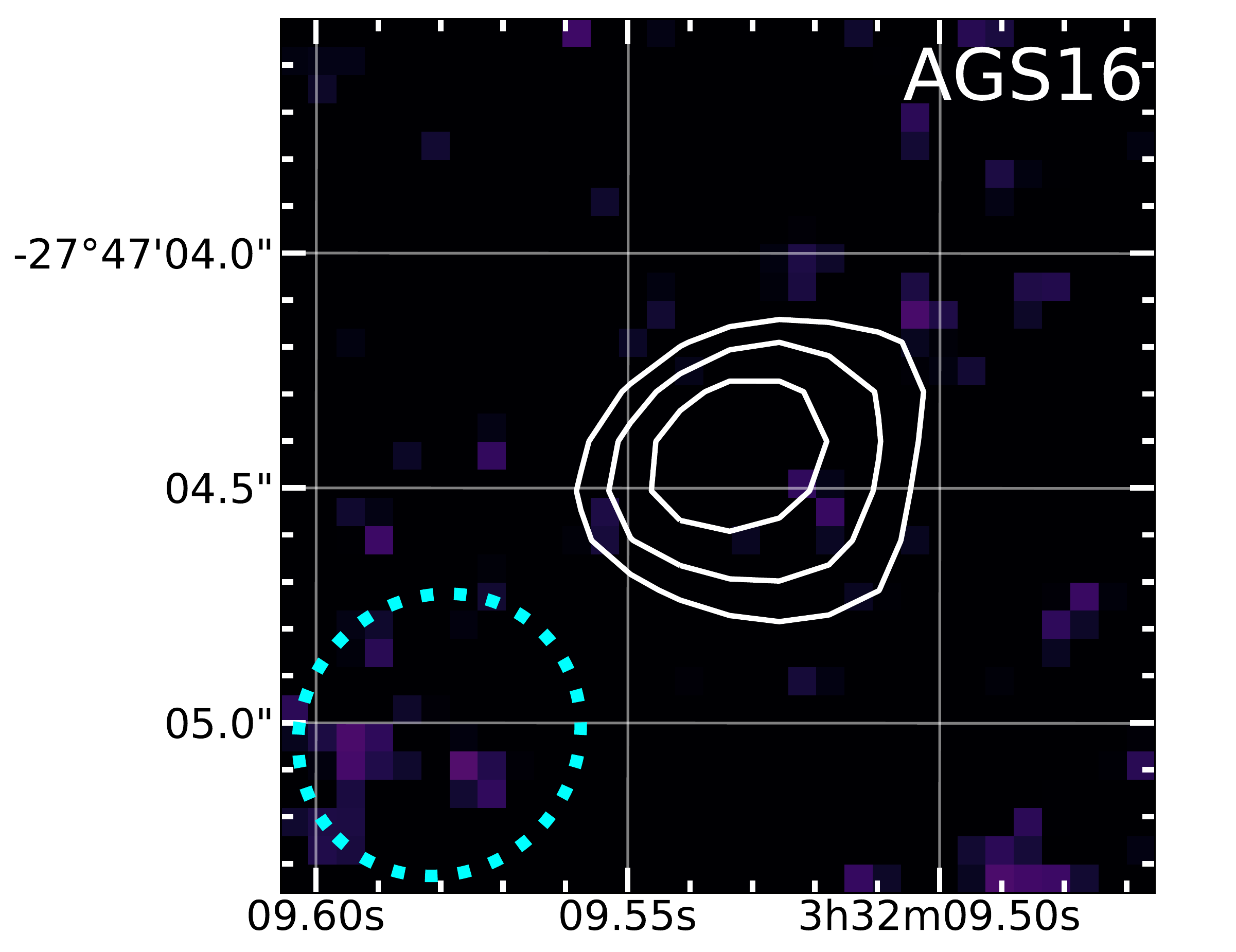}
}
\end{minipage}
\begin{minipage}[t]{0.9\textwidth}
\resizebox{\hsize}{!} {
\includegraphics[width=3.08cm,clip]{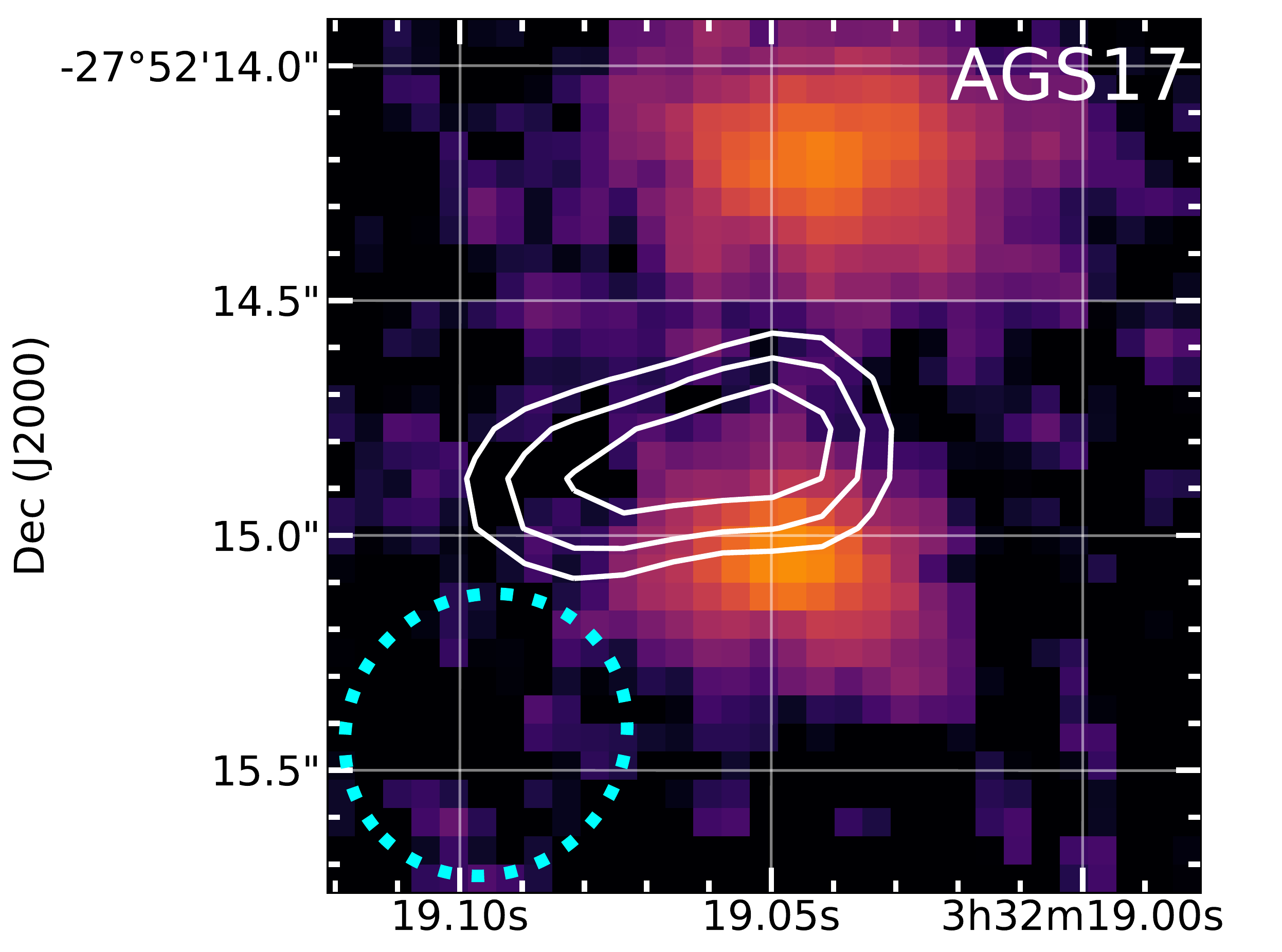}
\includegraphics[width=3cm,clip]{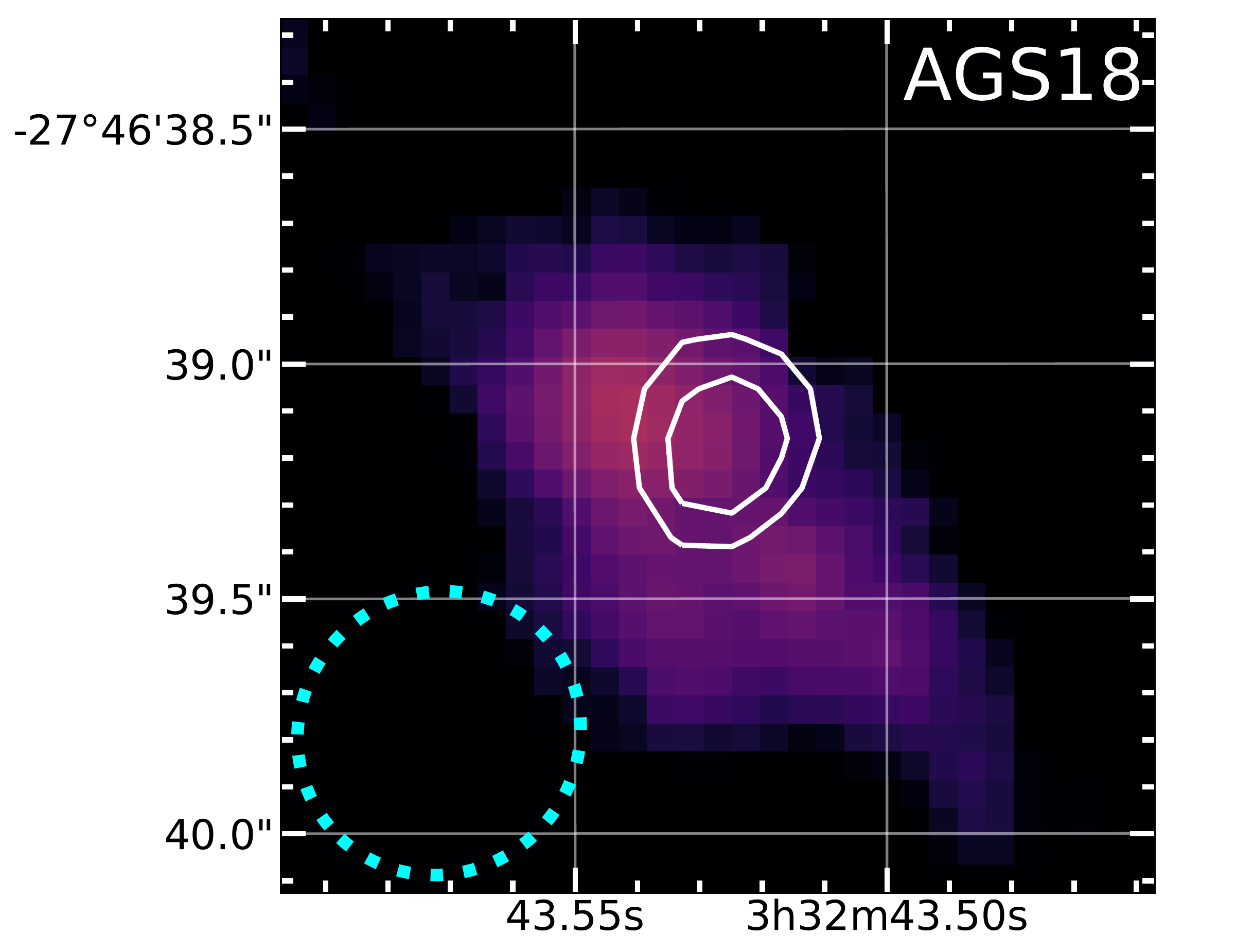}
\includegraphics[width=3cm,clip]{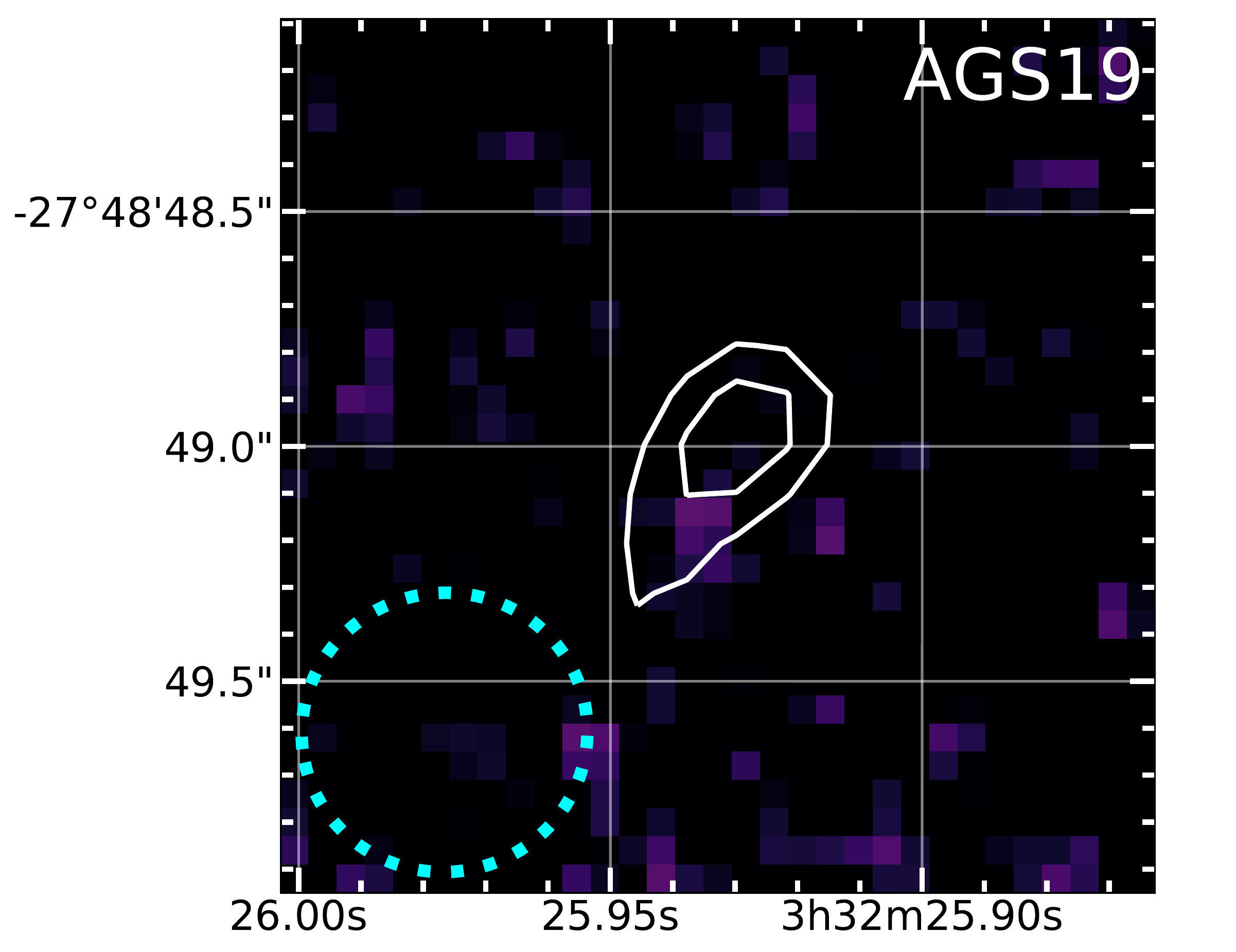}
\includegraphics[width=3cm,clip]{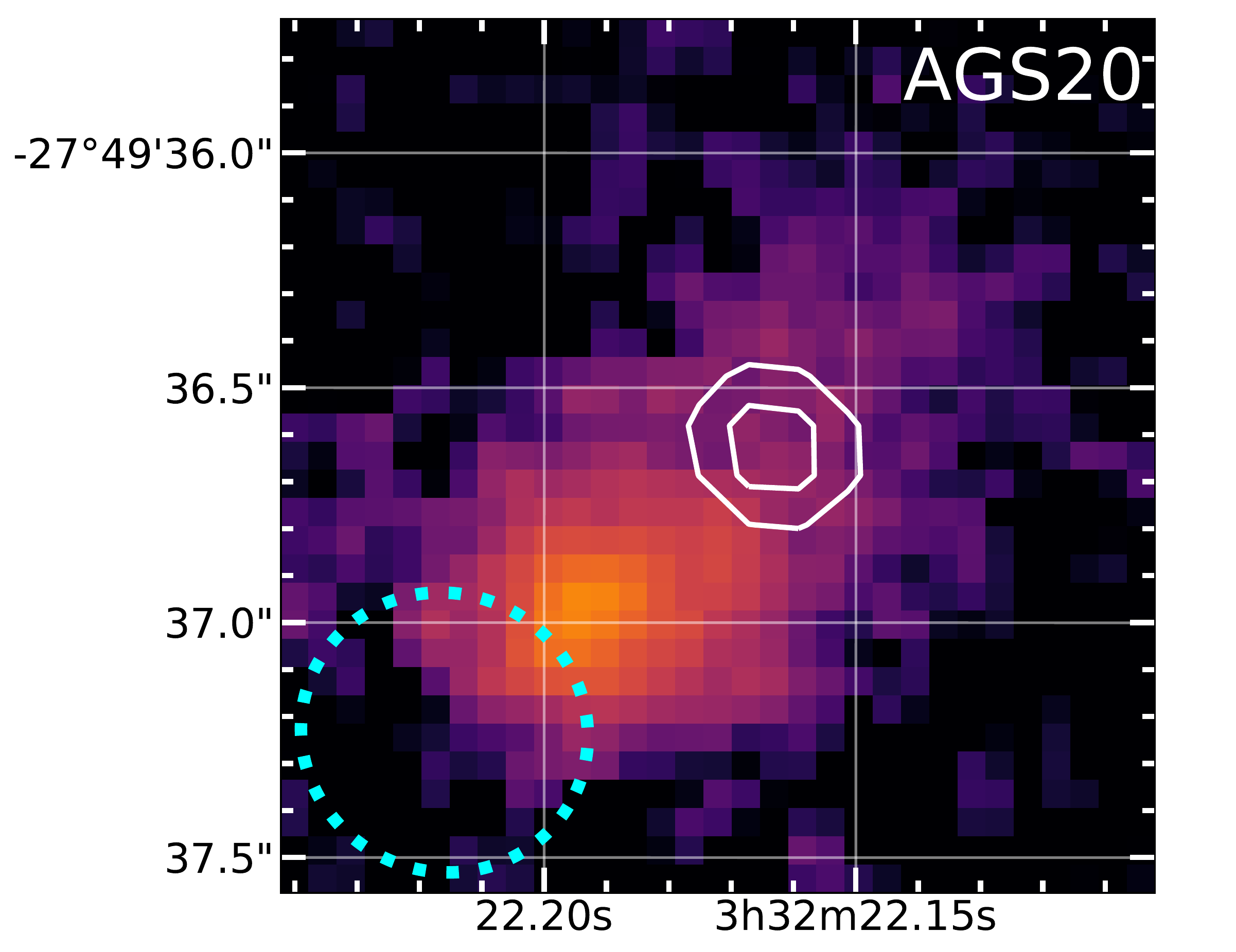}
}
\end{minipage}
\begin{minipage}[t]{0.9\textwidth}
\resizebox{\hsize}{!} {

\includegraphics[width=3.3cm,clip]{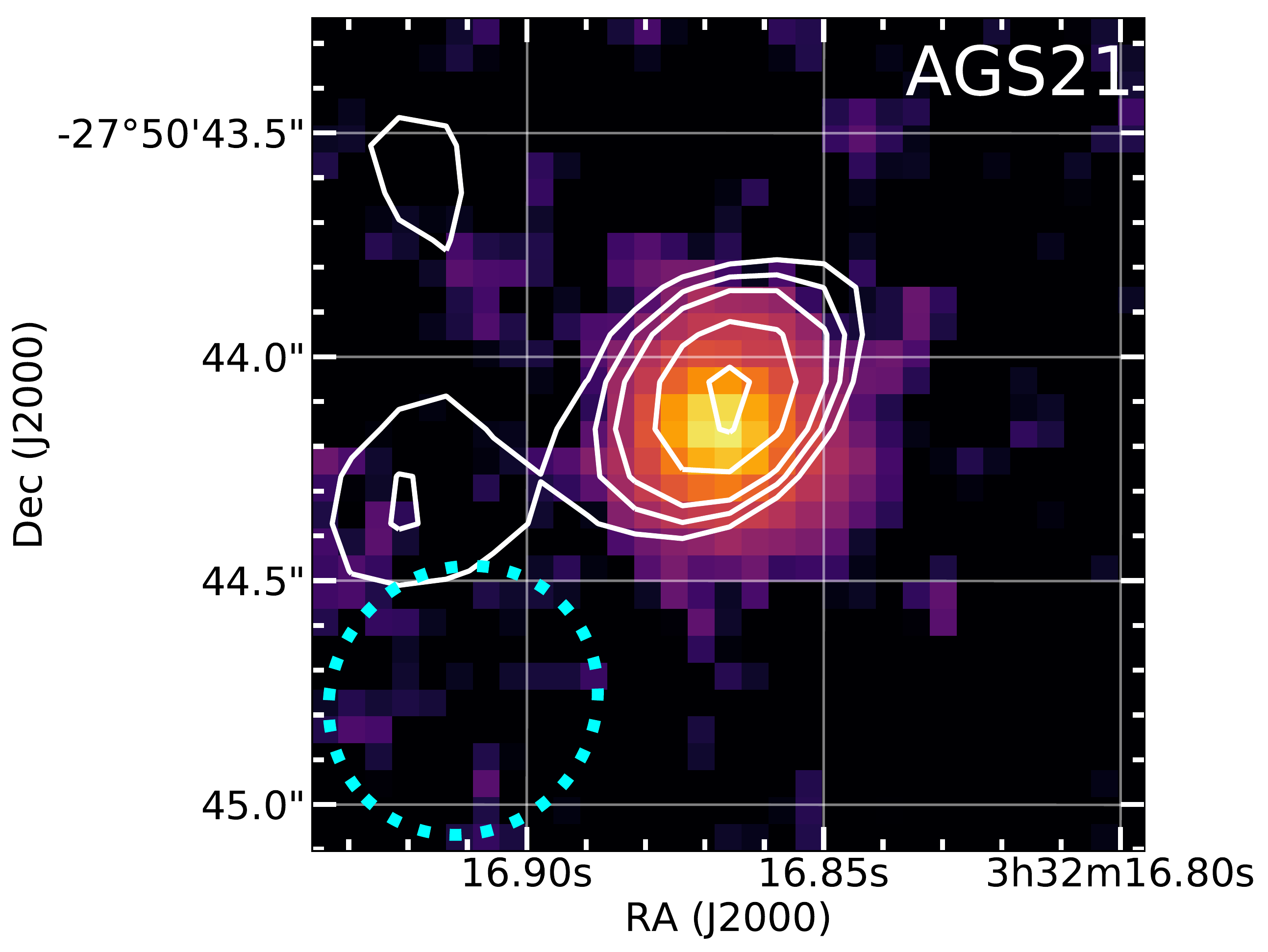}
\includegraphics[width=3.1cm,clip]{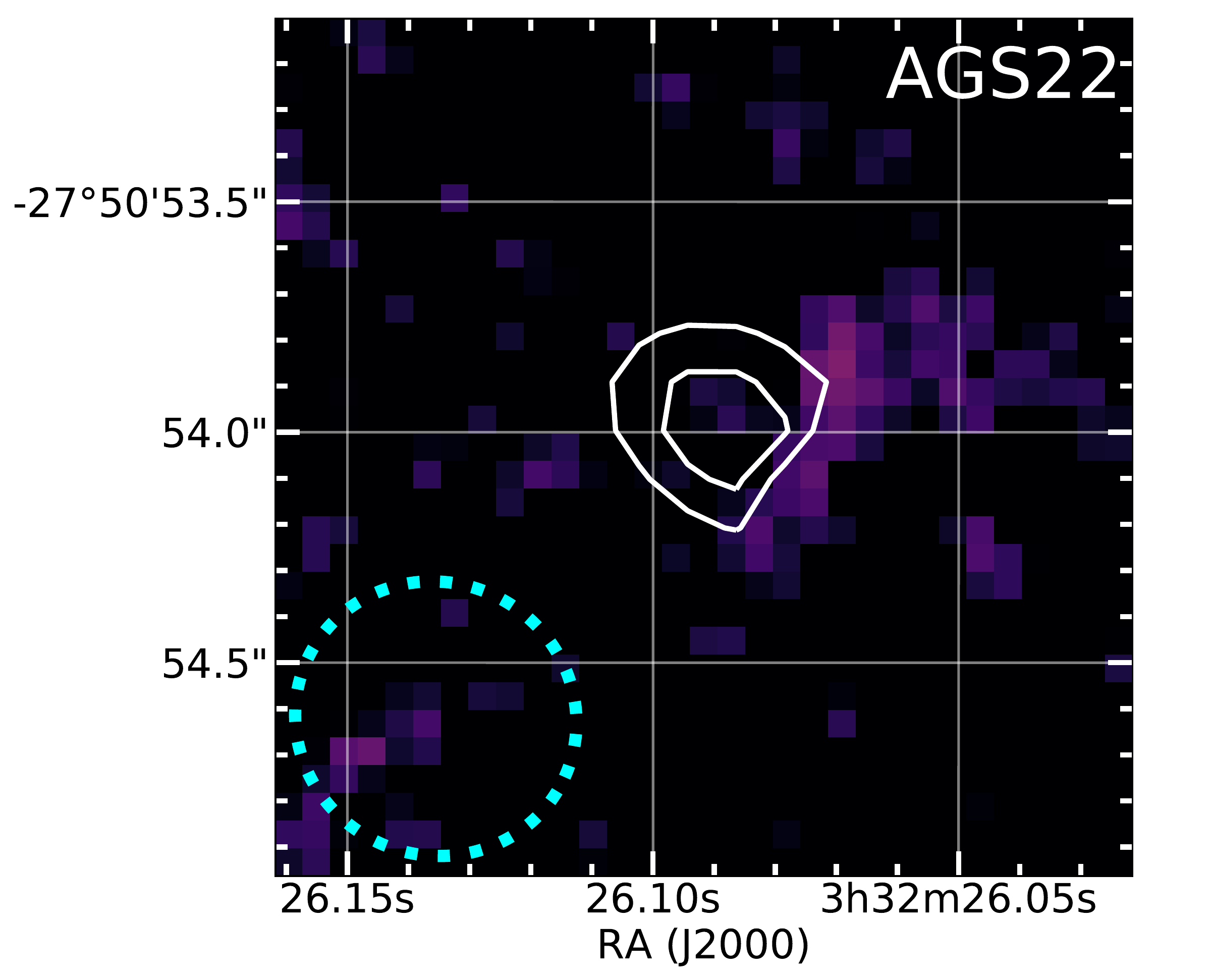}
\includegraphics[width=3.1cm,clip]{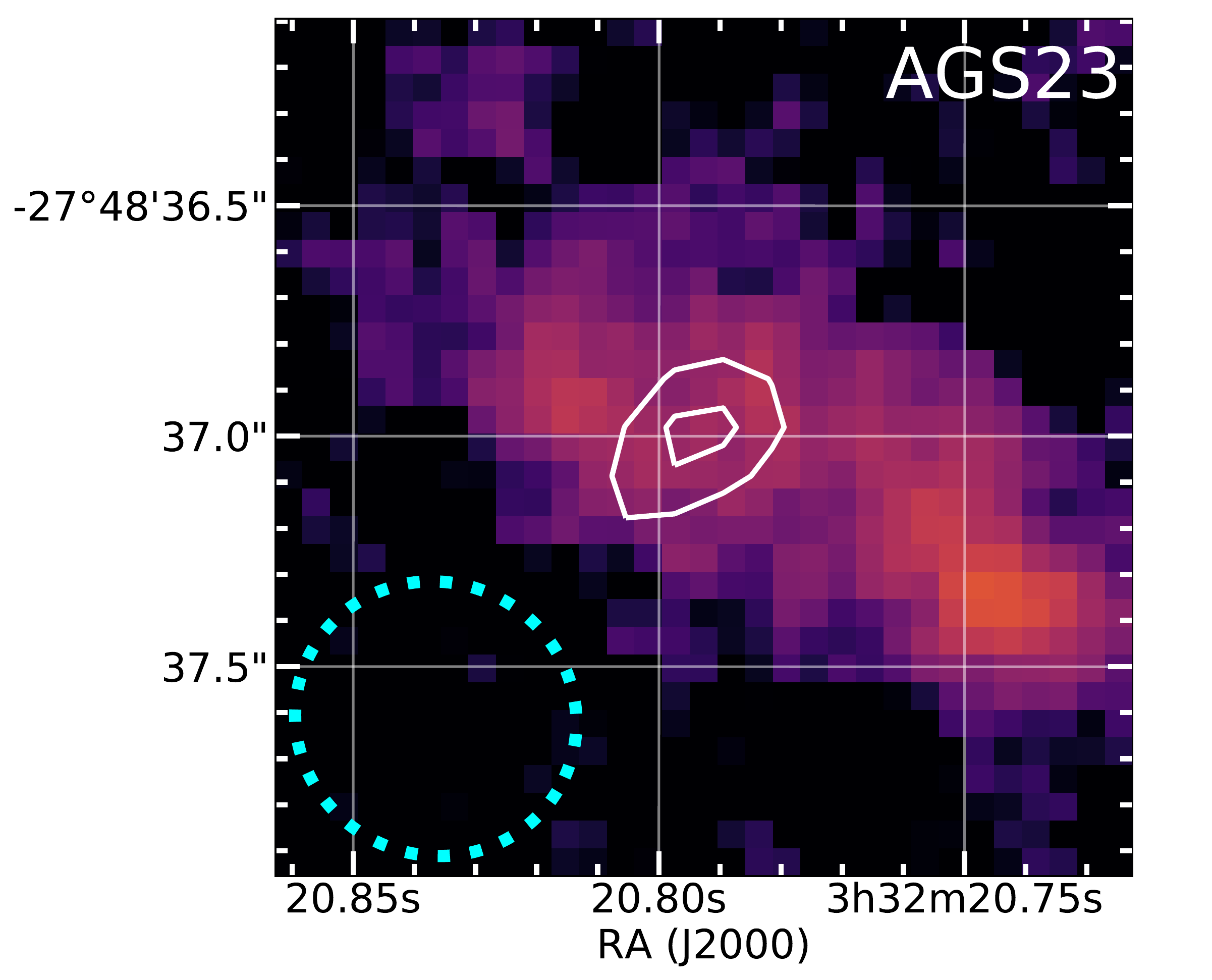}
\includegraphics[width=3.14cm,clip]{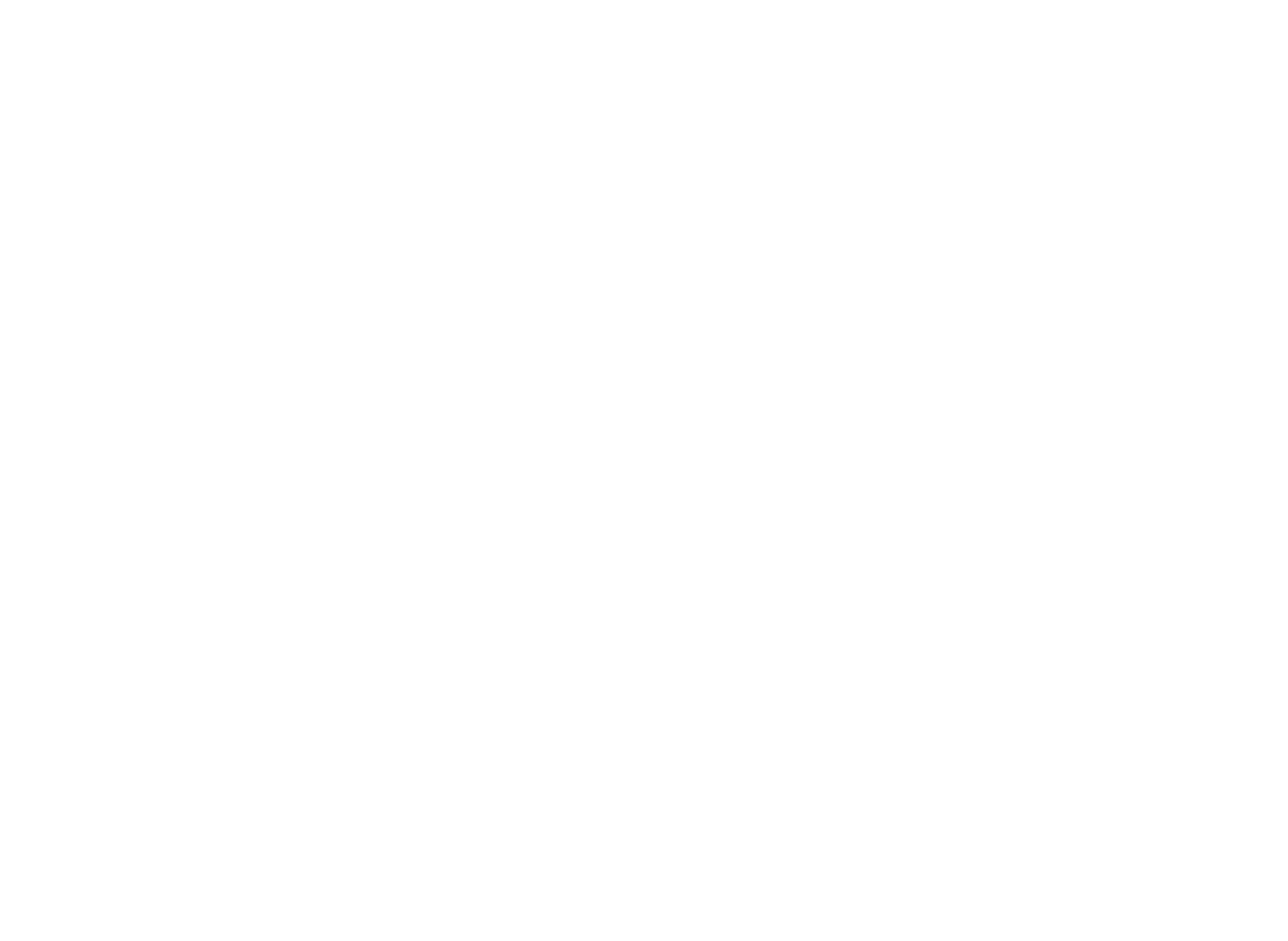}
}
\end{minipage}
\caption{Postage stamps of 1.8\,$\times$\,1.8 arcsec. ALMA contours (4, 4.5 then 5 to 10-$\sigma$ with a step of 1-$\sigma$) at 1.1mm (white lines) are overlaid on F160W HST/WFC3 images. The images are centred on the ALMA detections. The shape of the synthesized beam is given in the bottom left corner. Astrometry corrections described in Sect.~\ref{sec:Astrometric_correction} have been applied to the HST images. In some cases (AGS1, AGS3, AGS6, AGS13, AGS21 for example), the position of the dust radiation matches that of the stellar emission; in other cases, (AGS4, AGS17 for example), a displacement appears between both two wavelengths. Finally, in some cases (AGS11, AGS14, AGS16, and AGS19) there are no optical counterparts. We will discuss the possible explanations for this in Sect.~\ref{sec:HST-dark}.}
\label{ALMA_contours_H-band}
\end{figure*}

This accurate subtraction of the global systematic offset as well as the local offset does not however guarantee a perfect overlap between ALMA and HST emission. The location of the dust emission may not align perfectly with the starlight from a galaxy, due to the difference in ALMA and HST resolutions, as well as the physical offsets between dust and stellar emission that may exist. In Fig.~\ref{ALMA_contours_H-band}, we show the ALMA contours (4 to 10\,$\sigma$) overlaid on the F160W HST-WFC3 images after astrometric correction. In some cases (AGS1, AGS3, AGS6, AGS13, AGS21 for example), the position of the dust radiation matches that of the stellar emission; in other cases, (AGS4, AGS17 for example), a displacement appears between both two wavelengths. Finally, in some cases (AGS11, AGS14, AGS16, and AGS19) there are no optical counterparts. We will discuss the possible explanations for this in Sect.~\ref{sec:HST-dark}.
 
\begin{figure}
   \centering
   \includegraphics[width=\hsize]{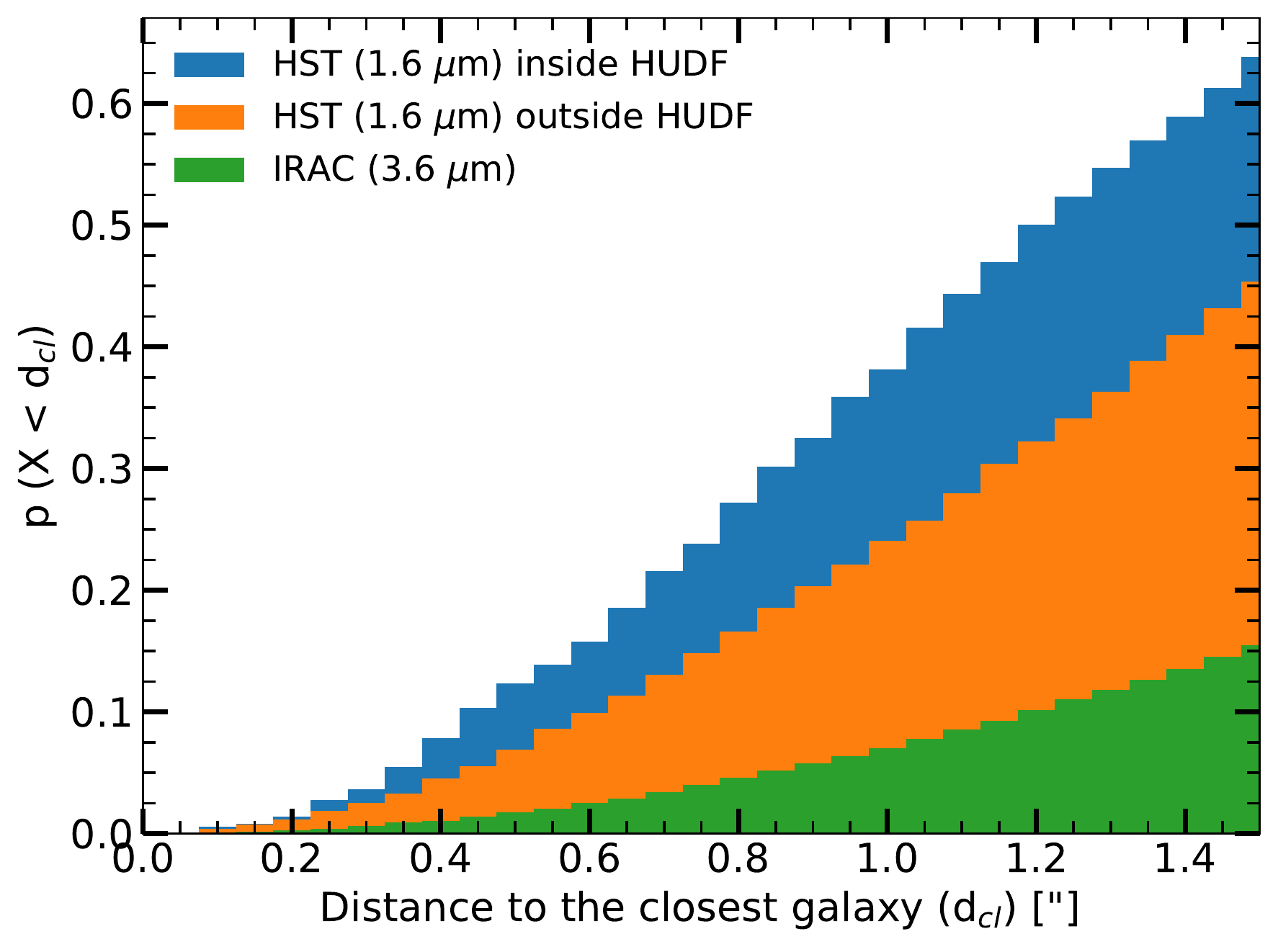}
      \caption{Probability of a randomly selected position in the area defined by this survey to have at least one HST (blue, orange) or $IRAC$ (green) neighbour as a function of distance. We compute this probability by Monte Carlo simulation using the distribution of galaxies listed in the CANDELS/GOODS--\textit{South} catalogue. Due to the presence of the HUDF within the GOODS--\textit{South} Field, we cannot consider that the density of HST galaxies is uniform, and we consider these two fields separately (blue inside and orange outside the HUDF).}
         \label{distance_closest_galaxy}
\end{figure}

\subsection{Identification of counterparts}\label{sec:galaxies_identification}
We searched for optical counterparts in the CANDELS/GOODS--\textit{South} catalogue, within a radius of 0\arcsec6 from the millimetre position after having applied the astrometric corrections to the source positions described in Sect.~\ref{sec:Astrometric_correction}. The radius of the cross-matching has been chosen to correspond to the synthesized beam (0\arcsec60) of the tapered ALMA map used for galaxy detection. Following \cite{Condon1997}, the maximal positional accuracy of the detection in the 1.1mm map is given by $\theta_{\text{beam}}$/(2$\times$SNR). In the 0\arcsec60-mosaic, the positional accuracy therefore ranges between 26.5 mas and 62.5 mas for our range of SNR (4.8-11.3), corresponding to physical sizes between 200 and 480 pc  at $z$\,=\,3. 

Despite the high angular resolution of ALMA, the chance of an ALMA-HST coincidence is not negligible, because of the large projected source density of the CANDELS/GOODS--\textit{South} catalogue. Fig.~\ref{distance_closest_galaxy} shows a Monte Carlo simulation performed to estimate this probability. We separate here the deeper Hubble Ultra Deep Field (blue histogram) from the rest of the CANDELS-Deep area (orange histogram). We randomly define a position within GOODS--\textit{South} and then measure the distance to its closest HST neighbour using the source positions listed in \cite{Guo2013}. We repeat this procedure 100\,000 times inside and outside the HUDF. The probability for a position randomly selected in the GOODS--\textit{South} field to fall within 0.6 arcsec of an HST source is 9.2\% outside the HUDF, and 15.8\% inside the HUDF. We repeat this exercise to test the presence of an IRAC counterpart with the \cite{Ashby2015} catalogue (green histogram). The probability to randomly fall on an IRAC source is only 2.1\%.

With the detection threshold determined in Sect.~\ref{sec:Source_Detection}, 80\% of the millimetre galaxies detected have an HST-WFC3 counterpart, and 4 galaxies remain without an optical counterpart. We cross-matched our detections with the ZFOURGE catalogue.

\begin{figure*}
\centering

\begin{minipage}[t]{0.9\textwidth}
\resizebox{\hsize}{!} {
\includegraphics[width=3cm,clip]{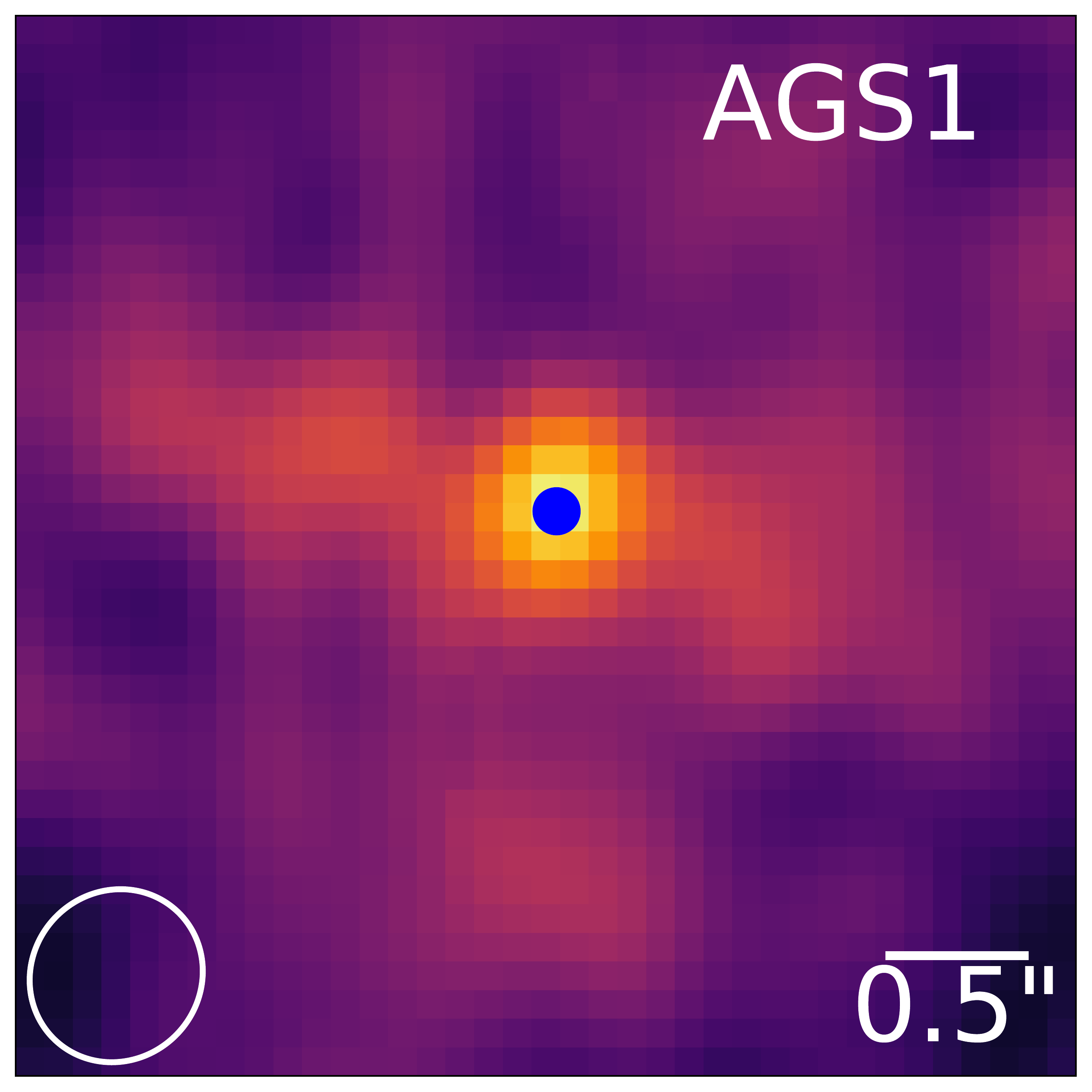}
\includegraphics[width=3cm,clip]{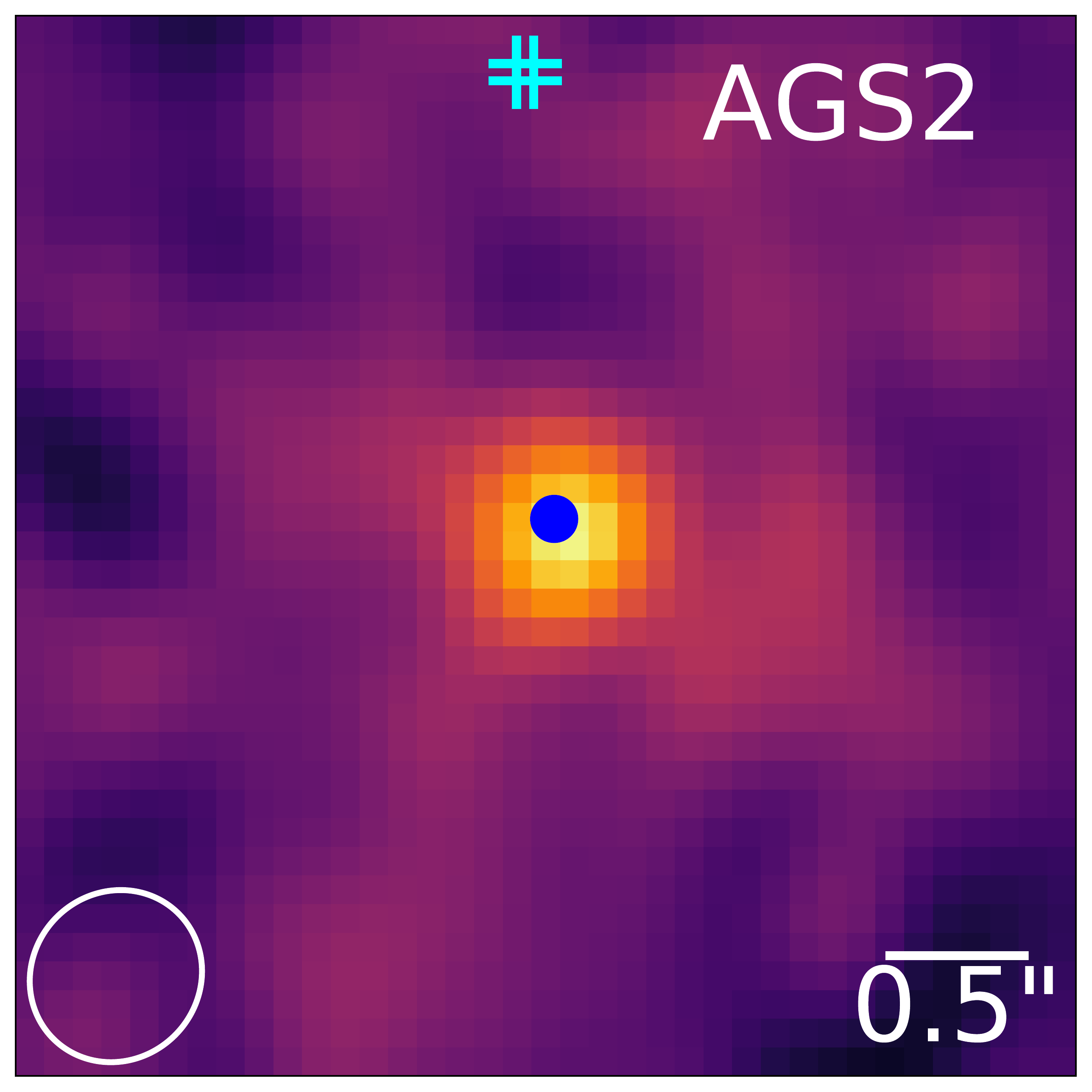}
\includegraphics[width=3cm,clip]{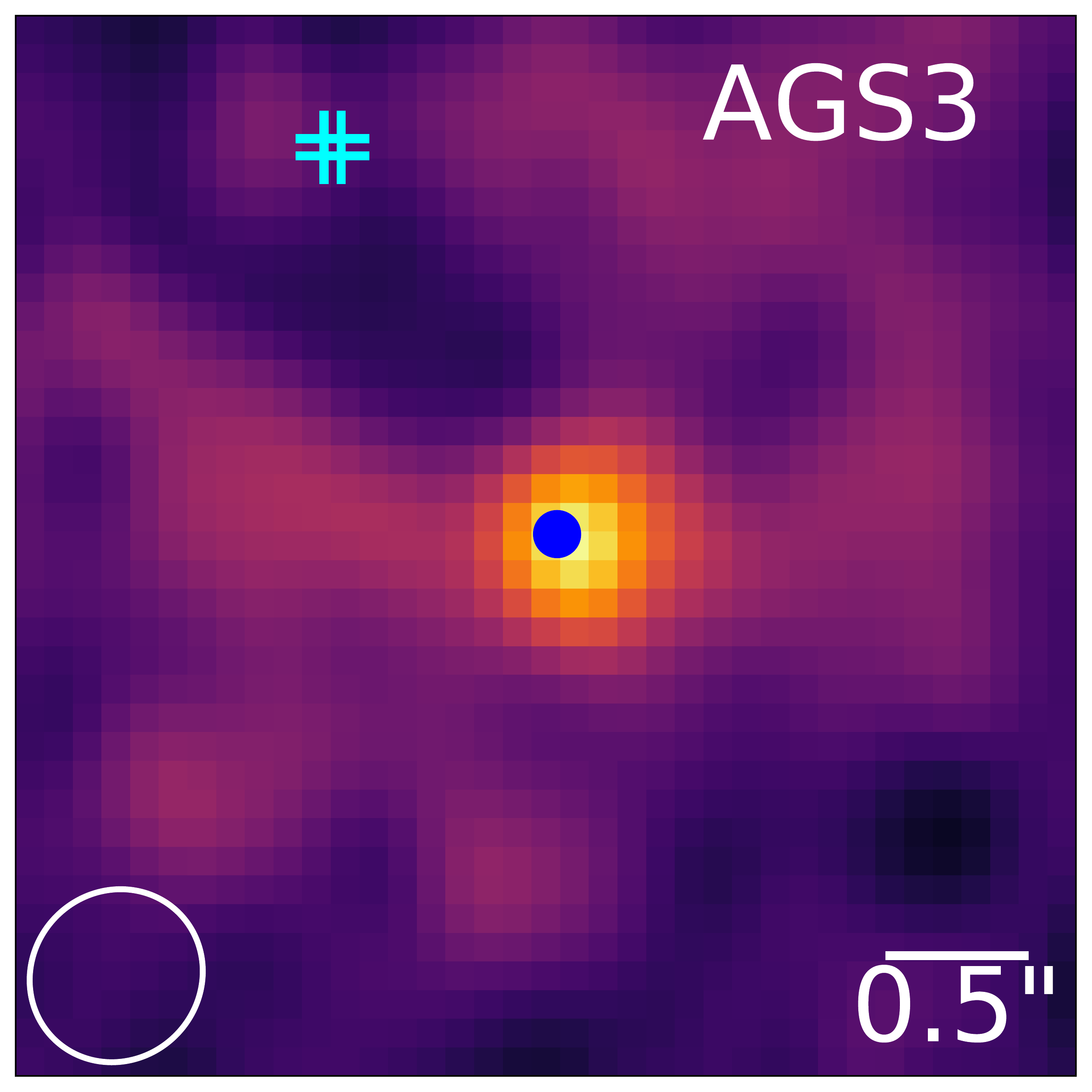}
\includegraphics[width=3cm,clip]{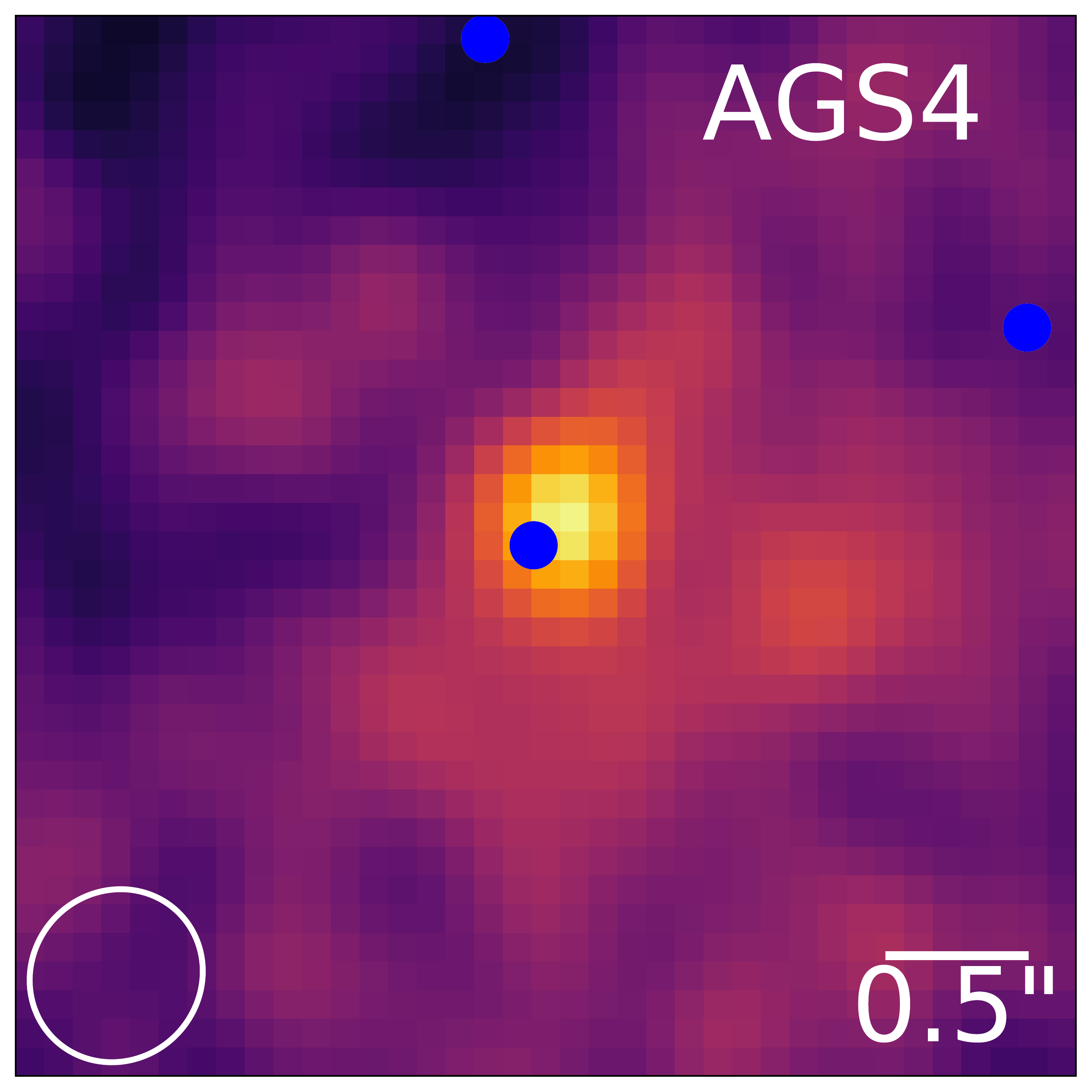}
\includegraphics[width=3cm,clip]{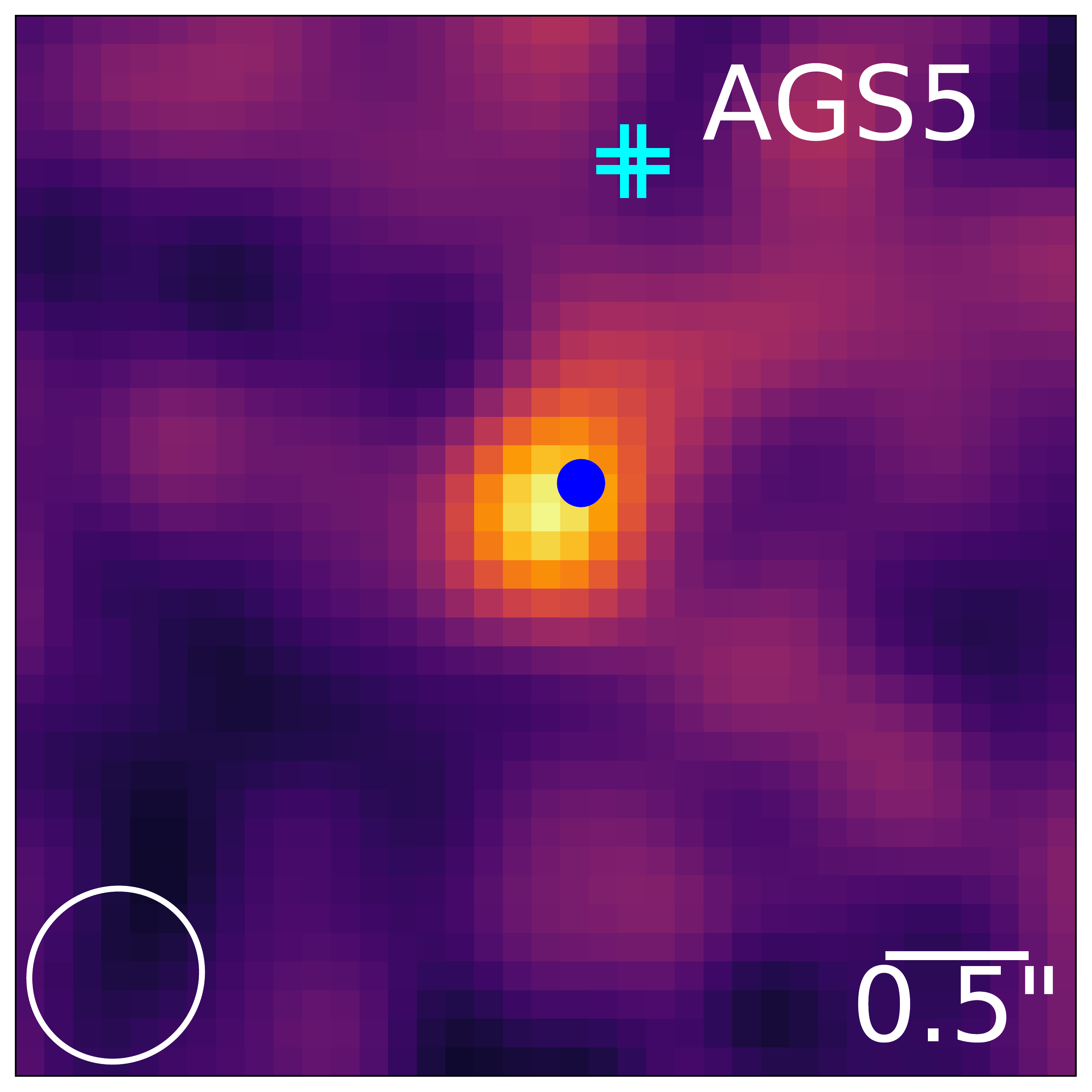}
\includegraphics[width=3cm,clip]{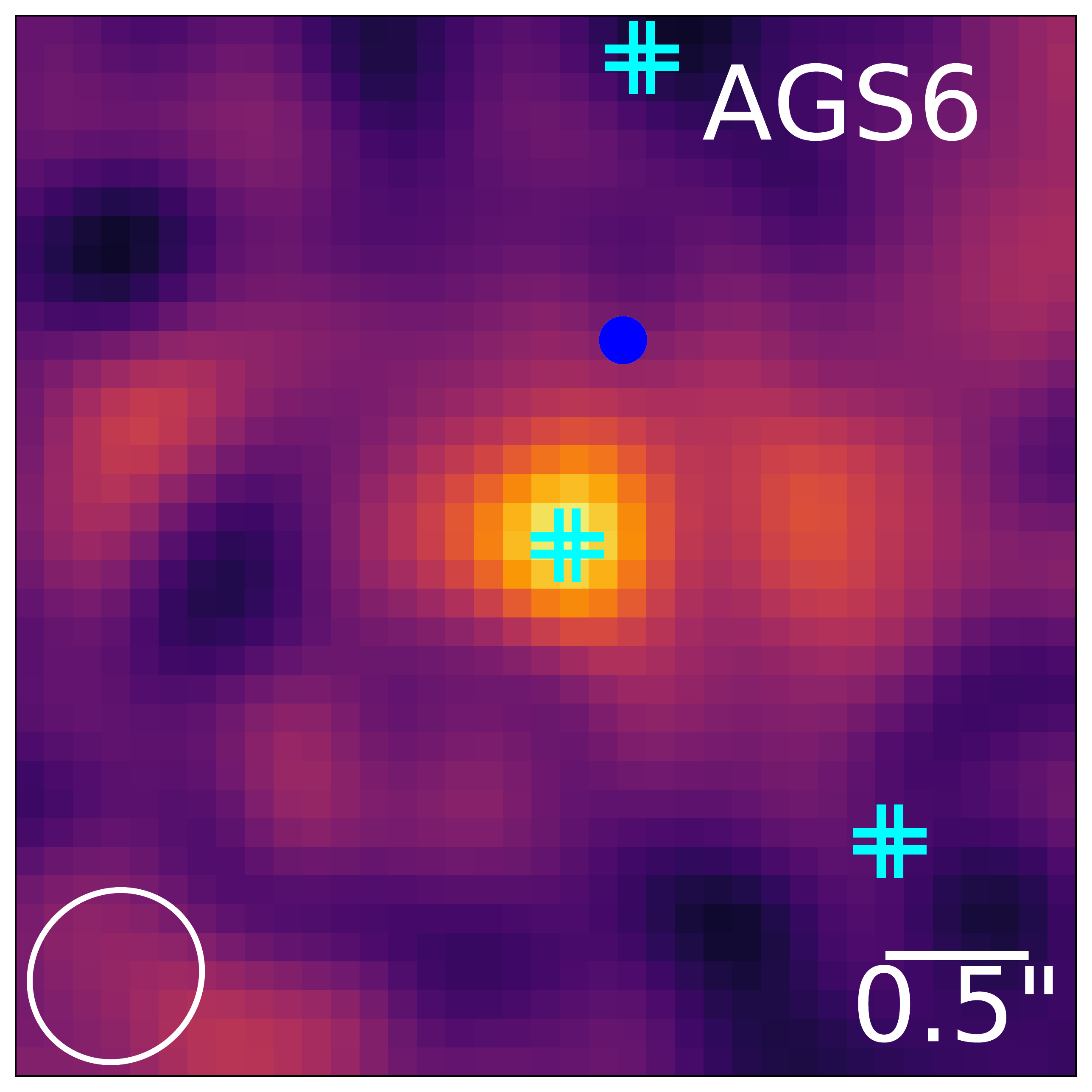}
}
\end{minipage}
\begin{minipage}[t]{0.9\textwidth}
\resizebox{\hsize}{!} {
\includegraphics[width=3cm,clip]{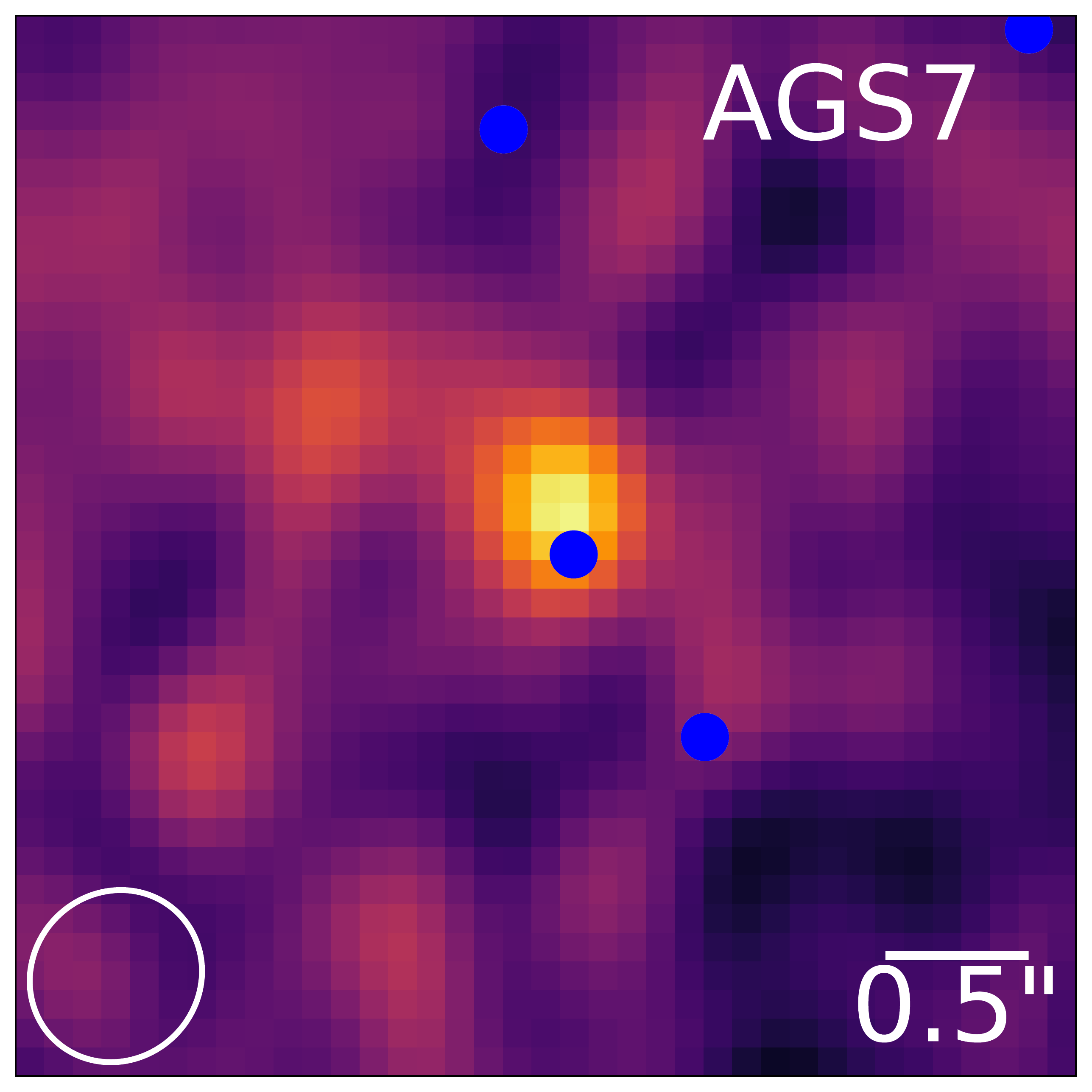}
\includegraphics[width=3cm,clip]{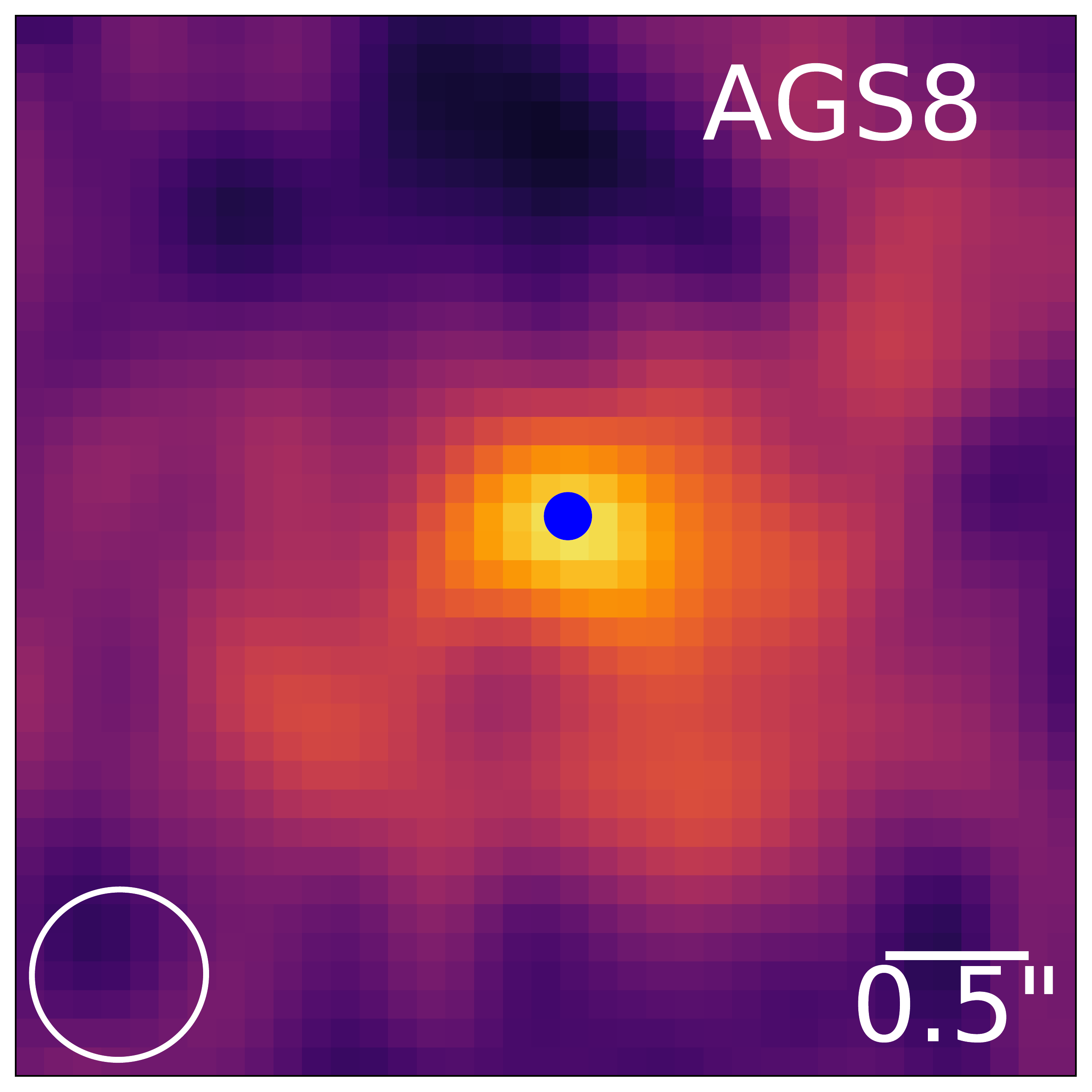}
\includegraphics[width=3cm,clip]{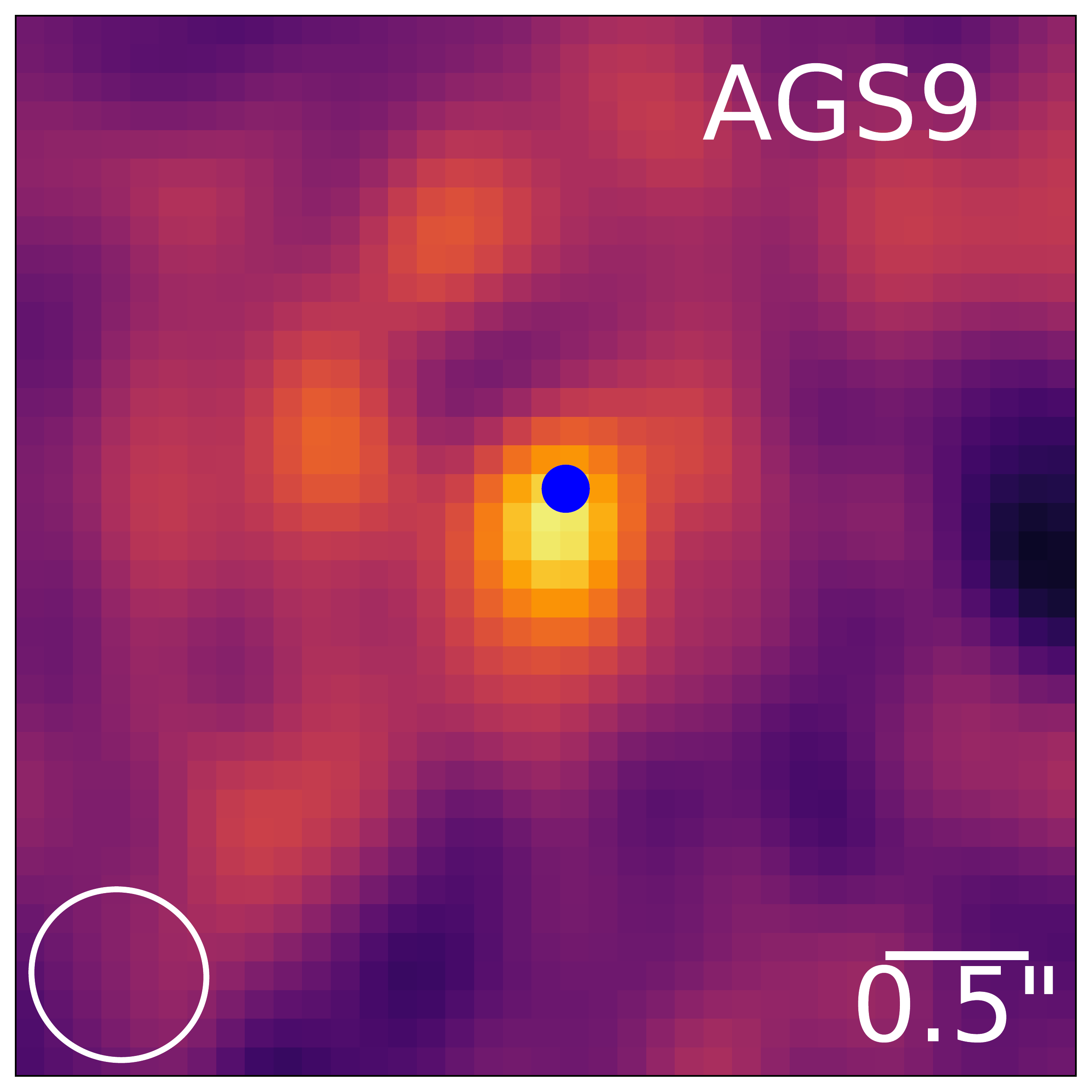}
\includegraphics[width=3cm,clip]{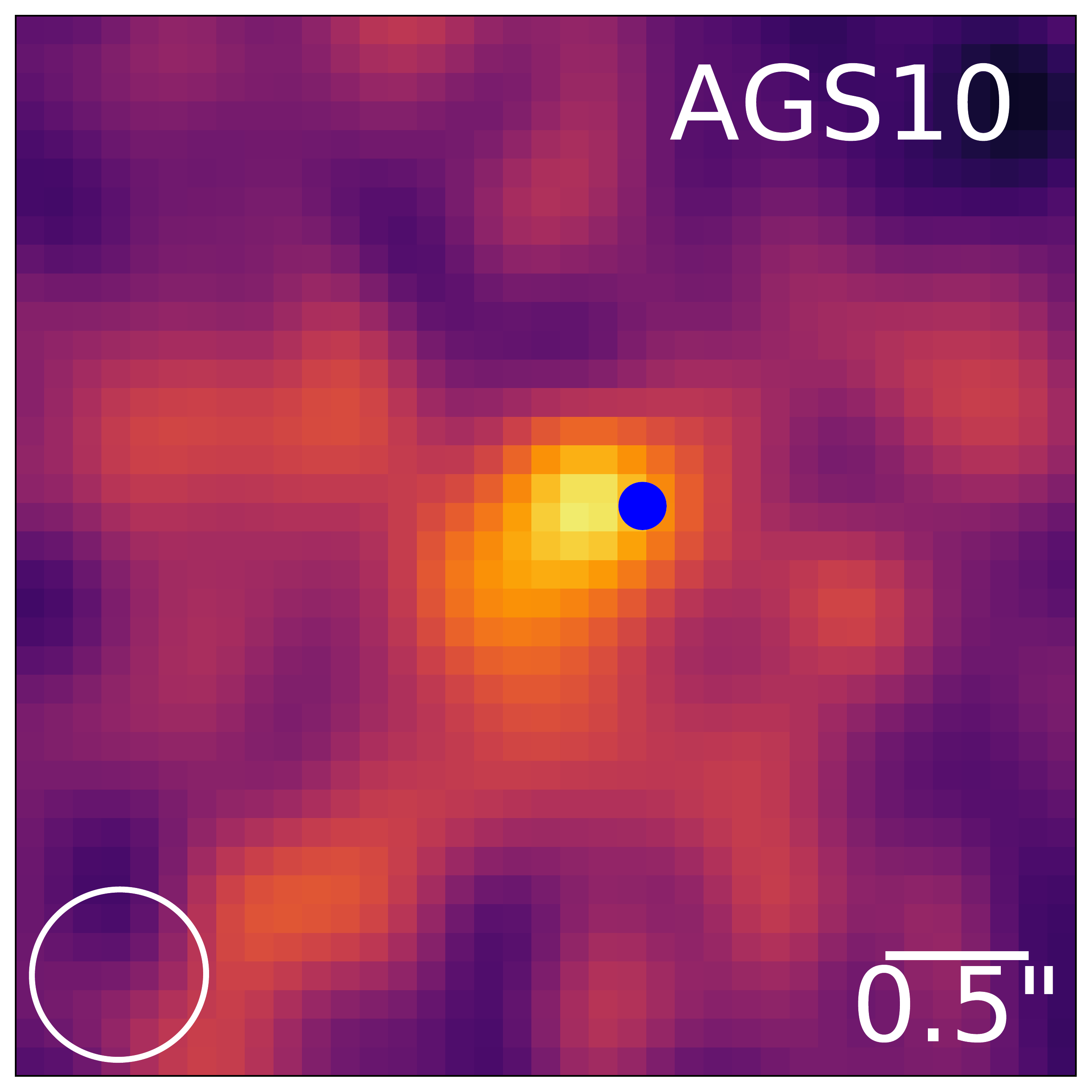}
\includegraphics[width=3cm,clip]{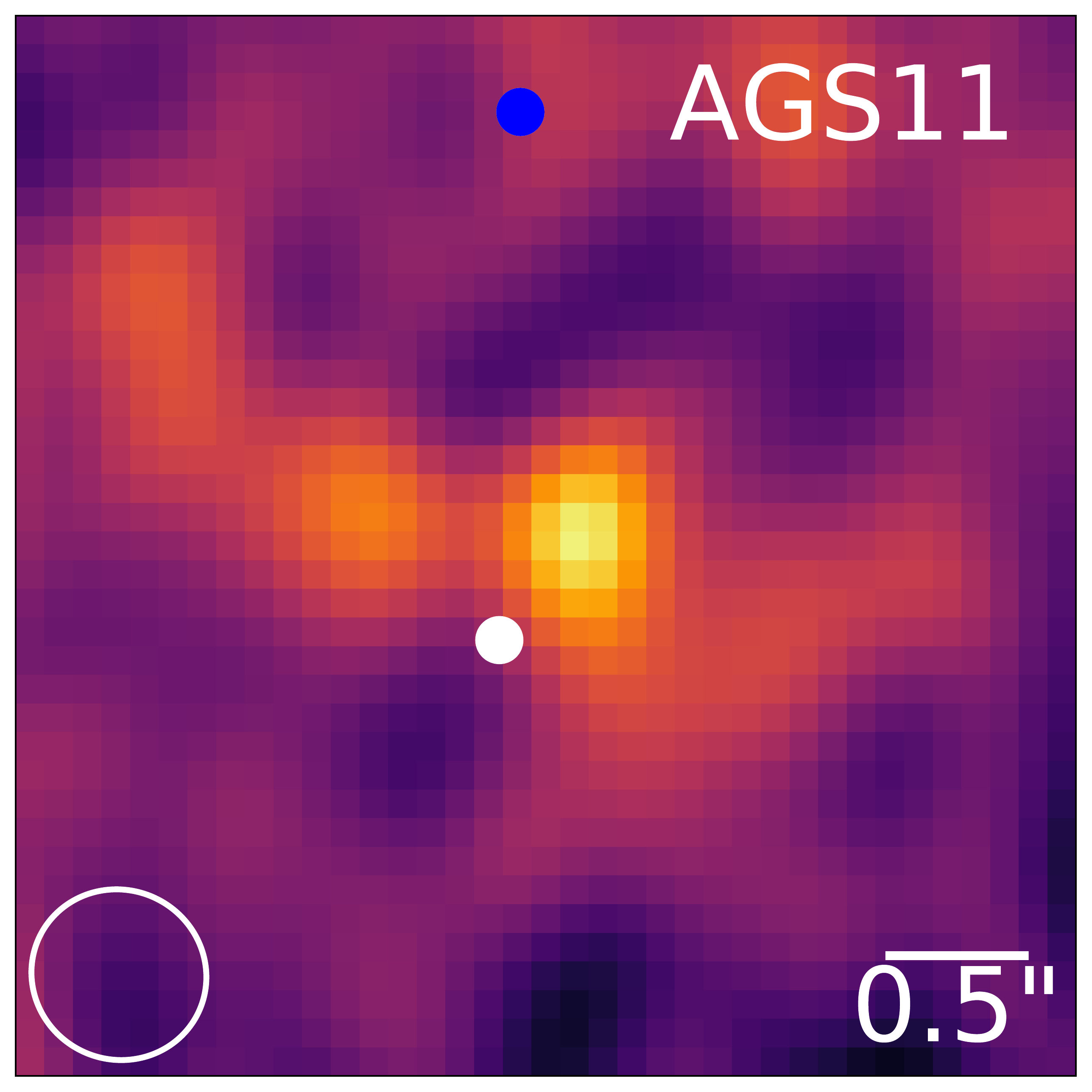}
\includegraphics[width=3cm,clip]{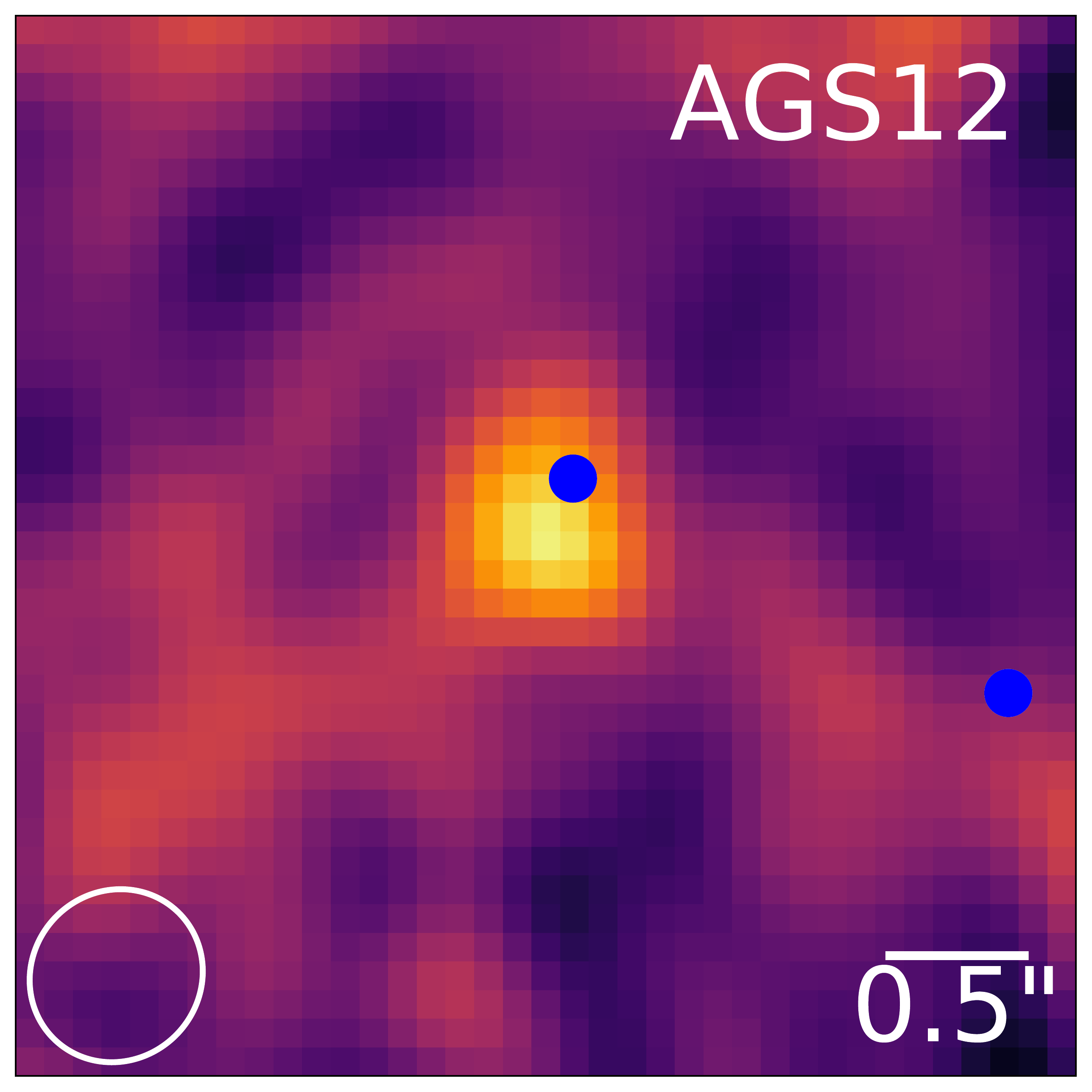}
}
\end{minipage}

\begin{minipage}[t]{0.9\textwidth}
\resizebox{\hsize}{!} {
\includegraphics[width=3cm,clip]{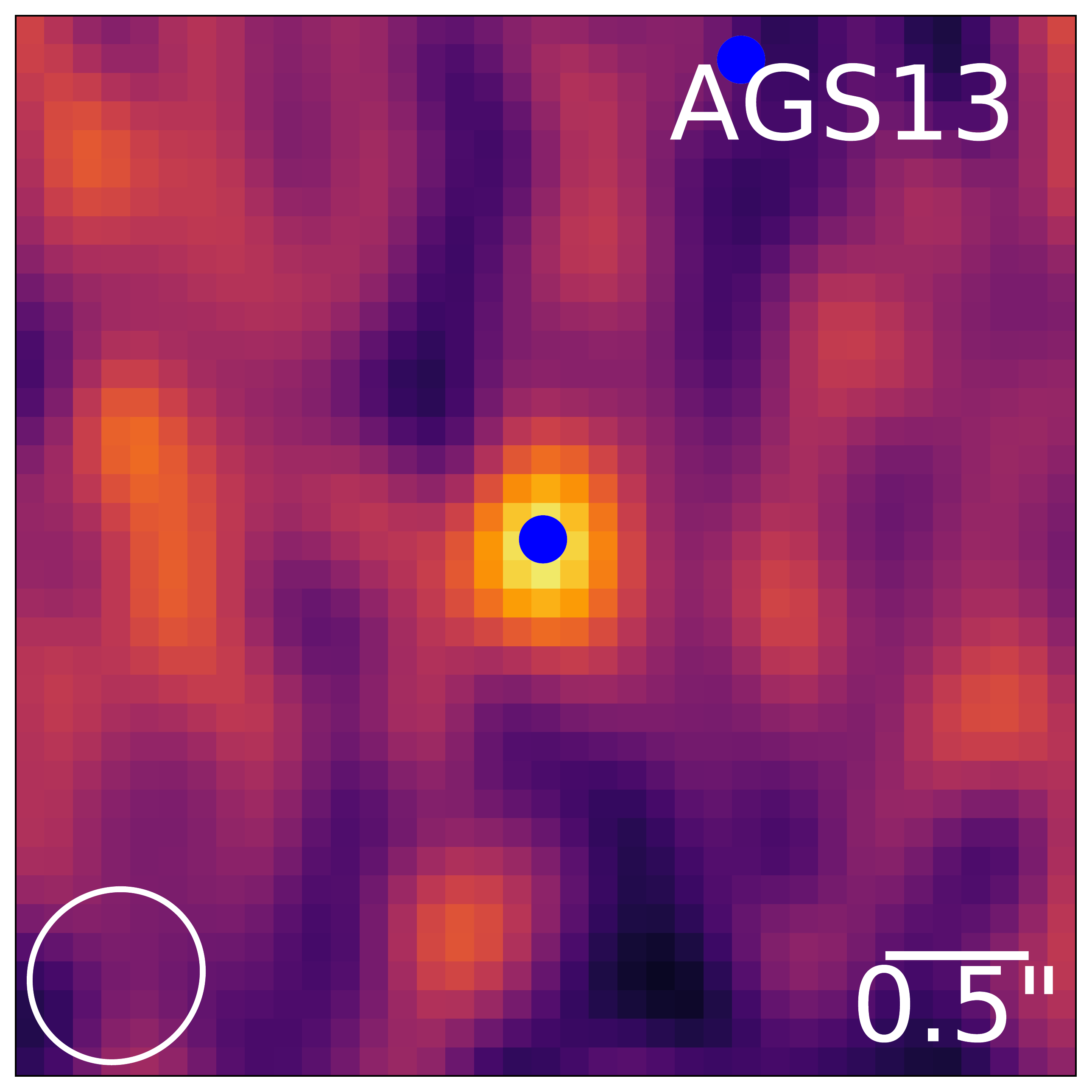}
\includegraphics[width=3cm,clip]{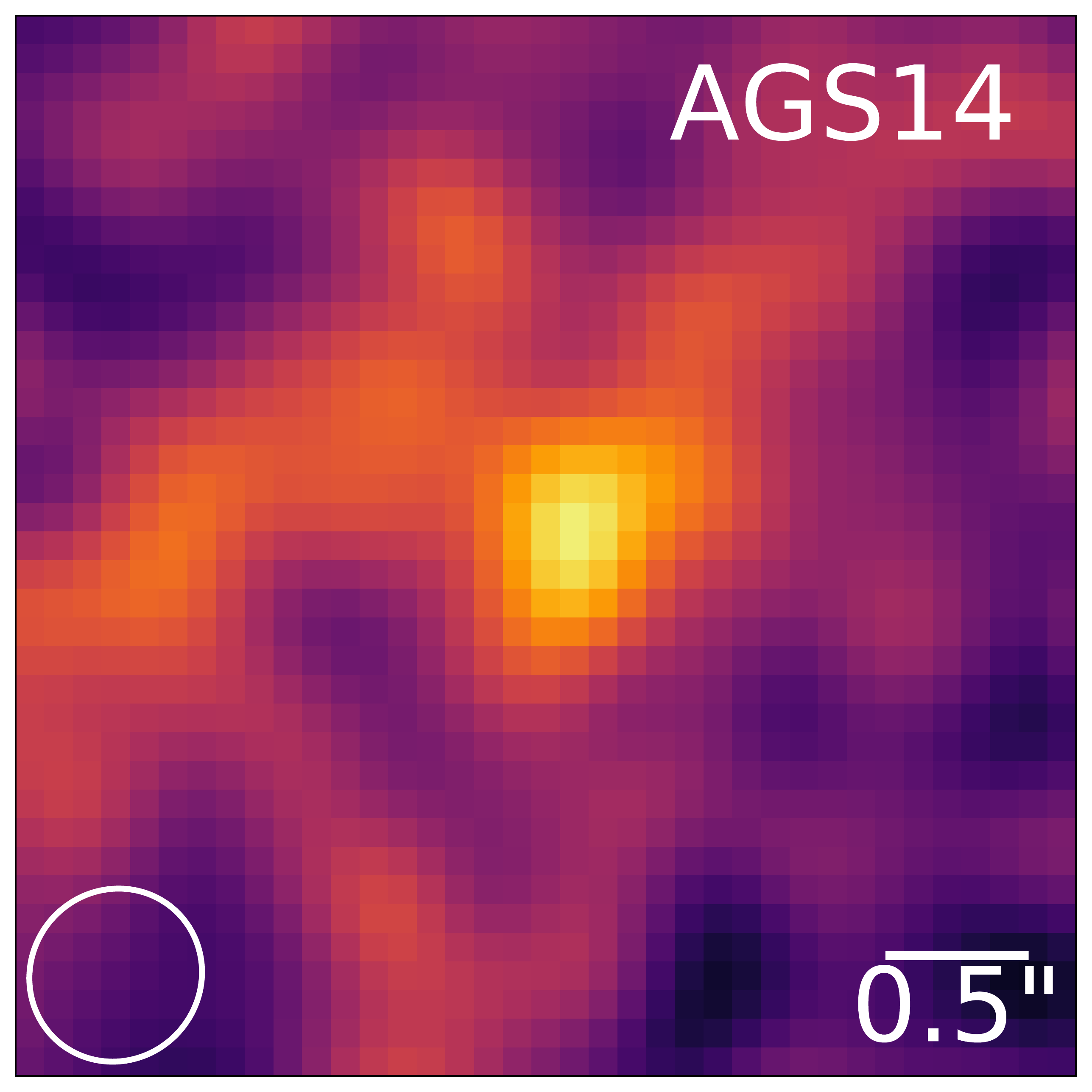}
\includegraphics[width=3cm,clip]{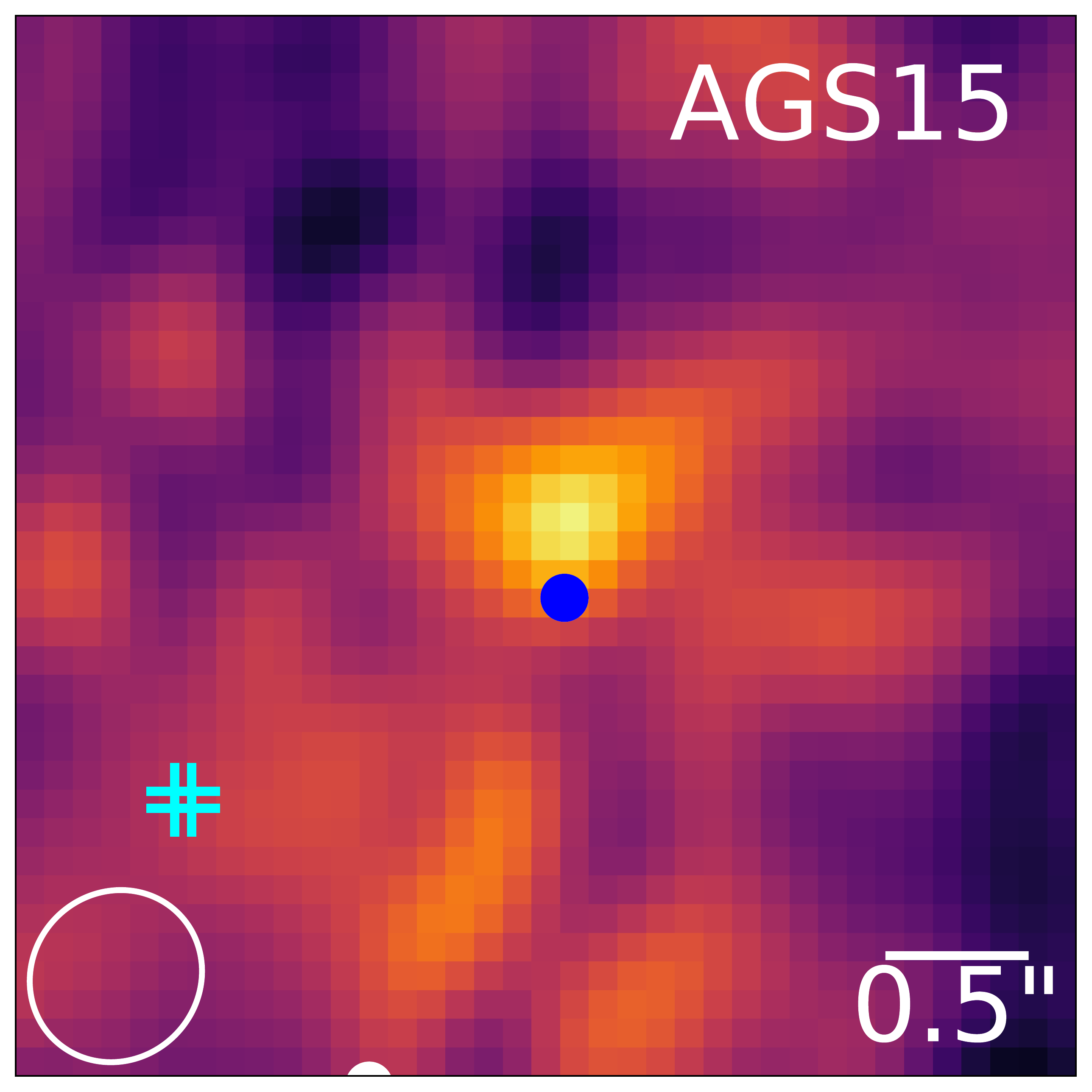}
\includegraphics[width=3cm,clip]{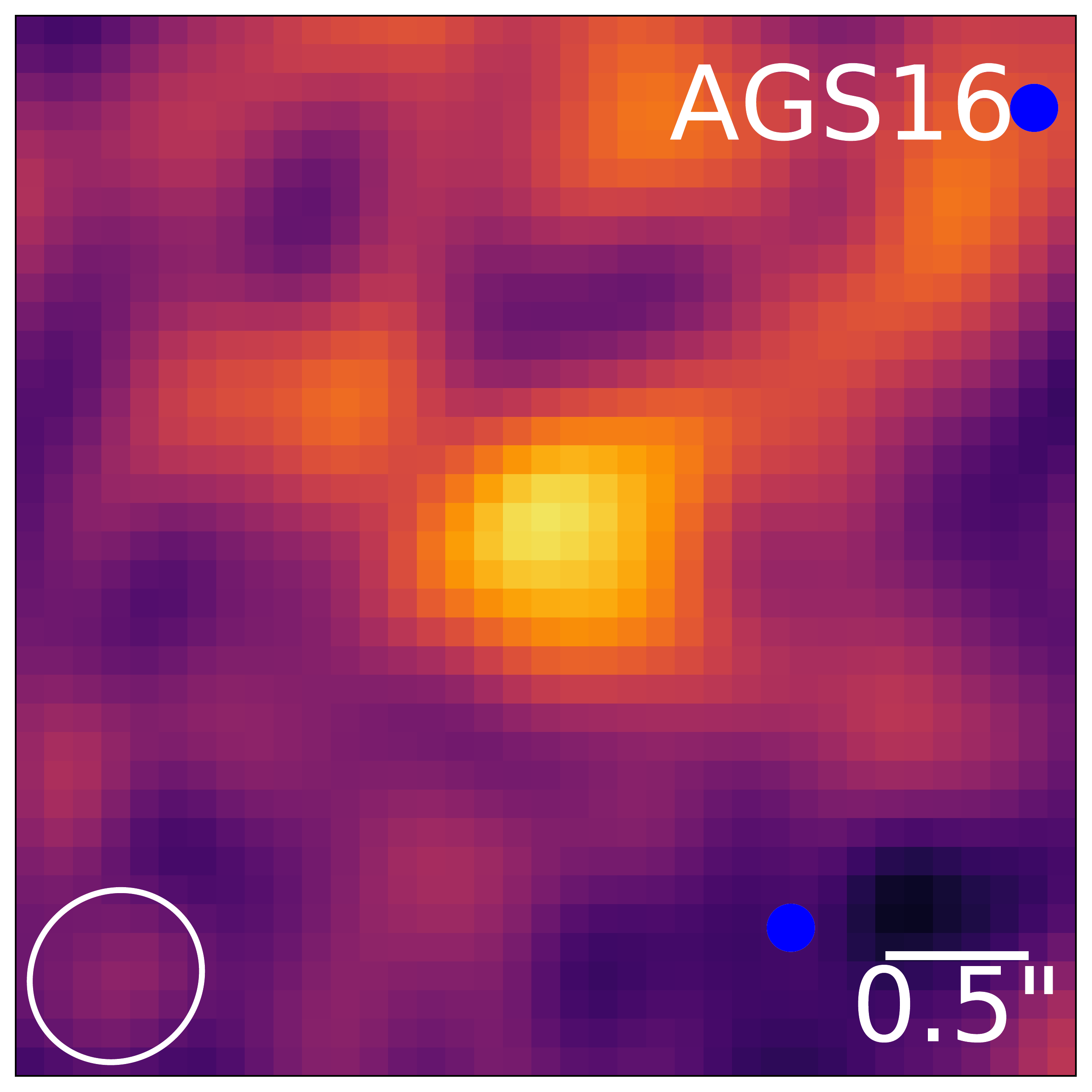}
\includegraphics[width=3cm,clip]{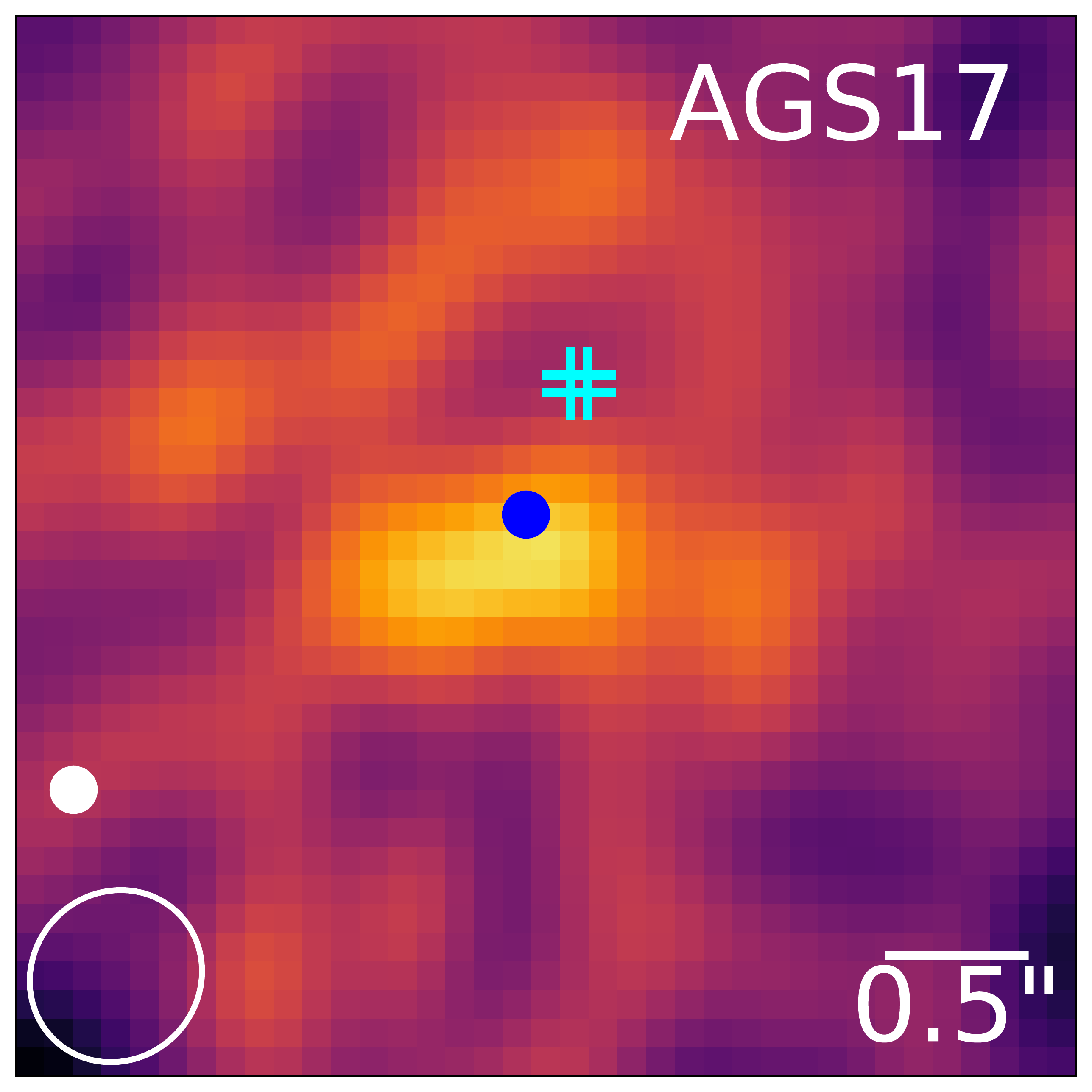}
\includegraphics[width=3cm,clip]{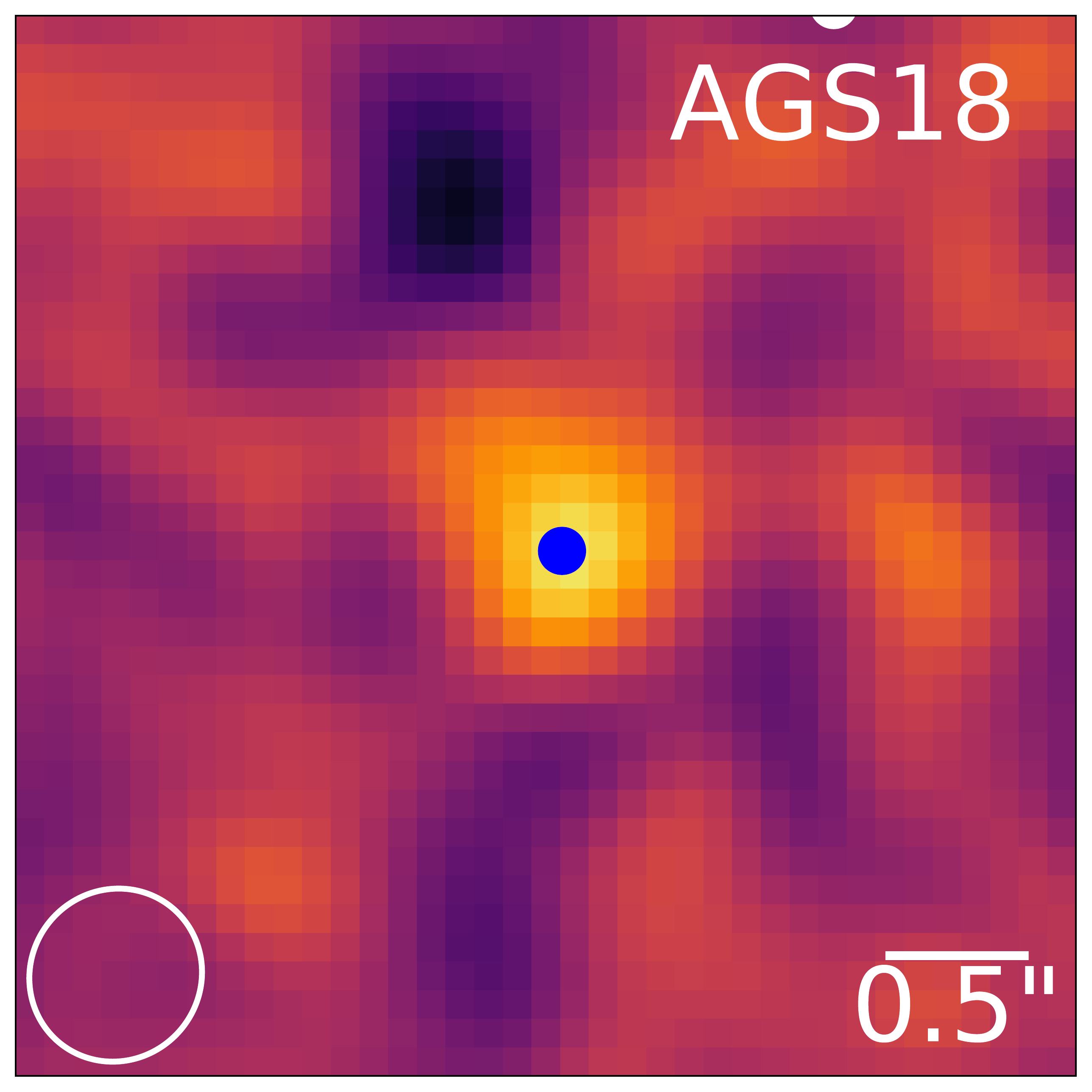}
}
\end{minipage}
\begin{minipage}[t]{0.9\textwidth}
\resizebox{\hsize}{!} {
\includegraphics[width=3cm,clip]{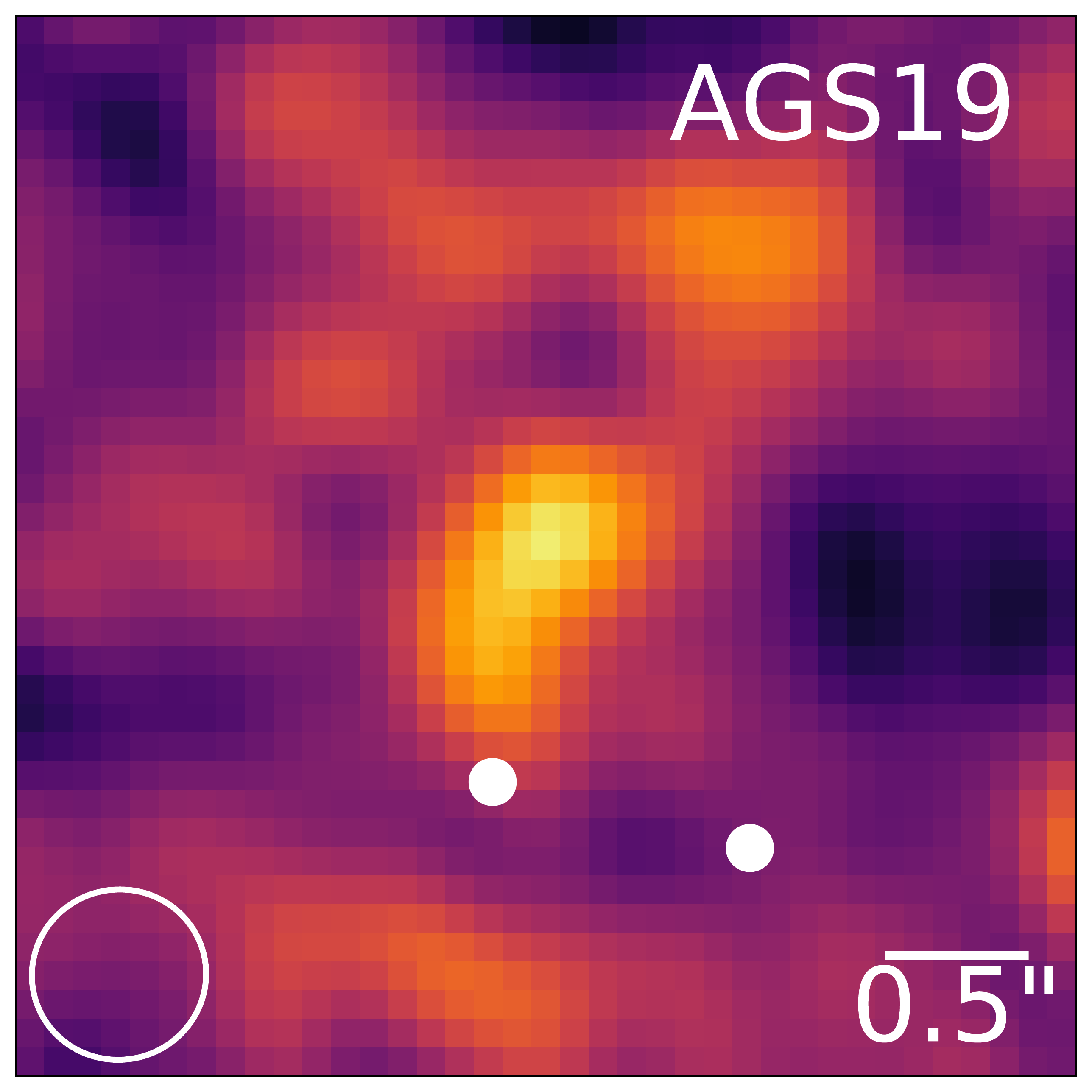}
\includegraphics[width=3cm,clip]{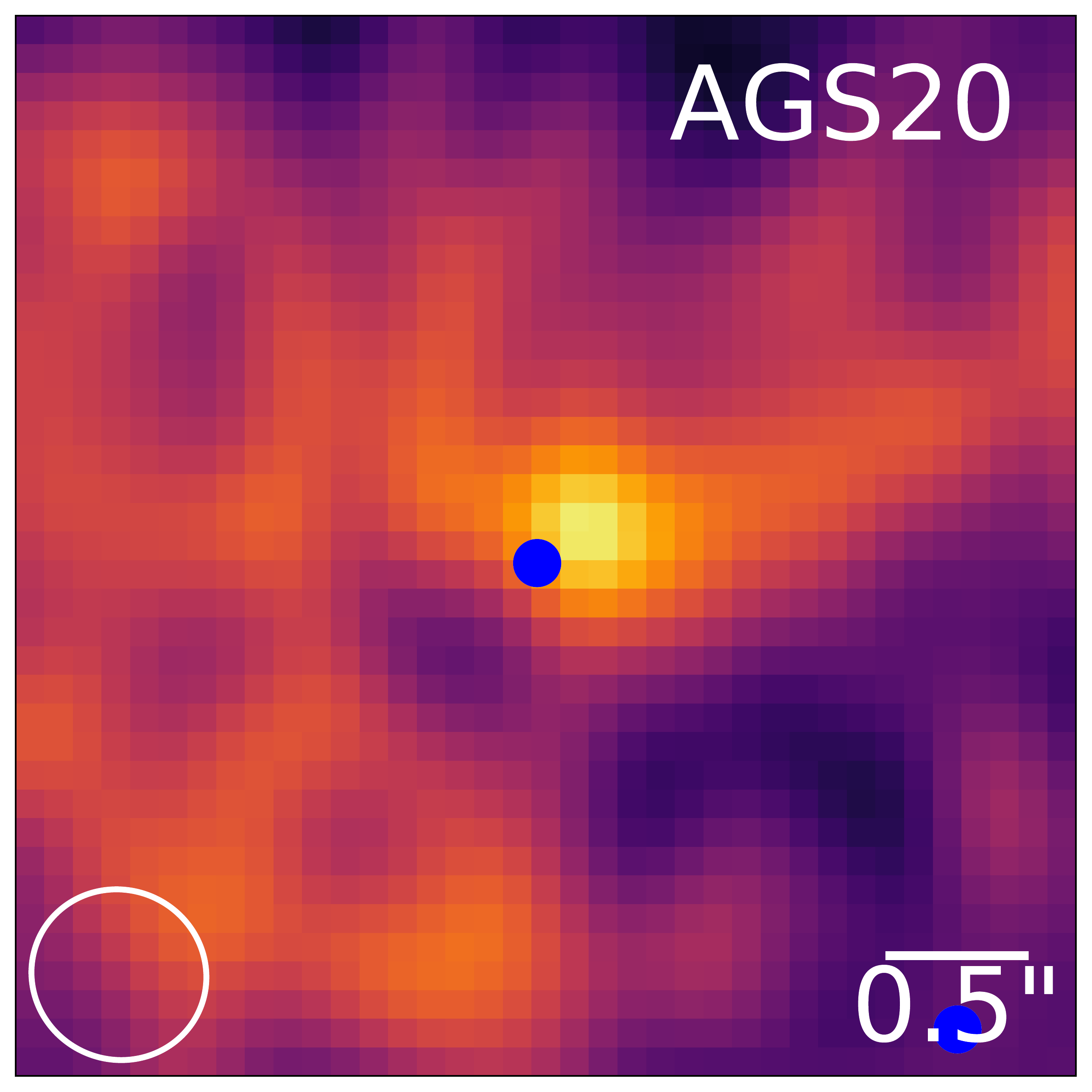}
\includegraphics[width=3cm,clip]{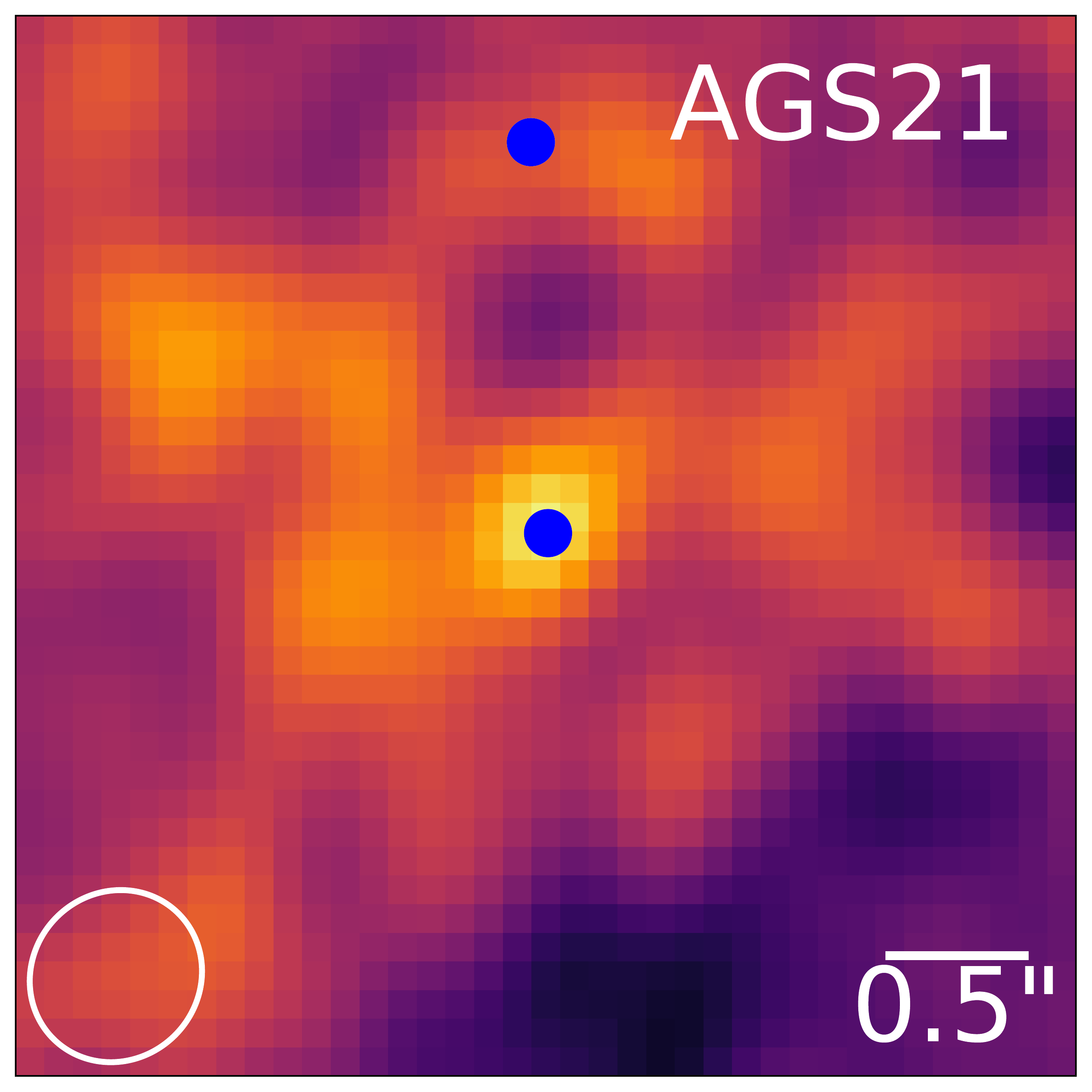}
\includegraphics[width=3cm,clip]{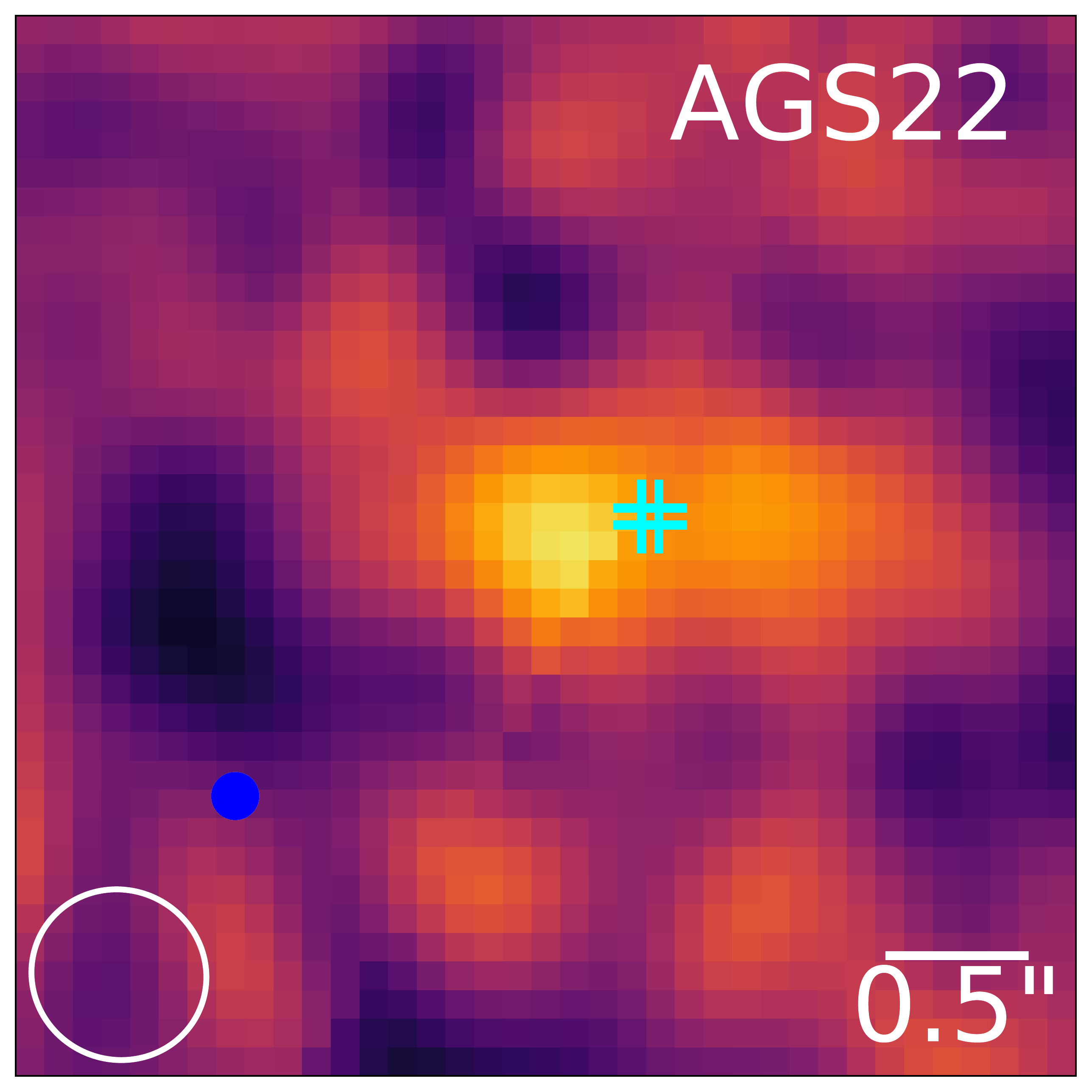}
\includegraphics[width=3cm,clip]{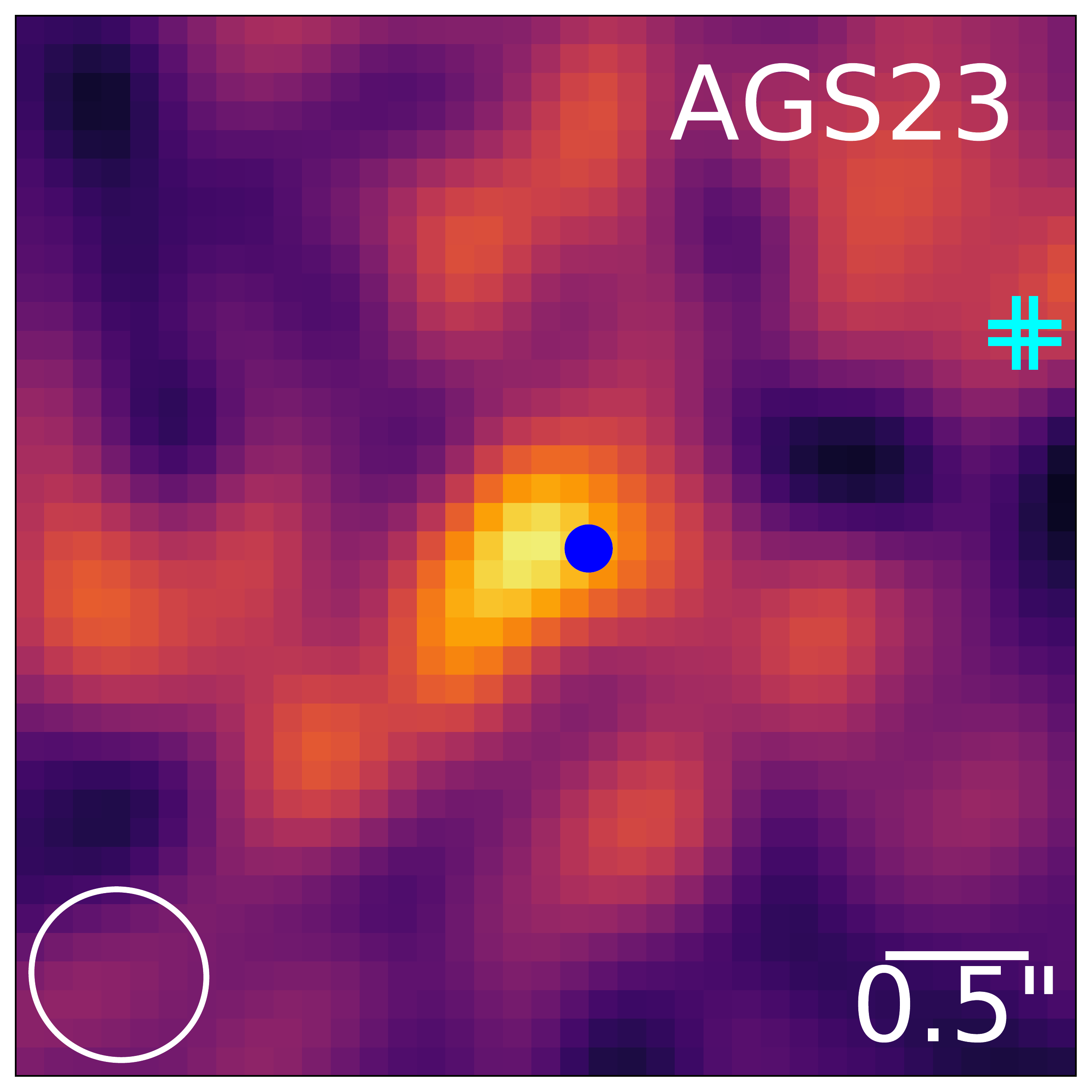}
\includegraphics[width=3cm,clip]{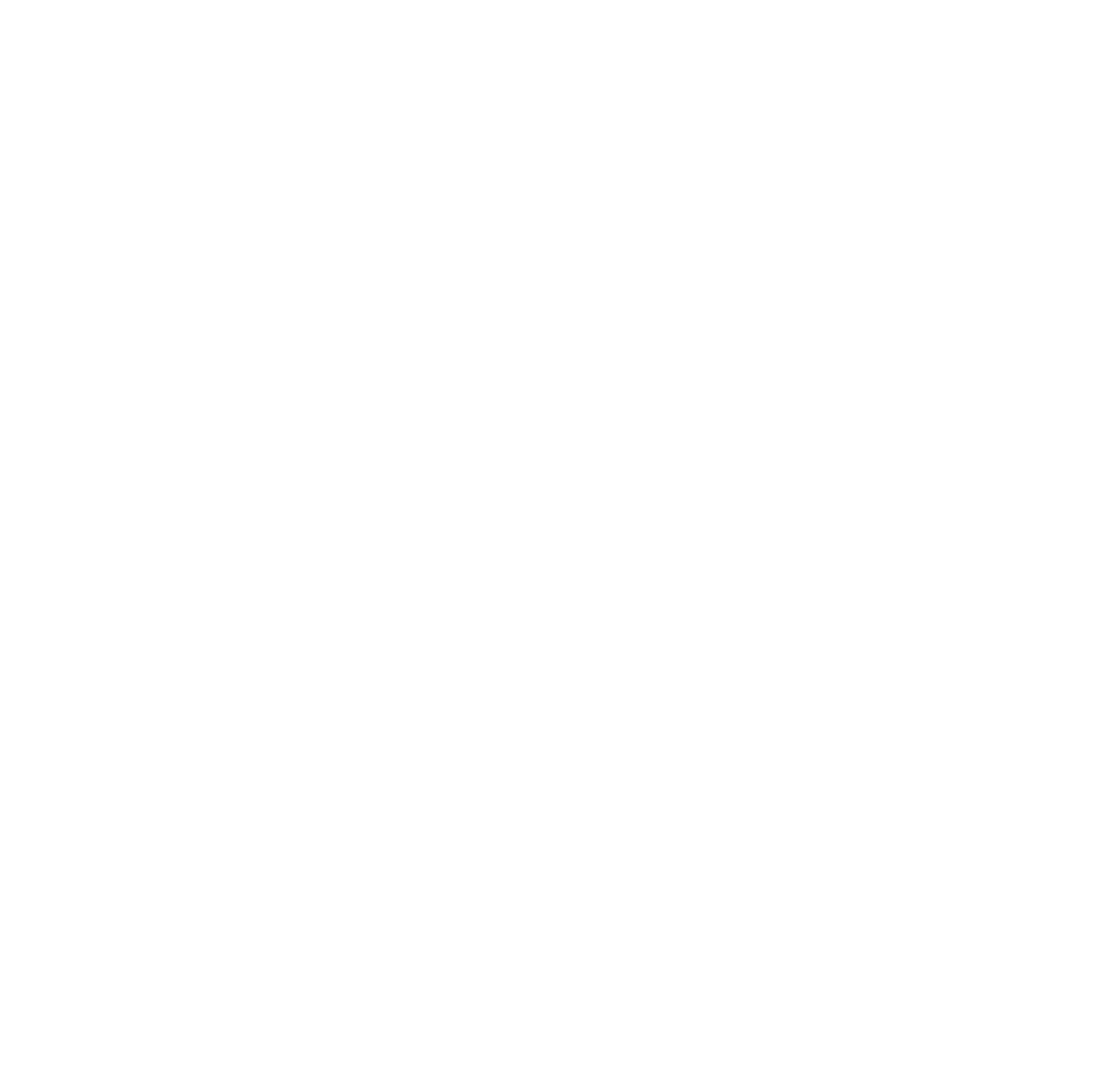}
}
\end{minipage}
\caption{ALMA 1.1 mm continuum maps for the 23 detections tapered at 0.60 arcsec. Each 3\arcsec5\,$\times$\,3\arcsec5 image is centred on the position of the ALMA detection. Cyan double crosses show sources from the GOODS--\textit{S} CANDELS catalogue. White circles show sources from the ZFOURGE catalogue. Blue circles show common sources from both optical catalogues (i.e. sources with an angular separation lower than 0\arcsec4). The shape of the synthesized beam is given in the bottom left corner.}
\label{compare_CANDELS/ZFOURGE}
\end{figure*}

Fig.~\ref{compare_CANDELS/ZFOURGE} shows 3\arcsec5\,$\times$\,3.\arcsec5 postage stamps of the ALMA-detected galaxies, overlaid with the positions of galaxies from the CANDELS/GOODS--\textit{South} catalogue (magenta double crosses), ZFOURGE catalogue (white circles) or both catalogues (i.e. sources with an angular separation lower than 0\arcsec4, blue circles). These are all shown after astrometric correction. Based on the ZFOURGE catalogue, we find optical counterparts for one galaxy that did not have an HST counterpart: AGS11, a photometric redshift has been computed in the ZFOURGE catalogue for this galaxy.

The redshifts of AGS4 and AGS17 as given in the CANDELS catalogue are unexpectedly low ($z$\,=\,0.24 and $z$\,=\,0.03, respectively), but the redshifts for these galaxies given in the ZFOURGE catalogue ($z$\,=\,3.76 and $z$\,=\,1.85, respectively) are more compatible with the expected redshifts for galaxies detected with ALMA. These galaxies, missed by the HST or incorrectly listed as local galaxies are particularly interesting galaxies (see Sect.~\ref{sec:HST-dark}). AGS6 is not listed in the ZFOURGE catalogue, most likely because it is close (< 0\arcsec7) to another bright galaxy (ID$_{CANDELS}$\,=\,15768). These galaxies are blended in the ZFOURGE ground-based $K_s$-band images. AGS6 has previously been detected at 1.3 mm in the HUDF, so we adopt the redshift and stellar mass found by \cite{Dunlop2017}. The consensus CANDELS zphot from \cite{Santini2015} is $z$\,=\,3.06 (95\% confidence: 2.92\,$<$\, $z$\,$<$\, 3.40), consistent with the value in \cite{Dunlop2017}.

\begin{figure}
   \centering
   \includegraphics[width=\hsize]{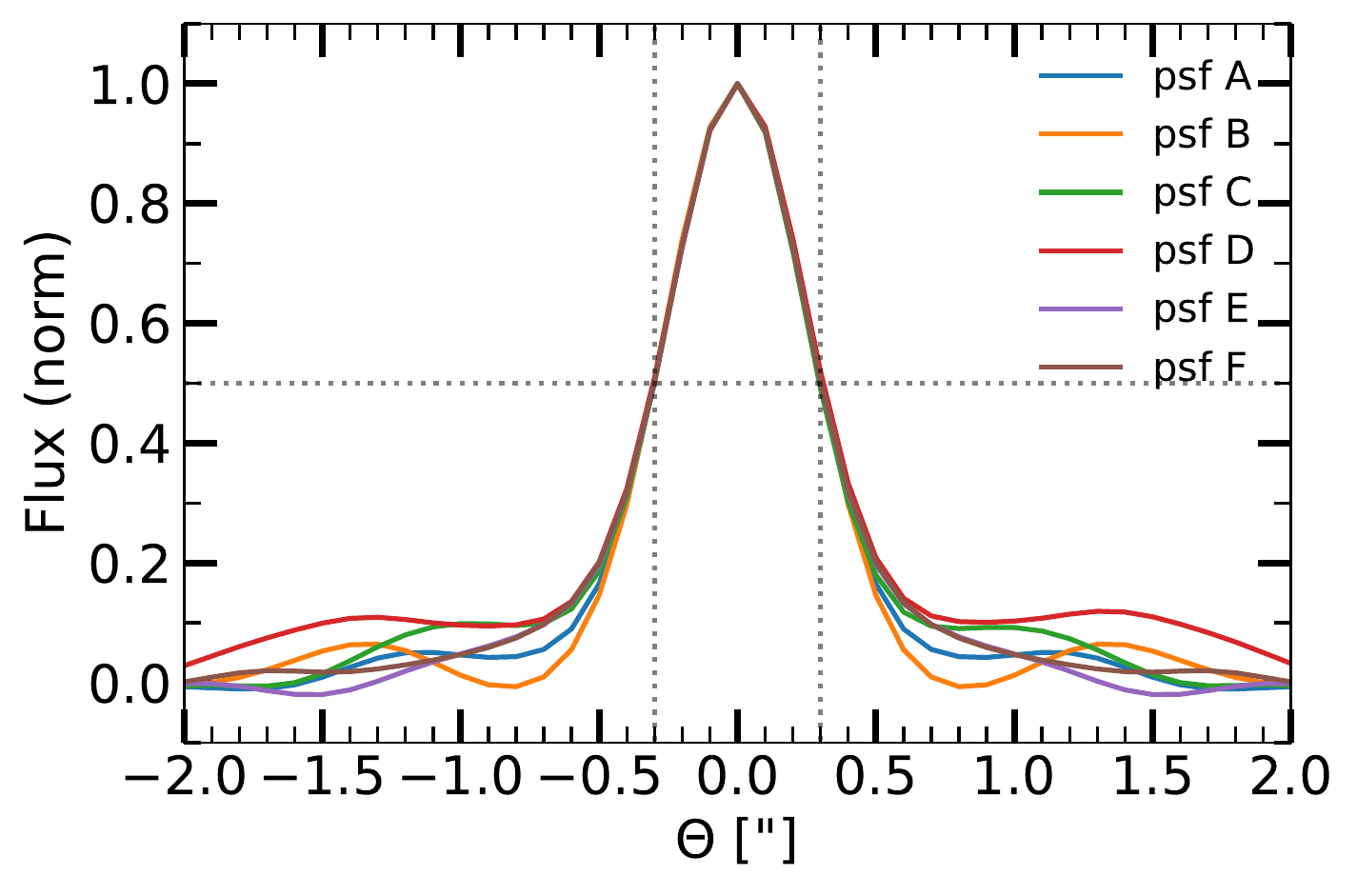}
      \caption{East-West profile of the PSFs corresponding to the 6 different parallel slices composing the ALMA image in the 0\arcsec60-mosaic (see Fig.~\ref{map_detection}).}
         \label{profile_psf}
\end{figure}

\begin{figure}
   \centering
   \includegraphics[width=\hsize]{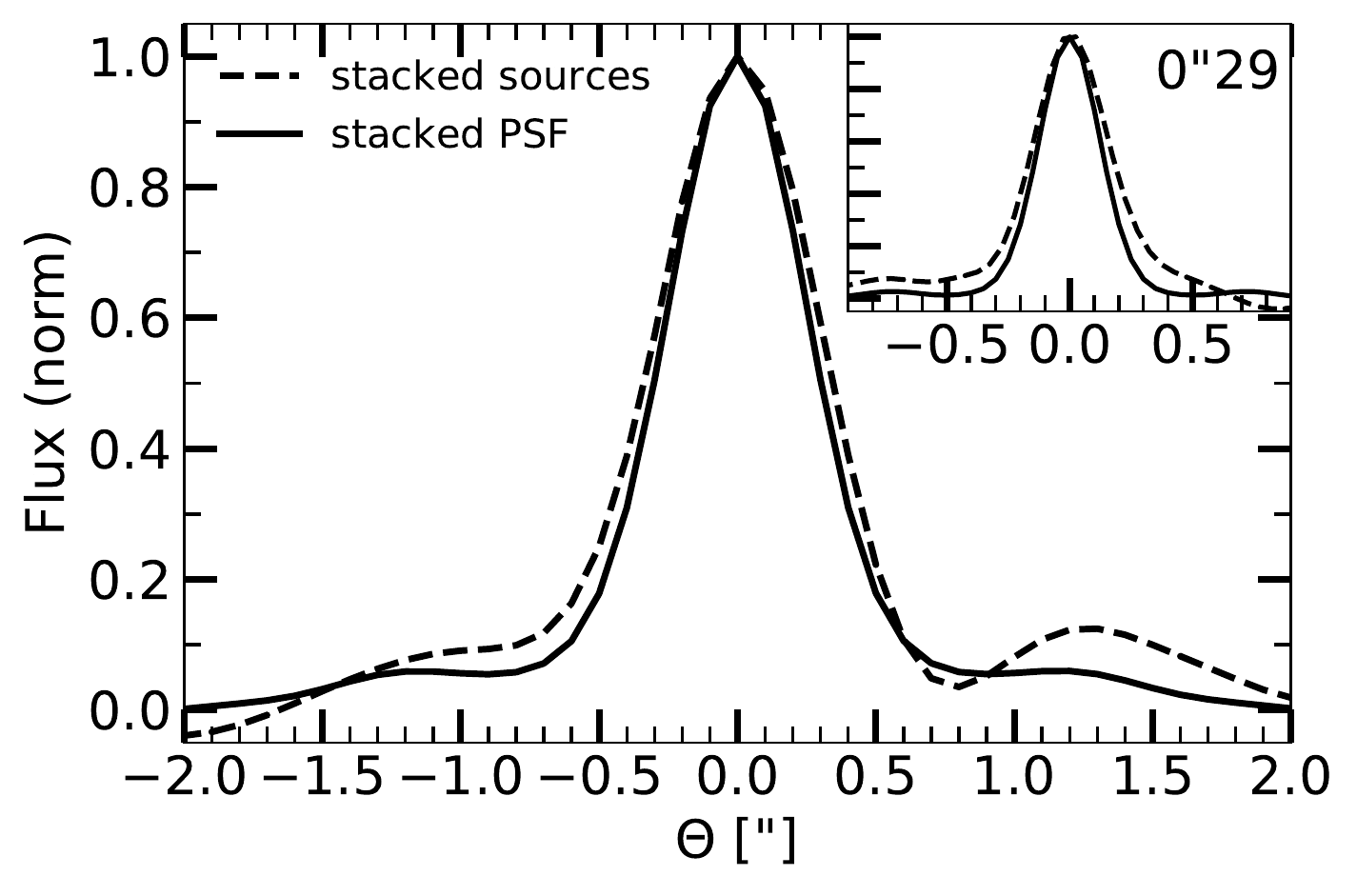}
      \caption{Comparison between the stacked PSF (black solid line) and the stack of the 23 ALMA-detections (black dashed line) in the 0\arcsec60-mosaic. As each slice has a specific PSF, we stack the PSF corresponding to the position of each detection. The fluxes of each detection have been normalized, so that the brightest sources do not skew the results. Fluxes of the PSF and ALMA detections are normalized to 1. Flux profiles are taken across the East-West direction. The result is consistent with unresolved or marginally resolved sources at this resolution. The insert in the top-right corner shows the same procedure for the 15 sources detected in the 0\arcsec29-mosaic (see Table~\ref{catalogue}).}
         \label{compare_stack_psf_sources_23_peak}
\end{figure}

\subsection{Galaxy sizes}\label{sec:galaxy_sizes}

Correctly estimating the size of a source is an essential ingredient for measuring its flux. As a first step, it is imperative to know if the detections are resolved or unresolved. In this section, we discuss our considerations regarding the sizes of our galaxies. The low number of galaxies with measured ALMA sizes in the literature makes it difficult to constrain the size distribution of dust emission in galaxies. Recent studies \citep[e.g.][]{Barro2016, Rujopakarn2016, Elbaz2017, Ikarashi2017, Fujimoto2017} with sufficient resolution to measure ALMA sizes of galaxies suggest that dust emission takes place within compact regions of the galaxy.

Two of our galaxies (AGS1 and AGS3) have been observed in individual pointings (ALMA Cycle 1; P.I. R.Leiton, presented in \citealt{Elbaz2017}) at 870\,$\mu$m with a long integration time (40-50 min on source). These deeper observations give more information on the nature of the galaxies, in particular on their morphology.. Due to their high SNR ($\sim$100) the sizes of the dust emission could be measured accurately: R$_{1/2maj}$\,=\,120$\pm$4 and 139$\pm$6 mas for AGS1 and AGS3 respectively, revealing extremely compact star forming regions corresponding to circularized effective radii of $\sim$1 kpc at redshift $z$ $\sim$2. The Sersic indices are 1.27$\pm$0.22 and 1.15$\pm$0.22 for AGS1 and AGS3 respectively: the dusty star forming regions therefore seem to be disk-like. Based on their sizes, their stellar masses ($>$ 10$^{11}$ M$_\sun$), their SFRs ($>$ 10$^{3}$ M$_\sun$yr$^{-1}$) and their redshifts ($z$ $\sim$2), these very compact galaxies are ideal candidate progenitors of compact quiescent galaxies at z$\sim$2 (\citealt{Barro2013,Williams2014,VanderWel2014,Kocevski2017}, see also \citealt{Elbaz2017}).

\begin{figure*}[h!]
\centering
\begin{minipage}[t]{1.\textwidth}
\resizebox{\hsize}{!} { 
\includegraphics[width=3cm,clip]{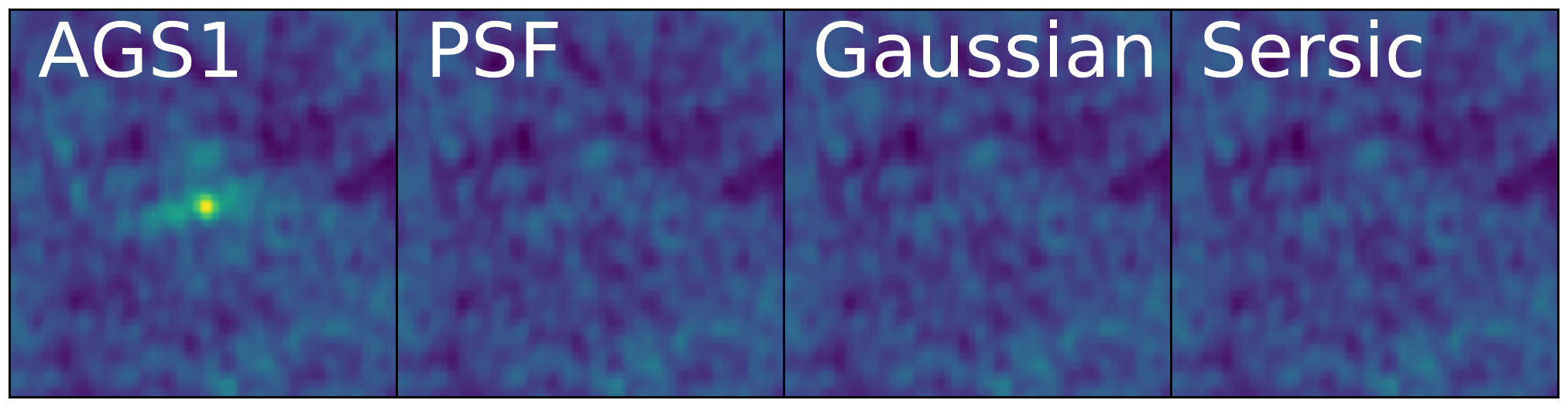} 
\includegraphics[width=3cm,clip]{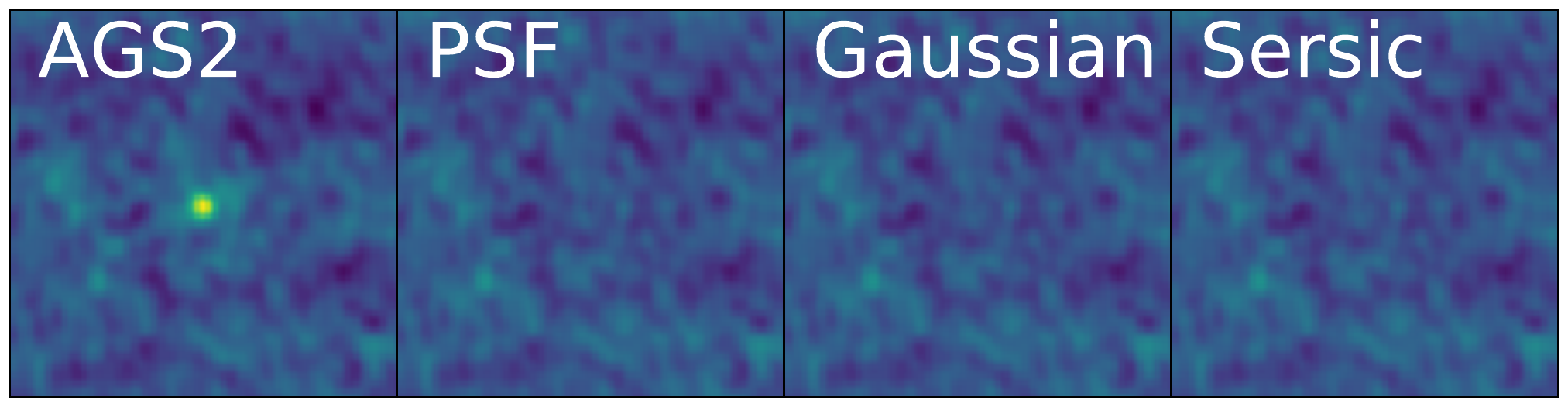} 
\includegraphics[width=3cm,clip]{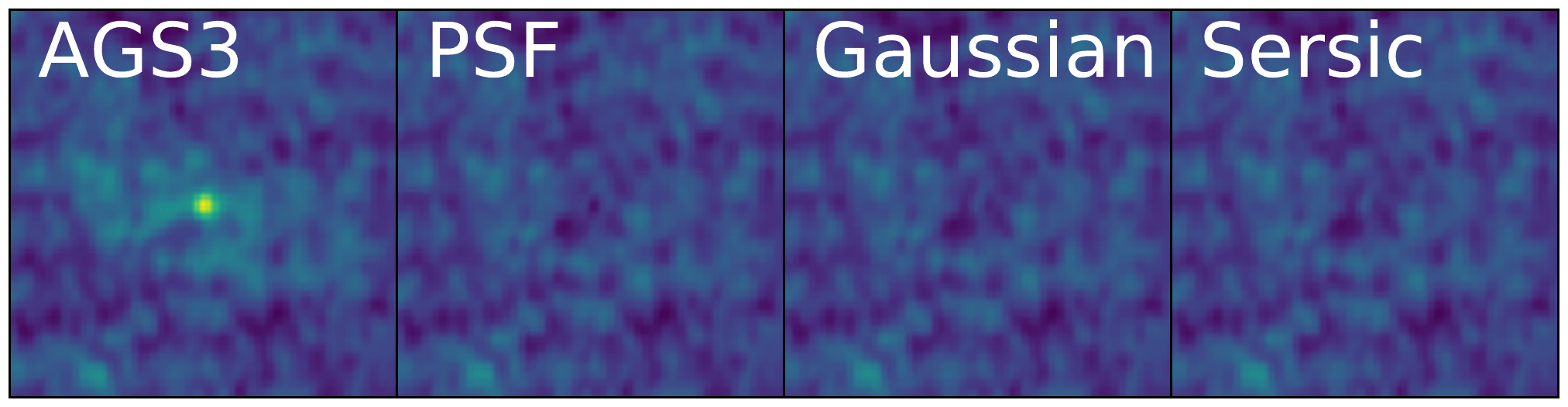}
}
\end{minipage}
\begin{minipage}[t]{1.\textwidth}
\resizebox{\hsize}{!} { 
\includegraphics[width=3cm,clip]{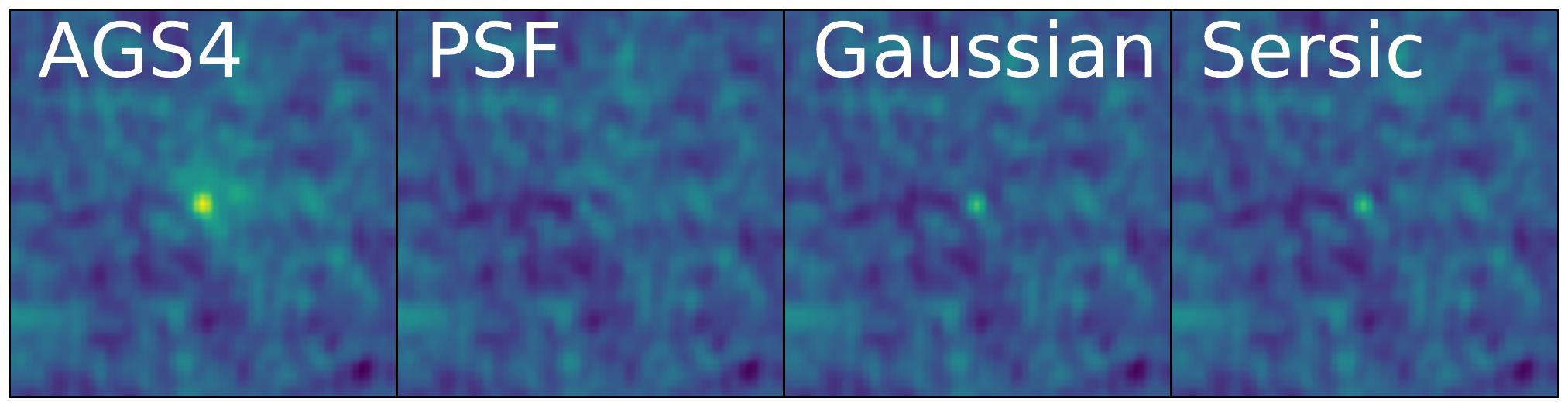} 
\includegraphics[width=3cm,clip]{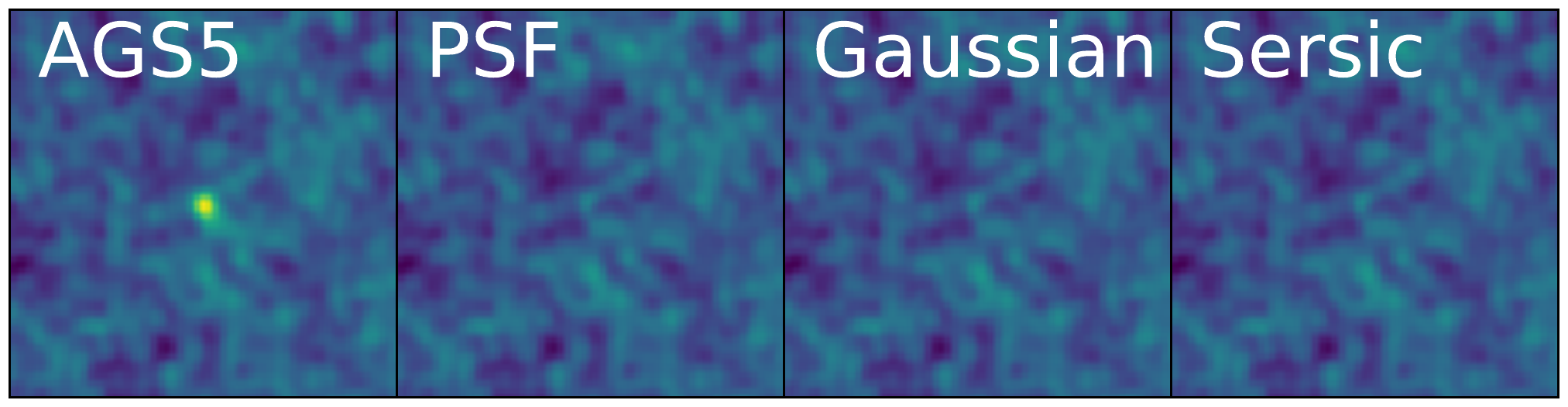} 
\includegraphics[width=3cm,clip]{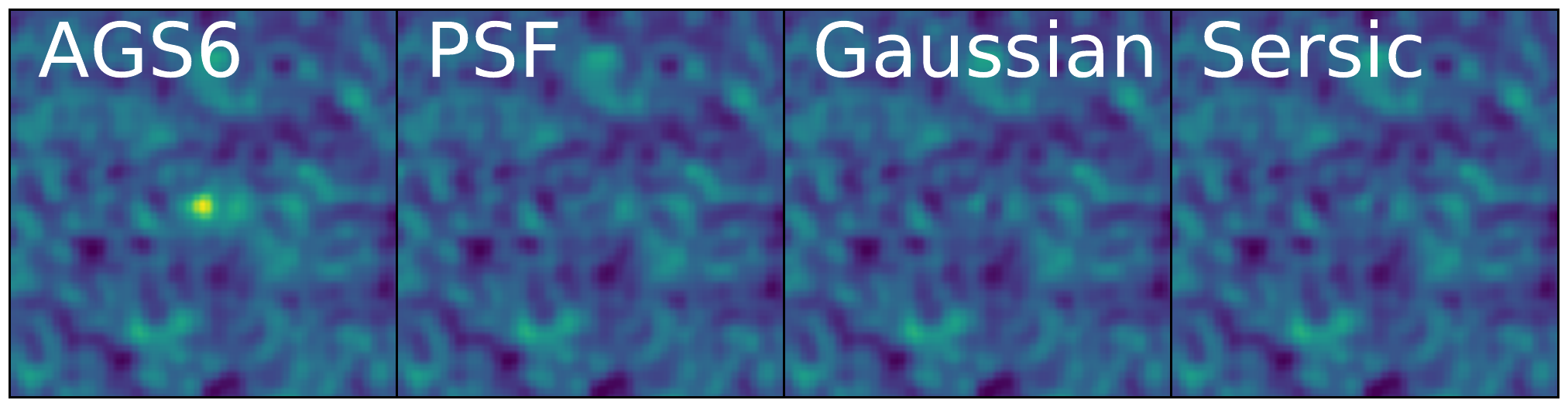}
}
\end{minipage}
      \caption{10\arcsec\,$\times$\,10\arcsec postage stamps, centred on the galaxy detections. Left to right: the source in the 0\arcsec60-mosaic map, and the residuals obtained after PSF, Gaussian and Sersic flux fitting. The residuals are very similar between the three different extraction methods. Only the 6 brightest galaxies are shown.}
         \label{residuals}
\end{figure*}
Size measurements of galaxies at (sub)millimetre wavelengths have previously been made as part of several different studies. \cite{Ikarashi2015} measured sizes for 13 AzTEC-selected SMGs. The Gaussian FWHM range between 0\arcsec10 and 0\arcsec38 with a median of $0\arcsec20^{+0\arcsec03}_{-0\arcsec05}$ at 1.1 mm. \cite{Simpson2015b} derived a median intrinsic angular size of FWHM\,=\,0\arcsec30$\pm$0\arcsec04 for their 23 detections with a SNR\,$>$\, 10 in the Ultra Deep Survey (UDS) for a resolution of 0\arcsec3 at 870\,$\mu$m. \mbox{\cite{Tadaki2017}} found a median FWHM of 0\arcsec11$\pm$0.02 for 12 sources in a 0\arcsec2-resolution survey at 870\,$\mu$m. \cite{Barro2016} use a high spatial resolution (FWHM $\sim$0\arcsec14) to measure a median Gaussian FWHM of 0\arcsec12 at 870\,$\mu$m, with an average Sersic index of 1.28. For \cite{Hodge2016}, the median major axis size of the Gaussian fit is FWHM\,=\,0\arcsec42$\pm$0\arcsec04 with a median axis ratio $b/a$\,=\,0.53$\pm$0.03 for 16 luminous ALESS SMGs, using high-resolution ($\sim$0\arcsec16) data at 870\,$\mu$m. \cite{Rujopakarn2016} found a median circular FWHM at 1.3 mm of 0\arcsec46 from the ALMA image of the HUDF \citep{Dunlop2017}. \cite{Gonzalez-Lopez2017} studied 12 galaxies at S/N $\ge$ 5,  using 3 different beam sizes (0\arcsec63\,$\times$\,0\arcsec49), (1\arcsec52\,$\times$\,0\arcsec85) and (1\arcsec22\,$\times$\,1\arcsec08). They found effective radii spanning\,$<$\,0\arcsec05 to 0\arcsec37$\pm$0\arcsec21 in the ALMA Frontier Fields survey at 1.1mm. \cite{Ikarashi2017} obtained ALMA millimetre-sizes of 0\arcsec08 -- 0\arcsec68 (FWHM) for 69 ALMA-identified AzTEC SMGs with an SNR greater than 10. These galaxies have a median size of 0\arcsec31. These studies are all broadly in agreement, revealing compact galaxy sizes in the sub(millimetre) regime of typically 0\arcsec3$\pm$0\arcsec1.

Size measurements require a high SNR detection to ensure a reliable result. The SNR range of our detections is 4.8-11.3. Following \cite{Marti-Vidal2012}, the reliable size measurement limit for an interferometer is:

\begin{equation}
\theta_{min}\,=\,\beta \left(   \frac{\lambda_c}{2 \ \text{SNR}^2}    \right)^{1/4} \times \theta_{beam}\,\simeq\,0.88\frac{\theta_{beam}}{\sqrt{\text{SNR}}}
\end{equation}

\noindent where $\lambda_c$ is the value of the log-likelihood, corresponding to the cutoff of a Gaussian distribution to have a false detection and $\beta$ is a coefficient related to the intensity profile of the source model and the density of the visibilities in Fourier space. This coefficient usually takes values in the range 0.5-1. We assume $\lambda_c$\,=\,3.84 corresponding to a 2\,$\sigma$ cut-off, and $\beta$\,=\,0.75. For $\theta_{beam}$\,=\,0\arcsec60 and a range of SNR between 4.8 and 11.3, the minimum detectable size (FWHM) therefore varies between 0\arcsec16 and 0\arcsec24. Using the 0\arcsec60-mosaic map, the sizes of a large number of detections found in previous studies could therefore not be reliably measured.

To quantitatively test if the millimetre galaxies are resolved in our survey we perform several tests.

The first test is to stack the 23 ALMA-detections and compare the obtained flux profile with the profile of the PSF. However, in the mosaic map, each slice has its own PSF. We therefore also need to stack the PSFs at these 23 positions in order to obtain a global PSF for comparison. Fig.~\ref{profile_psf} shows the different PSFs used in this survey in the 0\arcsec60-mosaic. The FWHM of each PSF is identical, the differences are only in the wings. The stack of the 23 PSFs for the 23 detections and the result of the source stacking in the 0\arcsec60-mosaic is shown in Fig.~\ref{compare_stack_psf_sources_23_peak}. The flux of each detection is normalized so that all sources have the same weight, and the stacking is not skewed by the brightest sources. 

Size stacking to measure the structural parameters of galaxies is at present a relatively unexplored area. This measurement could suffer from several sources of bias. The uncertainties on the individual ALMA peak positions could increase the measured size in the stacked image, for example. On the other hand, due to the different inclination of each galaxy, the stacked galaxy could appear more compact than the individual galaxies \mbox{\citep[eg.][]{Hao2006,Padilla2008,Li2016}}. Alternatively, some studies \citep[eg.][]{VanDokkum2010} indicate that size stacking gives reasonably accurate mean galaxy radii. In our case, the result of the size stacking is consistent with unresolved sources or marginally resolved at this resolution which corresponds to a physical diameter of 4.6 kpc at $z$\,=\,3.

   \begin{figure}
   \centering
   \includegraphics[width=1.\hsize]{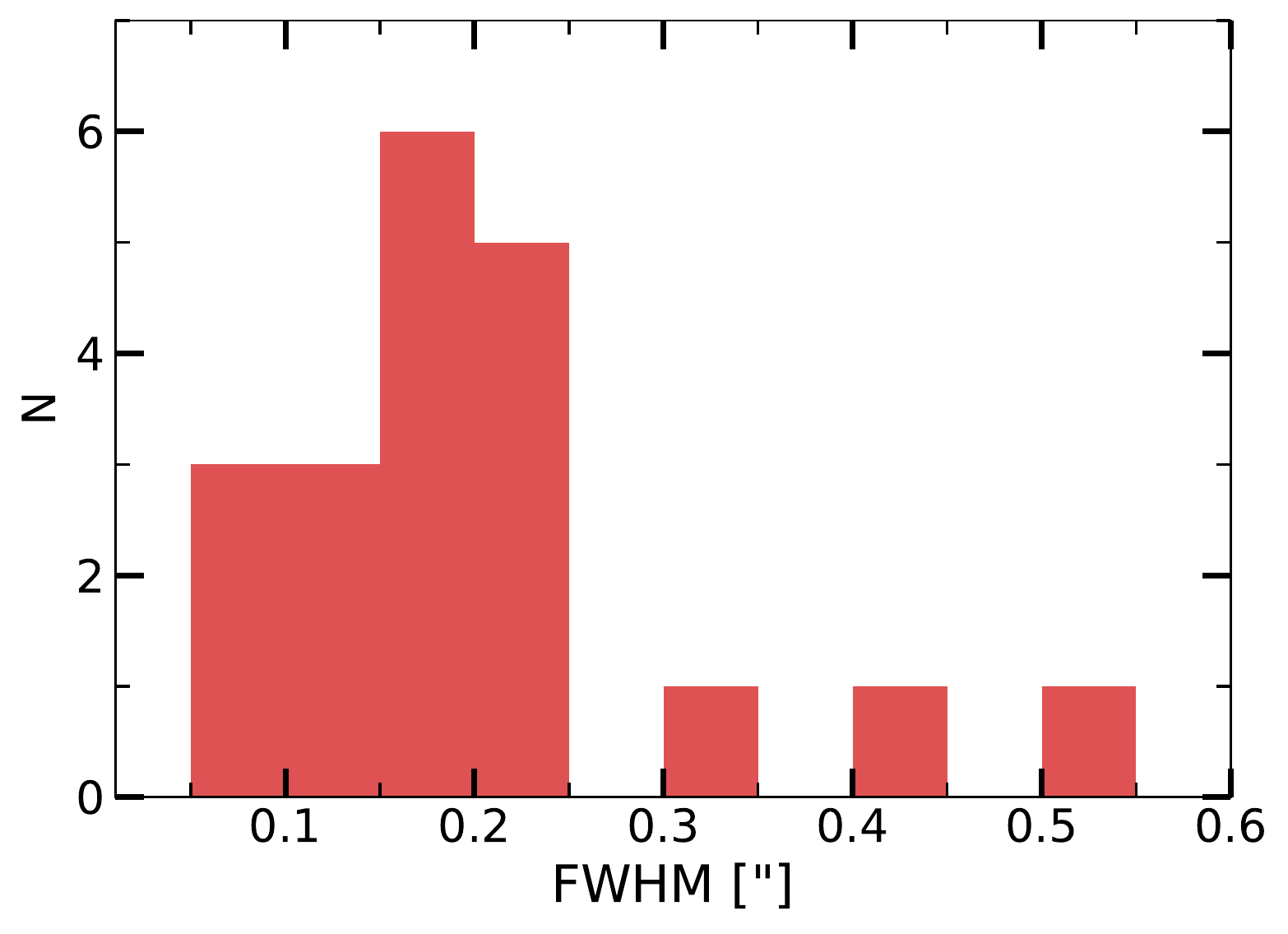}
      \caption{Size distribution histogram for the 20 robust detections. These sizes are computed by fitting the ALMA detections with a circular Gaussian in the \textit{uv}-plane using \texttt{uvmodelfit} in CASA. 85\% of the sources exhibit a FWHM smaller than 0\arcsec25.}
         \label{size_distribution}
   \end{figure}

The second test is to extract the flux for each galaxy using PSF-fitting. We use \texttt{Galfit} \citep{Peng2010} on the 0\arcsec60-mosaic. The residuals of this PSF-extraction are shown for the 6 brightest galaxies in Fig.~\ref{residuals}. The residuals of 21/23 detections do not have a peak greater than 3\,$\sigma$ in a radius of 1\arcsec around the source. Only sources AGS10 and AGS21 present a maximum in the residual map at $\sim$3.1\,$\sigma$.

We compare the PSF flux extraction method with Gaussian and Sersic shapes. As our sources are not detected with a particularly high SNR, and in order to limit the number of degrees of freedom, the Sersic index was frozen to n\,=\,1 (exponential disk profile, in good agreement with \citealt{Hodge2016} and \citealt{Elbaz2017} for example), assuming that the dust emission is disk-like. Fig.~\ref{residuals} shows the residuals for the 3 different extraction profiles. The residuals are very similar between the point source, Gaussian and Sersic profiles, suggesting that the approximation that the sources are not resolved is appropriate, and does not result in significant flux loss. We also note that, for several galaxies, due to large size uncertainties, the Gaussian and Sersic fits give worse residuals than the PSF fit (AGS4 for example).

For the third test, we take advantage of the different tapered maps. We compare the peak flux for each detection between the 0\arcsec60-mosaic map and the 0\arcsec29-mosaic map. The median ratio is $S_{peak}^{0\arcsec29}/S_{peak}^{0\arcsec60}$\,=\,0.87$\pm$0.16. This small decrease, of only 10\% in the peak flux density between the two tapered maps suggests that the flux of the galaxies is only slightly more resolved in the 0\arcsec29-mosaic map.

In order to test the impact of our hypothesis that the sources can be considered as point-like in the mosaic tapered at 0\arcsec60, we fitted their light profiles with a circular Gaussian in the \textit{uv}-plane using \texttt{uvmodelfit} in CASA (we also tested the use of an asymmetric Gaussian but the results remained similar although with a lower precision due to the larger number of free parameters in the fit). The sizes that we obtained confirm our hypothesis that our galaxies are particularly compact since 85\% of the sources (17 out of 20 robust detections) exhibit a FWHM smaller than 0\arcsec25 (i.e. the half-light radius is twice smaller than this value). The median size of our sample of 20 galaxies is 0\arcsec18 (see the distribution of sizes in Fig.~\ref{size_distribution}). This analysis shows that two sources are outliers with sizes of 0\arcsec41$\pm$0\arcsec03 and 0\arcsec50$\pm$0\arcsec08, for AGS17 and AGS18 respectively. For these two sources, the assumption of point-like sources is not valid and leads to an underestimate of the actual flux densities by a factor of 2.3 and 1.7 respectively. This correction has been applied to the list of peak flux densities provided in Table~\ref{catalogue}.

Having performed these tests, we conclude that for all of the detections, except AGS17 and AGS18, the approximation that these sources appear point-like in the 0\arcsec60-mosaic map is justified. For the two remaining sources, we apply a correction given above. Our photometry is therefore performed under this assumption.

\section{Number counts}\label{sec:Number_counts}

\subsection{Completeness}\label{sec:completeness}

We assess the accuracy of our catalogue by performing completeness tests. The completeness is the probability for a source to be detected in the map given factors such as the depth of the observations. We computed the completeness of our observations using Monte Carlo simulations performed on the 0\arcsec60-mosaic map. We injected 50 artificial sources in each slice. Each source was convolved with the PSF and randomly injected on the dirty map tapered at 0\arcsec60. In total, for each simulation run, 300 sources with the same flux were injected into the total map. In view of the size of the map, the number of independent beams and the few number of sources detected in our survey, we can consider, to first order, that our dirty map can be used as a blank map containing only noise, and that the probability to inject a source exactly at the same place as a detected galaxy is negligible. The probability that at least two point sources, randomly injected, are located within the same beam ($p_b$) is:
\begin{equation}
p_b\,=\,1- \prod_{k\,=\,0}^{k\,=\,n - 1} \frac{N_b - k}{N_b}
\end{equation}
\noindent where $N_b$ is the number of beams and $n$ is the number of injected sources. For each one of the six slices of the survey, we count $\sim$100\,000 independent beams. The probability of having source blending for 50 simulated sources in one map is $\sim$1\%.

We then count the number of injected sources detected with $\sigma_p$\,=\,4.8\,$\sigma$ and  $\sigma_f$\,=\,2.7\,$\sigma$, corresponding to the thresholds of our main catalogue. We inject 300 artificial sources of a given flux, and repeat this procedure 100 times for each flux density. Our simulations cover the range S$_\nu$\,=\,0.5-2.4\,mJy in steps of 0.1\,mJy. Considering the resolution of the survey, it would be reasonable to expect that a non-negligible number of galaxies are not seen as point sources but extended sources (see Sect.~\ref{sec:galaxy_sizes}). We simulate different sizes of galaxies with Gaussian FWHM between 0\arcsec2 and 0\arcsec9 in steps of 0\arcsec1, as well as point-source galaxies, to better understand the importance of the galaxy size in the detectability process. We match the recovered source with the input position within a radius of 0\arcsec6.

   \begin{figure}
   \centering
   \includegraphics[width=\hsize]{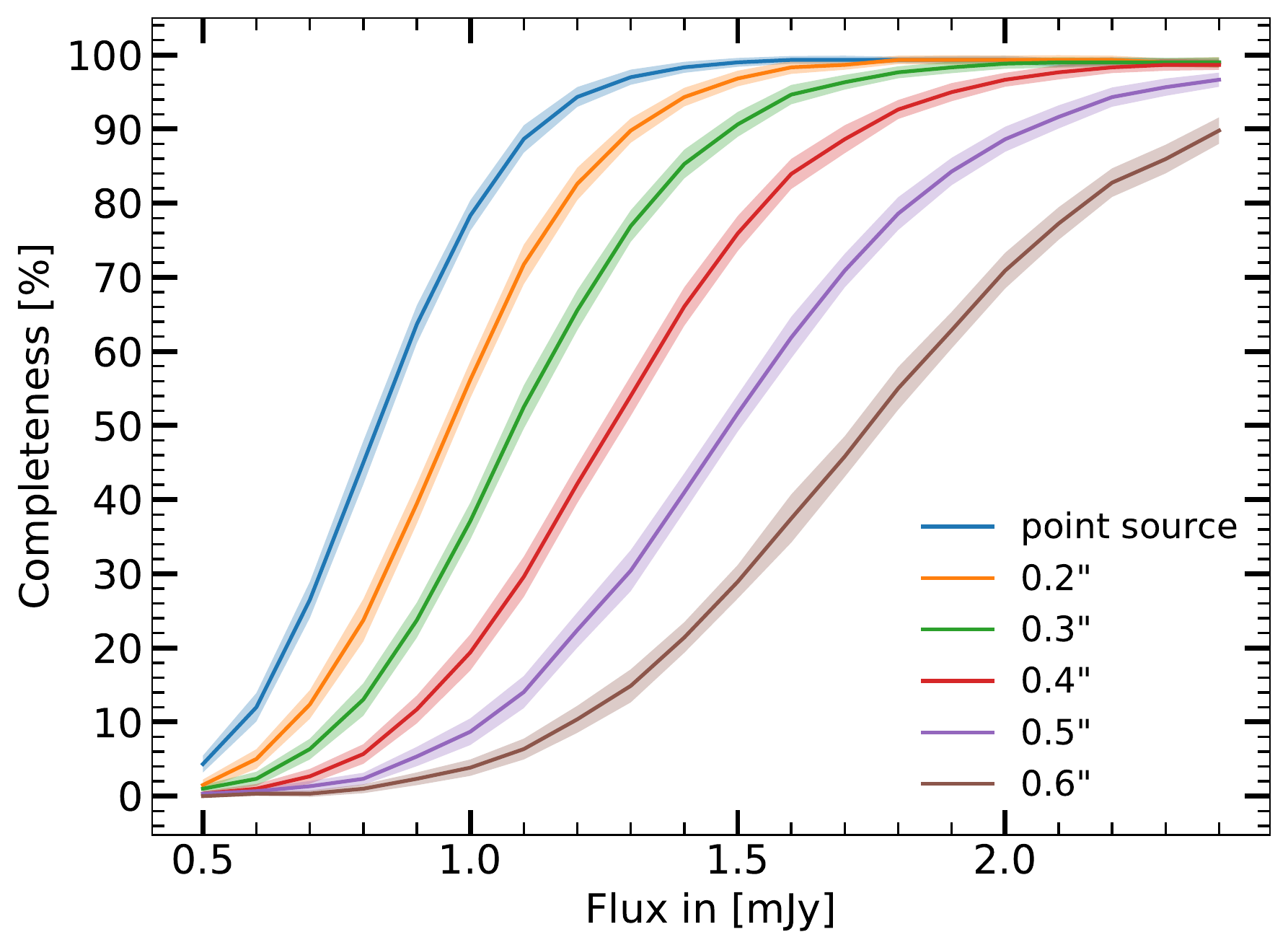}
      \caption{Median source detection completeness for simulated point-like and Gaussian galaxies as a function of integrated flux, for different FWHMs (see figure's legend). The shaded regions correspond to the standard deviation of 100 runs, each containing 300 simulated sources.}
         \label{Completeness}
   \end{figure}

Fig.~\ref{Completeness} shows the resulting completeness as a function of input flux, for different FWHM Gaussian sizes convolved by the PSF and injected into the map.

As a result of our simulations, we determine that at 1.2\,mJy, our sample is 94$\pm$1\% complete for point sources. This percentage drastically decreases for larger galaxy sizes. For the same flux density, the median detection rate drops to 61$\pm$3\% for a galaxy with FWHM $\sim$0\arcsec3, and to 9$\pm$1\% for a FWHM $\sim$0\arcsec6 galaxy. This means, that for a galaxy with an intrinsic flux density of 1.2\,mJy, we are more than ten times more likely to detect a point source galaxy than a galaxy with FWHM $\sim$0\arcsec6. 

The size of the millimetre emission area plays an essential role in the flux measurement and completeness evaluation. We took the hypothesis that ALMA sizes are 1.4 times smaller than the size measured in HST $H$-band (as derived by \citealt{Fujimoto2017} using 1034 ALMA galaxies). We are aware that this size ratio is poorly constrained at the present time, but such relation has been observed in several studies (see Sect.~\ref{sec:galaxy_sizes}). For example, of the 12 galaxies presented by \cite{Laporte2017}, with fluxes measured using ALMA at 1.1mm \citep{Gonzalez-Lopez2017}, 7 of them have a size measured by HST F140W/WFC3 similar to the size measured in the ALMA map. On the other hand, for the remaining 5 galaxies, their sizes are approximately two times more compact at millimetre wavelengths than at optical wavelengths. This illustrates the dispersion of this ratio.

\subsection{Effective area}

\begin{figure}
\centering
\resizebox{\hsize}{!} {
\includegraphics[width=5cm,clip]{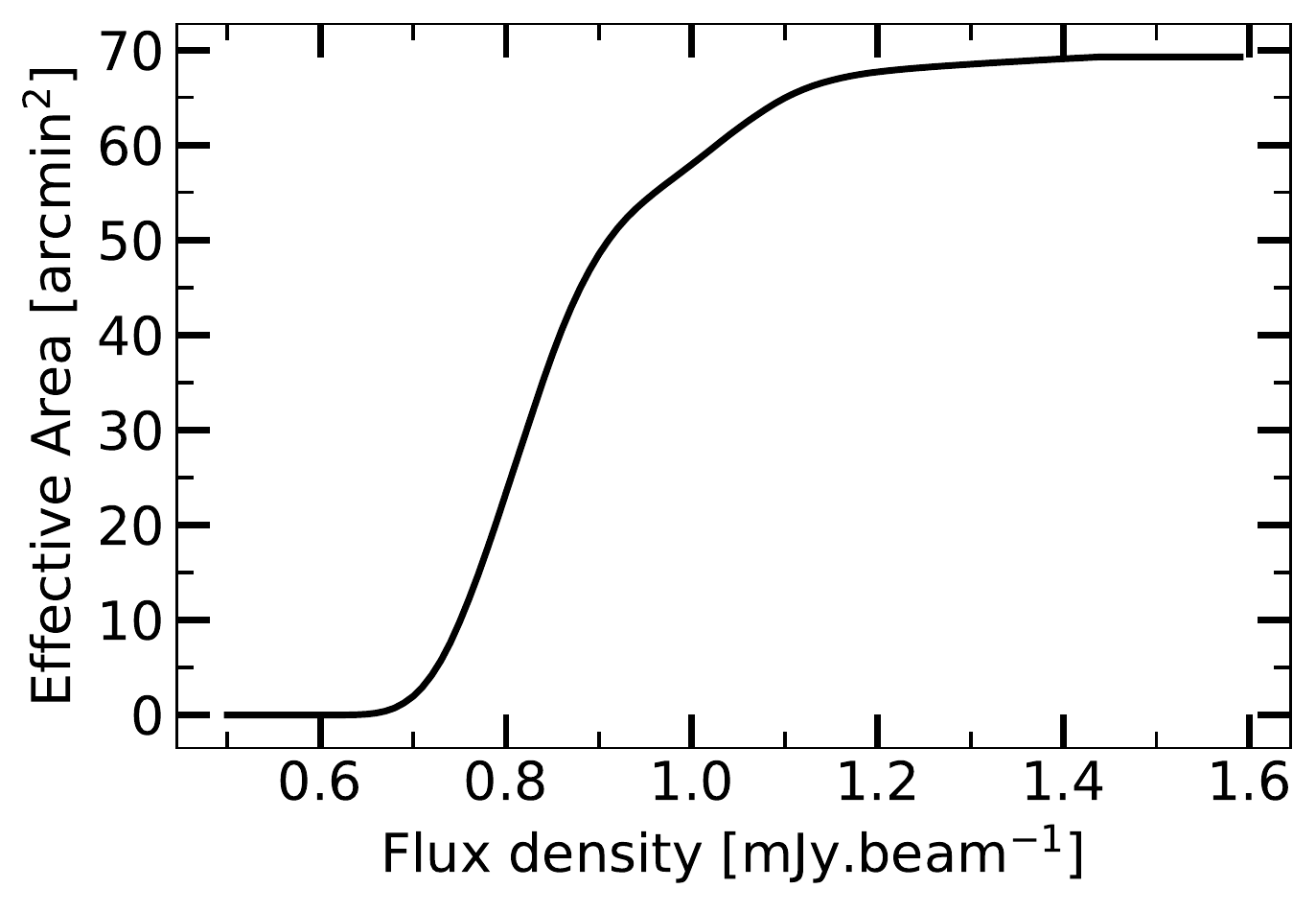}
}
\caption{Effective area as a function of flux density, where a source with a given flux can be detected with an SNR\,$>$\, 4.8\,$\sigma$. Ninety percent of the survey area reaches a sensitivity of at least 1.06\,mJy.beam$^{-1}$.}
   \label{Effective_area}
\end{figure}

As the sensitivity of our 1.1mm ALMA map is not uniform, we define an effective area where a source with a given flux can be detected with an SNR\,$>$\, 4.8\,$\sigma$, as shown in Fig.~\ref{Effective_area}. Our map is composed of 6 different slices - one of them, slice B, presents a noise 30\% greater than the mean of the other 5, whose noise levels are comparable. The total survey area is 69.46 arcmin$^2$, with 90\% of the survey area reaching a sensitivity of at least 1.06\,mJy.beam$^{-1}$. We consider the relevant effective area for each flux density in order to compute the number counts. We consider the total effective area over all slices in the number counts computation.

\subsection{Flux Boosting and Eddington bias}

In this section, we evaluate the effect of flux boosting. Galaxies detected with a relatively low SNR tend to be boosted by noise fluctuations (see \citealt{Hogg1998,Coppin2005, Scott2002}). To estimate the effect of flux boosting, we use the same set of simulations that we used for completeness estimations. 

\begin{figure}
 \centering
   \includegraphics[width=\hsize]{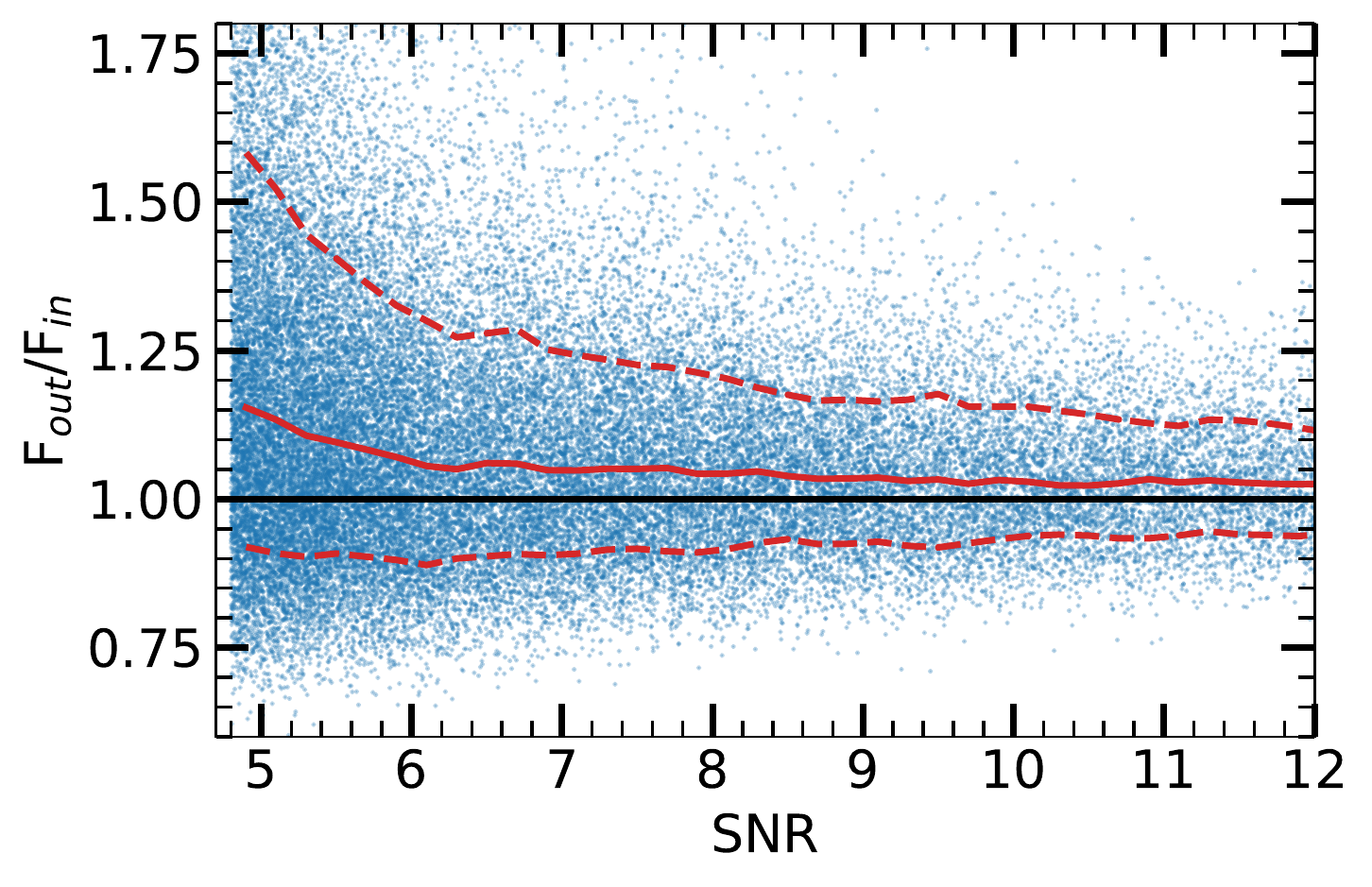}
      \caption{Flux boosting as a function of measured SNR estimated from simulations. The median of the boosting is shown by a solid red line. The 1\,$\sigma$ confidence intervals (dashed red lines) are overplotted. The solid black horizontal line corresponds to F$_\text{out}$\,=\,F$_\text{in}$ (see text for details). We use the same set of simulations that we used for the completeness analysis.}
         \label{Flux_boosting}
   \end{figure}

The results of our simulations are shown in Fig.~\ref{Flux_boosting}. The boosting effect is shown as the ratio between the input and output flux densities as a function of the measured SNR. For point sources, we observe the well-known flux boosting effect for the lowest SNRs. This effect is not negligible for the faintest sources in our survey. At 4.8\,$\sigma$, the flux boosting is $\sim$15\%, and drops below 10\% for an SNR greater than 5.2. We estimate the de-boosted flux by dividing the measured flux by the median value of the boosting effect as a function of SNR (red line in Fig.~\ref{Flux_boosting}).

We also correct for the effects of the Eddington bias \citep{Eddington1913}. As sources with lower luminosities are more numerous than bright sources, Gaussian distributed noise gives rise to an overestimation of the number counts in the lowest flux bins. We simulate a realistic number of sources (the slope of the number counts is computed using the coefficients given in Table~\ref{best_fit_powerlaw_shechter}) and add Gaussian noise to each simulated source. The correction factor for each flux bin is therefore the ratio between the flux distribution before and after adding the noise.

\subsection{Cumulative and differential number counts}\label{sec:Cumulative_number_counts}

We use sources with an SNR greater than 4.8 from the main catalogue to create cumulative and differential number counts. We need to take into account the contamination by spurious sources, completeness effects, and flux boosting in order to compute these number counts. 

The contribution of a source with flux density $S_i \pm dS_i$ to the cumulative number count is given by:

\begin{equation}
\label{eq_cumulative_number_counts}
\frac{dN(S_i)}{dS_i}\,=\,\frac{p_c(S_i)}{A_{eff}(S_i)C(S_i,R_{\text{ALMA}}^{\text{circ}})}\times\frac{dN_{obs}(S_i)}{dS_i}
\end{equation}
\noindent where $p_c(S_i)$ is the purity criterion as defined in Eq.~\ref{quality_criteria} at the flux density $S_i$, $A_{eff}(S_i)$ and $C(S_i,R_{\text{ALMA}}^{\text{circ}})$ are the effective area and the completeness for the flux interval $dS_i$, as shown in Fig.~\ref{Completeness} and Fig.~\ref{Effective_area}.
The completeness is strongly correlated with the sizes of the galaxies. To estimate the completeness, galaxies that do not have measured sizes in the $H$-band \citep{van_der_Wel2012} are considered as point sources otherwise we use $R_{\text{ALMA}}^{\text{circ}}$\,=\,$R_{\text{H}}^{\text{circ}} / 1.4$ (see Sect.~\ref{sec:completeness}).

The cumulative number counts are given by the sum over all of the galaxies with a flux density higher than $S$:
\begin{equation}
N(>S)\,=\,\sum^{S_i>S} \frac{p_c(S_i)}{A_{eff}(S_i)C(S_i,R_{\text{ALMA}}^{\text{circ}})}\times\frac{dN_{obs}(S_i)}{dS_i}\times dS_i
\end{equation}
Errors are computed by Monte-Carlo simulations, added in quadrature to the Poisson uncertainties. The derived number counts are provided in Tab.~\ref{Table_count}. AGS19 is located at a position where the noise is artificially low, and has therefore not been taken into account

\begin{table}
\centering          
\begin{tabular}{c c c | c c c}  
\hline
$S_\nu$ &N($>S_\nu$) & N$_{\text{cum}}$ & $S_\nu$& $dN/dS_\nu$  & N$_{\text{diff}}$ \\
mJy & deg$^{-2}$ & &mJy&\,mJy$^{-1}$deg$^{-2}$ & \\
(1)&(2)&(3)&(4)&(5)&(6)\\
\hline    
\hline    
 0.70 & $2772_{-2641}^{+1776}$	&	19 	&   0.80 	& $8257_{-8023}^{+26121}$ 	&    7\\
 0.88 & $950_{-775}^{+575}$ 		&  	13 	&   1.27 	& $1028_{- 638}^{+6547} $	&    6\\
 1.11 & $524_{-188}^{+ 530}$ 		&  	11 	&   2.01 	&  $327_{- 160}^{+ 148}$ 		&    6\\
 1.40 &$327_{-124}^{+277}$		&	7	& 		& 						& 	\\
 1.76 &$209_{-119}^{+178}$		&	4	& 		& 						& 	\\
\hline 
   \end{tabular}
\caption{Number counts at 1.1mm derived from\,$>$\, 4.8\,$\sigma$ detections (main catalogue). Columns: (1) Flux Density; (2) Cumulative number counts; (3) Number of entries per bin for cumulative number counts; (4) Centre of the flux density bin; (5) Differential number counts; (6) Number of entries per bin for differential number counts. Flux density bins, $\Delta$log$S_\nu$\,=\,0.20 dex wide for differential number counts. The uncertainties are computed by Monte-Carlo simulations, added in quadrature to the Poisson uncertainties.}
\label{Table_count}    
\end{table}

\begin{figure*}
\centering
\begin{minipage}[t]{1.0\textwidth}
\resizebox{\hsize}{!} {
\includegraphics[width=5.2cm,clip]{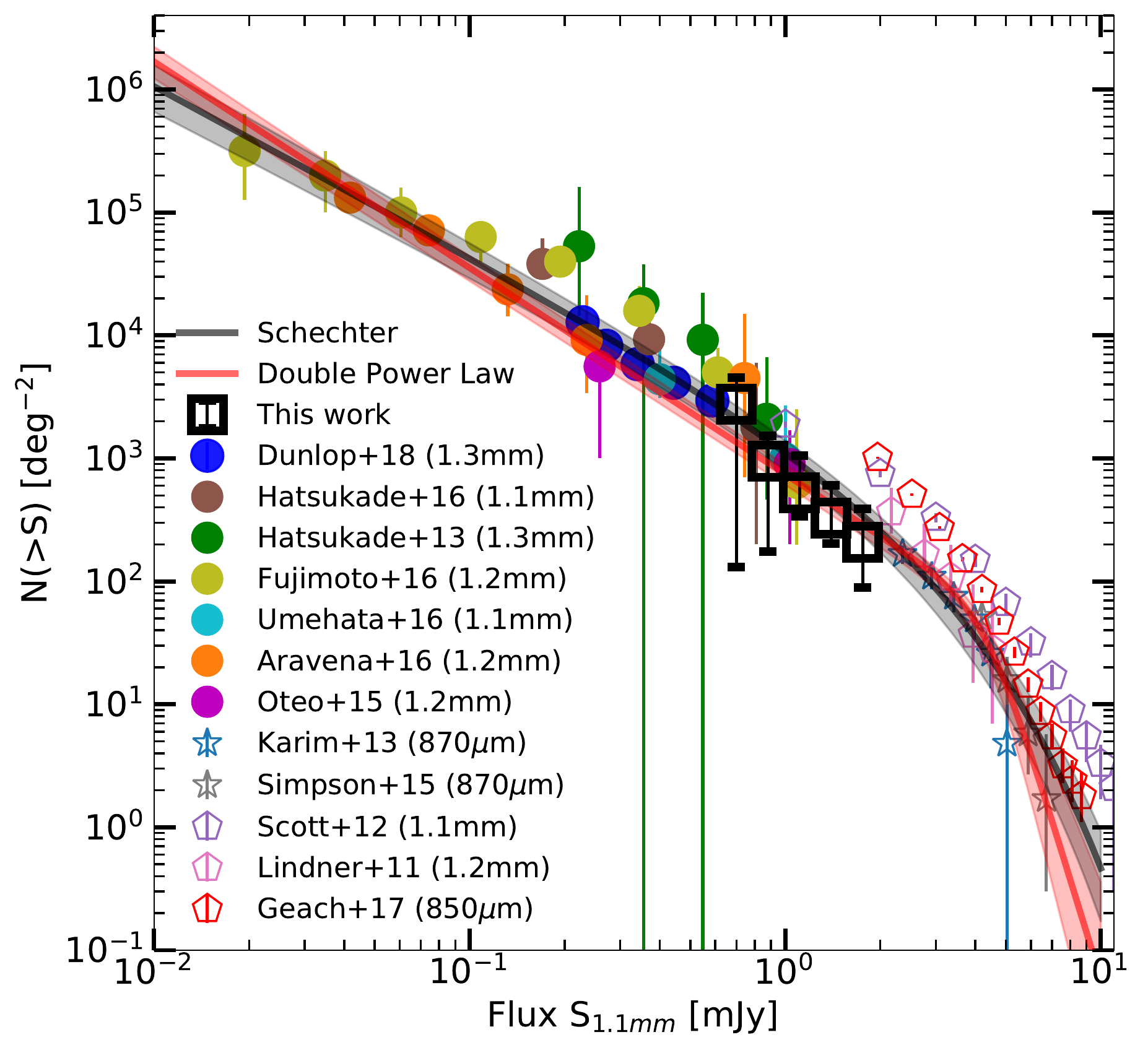}\\
\includegraphics[width=5cm,clip]{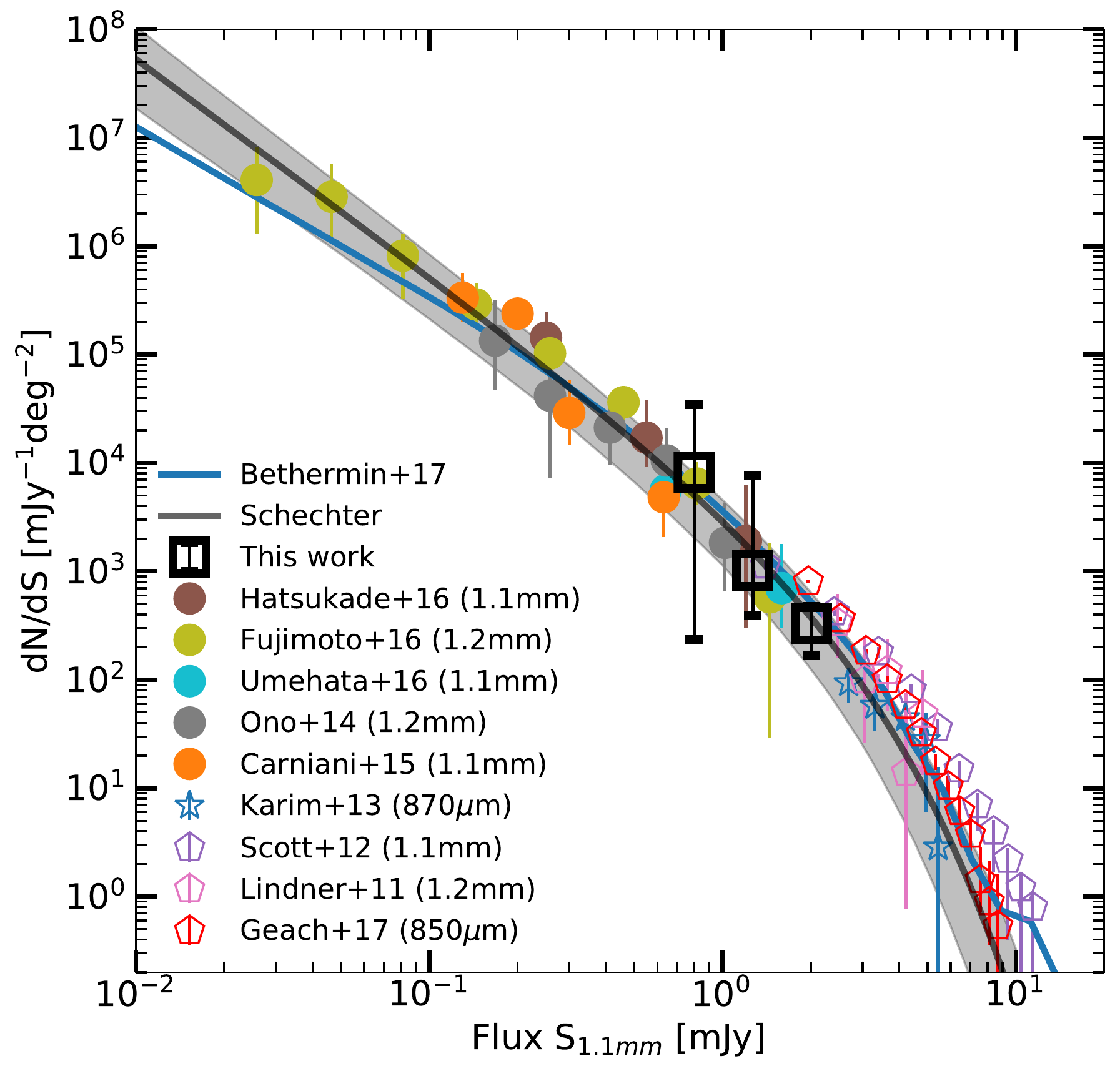}
}
\end{minipage}

\caption{1.1-mm cumulative (left panel) and differential (right panel) number counts derived using the corrections described in Sect.~\ref{sec:Cumulative_number_counts}, for the sources detected at $> $ 4.8\,$\sigma$ in the main catalogue. AGS19 is located at a position where the noise is artificially low, and has therefore not been taken into account. Previous (sub)millimetre cumulative number counts are also shown \citep{Lindner2011, Scott2012, Karim2013, Hatsukade2013, Ono2014, Simpson2015, Oteo2015, Carniani2015, Hatsukade2016, Aravena2016,Fujimoto2016,Umehata2017,Geach2017, Dunlop2017}. The different fluxes are scaled to 1.1 mm flux densities using S$_{1.1\text{mm}}$/S$_{1.2\text{mm}}$\,=\,1.29, S$_{1.1\text{mm}}$/S$_{1.3\text{mm}}$\,=\,1.48, S$_{1.1\text{mm}}$/S$_{870\mu m}$\,=\,0.56. From the \cite{Umehata2017} study, we use only sources which do not have $z$\,=\,3.09, (i.e. we are excluding the protocluster members). Results from single-dish surveys are shown with unfilled pentagon markers and are only indicative, they are not considered for model fitting. The grey curve shows the best-fit Schechter function (with 1-$\sigma$) uncertainties, the red curve shows the best-fit DPL function (with 1-$\sigma$).}
   \label{cumulative_number_count}
\end{figure*}

In Fig.~\ref{cumulative_number_count}, we compare our results with previous studies \citep{Lindner2011,Scott2012,Karim2013,Hatsukade2013, Simpson2015, Oteo2015, Hatsukade2016, Aravena2016,Fujimoto2016,Umehata2017, Geach2017, Dunlop2017}. To standardize these previous studies, the different flux densities are scaled to 1.1 mm using a Modified Black Body (MBB) model, assuming a dust emissivity index $\beta$\,=\,1.5 \citep[e.g.][]{Gordon2010}, a dust temperature $T_d$\,=\,35 K  \citep[eg.][]{Chapman2005, Kovacs2006, Coppin2008}, and a redshift of $z$\,=\,2.5 \citep[e.g.][]{Wardlow2011,Yun2012}. These values have also been chosen to be consistent with \cite{Hatsukade2016}. The different fluxes are therefore scaled to 1.1 mm using the relations S$_{1.1\text{mm}}$/S$_{1.2\text{mm}}$\,=\,1.29 , S$_{1.1\text{mm}}$/S$_{1.3\text{mm}}$\,=\,1.48 and S$_{1.1\text{mm}}$/S$_{870\mu m}$\,=\,0.56. It is a real challenge to standardize these previous studies because instruments, observational techniques or resolution often vary between studies. Some of these counts have been computed from individual pointings, by brightness selection, or by serendipitous detections. Observations with a single dish or a low resolution can also overestimate the number counts for the brightest galaxies, because of blending effects (see \citealt{Ono2014}). Another non-negligible source of error can come from an inhomogeneous distribution of bright galaxies. An underdensity by a factor of two of submillimetre galaxies with far infrared luminosities greater than $2 \times 10^{12}$L$_\odot$ in the Extended \textit{Chandra} Deep Field South (ECDFS) compared to other deep fields has been revealed by \cite{Weiss2009}.

Despite those potential caveats, the results from our ALMA survey in the GOODS--\textit{South} field are in good agreement with previous studies for flux densities below 1\,mJy. For values above this flux density, two different trends coexist as illustrated in Fig.~\ref{cumulative_number_count}: our counts are similar to those found by \cite{Karim2013}, but below the trend characterised by \cite{Scott2012}. These two previous studies have been realized under different conditions. The effects of blending, induced by the low resolution of a single dish observation, as with \cite{Scott2012}, tend to overestimate the number counts at the bright-end \citep[][]{Ono2014, Karim2013, Bethermin2017}. We indicate these points on the  Fig.~\ref{cumulative_number_count} on an indicative basis only. 

The differences in wavelength between the different surveys, even after applying the scaling corrections above, can also induce scatter in the results, especially for wavelengths far from 1.1mm.
The cumulative source counts from the 20 detections in this study and the results from other multi-dish blank surveys are fitted with a Double Power Law (DPL) function \citep[e.g.][]{Scott2002} given by:

\begin{equation}
\label{DPL}
N(>S)\,=\,\frac{N_0}{S_0}\left[ \left( \frac{S}{S_0} \right) ^\alpha + \left( \frac{S}{S_0} \right) ^\beta  \right]^{-1}
\end{equation}
and a modified Schechter \citep{Schechter1976} function \citep[e.g.][]{Knudsen2008}:

\begin{equation}
\label{Schechter}
N(>S) \,=\,\frac{N_0}{S_0} \left( \frac{S}{S_0} \right) ^\alpha \text{exp}\left( -\frac{S}{S_0} \right) d\left( \frac{S}{S_0} \right)
\end{equation}
\noindent where $N_0$ is the normalization, $S_0$ the characteristic flux density and $\alpha$ is the faint-end slope. $\beta$ is the bright-end slope of the number of counts in Eq.~\ref{DPL}. We use a least squares method with the Trust Region Reflective algorithm for these two fitted-functions. The best-fit parameters are given in Table~\ref{best_fit_powerlaw_shechter}.

\begin{table}
\centering          
\begin{tabular}{c c c c c }   
\hline    

			& $N_0$ 					& $S_0$				& $\alpha$				&$\beta$\\
          		& 10$^{2}$deg$^{-2}$		&mJy				&						&		\\
\hline    
\hline    
\multicolumn{5}{c}{Cumulative number counts}\\
DPL			&	2.8$\pm$0.2			&	$4.4_{-0.5}^{+0.3}$	&	$8.45_{-1.07}^{+0.28}$	&1.68$\pm$0.02\\
Schechter 	&	$14.3_{-2.3}^{+1.4}$		&	$2.0\pm0.3$		&	-1.38$\pm$0.05			&\\
\hline 
\multicolumn{5}{c}{Differential number counts}\\
Schechter 	&	$35.2_{-10.8}^{+4.6}$	&	$1.6_{-0.4}^{+0.3}$	&	-1.99$\pm$0.07			&\\
\hline
   \end{tabular}
\caption{Best-fit parameters for the cumulative and differential  number counts for a double power law function (Eq.~\ref{DPL}) and a Schechter function (Eq.~\ref{Schechter}).}
\label{best_fit_powerlaw_shechter}    

\end{table}

One of the advantages of using differential number counts compared to cumulative number counts is the absence of correlation of the counts between the different bins. However, the differential number counts are sensitive to the lower number of detections per flux density bin. Here we use $\Delta$log$S_\nu$\,=\,0.2 dex flux density bins.

We compare our results with an empirical model that predicts the number counts at far-IR and millimetre wavelengths, developed by \cite{Bethermin2017}. This simulation, called SIDES (Simulated Infrared Dusty Extragalactic Sky), updates the \cite{Bethermin2012} model. These predictions are based on the redshift evolution of the galaxy properties, using a two star-formation mode galaxy evolution model (see also \citealt{Sargent2012}).
 
The \cite{Bethermin2017} prediction is in good agreement with the number counts derived in this study, for the two bins with the lowest fluxes. For the highest-flux bin, the model is slightly above the data ($\sim$1$\sigma$ above the best Schechter fit for fluxes greater than 1\,mJy). However, both the \cite{Bethermin2017} model and our data points are below the single-dish measurements for fluxes greater than 1\,mJy. This disagreement between interferometric and single-dish counts is expected, because the boosting of the flux of single-dish sources by their neighbour in the beam \citep{Karim2013,Hodge2013,Scudder2016}. \cite{Bethermin2017} derived numbers counts from a simulated single-dish map based on their model and found a nice agreement with single-dish data, while the intrinsic number counts in the simulation are much lower and compatible with our interferometric study.

Cosmic variance was not taken into account in the calculation of the errors. Above $z$ = 1.8 and up to the redshift of the farthest galaxy in our catalog at $z$ = 4.8, the strong negative K-correction at this wavelength ensures that the selection of galaxies is not redshift-biased. The cosmic variance, although significant for massive galaxies in a small solid angle, is counterbalanced by the negative K-correction, which makes the redshift interval of our sources ($\Delta z$ = 3 in Eq. 12  in \citealt{Moster2011}) relatively large, spanning a comoving volume of 1400 Gpc$^3$. Based on \cite{Moster2011}, the cosmic variance for our sources is $\sim$15\,\%,  which does not significantly affect the calculation of the errors on our number counts.

\subsection{Contribution to the cosmic infrared background}

\begin{figure}
\centering
\resizebox{\hsize}{!} {
\includegraphics[width=5cm,clip]{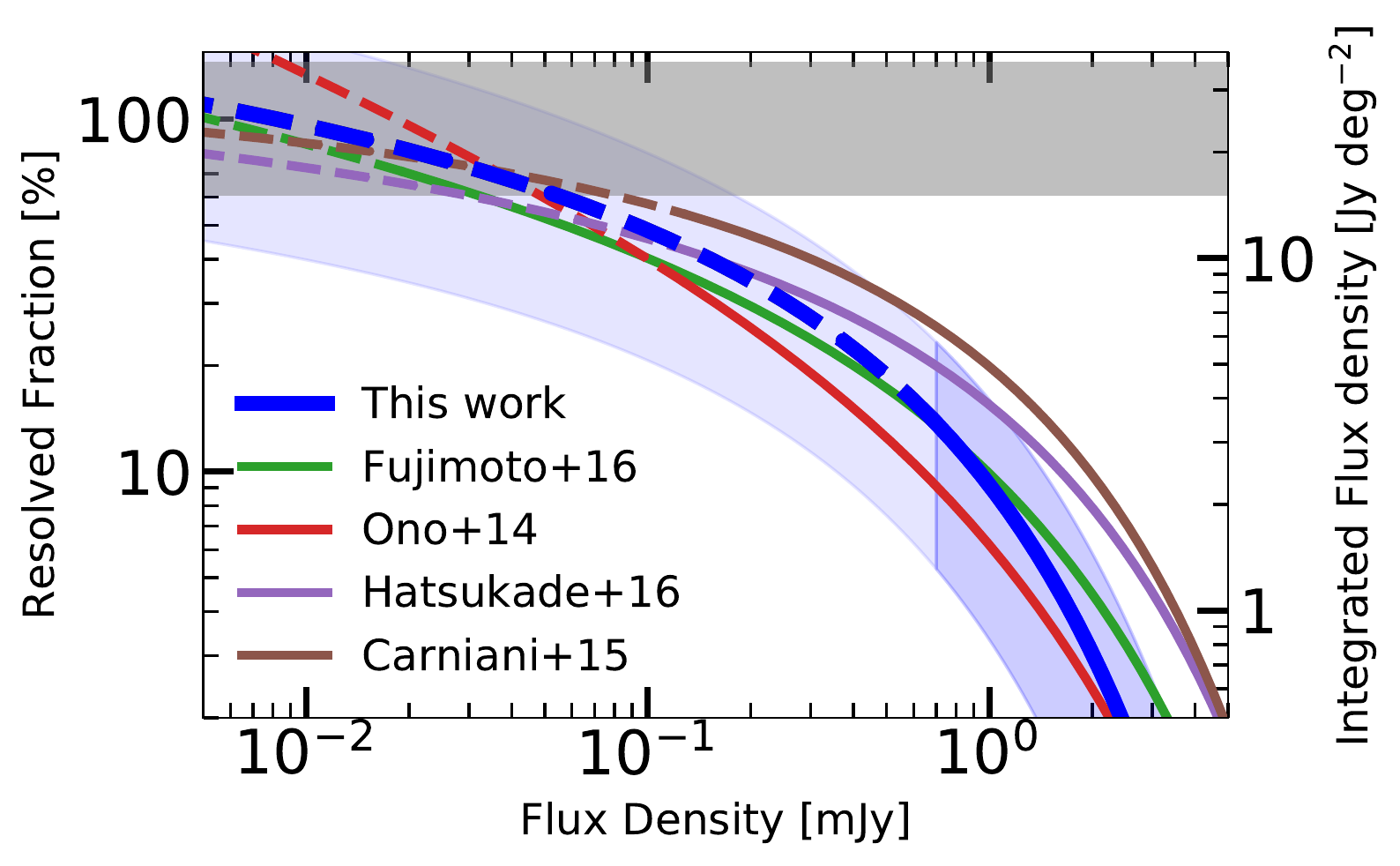}
}
\caption{Resolved 1.1 mm EBL computed from the best-fit Schechter function to the differential number counts. The green, red, purple and brown lines are from the number counts estimated by \cite{Fujimoto2016}, \cite{Ono2014}, \cite{Hatsukade2016} and \cite{Carniani2015} respectively. The blue line and shaded region show the results from this work and the associated uncertainty. The solid lines represent the model above the detection limits, and the dashed lines show the extrapolation below these limits. The grey shaded region shows the 1.1 mm cosmic infrared background measured by COBE \citep{Fixsen1998}.}
   \label{fig:EBL}
\end{figure}

The extragalactic background light (EBL) is the integrated intensity of all of the light emitted throughout cosmic time. Radiation re-emitted by dust comprises a significant fraction of the EBL, because this re-emitted radiation, peaking around 100 $\mu$m, has an intensity comparable to optical background \citep{Dole2006}. The contribution of our ALMA sources to the EBL is derived by integrating the derived number counts down to a certain flux density limit. Using the 20 ($>$4.8\,$\sigma$) sources detected, we compute the fraction of the 1.1 mm EBL resolved into discrete sources. The integrated flux density is given by:
\begin{equation}
\label{eq:EBL}
I(S>S_{lim})\,=\,\int_{S_{lim}}^{\inf}\frac{dN(S)}{dS}SdS
\end{equation}
We use the set of parameters given in Table~\ref{best_fit_powerlaw_shechter} on the differential number counts. We compare our results with observations from the Far Infrared Absolute Spectrophotometer (FIRAS) on the COsmic Background Explorer (COBE), knowing that uncertainties exist on the COBE measurements \citep[e.g.][]{Yamaguchi2016}. We use the equation given in \cite{Fixsen1998} to compute the total energy of the EBL:
\begin{equation}
I_\nu\,=\,(1.3\pm0.4)\times10^{-5}\left(\frac{\nu}{\nu_0}\right)^{0.64\pm0.12}P_\nu(18.5+1.2 K)
\end{equation}
\noindent where $\nu_0$\,=\,100 cm$^{-1}$, and $P_\nu$ is the familiar Planck function with $I_\nu$ in erg.s$^{-1}$cm$^{-1}$Hz$^{-1}$sr$^{-1}$.
From this equation, we find that at 1.1 mm, the energy of the EBL is 2.87 nW.m$^{-2}$sr$^{-1}$. 
From Eq.~\ref{eq:EBL} we can estimate the integrated EBL light. Fig.~\ref{fig:EBL} shows this total integrated flux density. For our data, the lowest flux density bin for differential counts $S_{lim}$ is 0.8\,mJy, and we extrapolate to lower flux densities. We have resolved only 13.5$_{-8.6}^{+9.0}$\% of the EBL into individual galaxies at 0.8\,mJy. This result is in good agreement with studies such as \cite{Fujimoto2016}. In order to have the majority of the EBL resolved (e.g. \mbox{\citealt{Hatsukade2013}}; \citealt{Ono2014, Carniani2015, Fujimoto2016}), we would need to detect galaxies down to 0.1\,mJy (about 50 \% of the EBL is resolved at this value).

The extrapolation of the integrated flux density below $S_{lim}$ suggests a flattening of the number counts. The population of galaxies that dominate this background is composed of the galaxies undetected in our survey, with a flux density below our detection limit. 

\section{Galaxy properties}\label{sec:Properties}
We now study the physical properties of the ALMA detected sources, taking advantage of the wealth of ancillary data available for the GOODS--\textit{South} field.

\subsection{Redshift distribution}

   \begin{figure}
   \centering
   \includegraphics[width=\hsize]{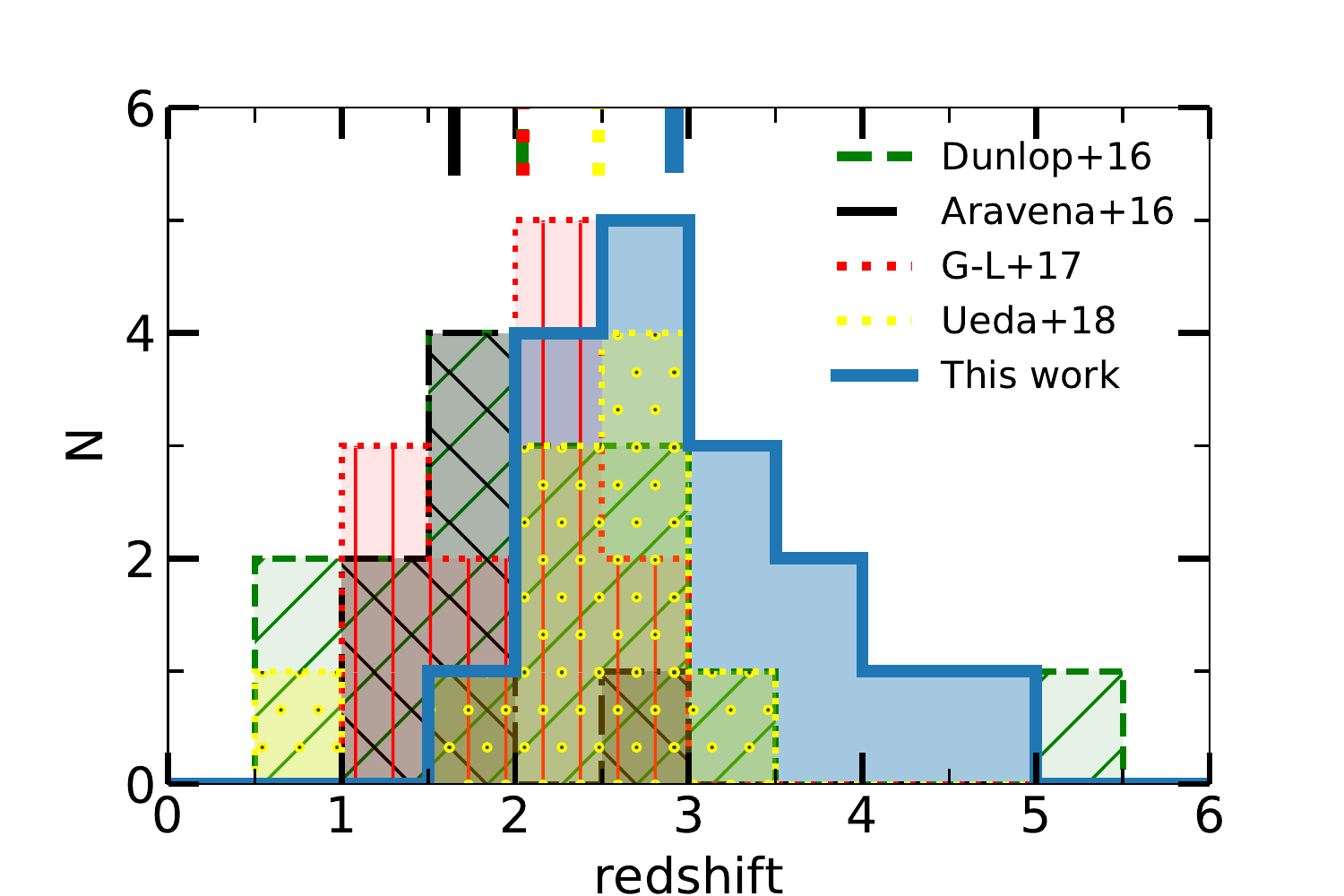}
      \caption{Redshift distributions (photometric or spectroscopic) for millimetre-selected galaxies. The blue solid line shows the redshift distribution of our ALMA GOODS--\textit{South} blind survey. The green dashed line shows the Hubble Ultra Deep Field Survey redshift distribution \citep{Dunlop2017}, the black dash-dotted line shows the ASPECS sample \citep{Aravena2016}, the red dotted line shows the ALMA Frontier Fields Survey \citep{Gonzalez-Lopez2017} and the yellow dotted line shows the ASAGAO survey \citep{Ueda2018}. Short coloured lines at the top of the figure indicate the median redshifts for these four studies.}
         \label{redshift_distribution}
   \end{figure}

Among the 17 ALMA detected sources for which redshifts have been computed, six have a spectroscopic redshift (AGS1, AGS2, AGS3, AGS9, AGS12, AGS13 and AGS18) determined by \mbox{\cite{Kurk2013}}, and recently confirmed by \cite{Barro2017}, \cite{Momcheva2016}, \cite{Vanzella2008}, Mobasher (private communication), \cite{Inami2017}, \cite{Kriek2008} and \citealt{Dunlop2017} -- from a private communication of Brammer -- respectively. The redshift distribution of these 17 ALMA sources is presented in Fig.~\ref{redshift_distribution}, compared to the distributions of four other deep ALMA blind surveys \citep{Dunlop2017,Aravena2016,Gonzalez-Lopez2017, Ueda2018}.
Of the 17 sources, 15 are in the redshift range $z$\,=\,1.9 $-$ 3.8. Only two galaxies (AGS4 and AGS11) have a redshift greater than 4 ($z_{\text{phot}}$\,=\,4.32 and 4.82 respectively). We discuss these galaxies further in Sect.~\ref{sec:HST-dark}. The mean redshift of the sample is $z$\,=\,3.03$\pm$0.17, where the error is computed by bootstrapping. This mean redshift is significantly higher than those found by \cite{Dunlop2017}, \cite{Aravena2016}, \cite{Gonzalez-Lopez2017} and \cite{Ueda2018} who find distributions peaking at 2.13, 1.67, 1.99 and 2.28 respectively. The median redshift of our sample is 2.92$\pm$0.20, which is a little higher than the value expected from the models of \cite{Bethermin2015}, which predict a median redshift of 2.5 at 1.1\,mm, considering our flux density limit of $\sim$874\,$\mu$Jy (4.8\,$\sigma$).

Our limiting sensitivity is shallower than that of previous blind surveys: 0.184\,mJy here compared with 13\,$\mu$Jy in \cite{Aravena2016}, 35\,$\mu$Jy in \cite{Dunlop2017}, (55-71)\,$\mu$Jy in \cite{Gonzalez-Lopez2017} and 89\,$\mu$Jy in \cite{Ueda2018}. However our survey covers a larger region on the sky: 69 arcmin$^2$ here, compared to 1 arcmin$^2$, 4.5 arcmin$^2$, 13.8 arcmin$^2$ and 26 arcmin$^2$ for these four surveys respectively. The area covered by our survey is therefore a key parameter in the detection of high redshift galaxies due to a tight link between 1.1mm luminosity and stellar mass as, we will show in the next section. The combination of two effects: a shallower survey allowing us to detect brighter SMGs, which are more biased toward higher redshifts \citep[e.g.][]{Pope2005}, as well as a larger survey allowing us to reach more massive galaxies, enables us to open the parameter space at redshifts greater than 3, as shown in Fig.~\ref{redshift_distribution}. This redshift space is partly or totally missed in smaller blind surveys.

We emphasize that the two HST--dark galaxies (see Sect.~\ref{sec:HST-dark}) for which the mass and redshift could be determined (AGS4 and AGS11) are the two most distant galaxies in our sample, with redshifts greater than 4.

\subsection{Stellar Masses}\label{sec::stellar_masses}

Over half (10/17) of our galaxies have a stellar mass greater than 10$^{11}$ M$_\sun$ (median mass of $\text{M}_{\star}$\,=\,1.1\,$\times$\,10$^{11}$ M$_\sun$).
The population of massive and compact star-forming galaxies at $z$$\sim$2 has been documented at length \citep[e.g.][]{Daddi2005,van_Dokkum2015}, but their high redshift progenitors are to-date poorly detected in the UV. Our massive galaxies at redshifts greater than 3 might therefore give us an insight into these progenitors.

   \begin{figure*}
   \centering
   \includegraphics[width=150mm]{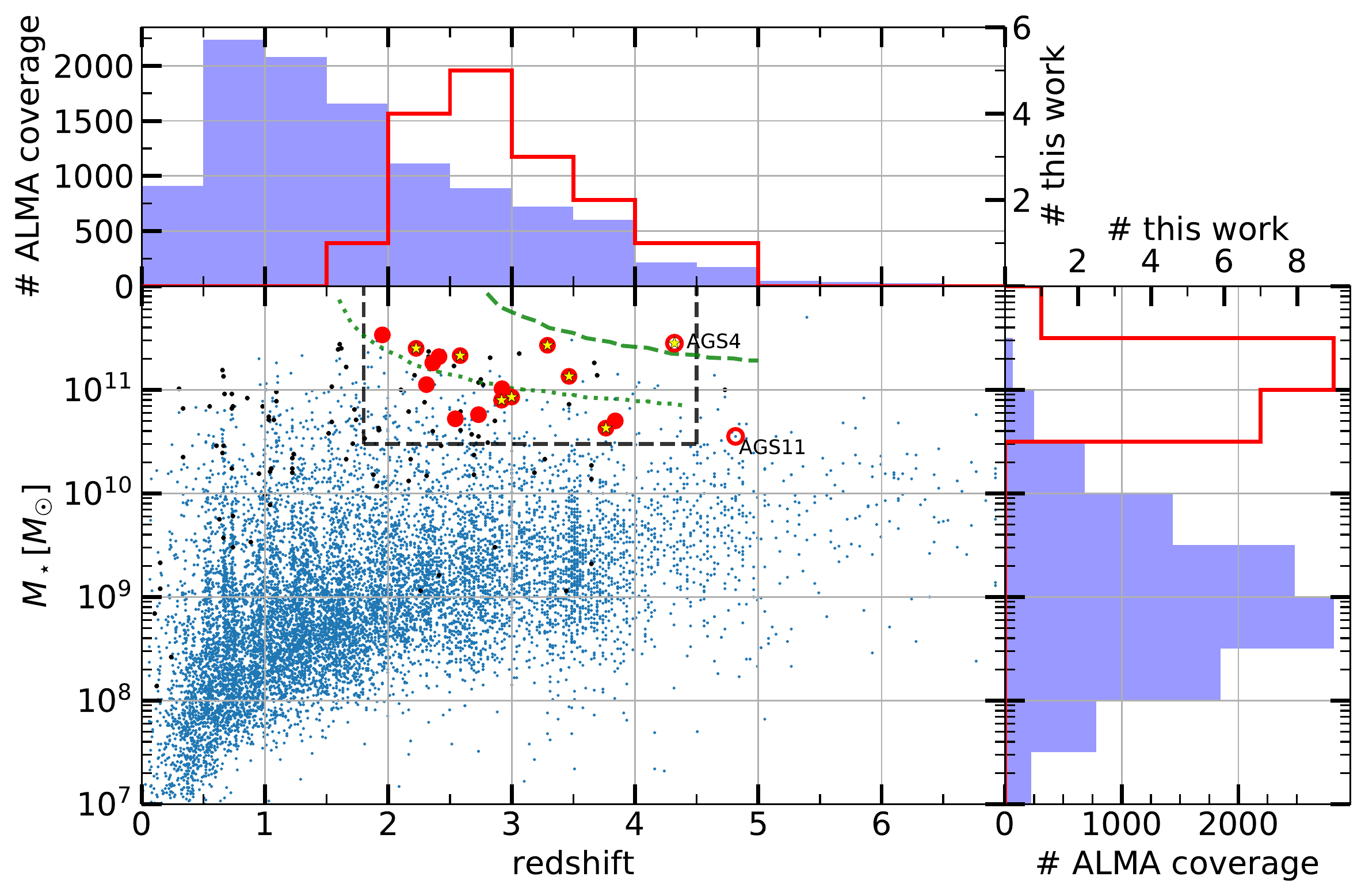}
      \caption{Stellar mass versus redshift for the galaxies detected in our ALMA GOODS--\textit{South} blind survey (red points). For comparison, the distribution of all of the galaxies, listed in the ZFOURGE catalogue, in the same field of view is given in blue. Only UVJ active galaxies are shown. The two HST-dark galaxies for which we have redshifts (AGS4 and AGS11) are represented by open circles. The redshift of AGS11 is however uncertain. The green dashed line shows the position that would be occupied by a typical star-forming galaxy -- lying on the median of the SFR-M$_{\star}$ star-formation main sequence (MS) -- that would produce a 1.1mm flux density equal to our average detection limit of 0.88\,mJy (4.8-$\sigma$) using the Spectral Energy Distribution (SED) library of \cite{Schreiber2017}. The dotted line illustrates the position of galaxies 3 times above the MS using the appropriate SEDs from the same library. Galaxies hosting an AGN that are undetected or detected by ALMA are identified with black dotes and yellow stars respectively. Inside the black dashed rectangle, 50\% of the galaxies detected by ALMA host an AGN, while only 14\% of the UVJ active galaxies undetected by ALMA host an AGN.}
         \label{mass_redshift_distribution}
   \end{figure*}

Fig.~\ref{mass_redshift_distribution} shows the stellar mass as a function of redshift for all of the UVJ active galaxies, listed in the ZFOURGE catalogue, in our ALMA survey field of view. Star forming galaxies (SFGs) have been selected by a UVJ colour-colour criterion as given by \cite{Williams2009} and applied at all redshifts and stellar masses as suggested by \cite{Schreiber2015}:

\begin{equation}
SFG =
\left\lbrace
\begin{array}{ccc}
U-V  &\,<\, & 1.3, \text{or}\\
V-J &\,>\, & 1.6, \text{or}\\
U-V &\,<\, & 0.88 \times (V-J) + 0.49
\end{array}\right.
\label{SFG_equation}
\end{equation}

\noindent All galaxies not fulfilling these colour criteria are considered as quiescent galaxies and are excluded from our comparison sample (9.3\% of the original sample). The ALMA detected galaxies in our survey are massive compared to typical SFGs detected in deep optical and near-IR surveys like CANDELS, in the same redshift range (2$<$ $z$\,$<$\,4), as shown in Fig.~\ref{mass_redshift_distribution}.

The high proportion of massive galaxies among the ALMA detected sources suggests that stellar mass can be a strong driver for a source to be detected by ALMA at high redshift \citep{Dunlop2017}. The strong link between detection and stellar mass is related to the underlying relation between stellar mass and star formation rate of SFGs \mbox{\citep[e.g.][]{Noeske2007,Elbaz2011}}. Almost one third (7/24) of the galaxies previously catalogued in the field of view of this study with M$_\star$\,$>$\, $10^{11}$ and 2\,$<$\, $z$\,$<$\, 3 are also detected with ALMA. The position of our galaxies along the main sequence of star formation will be studied in a following paper (Franco et al., in prep.).

We observe a lack of detections at redshift $z$\,<\,2, driven by both a strong positive K-correction favouring higher redshifts and a decrease in the star formation activity at low redshift. Indeed, the specific star formation rate (sSFR), defined as the ratio of galaxy SFR to stellar mass, drops quickly at lower redshifts ($z$\,<\,2), whereas this rate increases continuously at greater redshifts \mbox{\citep[e.g.][]{Schreiber2015}}.
In addition, very massive galaxies (stellar mass greater than $10^{11}$ M$_\sun$) are relatively rare objects in the smaller co-moving volumes enclosed by our survey at lower redshift. To detect galaxies with these masses, a survey has to be sufficiently large. The covered area is therefore a critical parameter for blind surveys to find massive high redshift galaxies. 

In order to estimate the selection bias relative to the position of our galaxies on the main sequence, we show in Fig.~\ref{mass_redshift_distribution} the minimum stellar mass as a function of redshift that our survey can detect, for galaxies on the MS of star formation (green dashed line), and for those with a SFR three times above the MS (green dotted line).

To determine this limit, we calculate the SFR of a given MS galaxy, based on the galaxy stellar mass and redshift as defined in \cite{Schreiber2015}. From this SFR and stellar mass, the galaxy SED can also be calculated using the \cite{Schreiber2017} library. We then integrate the flux of this SED around 1.1mm.
 
It can be seen that the stellar mass detection limit corresponding to MS galaxies lies at higher stellar mass than all of the galaxies detected by our ALMA survey (as well as all but one of the other star-forming galaxies present in the same region). This means that our survey is unable to detect star-forming galaxies below the main sequence.  We can quantify the offset of a galaxy from the main sequence, the so-called "starburstiness" \citep{Elbaz2011}, by the ratio SFR/SFR$_{MS}$, where SFR$_{MS}$ is the average SFR of "main sequence" galaxies computed from \cite{Schreiber2015}. We also indicate our detection limit for galaxies with SFR/SFR$_{MS}$\,=\,3. In this case, 7 of the 17 galaxies shown lie above the limit. To have been detected, these galaxies must therefore have SFRs at least larger than the SFR$_{MS}$, the other ten galaxies must have a SFR at least three times above the MS. This highlights that our survey is biased towards galaxies with high SFRs.

\subsection{AGN}\label{sec:AGN}

In this Section, we discuss the presence of AGN within the 20 most robust ALMA detections, i.e., rejecting the 3 spurious detections with no IRAC counterpart (AGS14, AGS16 and AGS19 marked with a star in Table~\ref{catalogue}) but including 3 of the supplementary sources (AGS21, AGS22 and AGS23). We find an X-ray counterpart for 65\% of them (13/20) in the 7 Msec X-ray survey of GOODS--\textit{South} with \textit{Chandra} \citep{Luo2017}. Most of these galaxies were classified as AGN in the catalogue of \cite{Luo2017} that identifies as AGN all galaxies with an intrinsic 0.5-7.0 keV luminosity higher than L$_{X,int}$\,=\,3\,$\times$\,10$^{42}$ erg.s$^{-1}$, among other criteria. However, our ALMA galaxies being biased towards highly star-forming galaxies, we decided to increase the minimum X-ray luminosity to a three times stronger X-ray luminosity threshold to avoid any contamination by star-formation. We also consider as AGN the galaxies exhibiting a hard X-ray spectrum. Hence, we adopt here the following criteria to identify AGN: either \textit{(i)} L$_{X,int}$\,$>$\, 10$^{43}$ erg.s$^{-1}$ (luminous X-ray sources) or \textit{(ii)} $\Gamma$\,$<$\,1.0 (hard X-ray sources).

As the redshifts adopted by \cite{Luo2017} are not always the same as ours, when necessary we scaled the X-luminosities to our redshifts using Eq.~1 from \cite{Alexander2003}, and assuming a photon index of $\Gamma$\,=\,2.

Using these conservative criteria, we find that 8 ALMA galaxies host an X-ray AGN (marked with a yellow star in Fig.~\ref{mass_redshift_distribution}). In order to compare the AGN fraction among ALMA detections with galaxies undetected by ALMA with similar masses and redshifts, we restrict our comparison to galaxies with M$_\star$\,$>$\, 3\,$\times$\,10$^{10}$ M$_\sun$ and 1.8\,$<$\, $z$\,$<$\, 4.5 (rectangle in black dotted lines in Fig.~\ref{mass_redshift_distribution}). In this area encompassing 16 ALMA detections, we find that 50\% of the ALMA sources host an AGN (8/16) as compared to only 14\% (23/160) of the star-forming galaxies undetected by ALMA located in this same area (selected using the UVJ criteria recalled in Eq.~\ref{SFG_equation} in the ZFOURGE catalogue).

The presence of a large fraction of AGN among the galaxies detected by ALMA may reflect the fact that the ALMA sources are experiencing a starburst (well above the MS marked with a green dashed line in Fig.~\ref{mass_redshift_distribution}), possibly triggered by a merger that may dramatically reduce the angular momentum of the gas and drive it towards the centre of the galaxies \cite[e.g.,][] {Rovilos2012, Gatti2015, Lamastra2013} or violent disk instabilities \citep{Bournaud2012}. In addition, the high AGN fraction may be driven by the link between the presence of an AGN and the compactness of their host galaxy. \cite{Elbaz2017}, \mbox{\cite{Chang2017}} and \mbox{\cite{Ueda2018}} suggest that the proportion of galaxies hosting an AGN increases with IR luminosity surface density. As discussed in Sect.~\ref{sec:completeness}, the size, and therefore the compactness of a galaxy, increases the likelihood of an ALMA detection at our angular resolution.
Alternatively ALMA might preferentially detect galaxies with a high gas, hence also dust, content, more prone to efficiently fuel the central black hole and trigger an AGN.

This fraction of galaxies with a high X-ray luminosity (L$_{X,int}$\,$>$\,10$^{43}$ erg.s$^{-1}$) seems to be significantly higher than that found in some other ALMA surveys, in particular  in \cite{Dunlop2017} (2/16) or \cite{Ueda2018} (4/12).

\section{HST-dark galaxies}\label{sec:HST-dark}
Some galaxies without $H$-band HST-WFC3 (1.6\,$\mu$m) counterparts have been discovered. We discuss below the possibility that these detections may be real HST-dark galaxies. Some ALMA detections previously attributed to an HST counterpart seem in fact to be either more distant galaxies, extremely close in the line of sight to another galaxy, hidden by a foreground galaxy, or too faint at optical rest-frame wavelengths to be detected by HST.

It is already known that some of the most luminous millimetre or submillimetre galaxies can be completely missed at optical wavelengths \mbox{\citep{Wang2016}}, even in the deepest optical surveys, due to dust extinction. Some of these galaxies can also be undetected in the NIR \mbox{\citep{Wang2009}}. 

Among the sources that do not have detections in the $H$-band of HST-WFC3, we distinguish the sources not detected by HST but detected by other instruments (we will discuss the importance of the IRAC filters), and sources undetected by HST and all of the other available instruments in the GOODS--\textit{South} Field (described in the Sect.~\ref{sec:Ancillary_data}).

Of the 20 galaxies detected in our main catalogue, 7 (35\%) do not present an obvious HST counterpart. This number is slightly higher than the expected number of spurious sources (4$\pm$2), predicted by the statistical analysis of our survey. 
To be more accurate, for three of these seven galaxies (AGS4, AGS15 and AGS17), an HST galaxy is in fact relatively close in the line of sight, but strong evidence, presented below, suggests that the HST galaxies are not the counterpart of the ALMA detections, and without the resolution of ALMA we would falsely associate the counterpart. For the 4 other ALMA detections without HST-WFC3 counterparts within a radius of 0\arcsec60, one of them (AGS11) has also been detected at other wavelengths. In this section, we will discuss 4 particularly interesting cases of HST-dark galaxies (AGS4, AGS11, AGS15 and AGS17), and discuss our reasons for classifying the other 3 as spurious sources.

\begin{figure*}
\centering
\begin{minipage}[t]{0.85\textwidth}
\resizebox{\hsize}{!} { 
\includegraphics[width=5.25cm,clip]{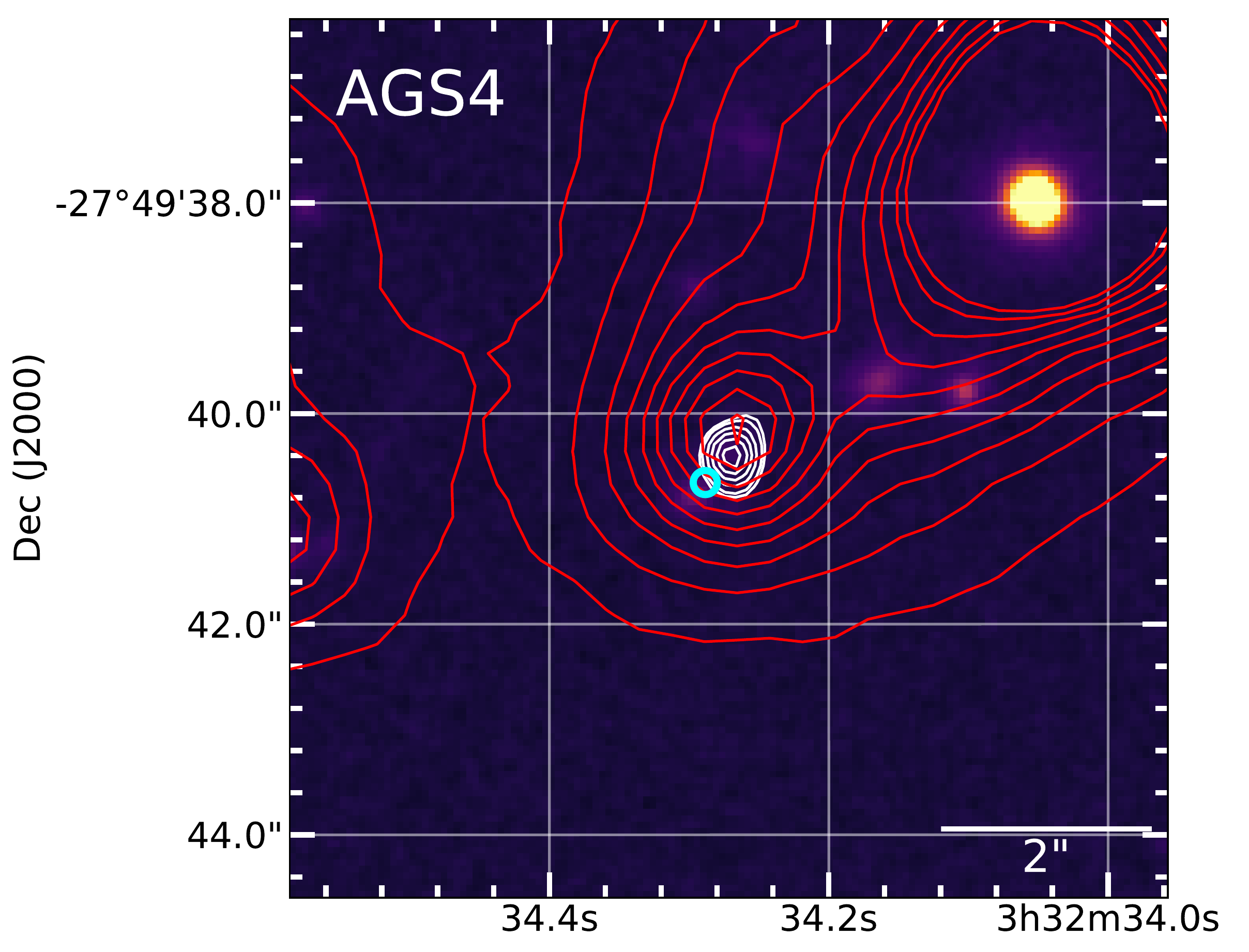} 
\includegraphics[width=5cm,clip]{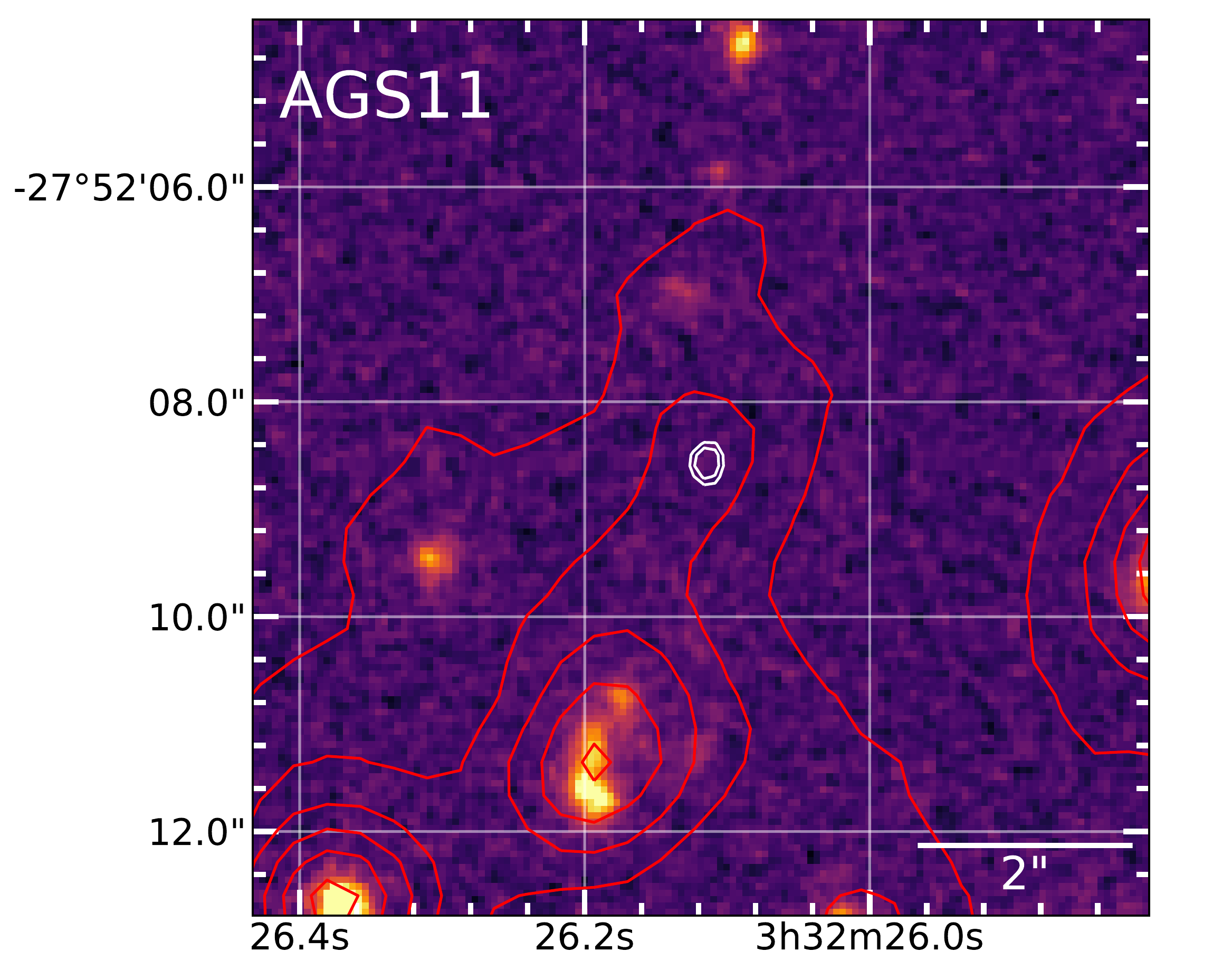} 
}
\end{minipage}
\begin{minipage}[t]{0.85\textwidth}
\resizebox{\hsize}{!} { 
\includegraphics[width=5.15cm,clip]{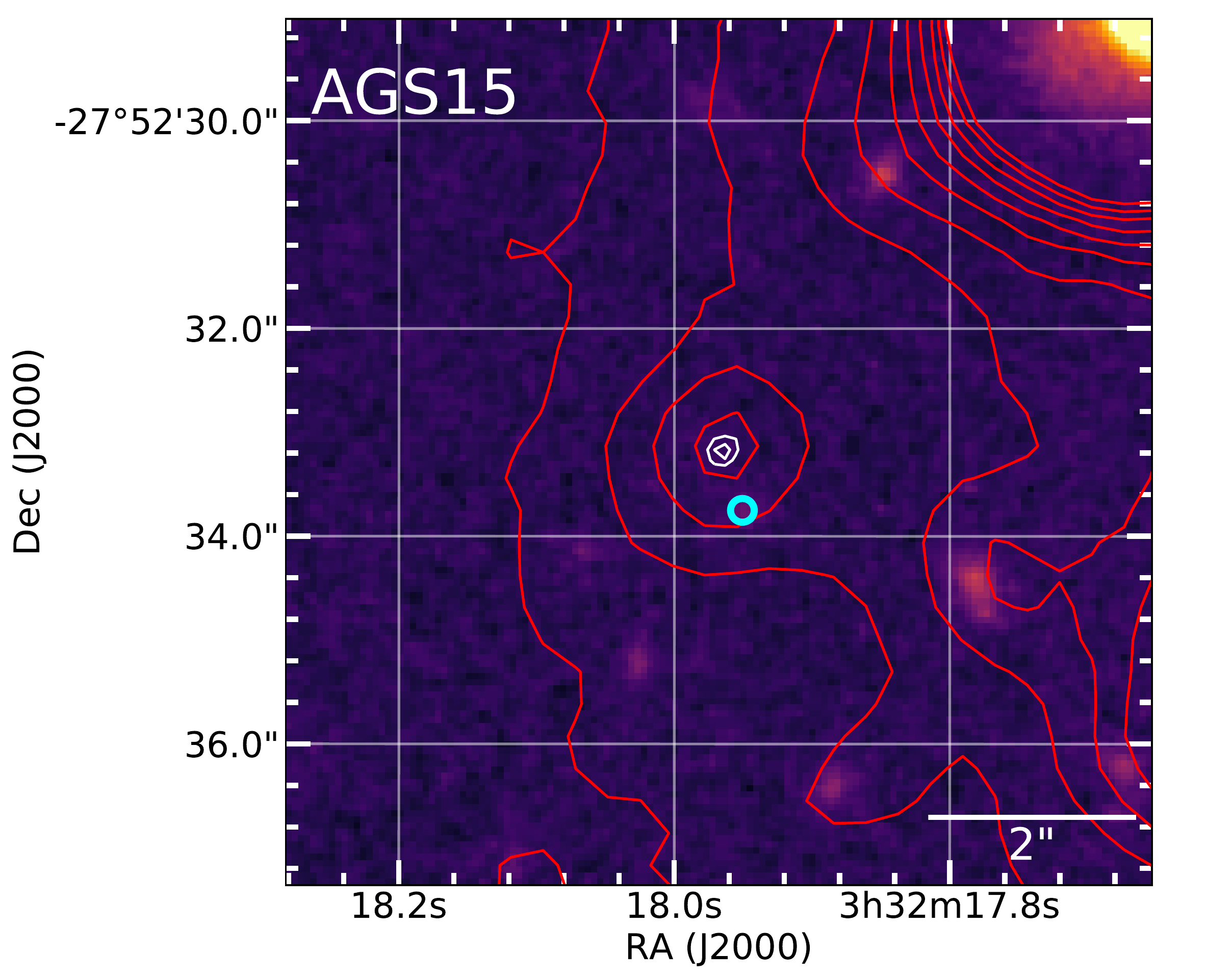} 
\includegraphics[width=5cm,clip]{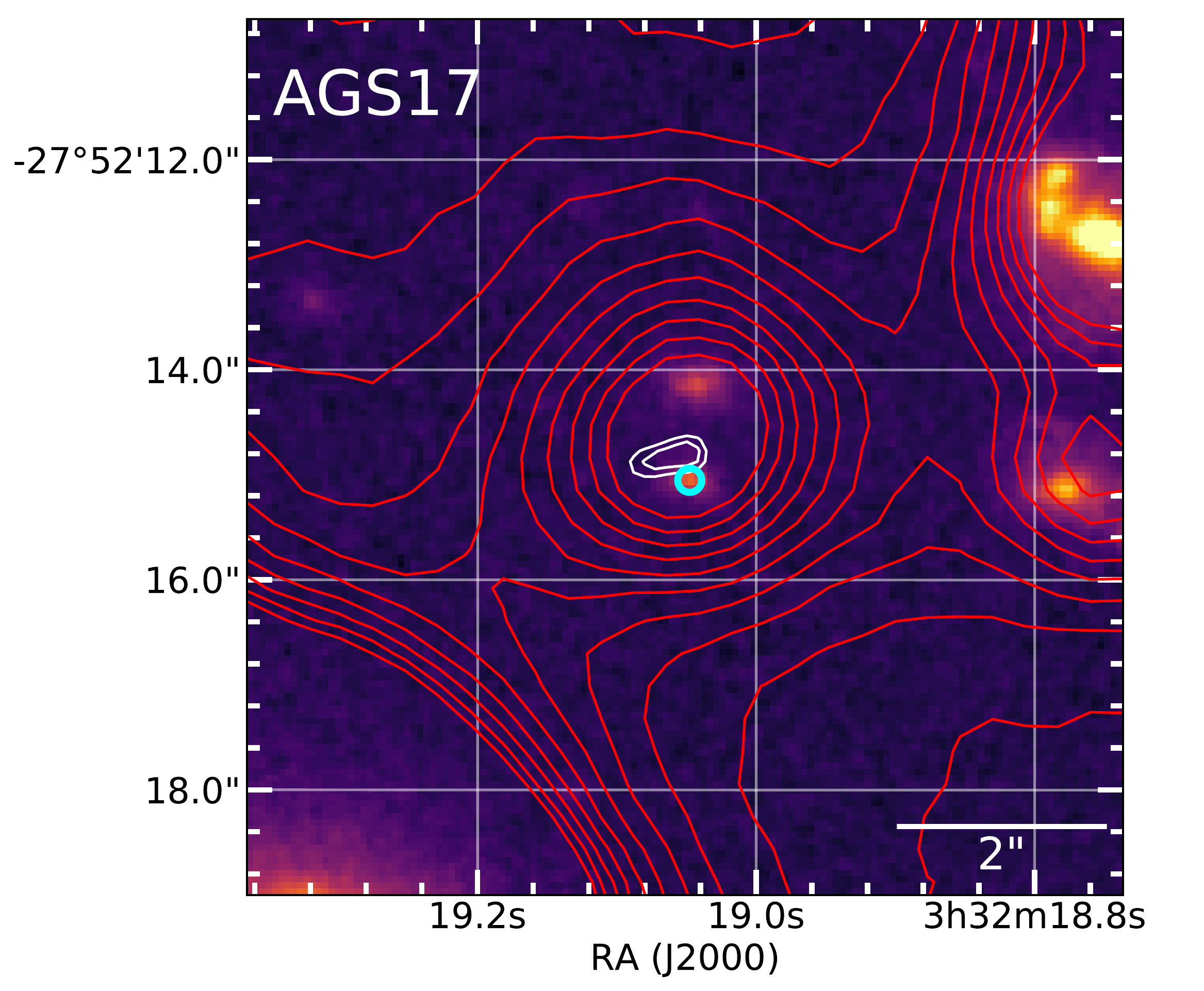} 

}
\end{minipage}
      \caption{IRAC 3.6\,$\mu$m (red contours, 3.\,$\mu$Jy to 30.\,$\mu$Jy in steps of of 3.0\,$\mu$Jy) and ALMA 1.1mm (white contours, 4, 4.5 then 5 to 10-$\sigma$ in steps of 1-$\sigma$) overlaid on 8\arcsec3\,$\times$\,8\arcsec3 HST $H$-band images. The position of the previously associated HST counterpart is shown by a cyan circle.}
         \label{IRAC_countours}
\end{figure*}

Our 4 HST-dark galaxies (AGS4, AGS11, AGS15 and AGS17) have at least one feature in common, the presence of an IRAC detection and the fact that this IRAC detection is closer on the sky than the unrelated HST detection (see Table~\ref{Hasardious_Association}). The IRAC detections come from the \mbox{\cite{Ashby2015}} catalogue, except for AGS15 where the position comes from the ZFOURGE catalogue, using the \mbox{\cite{Labbe2015}} survey. The offset between the IRAC and HST sources might suggest that they are different sources. Fig.~\ref{IRAC_countours} shows the IRAC contours at 3.6\,$\mu$m centred on the ALMA detection, superimposed over the HST $H$-band image.
The presence of IRAC detections at these distances from the ALMA galaxies is a very strong driver for the identification of sources. The probability of random IRAC association is between one and two orders of magnitude less likely than random HST association for this range of distances, as shown in Fig.~\ref{distance_closest_galaxy} and Table~\ref{Hasardious_Association}. The selection of ALMA candidates from galaxies detected in IRAC channels 1 and 2 but missed by HST-WFC3 at 1.6\,$\mu$m has already been experimented successfully by T. Wang et al. (in prep.), and seems to be a good indicator to detect HST-dark ALMA galaxies.

\begin{table}
\centering          
\begin{tabular}{l c c c c}  
\hline
ID&AGS4&AGS11&AGS15&AGS17\\
\hline    
\hline    

HST RaA [\%]& 4.52 & -  & 9.14 & 2.12 \\
IRAC RaA [\%]& 0.06  & 0.18 & 0.12 & 0.05 \\
\hline 
\end{tabular}
\caption{The probability of an HST or IRAC random association (RaA) between the ALMA detection and the closest HST and IRAC galaxies for the 4 HST-Dark galaxies discussed in Sect.~\ref{sec:HST-dark}}
\label{Hasardious_Association}  
\end{table}

As each of our HST-dark galaxies have different features, we will discuss each galaxy individually.

\begin{figure}
\centering
\begin{minipage}[t]{0.5\textwidth}
\resizebox{\hsize}{!} {
\includegraphics[width=3cm,clip]{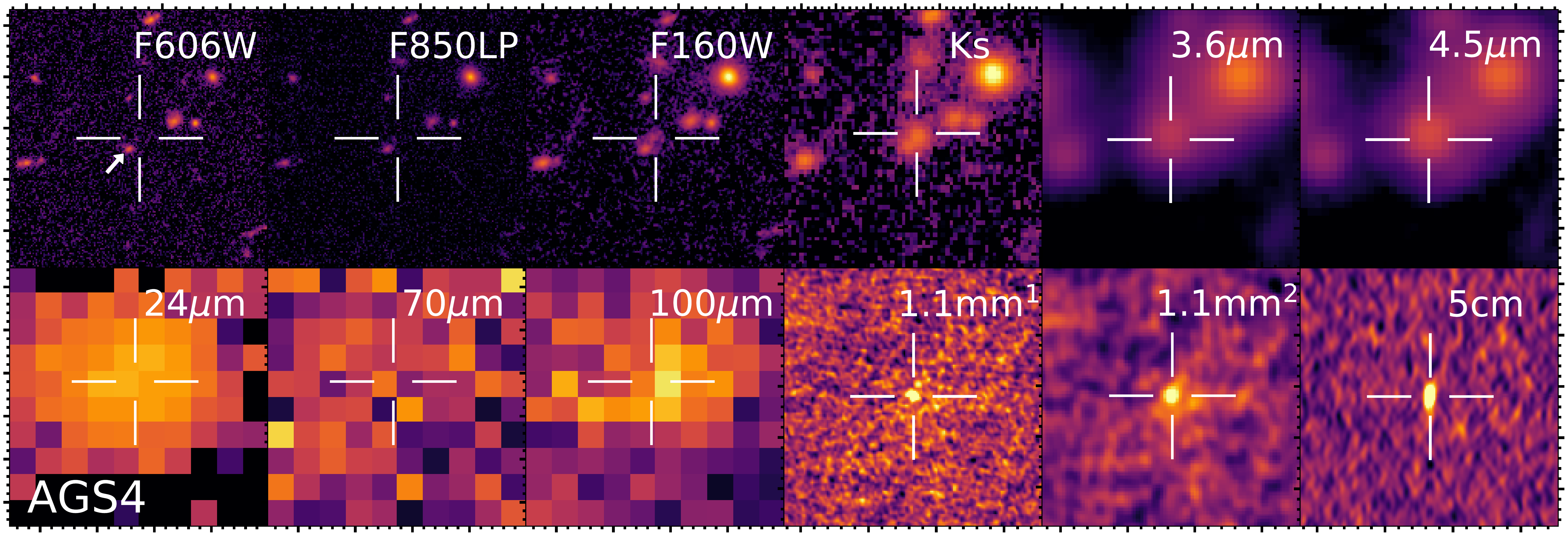}}
\end{minipage}
\begin{minipage}[t]{0.5\textwidth}
\resizebox{\hsize}{!} {
\includegraphics[width=3cm,clip]{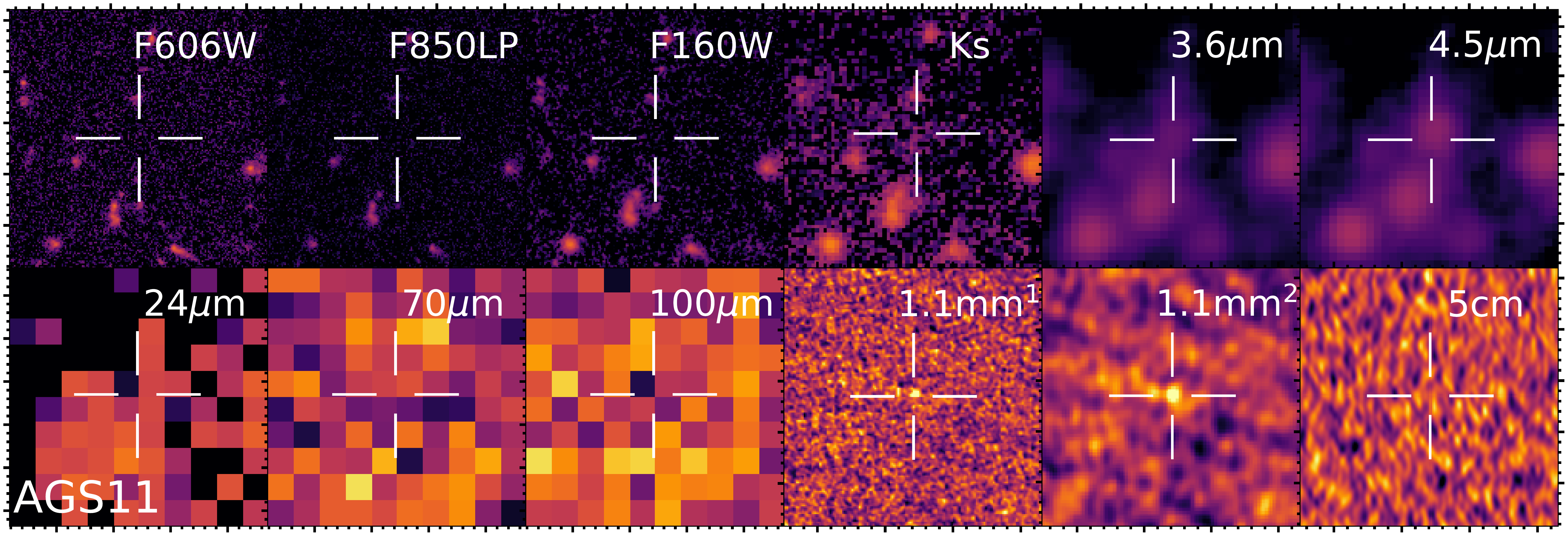}}
\end{minipage}
\begin{minipage}[t]{0.5\textwidth}
\resizebox{\hsize}{!} {
\includegraphics[width=3cm,clip]{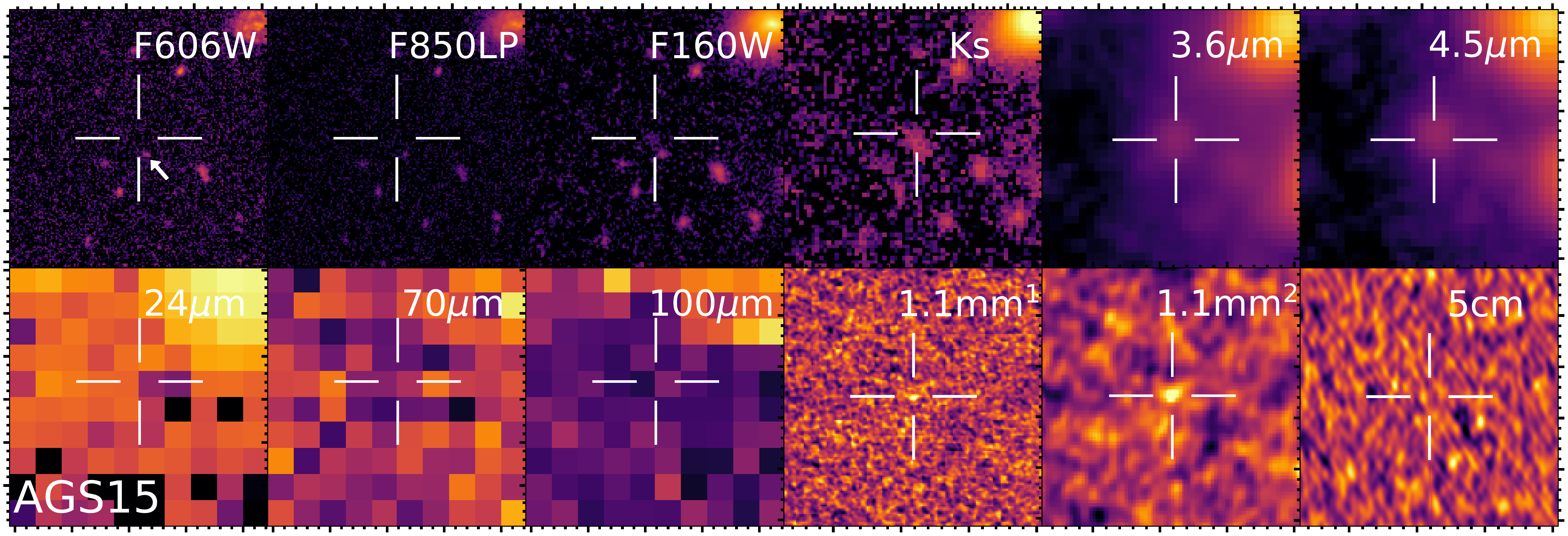}}
\end{minipage}
\begin{minipage}[t]{0.5\textwidth}
\resizebox{\hsize}{!} {
\includegraphics[width=3cm,clip]{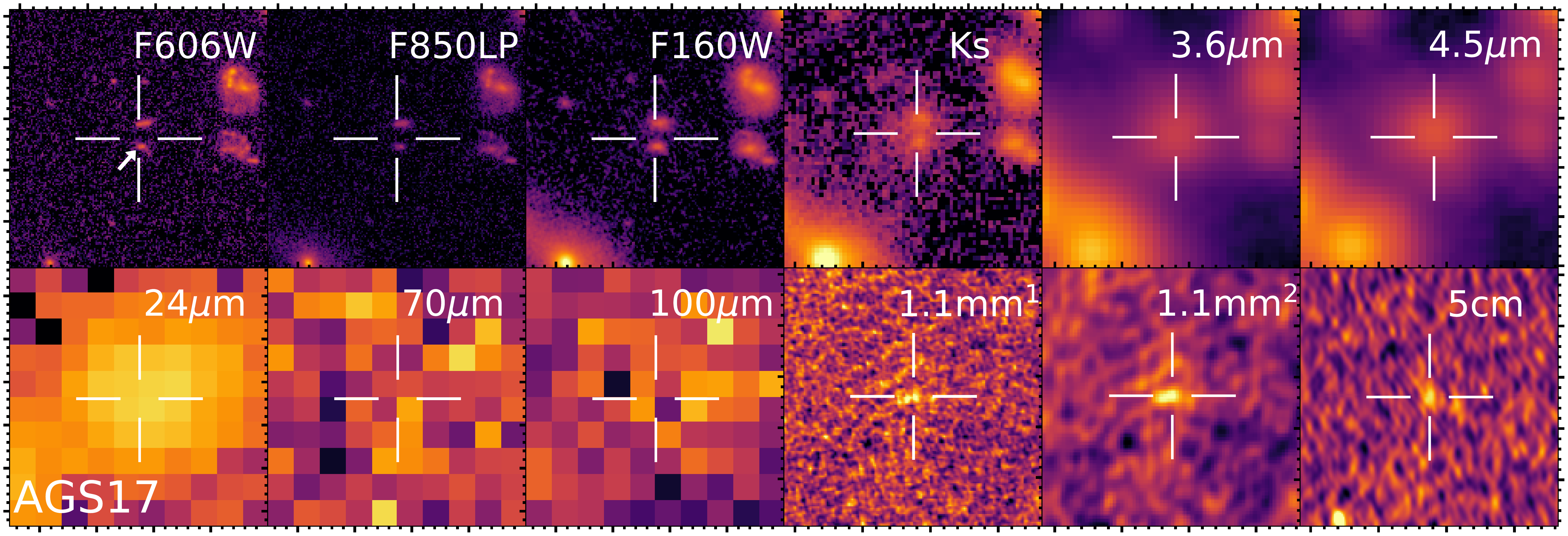}}
\end{minipage}
\caption{Postage stamps of 10\,$\times$\,10 arcsec from HST-WFC3 at 0.606\,$\mu$m to VLA at 5cm, for the four optically-dark galaxies discussed in Sect.~\ref{sec:HST-dark}. For the two ALMA images at 1.1 mm, those marked by $^1$ correspond to the non-tapered images, those marked by $^2$ correspond to the 0\arcsec60-mosaic images. The $K_s$-band thumbnail comes from the super-deep detection image described in Sect.~\ref{sec:Optical/near-infrared_imaging}. All images are centred on the ALMA detection. We indicate with white arrows the position of the previously associated HST counterpart.}
\label{multiwavelength-dark-HST}
\end{figure}

\begin{itemize}

\item AGS4 is a close neighbour of ID$_{CANDELS}$ 8923. AGS4 is the fourth brightest detection in our survey with an SNR greater than 9. The centre of the ALMA detection is located at only 0\arcsec38 from ID$_{CANDELS}$ 8923, its closest neighbouring galaxy. Before astrometric correction, this distance was only 0\arcsec21. This is therefore an example where the astrometric correction moves the ALMA galaxy away from the supposed counterpart. In Fig.~\ref{multiwavelength-dark-HST}, we can clearly see that the ALMA emission is offset from the observed $H$-band galaxy shown by the white arrow in Fig.~\ref{multiwavelength-dark-HST}. This offset could be explained physically, for example, as a region extremely obscured by dust, within the same galaxy, greatly extinguishing the optical rest-frame emission that is revealed by ALMA.
However, for AGS4, a series of clues suggest another explanation for this offset.

\begin{figure}
\centering

\includegraphics[width=8.7cm]{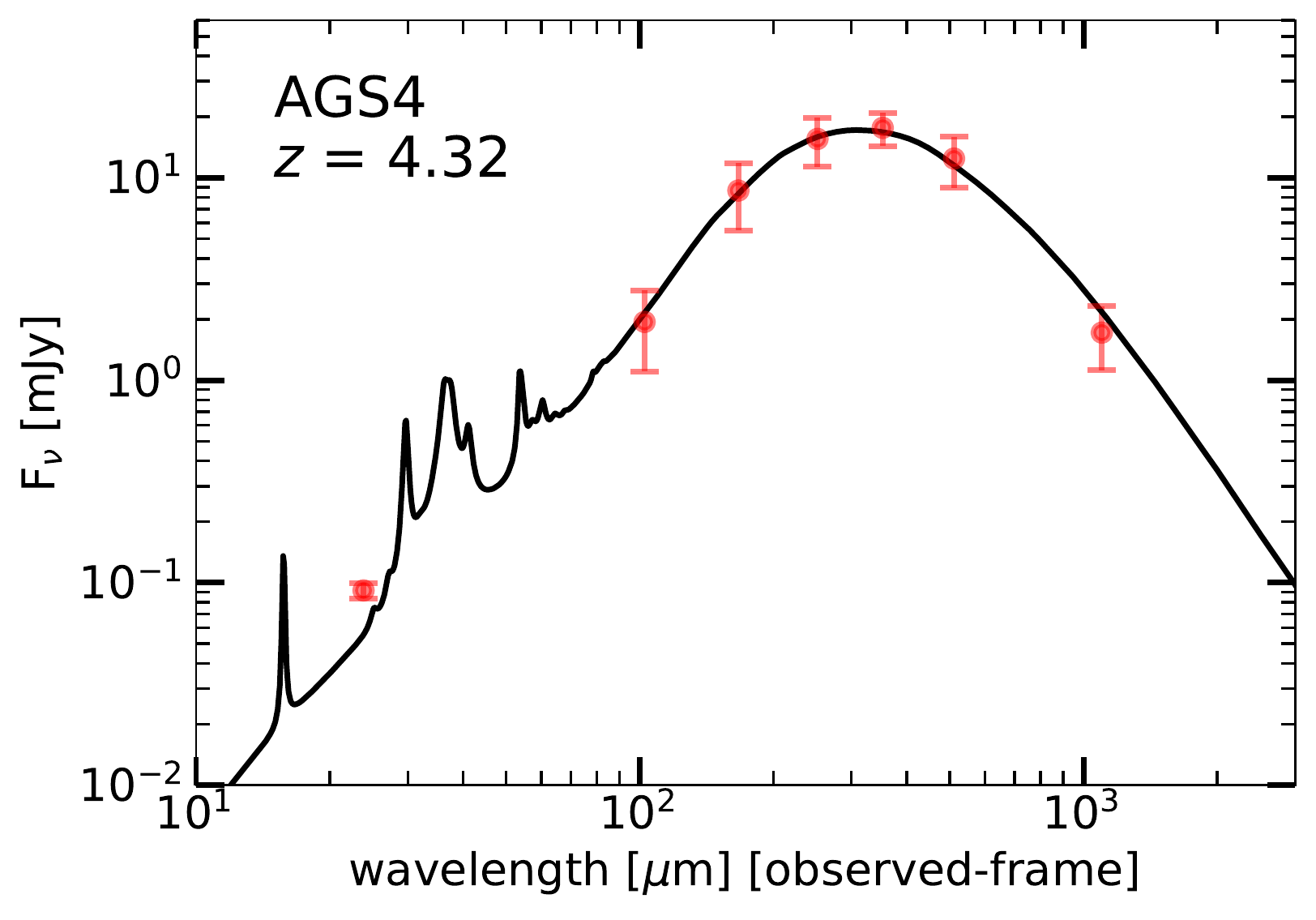}
\includegraphics[width=8.7cm]{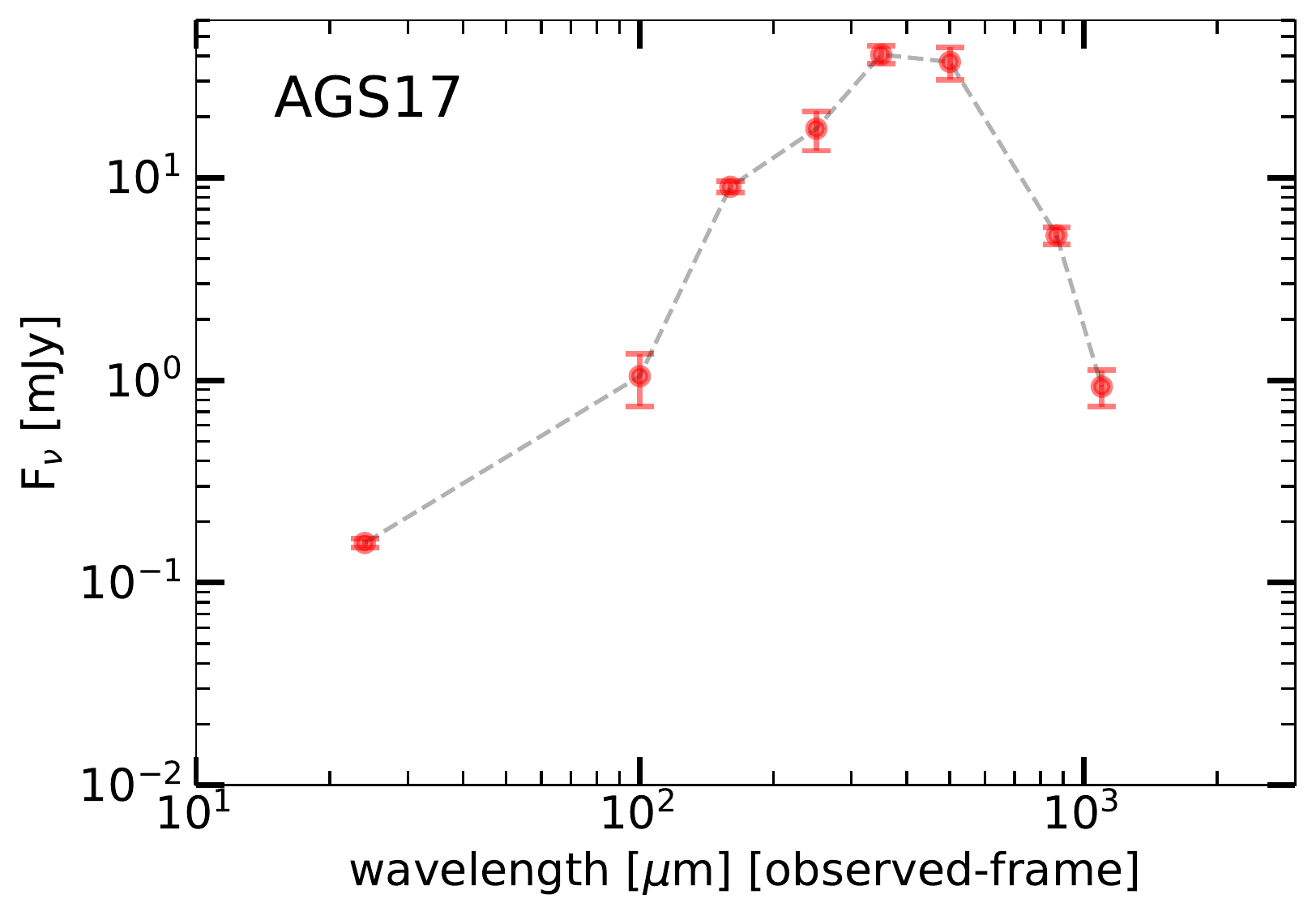}

\caption{Spectral energy distributions (SED) of the two optically-dark galaxies AGS4 and AGS17. The flux densities from 100 to 500\,$\mu$m are from GOODS-\textit{Herschel} \cite{Elbaz2011}. AGS4 (top) ($z$\,=\,4.32, see Sect.~\ref{sec:HST-dark}) is fitted with the model of \cite{Schreiber2017}. The SED of AGS17 (bottom), which has no known redshift, is simply presented with an interpolation between the observed flux densities to illustrate that it peaks around 400\,$\mu$m. This peak is inconsistent with the redshifts of the two optical sources with ID$_{CANDELS}$ 4414 ($z$\,=\,1.85) and 4436 ($z$\,=\,0.92)}
\label{SED}
\end{figure}

\begin{figure}
\centering
\includegraphics[width=\hsize]{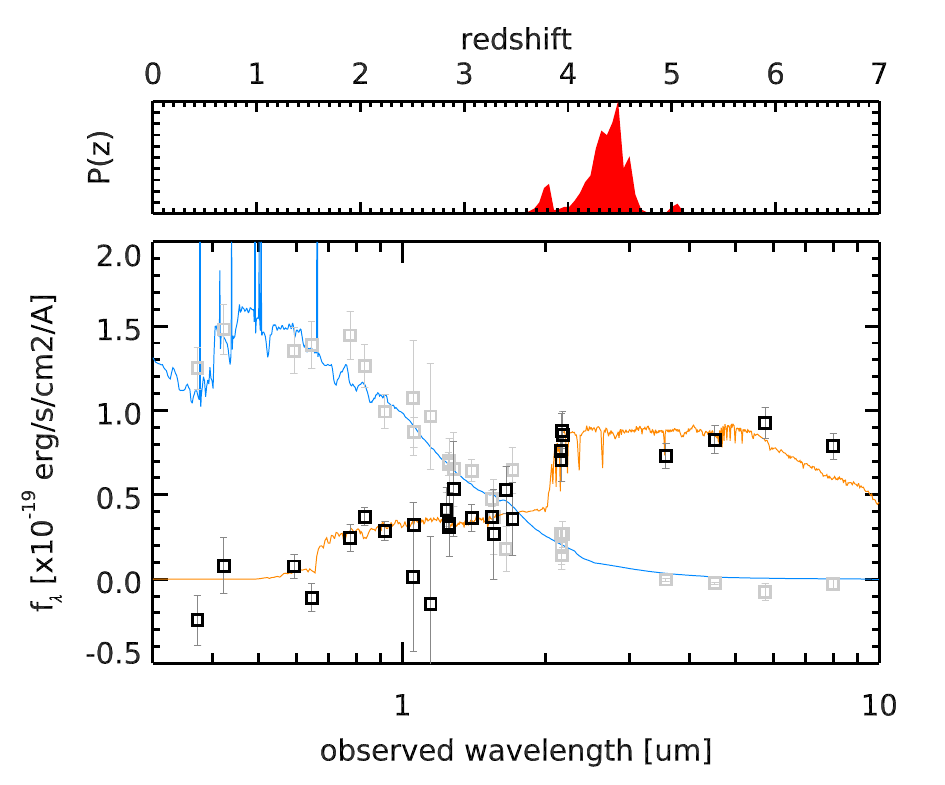}
\caption{Spectral energy distributions of AGS4 and ID$_{CANDELS}$ 8923. Aperture photometry allows the separation between the local galaxy detected by HST (blue, and indicated by a white arrow in Fig.~\ref{multiwavelength-dark-HST}, ID$_{CANDELS}$ 8923) and the distant galaxy detected by ALMA (orange). The top panel shows the photometric redshift probability distribution of AGS4. As the Balmer break is well established in the $K$-band, we consider that this redshift determination is robust, and we adopt the derived redshift $z_{AGS4}\,=\,4.32_{-0.21}^{+0.25}$ and stellar mass (10$^{11.45 \pm 0.2}$ M$_\sun$) values for AGS4.}
\label{AGS4}
\end{figure}

The first piece of evidence is the comparison between the IR SED at the position of the ALMA detection (the SEDs of all of the galaxies detected in this paper will be presented in a future publication, Franco et al., in prep.) and the redshift of the optical galaxy. The redshift of the optical galaxy is $z$\,=\,0.241, whereas the far IR SED peaks around 350\,$\mu$m (see Fig.~\ref{SED}). If AGS4 was a dusty star forming region on the outskirts of 8923, this infrared SED would suggest an abnormally cold dust temperature. It is therefore more probable that AGS4 is not part of 8923, and is a dusty distant galaxy. The fuzzy emission in the $H$-band HST image, exactly centred at the position of the ALMA detection (see Fig.~\ref{multiwavelength-dark-HST}) has not led to any detection in the CANDELS catalogue. In the $V$-band HST images, only ID$_{CANDELS}$ 8923 is present, seen to the South-East of the position of the ALMA detection (indicated by a white cross). No emission is visible at the exact position of the ALMA detection. In the $z$-band, a barely visible detection appears extremely close to the centre of the image.

The second clue is the detection of a galaxy with redshift $z$\,=\,3.76 in the \texttt{FourStar} Galaxy Evolution Survey, 0\arcsec16 from the ALMA detection. This redshift is much more consistent with the peak of the IR SED. The ZFOURGE survey is efficient at detecting galaxies with redshifts between 1 and 4 by using a $K_s$-band detection image (instead of $H$-band as used for the CANDELS survey), and also due to the high spectral resolution ($\lambda/\Delta\lambda \approx$ 10) of the medium-bandwidth filters which provide fine sampling of the Balmer/4000 $\AA$ spectral break at these redshifts \mbox{\citep{Tomczak2016}}. Furthermore, the stellar mass derived in the ZFOURGE catalogue (10$^{10.50}$ M$_\sun$ compared with 10$^{7.64}$ M$_\sun$ in the CANDELS catalogue) is more consistent with the expected mass of galaxies detected by ALMA. Indeed as shown in this paper, and as already shown by \cite{Dunlop2017}, ALMA tends to reveal the most massive dusty galaxies. 

The third piece of evidence is the presence, in the \textit{Spitzer}-CANDELS catalogue \citep{Ashby2015}, of a galaxy detected with the IRAC filers only 0\arcsec1 from our ALMA detection. This IRAC galaxy has a magnitude of 22.51 at 3.6\,$\mu$m, measured within an aperture of 2\arcsec4 radius.

We also note that \cite{Rujopakarn2016} detect a radio galaxy at SNR $\approx$ 17 only 55 mas from the centre of the ALMA detection shown in Fig.~\ref{multiwavelength-dark-HST} (the positional accuracy of this VLA image is 40 mas).
Additionally, AGS4 is detected in two of the three \textit{Chandra} bands: 0.5-7.0 keV (full band; FB) and 0.5-2.0 keV (soft band; SB), but not at 2-7 keV (hard band; HB) from the 7 Ms \textit{Chandra} observations of the GOODS--\textit{South} Field. The integrated X-ray flux is only 6.86\,$\times$\,10$^{40}$ erg.s$^{-1}$, but this galaxy is classified as an AGN in the 7 Ms catalogue.

The detection of a local galaxy at this position has been largely documented \citep[e.g.][]{Hsu2014,Skelton2014,Santini2015}. In contrast, some studies present the galaxy located as this location as a distant galaxy. \cite{Cardamone2010} take advantage of the 18-medium-band photometry from the Subaru telescope and the photometric redshift code \texttt{EAzY} \citep{Brammer2008} to derive a redshift $z$\,=\,3.60. \cite{Wuyts2008} find a redshift of $z$\,=\,3.52 also using \texttt{EAzY}. We can also add a redshift determination by \cite{Rafferty2011} using the Zurich Extragalactic Bayesian Redshift Analyzer (ZEBRA, \citealt{Feldmann2006}), at $z$\,=\,2.92. These determinations of high redshift by independent studies support the existence of a distant galaxy at this position.

Although close, the two sources (ID$_{CANDELS}$ 8923 and 8923b) were successfully de-blended using two light-profile models, determined by fitting the HST $H$-band image with \texttt{Galfit}. The two sources were then fitted simultaneously using these two models on all of the available images, fixing the profile to that observed in the $H$ band. The SEDs of these two galaxies are shown in Fig.~\ref{AGS4}, in blue for the HST galaxy and in orange for the ALMA galaxy, together with the photometric redshift probability distribution for AGS4. The redshifts were estimated using \texttt{EAZY}. For the blue HST galaxy we found $z$\,=\,0.09$_{-0.07}^{+0.06}$, in good agreement with that found by \cite{Skelton2014}. On the other hand, the redshift found for AGS4 is slightly higher than in ZFOURGE, with $z_{AGS4}\,=\,4.32_{-0.21}^{+0.25}$. However, we can also see a secondary peak in the redshift probability distribution, at the position of the ZFOURGE redshift. As the Balmer break is well established in the $K$-band, we consider that the redshift determination ($z_{AGS4}\,=\,4.32_{-0.21}^{+0.25}$) is robust and we adopt this redshift for AGS4. The stellar mass of the ALMA galaxy was then computed with \texttt{FAST} \citep{Kriek2009}, and we found 10$^{11.45 \pm 0.2}$ M$_\sun$ (probably slightly overestimated due to the presence of an AGN, suggested by a flux excess in the IRAC bands). The IR SED of this galaxy is shown in Fig.~\ref{SED}.

For the first time, thanks to ALMA, we can argue that there exists, at this position, not one but two galaxies, close to each other on the line of sight.

\item AGS11 is detected at 1.1 mm with a flux of 1.4\,mJy (S/N$\sim$8) without any counterpart in the deep HST image. However, the galaxy is also detected by IRAC, confirming the existence of a galaxy at this position. 
A galaxy was recently found, for the first time, in the ZFOURGE catalogue at 0\arcsec18 from the centre of the ALMA position. This galaxy was not detected directly in the Magellan image but in a super-deep combined $K_s$-band image at 4.5\,$\sigma$. From this position, the flux in the IRAC-bands have been extracted with SNRs of 26, 34, 8 and 8 at 3.6\,$\mu$m, 4.5\,$\mu$m, 5.8\,$\mu$m and 8.0\,$\mu$m respectively.

This HST-dark galaxy falls in a projected overdensity on the sky, consisting of sources in the redshift range 3.42 $\le$ $z$ $\le$ 3.56 and brighter than $K_s$\,$<$\, 24.9 \citep{Forrest2017}. This density has been computed by \cite{Forrest2017} using the 7$^{th}$ nearest-neighbour technique \citep{Papovich2010}. This overdensity, centred at RA\,=\,53.08$\degree$, DEC\,=\,-27.85$\degree$, extends beyond approximately 1.8 Mpc.

The redshift derived in the ZFOURGE catalogue is $z$\,=\,4.82, making it the farthest galaxy detected in this blind survey. However, we remain cautious regarding this redshift, as this entry has been flagged in the ZFOURGE catalogue (\texttt{use\,=\,0}) due to the SNR of this galaxy (4.7) being below the limit defining galaxies with good photometry (SNR $\ge$ 5). This galaxy is the only galaxy in our catalogue flagged in the ZFOURGE catalogue. For this reason we represent it with an empty circle Fig.~\ref{mass_redshift_distribution}. AGS11 has not been detected in the 7 Ms \textit{Chandra} survey.

The stellar mass, derived in the ZFOURGE catalogue, 3.55\,$\times$\,10$^{10}$ M$_\sun$, is consistent with the masses of all of the other ALMA galaxies found in this survey. What is particularly interesting in the multi-wavelength images of this galaxy is that AGS11 is detectable only by ALMA and in the IRAC-bands (in non-stacked images). Outside of these wavelengths, no emission is detectable.

\item AGS15 is at a distance of 0\arcsec59 from its possible HST counterpart (ID$_{CANDELS}$ 3818) after astrometric correction, corresponding to a physical distance of 4.33 kpc. This is the largest HST-ALMA offset in our entire catalogue. The IRAC position, in contrast, matches much more closely with the ALMA position, with an offset of only 0\arcsec14. The stellar mass of the optical galaxy, given by the ZFOURGE catalogue (7.24\,$\times$\,10$^{9}$ M$_\star$) would have made AGS15 a galaxy lying far from the median stellar mass (1.1\,$\times$\,10$^{11}$ M$_\sun$) of our survey. The redshift of ID$_{CANDELS}$ 3818 ($z$\,=\,3.46) is nevertheless consistent with the other redshifts found in this study.

\item AGS17 is a close neighbour (0\arcsec27) of ID$_{CANDELS}$ 4414 ($z$\,=\,1.85). AGS17 is one of the three galaxies detected by \cite{Hodge2013} at 870$\mu$m in the ALMA field of view (along with AGS8 and AGS15 previously discussed). The counterpart of AGS17 was attributed to ID$_{CANDELS}$ 4414 by \cite{Wiklind2014} with an offset between the ALMA detection and the corresponding F160W object of 0\arcsec32. 
Again, there are indications that the identification may be false:
the peak of the IR SED is $\sim$400\,$\mu$m (see Fig.~\ref{SED}), suggesting a more distant galaxy. 
To be detected with the flux densities reported in Table~\ref{sec:catalogue}, a galaxy at $z$\,=\,1.85 would have an extraordinarily high star formation rate ($\sim$820$\pm$240 M$_\sun$yr$^{-1}$), using the IR SEDs of \cite{Schreiber2017}.
If truly associated with the CANDELS counterpart, this galaxy would be an extreme starburst with an SFR 59$\pm$17 times greater than the SFR$_{MS}$. Galaxies with these properties cannot be ruled out, as galaxies with much higher star formation rates (and offsets from the main sequence) have already been observed \cite[e.g.][]{Pope2005,Fu2013}. However, such objects are relatively rare.
In addition, the stellar mass of ID$_{CANDELS}$ 4414 (10$^{10}$ M$_{\odot}$) is inconsistent with the trend of the other detections (more than one order of magnitude below the median stellar mass of our catalogue).

Another galaxy (ID$_{CANDELS}$ 4436) is relatively close (0\arcsec57) to the ALMA detection. The position of the ALMA detection, which is between ID$_{CANDELS}$ 4414 and ID$_{CANDELS}$ 4436, could be the signature of a major merger occurring between these two galaxies. The emission observed by ALMA could result in this case from the heating of the dust caused by the interaction of these two galaxies, but the redshift determination of 0.92$_{-0.18}^{+0.04}$ by \texttt{Le PHARE} \citep{Arnouts1999,Ilbert2006} dismisses this hypothesis. 

After subtraction of the 2 galaxies close on the line of sight (ID$_{CANDELS}$ 4414 and 4436) in the HAWK-I image, a diffuse source is revealed (half-light radius = 1\arcsec55$\pm$0\arcsec12, sersic index = 1.0). Lower resolution ALMA observations would be needed in order to correctly measure the total sub-mm flux of this extended source.

We also note the position of the IRAC source, located only 0\arcsec06 from the ALMA detection.
\end{itemize}

Of the total 23 detections in this survey, 7 do not show an HST $H$-band counterpart. This lack of counterpart could arise from an occultation of the optical counterpart by a foreground galaxy, faint emission at optical wavelengths, or a spurious ALMA detection.         

For the four galaxies previously discussed (AGS4, AGS11, AGS15 and AGS17), we observe a signal with IRAC at 3.6\,$\mu$m and 4.5\,$\mu$m, despite the limiting sensitivity of IRAC (26 AB mag at 3\,$\sigma$ for both 3.6 and 4.5\,$\mu$m; \citealt{Ashby2015}) being lower than HST-WFC3 (28.16 AB mag at 5\,$\sigma$ for F160W; \citealt{Guo2013}) in the respective images. Furthermore, two of the galaxies (AGS15 and AGS17) have already been detected at submillimetre wavelengths (870\,$\mu$m) by \cite{Hodge2013}.

The other three galaxies (AGS14, AGS16 and AGS19) are not detected at any other wavelength hence there is a high probability that they are spurious. This number is in good agreement with the expected number of spurious sources for our sample (4$\pm$2). 

Fig.~\ref{offset_plot} gives us a glimpse into how sources can be falsely associated with an HST galaxy. When the offset correction is applied, the three galaxies shown with magenta lines move further away from the centre position ($\Delta \delta$\,=\,0, $\Delta \alpha$\,=\,0), rather than closer to it. Another source also appears to show this behaviour: AGS20, seen in the lower left quadrant of Fig.~\ref{offset_plot}. The ALMA detection of AGS20 seems to be clearly offset from a HST galaxy, similar to AGS4. To ensure that there is not a more distant counterpart for AGS20 obscured by the HST source, we performed the same analysis as described in Sect.~\ref{sec:HST-dark} and illustrated in Fig.~\ref{AGS4}. The result of the decomposition suggests that the ALMA and HST sources are either two components of the same galaxy or two galaxies merging at this position. A spectroscopic analysis of AGS20 would allow for distinction between these two possibilities.

The IRAC detections seem to be particularly useful to confirm the existence of a source. In the main catalogue, except for the three galaxies that we consider as spurious, all others are also detected in the IRAC filters.

In conclusion, we have detected 20\% HST--dark galaxies (4 out of 20 robust detections) with a counterpart confirmed at least by IRAC. This proportion may depend in a manner that we cannot address here on the depth of the optical and millimetre images. 
Knowing that these HST--dark galaxies are dust, hence metal, rich they are likely progenitors of the most massive galaxies seen at $z$\,=\,0, hence potentially hosted by massive groups or clusters of galaxies. Two of these HST--dark galaxies have a tentative redshift of $z\,=\,$4.82 and $z$\,=\,3.76, we therefore expect these galaxies to be located on average within $z$$\sim$4--5. These two galaxies are already massive (10$^{10.55}$ M$_\sun$ and 10$^{10.50}$ M$_\sun$ respectively), suggesting that this population of galaxies is particularly interesting for understanding massive galaxy formation during the first billion years after the Big Bang. Spectroscopy with the JWST NIRSpec instrument will permit very sensitive spectroscopic detection of H$\alpha$ emission at $z\,<\,$6.6, and hence an important new tool to measure redshifts of these HST-dark galaxies. GOODS--\textit{South} will undoubtedly be a venue for extensive JWST spectroscopy, including Guaranteed Time Observations. Spectral scan observations with ALMA can also be a powerful tool to determine the distances, and hence physical properties, of this intriguing population of HST--dark galaxies.

\section{Summary and Conclusions}\label{sec:Conclusion}
The GOODS--ALMA survey covers an area of 69 arcmin$^2$ matching the deepest HST--WFC3 coverage of the GOODS--\textit{South} field at 1.1 mm and at a native resolution of $\sim$0\arcsec24. We used a 0\arcsec60 tapered mosaic due to the large number of independent beams at the native resolution. A comparison of the HST source positions with existing catalogues such as Pan-STARRS allowed us to correct the HST astrometry of the GOODS--\textit{South} field from both a global and local offset (equivalent to a distortion map, see also Dickinson et al. in prep.). We find a median offset between the HST and ALMA images of $-$94$\pm$42 mas in right ascension, $\alpha$, and 262$\pm$50 mas in declination, $\delta$. The main conclusions from our study are listed below.
\begin{enumerate}
\item 20 galaxies brighter than 0.7\,mJy at 1.1mm. We detect in total 20 sources above a detection threshold that guarantees an 80\% purity (less than 20\% chance to be spurious). Among these 20 galaxies (with an SNR\,$>$\,4.8), we expect 4$\pm$2 spurious galaxies from the analysis of the inverted map and we identify 3 probably spurious detections with no HST nor \textit{Spitzer}--IRAC counterpart, consistent with the expected number of spurious galaxies. An additional three sources with HST counterparts are detected either at high significance in the higher resolution map, or with different detection-algorithm parameters ensuring a purity greater than 80\%. Hence we identify in total 20 robust detections.

\item Pushing further in redshift the blind detection of massive galaxies with ALMA. The sources exhibit flux densities ranging from 0.6 to 2\,mJy, have a median redshift (and rms) of $z$\,=\,2.92$\pm$0.20 and stellar mass of M$_{\star}$\,=\,(1.1$\pm$0.4)\,$\times$\,10$^{11}$ M$_\sun$. By comparison with deeper but smaller ALMA extragalactic surveys (\citealt{Aravena2016,Dunlop2017,Gonzalez-Lopez2017, Ueda2018}), our redshift distribution is shifted to higher values even though our survey is shallower. This is due to the low surface density of massive, metal hence dust-rich, galaxies at high redshifts. The size of the ALMA survey is therefore a key parameter to detect high redshift galaxies.

\item 20\% HST--dark galaxies. The detection criteria of this main catalogue allowed us to identify sources with no HST counterparts. Out of the 20 galaxies listed above, and excluding the 3 candidate spurious detections, we identified 4 optically-dark or HST--dark galaxies with the request of 80\% purity and with a \textit{Spitzer}--IRAC counterpart at 3.6\,$\mu$m and 4.5\,$\mu$m, confirming the existence of a galaxy at the position of the ALMA detection. It is not the first time that such HST-dark sources have been found using e.g. infrared colour selections ($H$--dropouts, see e.g. \citealt{Wang2016}), but their identification in an unbiased survey at the depth of ALMA in the millimetre range allows us to determine that 20\% of the ALMA sources detected at 1.1mm above $\sim$0.7\,mJy are HST--dark (4 out of 20 sources in the main catalogue). Two of these sources are detected in the near-infrared in the ZFOURGE catalogue, with photometric redshift of $z_{\rm phot}$\,=\,4.32 (derived in this study; AGS4, also detected in the radio with VLA) and 4.82 (AGS11). The other two sources (AGS15 \& AGS17) were detected with the LABOCA ECDFS Submillimetre Survey (LESS) at 870\,$\mu$m and with ALMA after a follow-up at the same wavelength confirming that they were not the result of source blending \citep{Hodge2013}. 

\item Exceptionally high AGN fraction. We find a high proportion of AGNs in our ALMA 1.1mm sample with 40\% (8 out of 20 robust detections) detected in the 7Msec \textit{Chandra} X-ray survey of GOODS--\textit{South} in the 0.5-7.0 keV band with a X-luminosity greater than 10$^{43}$erg.sec$^{-1}$. Limiting our analysis to the ALMA sources with a redshift and stellar mass determination, we find that 50\% of the ALMA sources located in a well-defined stellar mass (M$_{\star}$$\,>\,$3\,$\times$\,10$^{10}$M$_{\sun}$) - redshift ($z$$\sim$1.8--4.5) range host an AGN as compared to 14\% only for the galaxies located within the same zone but undetected by ALMA.
This excess AGN contribution may be due to the fact that the ALMA galaxies are preferentially in a starburst mode due to our detection limit -- hence possibly experiencing a merger -- or/and that the high-resolution of ALMA favours unresolved, hence compact, sources knowing that the mechanism that leads to such compact star-formation may also trigger an AGN.

\item Alleviating the degeneracy of the bright end of the ALMA counts. The differential and cumulative number counts of our 20 primary detections allowed us to partly alleviate the degeneracy observed above 1\,mJy.beam$^{-1}$ in previous (sub)millimetre studies. We show that $\sim$15\% of the extragalactic background light is resolved into individual sources at 0.75\,mJy. By extrapolation, $\sim$50\% of the EBL is resolved at 0.1\,mJy.
\end{enumerate}

\begin{acknowledgements}
We wish to thank Ivo Labb\'{e} for sharing with us the IRAC/GREATS images and Bahram Mobasher for sharing the spectroscopic redshift of AGS4 with us.

This paper makes use of the following ALMA data: ADS/JAO.ALMA\#2015.1.00543.S. ALMA is a partnership of ESO (representing its member states), NSF (USA) and NINS (Japan), together with NRC (Canada), MOST and ASIAA (Taiwan), and KASI (Republic of Korea), in cooperation with the Republic of Chile. The Joint ALMA Observatory is operated by ESO, AUI/NRAO and NAOJ. RD and NN gratefully acknowledge the support provided by the BASAL Center for Astrophysics and Associated Technologies (CATA) through grant PFB-06 Etapa II. Support for BM was provided by the DFG priority program 1573 "The physics of the interstellar medium". DMA thanks the Science and Technology Facilities Council (STFC) for support from grant ST/L00075X/1. WR is supported by JSPS KAKENHI Grant Number JP15K17604, Thailand Research Fund/Office of the Higher Education Commission Grant Number MRG6080294, and Chulalongkorn University's CUniverse. NN acknowledges funding from Fondecyt 1171506.
\end{acknowledgements}

\bibliographystyle{aa}
\bibliography{biblio}

\begin{thebibliography}{158}
\expandafter\ifx\csname natexlab\endcsname\relax\def\natexlab#1{#1}\fi

\bibitem[{{Alexander} {et~al.}(2003){Alexander}, {Bauer}, {Brandt},
  {Hornschemeier}, {Vignali}, {Garmire}, {Schneider}, {Chartas}, \&
  {Gallagher}}]{Alexander2003}
{Alexander}, D.~M., {Bauer}, F.~E., {Brandt}, W.~N., {et~al.} 2003, \aj, 125,
  383

\bibitem[{{Alexander} {et~al.}(2008){Alexander}, {Brandt}, {Smail}, {Swinbank},
  {Bauer}, {Blain}, {Chapman}, {Coppin}, {Ivison}, \&
  {Men{\'e}ndez-Delmestre}}]{Alexander2008}
{Alexander}, D.~M., {Brandt}, W.~N., {Smail}, I., {et~al.} 2008, \aj, 135, 1968

\bibitem[{{Aravena} {et~al.}(2016){Aravena}, {Decarli}, {Walter}, {Da Cunha},
  {Bauer}, {Carilli}, {Daddi}, {Elbaz}, {Ivison}, {Riechers}, {Smail},
  {Swinbank}, {Weiss}, {Anguita}, {Assef}, {Bell}, {Bertoldi}, {Bacon},
  {Bouwens}, {Cortes}, {Cox}, {G{\'o}nzalez-L{\'o}pez}, {Hodge}, {Ibar},
  {Inami}, {Infante}, {Karim}, {Le Le F{\`e}vre}, {Magnelli}, {Ota}, {Popping},
  {Sheth}, {van der Werf}, \& {Wagg}}]{Aravena2016}
{Aravena}, M., {Decarli}, R., {Walter}, F., {et~al.} 2016, \apj, 833, 68

\bibitem[{{Arnouts} {et~al.}(1999){Arnouts}, {Cristiani}, {Moscardini},
  {Matarrese}, {Lucchin}, {Fontana}, \& {Giallongo}}]{Arnouts1999}
{Arnouts}, S., {Cristiani}, S., {Moscardini}, L., {et~al.} 1999, \mnras, 310,
  540

\bibitem[{{Ashby} {et~al.}(2015){Ashby}, {Willner}, {Fazio}, {Dunlop}, {Egami},
  {Faber}, {Ferguson}, {Grogin}, {Hora}, {Huang}, {Koekemoer}, {Labb{\'e}}, \&
  {Wang}}]{Ashby2015}
{Ashby}, M.~L.~N., {Willner}, S.~P., {Fazio}, G.~G., {et~al.} 2015, \apjs, 218,
  33

\bibitem[{{Ashby} {et~al.}(2013){Ashby}, {Willner}, {Fazio}, {Huang}, {Arendt},
  {Barmby}, {Barro}, {Bell}, {Bouwens}, {Cattaneo}, {Croton}, {Dav{\'e}},
  {Dunlop}, {Egami}, {Faber}, {Finlator}, {Grogin}, {Guhathakurta},
  {Hernquist}, {Hora}, {Illingworth}, {Kashlinsky}, {Koekemoer}, {Koo},
  {Labb{\'e}}, {Li}, {Lin}, {Moseley}, {Nandra}, {Newman}, {Noeske}, {Ouchi},
  {Peth}, {Rigopoulou}, {Robertson}, {Sarajedini}, {Simard}, {Smith}, {Wang},
  {Wechsler}, {Weiner}, {Wilson}, {Wuyts}, {Yamada}, \& {Yan}}]{Ashby2013}
{Ashby}, M.~L.~N., {Willner}, S.~P., {Fazio}, G.~G., {et~al.} 2013, \apj, 769,
  80

\bibitem[{{Barger} {et~al.}(1998){Barger}, {Cowie}, {Sanders}, {Fulton},
  {Taniguchi}, {Sato}, {Kawara}, \& {Okuda}}]{Barger1998}
{Barger}, A.~J., {Cowie}, L.~L., {Sanders}, D.~B., {et~al.} 1998, \nat, 394,
  248

\bibitem[{{Barro} {et~al.}(2013){Barro}, {Faber}, {P{\'e}rez-Gonz{\'a}lez},
  {Koo}, {Williams}, {Kocevski}, {Trump}, {Mozena}, {McGrath}, {van der Wel},
  {Wuyts}, {Bell}, {Croton}, {Ceverino}, {Dekel}, {Ashby}, {Cheung},
  {Ferguson}, {Fontana}, {Fang}, {Giavalisco}, {Grogin}, {Guo}, {Hathi},
  {Hopkins}, {Huang}, {Koekemoer}, {Kartaltepe}, {Lee}, {Newman}, {Porter},
  {Primack}, {Ryan}, {Rosario}, {Somerville}, {Salvato}, \& {Hsu}}]{Barro2013}
{Barro}, G., {Faber}, S.~M., {P{\'e}rez-Gonz{\'a}lez}, P.~G., {et~al.} 2013,
  \apj, 765, 104

\bibitem[{{Barro} {et~al.}(2017){Barro}, {Kriek}, {P{\'e}rez-Gonz{\'a}lez},
  {Diaz-Santos}, {Price}, {Rujopakarn}, {Pandya}, {Koo}, {Faber}, {Dekel},
  {Primack}, \& {Kocevski}}]{Barro2017}
{Barro}, G., {Kriek}, M., {P{\'e}rez-Gonz{\'a}lez}, P.~G., {et~al.} 2017,
  \apjl, 851, L40

\bibitem[{{Barro} {et~al.}(2016){Barro}, {Kriek}, {P{\'e}rez-Gonz{\'a}lez},
  {Trump}, {Koo}, {Faber}, {Dekel}, {Primack}, {Guo}, {Kocevski},
  {Mu{\~n}oz-Mateos}, {Rujopakarn}, \& {Seth}}]{Barro2016}
{Barro}, G., {Kriek}, M., {P{\'e}rez-Gonz{\'a}lez}, P.~G., {et~al.} 2016,
  \apjl, 827, L32

\bibitem[{{B{\'e}thermin} {et~al.}(2015){B{\'e}thermin}, {De Breuck},
  {Sargent}, \& {Daddi}}]{Bethermin2015}
{B{\'e}thermin}, M., {De Breuck}, C., {Sargent}, M., \& {Daddi}, E. 2015, \aap,
  576, L9

\bibitem[{{B{\'e}thermin} {et~al.}(2012){B{\'e}thermin}, {Le Floc'h}, {Ilbert},
  {Conley}, {Lagache}, {Amblard}, {Arumugam}, {Aussel}, {Berta}, {Bock},
  {Boselli}, {Buat}, {Casey}, {Castro-Rodr{\'{\i}}guez}, {Cava}, {Clements},
  {Cooray}, {Dowell}, {Eales}, {Farrah}, {Franceschini}, {Glenn}, {Griffin},
  {Hatziminaoglou}, {Heinis}, {Ibar}, {Ivison}, {Kartaltepe}, {Levenson},
  {Magdis}, {Marchetti}, {Marsden}, {Nguyen}, {O'Halloran}, {Oliver}, {Omont},
  {Page}, {Panuzzo}, {Papageorgiou}, {Pearson}, {P{\'e}rez-Fournon}, {Pohlen},
  {Rigopoulou}, {Roseboom}, {Rowan-Robinson}, {Salvato}, {Schulz}, {Scott},
  {Seymour}, {Shupe}, {Smith}, {Symeonidis}, {Trichas}, {Tugwell}, {Vaccari},
  {Valtchanov}, {Vieira}, {Viero}, {Wang}, {Xu}, \& {Zemcov}}]{Bethermin2012}
{B{\'e}thermin}, M., {Le Floc'h}, E., {Ilbert}, O., {et~al.} 2012, \aap, 542,
  A58

\bibitem[{{B{\'e}thermin} {et~al.}(2017){B{\'e}thermin}, {Wu}, {Lagache},
  {Davidzon}, {Ponthieu}, {Cousin}, {Wang}, {Dor{\'e}}, {Daddi}, \&
  {Lapi}}]{Bethermin2017}
{B{\'e}thermin}, M., {Wu}, H.-Y., {Lagache}, G., {et~al.} 2017, \aap, 607, A89

\bibitem[{{Blain} {et~al.}(2002){Blain}, {Smail}, {Ivison}, {Kneib}, \&
  {Frayer}}]{Blain2002}
{Blain}, A.~W., {Smail}, I., {Ivison}, R.~J., {Kneib}, J.-P., \& {Frayer},
  D.~T. 2002, \physrep, 369, 111

\bibitem[{{Bournaud} {et~al.}(2012){Bournaud}, {Juneau}, {Le Floc'h},
  {Mullaney}, {Daddi}, {Dekel}, {Duc}, {Elbaz}, {Salmi}, \&
  {Dickinson}}]{Bournaud2012}
{Bournaud}, F., {Juneau}, S., {Le Floc'h}, E., {et~al.} 2012, \apj, 757, 81

\bibitem[{{Brammer} {et~al.}(2008){Brammer}, {van Dokkum}, \&
  {Coppi}}]{Brammer2008}
{Brammer}, G.~B., {van Dokkum}, P.~G., \& {Coppi}, P. 2008, \apj, 686, 1503

\bibitem[{{Bruzual} \& {Charlot}(2003)}]{Bruzual2003}
{Bruzual}, G. \& {Charlot}, S. 2003, \mnras, 344, 1000

\bibitem[{{Calzetti} {et~al.}(2000){Calzetti}, {Armus}, {Bohlin}, {Kinney},
  {Koornneef}, \& {Storchi-Bergmann}}]{Calzetti2000}
{Calzetti}, D., {Armus}, L., {Bohlin}, R.~C., {et~al.} 2000, \apj, 533, 682

\bibitem[{{Caputi} {et~al.}(2012){Caputi}, {Dunlop}, {McLure}, {Huang},
  {Fazio}, {Ashby}, {Castellano}, {Fontana}, {Cirasuolo}, {Almaini}, {Bell},
  {Dickinson}, {Donley}, {Faber}, {Ferguson}, {Giavalisco}, {Grogin},
  {Kocevski}, {Koekemoer}, {Koo}, {Lai}, {Newman}, \&
  {Somerville}}]{Caputi2012}
{Caputi}, K.~I., {Dunlop}, J.~S., {McLure}, R.~J., {et~al.} 2012, \apjl, 750,
  L20

\bibitem[{{Caputi} {et~al.}(2015){Caputi}, {Ilbert}, {Laigle}, {McCracken}, {Le
  F{\`e}vre}, {Fynbo}, {Milvang-Jensen}, {Capak}, {Salvato}, \&
  {Taniguchi}}]{Caputi2015}
{Caputi}, K.~I., {Ilbert}, O., {Laigle}, C., {et~al.} 2015, \apj, 810, 73

\bibitem[{{Cardamone} {et~al.}(2010){Cardamone}, {van Dokkum}, {Urry},
  {Taniguchi}, {Gawiser}, {Brammer}, {Taylor}, {Damen}, {Treister}, {Cobb},
  {Bond}, {Schawinski}, {Lira}, {Murayama}, {Saito}, \&
  {Sumikawa}}]{Cardamone2010}
{Cardamone}, C.~N., {van Dokkum}, P.~G., {Urry}, C.~M., {et~al.} 2010, \apjs,
  189, 270

\bibitem[{{Carniani} {et~al.}(2015){Carniani}, {Maiolino}, {De Zotti},
  {Negrello}, {Marconi}, {Bothwell}, {Capak}, {Carilli}, {Castellano},
  {Cristiani}, {Ferrara}, {Fontana}, {Gallerani}, {Jones}, {Ohta}, {Ota},
  {Pentericci}, {Santini}, {Sheth}, {Vallini}, {Vanzella}, {Wagg}, \&
  {Williams}}]{Carniani2015}
{Carniani}, S., {Maiolino}, R., {De Zotti}, G., {et~al.} 2015, \aap, 584, A78

\bibitem[{{Casey} {et~al.}(2013){Casey}, {Chen}, {Cowie}, {Barger}, {Capak},
  {Ilbert}, {Koss}, {Lee}, {Le Floc'h}, {Sanders}, \& {Williams}}]{Casey2013}
{Casey}, C.~M., {Chen}, C.-C., {Cowie}, L.~L., {et~al.} 2013, \mnras, 436, 1919

\bibitem[{{Chabrier}(2003)}]{Chabrier2003}
{Chabrier}, G. 2003, \pasp, 115, 763

\bibitem[{{Chang} {et~al.}(2017){Chang}, {Le Floc'h}, {Juneau}, {da Cunha},
  {Salvato}, {Civano}, {Marchesi}, {Ilbert}, {Toba}, {Lim}, {Tang}, {Wang},
  {Ferraro}, {Urry}, {Griffiths}, \& {Kartaltepe}}]{Chang2017}
{Chang}, Y.-Y., {Le Floc'h}, E., {Juneau}, S., {et~al.} 2017, \apjs, 233, 19

\bibitem[{{Chapman} {et~al.}(2003){Chapman}, {Blain}, {Ivison}, \&
  {Smail}}]{Chapman2003}
{Chapman}, S.~C., {Blain}, A.~W., {Ivison}, R.~J., \& {Smail}, I.~R. 2003,
  \nat, 422, 695

\bibitem[{{Chapman} {et~al.}(2005){Chapman}, {Blain}, {Smail}, \&
  {Ivison}}]{Chapman2005}
{Chapman}, S.~C., {Blain}, A.~W., {Smail}, I., \& {Ivison}, R.~J. 2005, \apj,
  622, 772

\bibitem[{{Cimatti} {et~al.}(2008){Cimatti}, {Cassata}, {Pozzetti}, {Kurk},
  {Mignoli}, {Renzini}, {Daddi}, {Bolzonella}, {Brusa}, {Rodighiero},
  {Dickinson}, {Franceschini}, {Zamorani}, {Berta}, {Rosati}, \&
  {Halliday}}]{Cimatti2008}
{Cimatti}, A., {Cassata}, P., {Pozzetti}, L., {et~al.} 2008, \aap, 482, 21

\bibitem[{{Condon}(1997)}]{Condon1997}
{Condon}, J.~J. 1997, \pasp, 109, 166

\bibitem[{{Coppin} {et~al.}(2005){Coppin}, {Halpern}, {Scott}, {Borys}, \&
  {Chapman}}]{Coppin2005}
{Coppin}, K., {Halpern}, M., {Scott}, D., {Borys}, C., \& {Chapman}, S. 2005,
  \mnras, 357, 1022

\bibitem[{{Coppin} {et~al.}(2008){Coppin}, {Halpern}, {Scott}, {Borys},
  {Dunlop}, {Dunne}, {Ivison}, {Wagg}, {Aretxaga}, {Battistelli}, {Benson},
  {Blain}, {Chapman}, {Clements}, {Dye}, {Farrah}, {Hughes}, {Jenness}, {van
  Kampen}, {Lacey}, {Mortier}, {Pope}, {Priddey}, {Serjeant}, {Smail},
  {Stevens}, \& {Vaccari}}]{Coppin2008}
{Coppin}, K., {Halpern}, M., {Scott}, D., {et~al.} 2008, \mnras, 384, 1597

\bibitem[{{Daddi} {et~al.}(2010){Daddi}, {Bournaud}, {Walter}, {Dannerbauer},
  {Carilli}, {Dickinson}, {Elbaz}, {Morrison}, {Riechers}, {Onodera}, {Salmi},
  {Krips}, \& {Stern}}]{Daddi2010}
{Daddi}, E., {Bournaud}, F., {Walter}, F., {et~al.} 2010, \apj, 713, 686

\bibitem[{{Daddi} {et~al.}(2005){Daddi}, {Dickinson}, {Chary}, {Pope},
  {Morrison}, {Alexander}, {Bauer}, {Brandt}, {Giavalisco}, {Ferguson}, {Lee},
  {Lehmer}, {Papovich}, \& {Renzini}}]{Daddi2005}
{Daddi}, E., {Dickinson}, M., {Chary}, R., {et~al.} 2005, \apjl, 631, L13

\bibitem[{{Dole} {et~al.}(2006){Dole}, {Lagache}, {Puget}, {Caputi},
  {Fern{\'a}ndez-Conde}, {Le Floc'h}, {Papovich}, {P{\'e}rez-Gonz{\'a}lez},
  {Rieke}, \& {Blaylock}}]{Dole2006}
{Dole}, H., {Lagache}, G., {Puget}, J.-L., {et~al.} 2006, \aap, 451, 417

\bibitem[{{Dunlop} {et~al.}(2017){Dunlop}, {McLure}, {Biggs}, {Geach},
  {Micha{\l}owski}, {Ivison}, {Rujopakarn}, {van Kampen}, {Kirkpatrick},
  {Pope}, {Scott}, {Swinbank}, {Targett}, {Aretxaga}, {Austermann}, {Best},
  {Bruce}, {Chapin}, {Charlot}, {Cirasuolo}, {Coppin}, {Ellis}, {Finkelstein},
  {Hayward}, {Hughes}, {Ibar}, {Jagannathan}, {Khochfar}, {Koprowski},
  {Narayanan}, {Nyland}, {Papovich}, {Peacock}, {Rieke}, {Robertson},
  {Vernstrom}, {Werf}, {Wilson}, \& {Yun}}]{Dunlop2017}
{Dunlop}, J.~S., {McLure}, R.~J., {Biggs}, A.~D., {et~al.} 2017, \mnras, 466,
  861

\bibitem[{{Eddington}(1913)}]{Eddington1913}
{Eddington}, A.~S. 1913, \mnras, 73, 359

\bibitem[{{Elbaz} {et~al.}(2011){Elbaz}, {Dickinson}, {Hwang},
  {D{\'{\i}}az-Santos}, {Magdis}, {Magnelli}, {Le Borgne}, {Galliano},
  {Pannella}, {Chanial}, {Armus}, {Charmandaris}, {Daddi}, {Aussel}, {Popesso},
  {Kartaltepe}, {Altieri}, {Valtchanov}, {Coia}, {Dannerbauer}, {Dasyra},
  {Leiton}, {Mazzarella}, {Alexander}, {Buat}, {Burgarella}, {Chary}, {Gilli},
  {Ivison}, {Juneau}, {Le Floc'h}, {Lutz}, {Morrison}, {Mullaney}, {Murphy},
  {Pope}, {Scott}, {Brodwin}, {Calzetti}, {Cesarsky}, {Charlot}, {Dole},
  {Eisenhardt}, {Ferguson}, {F{\"o}rster Schreiber}, {Frayer}, {Giavalisco},
  {Huynh}, {Koekemoer}, {Papovich}, {Reddy}, {Surace}, {Teplitz}, {Yun}, \&
  {Wilson}}]{Elbaz2011}
{Elbaz}, D., {Dickinson}, M., {Hwang}, H.~S., {et~al.} 2011, \aap, 533, A119

\bibitem[{{Elbaz} {et~al.}(2017){Elbaz}, {Leiton}, {Nagar}, {Okumura},
  {Franco}, {Schreiber}, {Pannella}, {Wang}, {Dickinson}, {Diaz-Santos},
  {Ciesla}, {Daddi}, {Bournaud}, {Magdis}, {Zhou}, \& {Rujopakarn}}]{Elbaz2017}
{Elbaz}, D., {Leiton}, R., {Nagar}, N., {et~al.} 2017, ArXiv e-prints
  [\eprint[arXiv]{1711.10047}]

\bibitem[{{Fazio} {et~al.}(2004){Fazio}, {Hora}, {Allen}, {Ashby}, {Barmby},
  {Deutsch}, {Huang}, {Kleiner}, {Marengo}, {Megeath}, {Melnick}, {Pahre},
  {Patten}, {Polizotti}, {Smith}, {Taylor}, {Wang}, {Willner}, {Hoffmann},
  {Pipher}, {Forrest}, {McMurty}, {McCreight}, {McKelvey}, {McMurray}, {Koch},
  {Moseley}, {Arendt}, {Mentzell}, {Marx}, {Losch}, {Mayman}, {Eichhorn},
  {Krebs}, {Jhabvala}, {Gezari}, {Fixsen}, {Flores}, {Shakoorzadeh}, {Jungo},
  {Hakun}, {Workman}, {Karpati}, {Kichak}, {Whitley}, {Mann}, {Tollestrup},
  {Eisenhardt}, {Stern}, {Gorjian}, {Bhattacharya}, {Carey}, {Nelson},
  {Glaccum}, {Lacy}, {Lowrance}, {Laine}, {Reach}, {Stauffer}, {Surace},
  {Wilson}, {Wright}, {Hoffman}, {Domingo}, \& {Cohen}}]{Fazio2004}
{Fazio}, G.~G., {Hora}, J.~L., {Allen}, L.~E., {et~al.} 2004, \apjs, 154, 10

\bibitem[{{Feldmann} {et~al.}(2006){Feldmann}, {Carollo}, {Porciani}, {Lilly},
  {Capak}, {Taniguchi}, {Le F{\`e}vre}, {Renzini}, {Scoville}, {Ajiki},
  {Aussel}, {Contini}, {McCracken}, {Mobasher}, {Murayama}, {Sanders},
  {Sasaki}, {Scarlata}, {Scodeggio}, {Shioya}, {Silverman}, {Takahashi},
  {Thompson}, \& {Zamorani}}]{Feldmann2006}
{Feldmann}, R., {Carollo}, C.~M., {Porciani}, C., {et~al.} 2006, \mnras, 372,
  565

\bibitem[{{Fixsen} {et~al.}(1998){Fixsen}, {Dwek}, {Mather}, {Bennett}, \&
  {Shafer}}]{Fixsen1998}
{Fixsen}, D.~J., {Dwek}, E., {Mather}, J.~C., {Bennett}, C.~L., \& {Shafer},
  R.~A. 1998, \apj, 508, 123

\bibitem[{{Flewelling} {et~al.}(2016){Flewelling}, {Magnier}, {Chambers},
  {Heasley}, {Holmberg}, {Huber}, {Sweeney}, {Waters}, {Chen}, {Farrow},
  {Hasinger}, {Henderson}, {Long}, {Metcalfe}, {Nieto-Santisteban}, {Norberg},
  {Saglia}, {Szalay}, {Rest}, {Thakar}, {Tonry}, {Valenti}, {Werner}, {White},
  {Denneau}, {Draper}, {Hodapp}, {Jedicke}, {Kaiser}, {Kudritzki}, {Price},
  {Wainscoat}, {Chastel}, {McClean}, {Postman}, \& {Shiao}}]{Flewelling2016}
{Flewelling}, H.~A., {Magnier}, E.~A., {Chambers}, K.~C., {et~al.} 2016, ArXiv
  e-prints [\eprint[arXiv]{1612.05243}]

\bibitem[{{Fontana} {et~al.}(2014){Fontana}, {Dunlop}, {Paris}, {Targett},
  {Boutsia}, {Castellano}, {Galametz}, {Grazian}, {McLure}, {Merlin},
  {Pentericci}, {Wuyts}, {Almaini}, {Caputi}, {Chary}, {Cirasuolo},
  {Conselice}, {Cooray}, {Daddi}, {Dickinson}, {Faber}, {Fazio}, {Ferguson},
  {Giallongo}, {Giavalisco}, {Grogin}, {Hathi}, {Koekemoer}, {Koo}, {Lucas},
  {Nonino}, {Rix}, {Renzini}, {Rosario}, {Santini}, {Scarlata}, {Sommariva},
  {Stark}, {van der Wel}, {Vanzella}, {Wild}, {Yan}, \&
  {Zibetti}}]{Fontana2014}
{Fontana}, A., {Dunlop}, J.~S., {Paris}, D., {et~al.} 2014, \aap, 570, A11

\bibitem[{{Forrest} {et~al.}(2017){Forrest}, {Tran}, {Broussard}, {Allen},
  {Apfel}, {Cowley}, {Glazebrook}, {Kacprzak}, {Labb{\'e}}, {Nanayakkara},
  {Papovich}, {Quadri}, {Spitler}, {Straatman}, \& {Tomczak}}]{Forrest2017}
{Forrest}, B., {Tran}, K.-V.~H., {Broussard}, A., {et~al.} 2017, \apjl, 838,
  L12

\bibitem[{{Fu} {et~al.}(2013){Fu}, {Cooray}, {Feruglio}, {Ivison}, {Riechers},
  {Gurwell}, {Bussmann}, {Harris}, {Altieri}, {Aussel}, {Baker}, {Bock},
  {Boylan-Kolchin}, {Bridge}, {Calanog}, {Casey}, {Cava}, {Chapman},
  {Clements}, {Conley}, {Cox}, {Farrah}, {Frayer}, {Hopwood}, {Jia}, {Magdis},
  {Marsden}, {Mart{\'{\i}}nez-Navajas}, {Negrello}, {Neri}, {Oliver}, {Omont},
  {Page}, {P{\'e}rez-Fournon}, {Schulz}, {Scott}, {Smith}, {Vaccari},
  {Valtchanov}, {Vieira}, {Viero}, {Wang}, {Wardlow}, \& {Zemcov}}]{Fu2013}
{Fu}, H., {Cooray}, A., {Feruglio}, C., {et~al.} 2013, \nat, 498, 338

\bibitem[{{Fujimoto} {et~al.}(2016){Fujimoto}, {Ouchi}, {Ono}, {Shibuya},
  {Ishigaki}, {Nagai}, \& {Momose}}]{Fujimoto2016}
{Fujimoto}, S., {Ouchi}, M., {Ono}, Y., {et~al.} 2016, \apjs, 222, 1

\bibitem[{{Fujimoto} {et~al.}(2017){Fujimoto}, {Ouchi}, {Shibuya}, \&
  {Nagai}}]{Fujimoto2017}
{Fujimoto}, S., {Ouchi}, M., {Shibuya}, T., \& {Nagai}, H. 2017, \apj, 850, 83

\bibitem[{{Gaia Collaboration} {et~al.}(2016){Gaia Collaboration}, {Brown},
  {Vallenari}, {Prusti}, {de Bruijne}, {Mignard}, {Drimmel}, {Babusiaux},
  {Bailer-Jones}, {Bastian}, \& et~al.}]{Gaia_Collaboration2016}
{Gaia Collaboration}, {Brown}, A.~G.~A., {Vallenari}, A., {et~al.} 2016, \aap,
  595, A2

\bibitem[{{Gatti} {et~al.}(2015){Gatti}, {Lamastra}, {Menci}, {Bongiorno}, \&
  {Fiore}}]{Gatti2015}
{Gatti}, M., {Lamastra}, A., {Menci}, N., {Bongiorno}, A., \& {Fiore}, F. 2015,
  \aap, 576, A32

\bibitem[{{Geach} {et~al.}(2017){Geach}, {Dunlop}, {Halpern}, {Smail}, {van der
  Werf}, {Alexander}, {Almaini}, {Aretxaga}, {Arumugam}, {Asboth}, {Banerji},
  {Beanlands}, {Best}, {Blain}, {Birkinshaw}, {Chapin}, {Chapman}, {Chen},
  {Chrysostomou}, {Clarke}, {Clements}, {Conselice}, {Coppin}, {Cowley},
  {Danielson}, {Eales}, {Edge}, {Farrah}, {Gibb}, {Harrison}, {Hine}, {Hughes},
  {Ivison}, {Jarvis}, {Jenness}, {Jones}, {Karim}, {Koprowski}, {Knudsen},
  {Lacey}, {Mackenzie}, {Marsden}, {McAlpine}, {McMahon}, {Meijerink},
  {Micha{\l}owski}, {Oliver}, {Page}, {Peacock}, {Rigopoulou}, {Robson},
  {Roseboom}, {Rotermund}, {Scott}, {Serjeant}, {Simpson}, {Simpson}, {Smith},
  {Spaans}, {Stanley}, {Stevens}, {Swinbank}, {Targett}, {Thomson}, {Valiante},
  {Wake}, {Webb}, {Willott}, {Zavala}, \& {Zemcov}}]{Geach2017}
{Geach}, J.~E., {Dunlop}, J.~S., {Halpern}, M., {et~al.} 2017, \mnras, 465,
  1789

\bibitem[{{Gonz{\'a}lez-L{\'o}pez} {et~al.}(2017){Gonz{\'a}lez-L{\'o}pez},
  {Bauer}, {Romero-Ca{\~n}izales}, {Kneissl}, {Villard}, {Carvajal}, {Kim},
  {Laporte}, {Anguita}, {Aravena}, {Bouwens}, {Bradley}, {Carrasco}, {Demarco},
  {Ford}, {Ibar}, {Infante}, {Messias}, {Mu{\~n}oz Arancibia}, {Nagar},
  {Padilla}, {Treister}, {Troncoso}, \& {Zitrin}}]{Gonzalez-Lopez2017}
{Gonz{\'a}lez-L{\'o}pez}, J., {Bauer}, F.~E., {Romero-Ca{\~n}izales}, C.,
  {et~al.} 2017, \aap, 597, A41

\bibitem[{{Gordon} {et~al.}(2010){Gordon}, {Galliano}, {Hony}, {Bernard},
  {Bolatto}, {Bot}, {Engelbracht}, {Hughes}, {Israel}, {Kemper}, {Kim}, {Li},
  {Madden}, {Matsuura}, {Meixner}, {Misselt}, {Okumura}, {Panuzzo}, {Rubio},
  {Reach}, {Roman-Duval}, {Sauvage}, {Skibba}, \& {Tielens}}]{Gordon2010}
{Gordon}, K.~D., {Galliano}, F., {Hony}, S., {et~al.} 2010, \aap, 518, L89

\bibitem[{{Griffin} {et~al.}(2010){Griffin}, {Abergel}, {Abreu}, {Ade},
  {Andr{\'e}}, {Augueres}, {Babbedge}, {Bae}, {Baillie}, {Baluteau}, {Barlow},
  {Bendo}, {Benielli}, {Bock}, {Bonhomme}, {Brisbin}, {Brockley-Blatt},
  {Caldwell}, {Cara}, {Castro-Rodriguez}, {Cerulli}, {Chanial}, {Chen},
  {Clark}, {Clements}, {Clerc}, {Coker}, {Communal}, {Conversi}, {Cox},
  {Crumb}, {Cunningham}, {Daly}, {Davis}, {de Antoni}, {Delderfield}, {Devin},
  {di Giorgio}, {Didschuns}, {Dohlen}, {Donati}, {Dowell}, {Dowell}, {Duband},
  {Dumaye}, {Emery}, {Ferlet}, {Ferrand}, {Fontignie}, {Fox}, {Franceschini},
  {Frerking}, {Fulton}, {Garcia}, {Gastaud}, {Gear}, {Glenn}, {Goizel},
  {Griffin}, {Grundy}, {Guest}, {Guillemet}, {Hargrave}, {Harwit}, {Hastings},
  {Hatziminaoglou}, {Herman}, {Hinde}, {Hristov}, {Huang}, {Imhof}, {Isaak},
  {Israelsson}, {Ivison}, {Jennings}, {Kiernan}, {King}, {Lange}, {Latter},
  {Laurent}, {Laurent}, {Leeks}, {Lellouch}, {Levenson}, {Li}, {Li},
  {Lilienthal}, {Lim}, {Liu}, {Lu}, {Madden}, {Mainetti}, {Marliani}, {McKay},
  {Mercier}, {Molinari}, {Morris}, {Moseley}, {Mulder}, {Mur}, {Naylor},
  {Nguyen}, {O'Halloran}, {Oliver}, {Olofsson}, {Olofsson}, {Orfei}, {Page},
  {Pain}, {Panuzzo}, {Papageorgiou}, {Parks}, {Parr-Burman}, {Pearce},
  {Pearson}, {P{\'e}rez-Fournon}, {Pinsard}, {Pisano}, {Podosek}, {Pohlen},
  {Polehampton}, {Pouliquen}, {Rigopoulou}, {Rizzo}, {Roseboom}, {Roussel},
  {Rowan-Robinson}, {Rownd}, {Saraceno}, {Sauvage}, {Savage}, {Savini},
  {Sawyer}, {Scharmberg}, {Schmitt}, {Schneider}, {Schulz}, {Schwartz},
  {Shafer}, {Shupe}, {Sibthorpe}, {Sidher}, {Smith}, {Smith}, {Smith},
  {Spencer}, {Stobie}, {Sudiwala}, {Sukhatme}, {Surace}, {Stevens}, {Swinyard},
  {Trichas}, {Tourette}, {Triou}, {Tseng}, {Tucker}, {Turner}, {Vaccari},
  {Valtchanov}, {Vigroux}, {Virique}, {Voellmer}, {Walker}, {Ward}, {Waskett},
  {Weilert}, {Wesson}, {White}, {Whitehouse}, {Wilson}, {Winter}, {Woodcraft},
  {Wright}, {Xu}, {Zavagno}, {Zemcov}, {Zhang}, \& {Zonca}}]{Griffin2010}
{Griffin}, M.~J., {Abergel}, A., {Abreu}, A., {et~al.} 2010, \aap, 518, L3

\bibitem[{{Grogin} {et~al.}(2011){Grogin}, {Kocevski}, {Faber}, {Ferguson},
  {Koekemoer}, {Riess}, {Acquaviva}, {Alexander}, {Almaini}, {Ashby}, {Barden},
  {Bell}, {Bournaud}, {Brown}, {Caputi}, {Casertano}, {Cassata}, {Castellano},
  {Challis}, {Chary}, {Cheung}, {Cirasuolo}, {Conselice}, {Roshan Cooray},
  {Croton}, {Daddi}, {Dahlen}, {Dav{\'e}}, {de Mello}, {Dekel}, {Dickinson},
  {Dolch}, {Donley}, {Dunlop}, {Dutton}, {Elbaz}, {Fazio}, {Filippenko},
  {Finkelstein}, {Fontana}, {Gardner}, {Garnavich}, {Gawiser}, {Giavalisco},
  {Grazian}, {Guo}, {Hathi}, {H{\"a}ussler}, {Hopkins}, {Huang}, {Huang},
  {Jha}, {Kartaltepe}, {Kirshner}, {Koo}, {Lai}, {Lee}, {Li}, {Lotz}, {Lucas},
  {Madau}, {McCarthy}, {McGrath}, {McIntosh}, {McLure}, {Mobasher},
  {Moustakas}, {Mozena}, {Nandra}, {Newman}, {Niemi}, {Noeske}, {Papovich},
  {Pentericci}, {Pope}, {Primack}, {Rajan}, {Ravindranath}, {Reddy}, {Renzini},
  {Rix}, {Robaina}, {Rodney}, {Rosario}, {Rosati}, {Salimbeni}, {Scarlata},
  {Siana}, {Simard}, {Smidt}, {Somerville}, {Spinrad}, {Straughn}, {Strolger},
  {Telford}, {Teplitz}, {Trump}, {van der Wel}, {Villforth}, {Wechsler},
  {Weiner}, {Wiklind}, {Wild}, {Wilson}, {Wuyts}, {Yan}, \& {Yun}}]{Grogin2011}
{Grogin}, N.~A., {Kocevski}, D.~D., {Faber}, S.~M., {et~al.} 2011, \apjs, 197,
  35

\bibitem[{{Guo} {et~al.}(2013){Guo}, {Ferguson}, {Giavalisco}, {Barro},
  {Willner}, {Ashby}, {Dahlen}, {Donley}, {Faber}, {Fontana}, {Galametz},
  {Grazian}, {Huang}, {Kocevski}, {Koekemoer}, {Koo}, {McGrath}, {Peth},
  {Salvato}, {Wuyts}, {Castellano}, {Cooray}, {Dickinson}, {Dunlop}, {Fazio},
  {Gardner}, {Gawiser}, {Grogin}, {Hathi}, {Hsu}, {Lee}, {Lucas}, {Mobasher},
  {Nandra}, {Newman}, \& {van der Wel}}]{Guo2013}
{Guo}, Y., {Ferguson}, H.~C., {Giavalisco}, M., {et~al.} 2013, \apjs, 207, 24

\bibitem[{{Hainline} {et~al.}(2011){Hainline}, {Blain}, {Smail}, {Alexander},
  {Armus}, {Chapman}, \& {Ivison}}]{Hainline2011}
{Hainline}, L.~J., {Blain}, A.~W., {Smail}, I., {et~al.} 2011, \apj, 740, 96

\bibitem[{{Hales} {et~al.}(2012){Hales}, {Murphy}, {Curran}, {Middelberg},
  {Gaensler}, \& {Norris}}]{Hales2012}
{Hales}, C.~A., {Murphy}, T., {Curran}, J.~R., {et~al.} 2012, {BLOBCAT:
  Software to Catalog Blobs}, AstrophyFsics Source Code Library

\bibitem[{{Hao} {et~al.}(2006){Hao}, {Mao}, {Deng}, {Xia}, \& {Wu}}]{Hao2006}
{Hao}, C.~N., {Mao}, S., {Deng}, Z.~G., {Xia}, X.~Y., \& {Wu}, H. 2006, \mnras,
  370, 1339

\bibitem[{{Hatsukade} {et~al.}(2016){Hatsukade}, {Kohno}, {Umehata},
  {Aretxaga}, {Caputi}, {Dunlop}, {Ikarashi}, {Iono}, {Ivison}, {Lee},
  {Makiya}, {Matsuda}, {Motohara}, {Nakanishi}, {Ohta}, {Tadaki}, {Tamura},
  {Wang}, {Wilson}, {Yamaguchi}, \& {Yun}}]{Hatsukade2016}
{Hatsukade}, B., {Kohno}, K., {Umehata}, H., {et~al.} 2016, \pasj, 68, 36

\bibitem[{{Hatsukade} {et~al.}(2013){Hatsukade}, {Ohta}, {Seko}, {Yabe}, \&
  {Akiyama}}]{Hatsukade2013}
{Hatsukade}, B., {Ohta}, K., {Seko}, A., {Yabe}, K., \& {Akiyama}, M. 2013,
  \apjl, 769, L27

\bibitem[{{Hodge} {et~al.}(2013){Hodge}, {Karim}, {Smail}, {Swinbank},
  {Walter}, {Biggs}, {Ivison}, {Weiss}, {Alexander}, {Bertoldi}, {Brandt},
  {Chapman}, {Coppin}, {Cox}, {Danielson}, {Dannerbauer}, {De Breuck},
  {Decarli}, {Edge}, {Greve}, {Knudsen}, {Menten}, {Rix}, {Schinnerer},
  {Simpson}, {Wardlow}, \& {van der Werf}}]{Hodge2013}
{Hodge}, J.~A., {Karim}, A., {Smail}, I., {et~al.} 2013, \apj, 768, 91

\bibitem[{{Hodge} {et~al.}(2016){Hodge}, {Swinbank}, {Simpson}, {Smail},
  {Walter}, {Alexander}, {Bertoldi}, {Biggs}, {Brandt}, {Chapman}, {Chen},
  {Coppin}, {Cox}, {Dannerbauer}, {Edge}, {Greve}, {Ivison}, {Karim},
  {Knudsen}, {Menten}, {Rix}, {Schinnerer}, {Wardlow}, {Weiss}, \& {van der
  Werf}}]{Hodge2016}
{Hodge}, J.~A., {Swinbank}, A.~M., {Simpson}, J.~M., {et~al.} 2016, \apj, 833,
  103

\bibitem[{{Hogg} \& {Turner}(1998)}]{Hogg1998}
{Hogg}, D.~W. \& {Turner}, E.~L. 1998, \pasp, 110, 727

\bibitem[{{Holland} {et~al.}(1999){Holland}, {Robson}, {Gear}, {Cunningham},
  {Lightfoot}, {Jenness}, {Ivison}, {Stevens}, {Ade}, {Griffin}, {Duncan},
  {Murphy}, \& {Naylor}}]{Holland1999}
{Holland}, W.~S., {Robson}, E.~I., {Gear}, W.~K., {et~al.} 1999, \mnras, 303,
  659

\bibitem[{{Hsieh} {et~al.}(2012){Hsieh}, {Wang}, {Hsieh}, {Lin}, {Yan}, {Lim},
  \& {Ho}}]{Hsieh2012}
{Hsieh}, B.-C., {Wang}, W.-H., {Hsieh}, C.-C., {et~al.} 2012, \apjs, 203, 23

\bibitem[{{Hsu} {et~al.}(2014){Hsu}, {Salvato}, {Nandra}, {Brusa}, {Bender},
  {Buchner}, {Donley}, {Kocevski}, {Guo}, {Hathi}, {Rangel}, {Willner},
  {Brightman}, {Georgakakis}, {Budav{\'a}ri}, {Szalay}, {Ashby}, {Barro},
  {Dahlen}, {Faber}, {Ferguson}, {Galametz}, {Grazian}, {Grogin}, {Huang},
  {Koekemoer}, {Lucas}, {McGrath}, {Mobasher}, {Peth}, {Rosario}, \&
  {Trump}}]{Hsu2014}
{Hsu}, L.-T., {Salvato}, M., {Nandra}, K., {et~al.} 2014, \apj, 796, 60

\bibitem[{{Huang} {et~al.}(2011){Huang}, {Zheng}, {Rigopoulou}, {Magdis},
  {Fazio}, \& {Wang}}]{Huang2012}
{Huang}, J.-S., {Zheng}, X.~Z., {Rigopoulou}, D., {et~al.} 2011, \apjl, 742,
  L13

\bibitem[{{Hughes} {et~al.}(1998){Hughes}, {Serjeant}, {Dunlop},
  {Rowan-Robinson}, {Blain}, {Mann}, {Ivison}, {Peacock}, {Efstathiou}, {Gear},
  {Oliver}, {Lawrence}, {Longair}, {Goldschmidt}, \& {Jenness}}]{Hughes1998}
{Hughes}, D.~H., {Serjeant}, S., {Dunlop}, J., {et~al.} 1998, \nat, 394, 241

\bibitem[{{Ikarashi} {et~al.}(2017){Ikarashi}, {Caputi}, {Ohta}, {Ivison},
  {Lagos}, {Bisigello}, {Hatsukade}, {Aretxaga}, {Dunlop}, {Hughes}, {Iono},
  {Izumi}, {Kashikawa}, {Koyama}, {Kawabe}, {Kohno}, {Motohara}, {Nakanishi},
  {Tamura}, {Umehata}, {Wilson}, {Yabe}, \& {Yun}}]{Ikarashi2017}
{Ikarashi}, S., {Caputi}, K.~I., {Ohta}, K., {et~al.} 2017, \apjl, 849, L36

\bibitem[{{Ikarashi} {et~al.}(2015){Ikarashi}, {Ivison}, {Caputi}, {Aretxaga},
  {Dunlop}, {Hatsukade}, {Hughes}, {Iono}, {Izumi}, {Kawabe}, {Kohno}, {Lagos},
  {Motohara}, {Nakanishi}, {Ohta}, {Tamura}, {Umehata}, {Wilson}, {Yabe}, \&
  {Yun}}]{Ikarashi2015}
{Ikarashi}, S., {Ivison}, R.~J., {Caputi}, K.~I., {et~al.} 2015, \apj, 810, 133

\bibitem[{{Ilbert} {et~al.}(2006){Ilbert}, {Arnouts}, {McCracken},
  {Bolzonella}, {Bertin}, {Le F{\`e}vre}, {Mellier}, {Zamorani}, {Pell{\`o}},
  {Iovino}, {Tresse}, {Le Brun}, {Bottini}, {Garilli}, {Maccagni}, {Picat},
  {Scaramella}, {Scodeggio}, {Vettolani}, {Zanichelli}, {Adami}, {Bardelli},
  {Cappi}, {Charlot}, {Ciliegi}, {Contini}, {Cucciati}, {Foucaud}, {Franzetti},
  {Gavignaud}, {Guzzo}, {Marano}, {Marinoni}, {Mazure}, {Meneux}, {Merighi},
  {Paltani}, {Pollo}, {Pozzetti}, {Radovich}, {Zucca}, {Bondi}, {Bongiorno},
  {Busarello}, {de La Torre}, {Gregorini}, {Lamareille}, {Mathez}, {Merluzzi},
  {Ripepi}, {Rizzo}, \& {Vergani}}]{Ilbert2006}
{Ilbert}, O., {Arnouts}, S., {McCracken}, H.~J., {et~al.} 2006, \aap, 457, 841

\bibitem[{{Inami} {et~al.}(2017){Inami}, {Bacon}, {Brinchmann}, {Richard},
  {Contini}, {Conseil}, {Hamer}, {Akhlaghi}, {Bouch{\'e}}, {Cl{\'e}ment},
  {Desprez}, {Drake}, {Hashimoto}, {Leclercq}, {Maseda}, {Michel-Dansac},
  {Paalvast}, {Tresse}, {Ventou}, {Kollatschny}, {Boogaard}, {Finley},
  {Marino}, {Schaye}, \& {Wisotzki}}]{Inami2017}
{Inami}, H., {Bacon}, R., {Brinchmann}, J., {et~al.} 2017, \aap, 608, A2

\bibitem[{{Karim} {et~al.}(2013){Karim}, {Swinbank}, {Hodge}, {Smail},
  {Walter}, {Biggs}, {Simpson}, {Danielson}, {Alexander}, {Bertoldi}, {de
  Breuck}, {Chapman}, {Coppin}, {Dannerbauer}, {Edge}, {Greve}, {Ivison},
  {Knudsen}, {Menten}, {Schinnerer}, {Wardlow}, {Wei{\ss}}, \& {van der
  Werf}}]{Karim2013}
{Karim}, A., {Swinbank}, A.~M., {Hodge}, J.~A., {et~al.} 2013, \mnras, 432, 2

\bibitem[{{Knudsen} {et~al.}(2008){Knudsen}, {van der Werf}, \&
  {Kneib}}]{Knudsen2008}
{Knudsen}, K.~K., {van der Werf}, P.~P., \& {Kneib}, J.-P. 2008, \mnras, 384,
  1611

\bibitem[{{Kocevski} {et~al.}(2017){Kocevski}, {Barro}, {Faber}, {Dekel},
  {Somerville}, {Young}, {Williams}, {McIntosh}, {Georgakakis}, {Hasinger},
  {Nandra}, {Civano}, {Alexander}, {Almaini}, {Conselice}, {Donley},
  {Ferguson}, {Giavalisco}, {Grogin}, {Hathi}, {Hawkins}, {Koekemoer}, {Koo},
  {McGrath}, {Mobasher}, {P{\'e}rez Gonz{\'a}lez}, {Pforr}, {Primack},
  {Santini}, {Stefanon}, {Trump}, {van der Wel}, {Wuyts}, \&
  {Yan}}]{Kocevski2017}
{Kocevski}, D.~D., {Barro}, G., {Faber}, S.~M., {et~al.} 2017, \apj, 846, 112

\bibitem[{{Koekemoer} {et~al.}(2011){Koekemoer}, {Faber}, {Ferguson}, {Grogin},
  {Kocevski}, {Koo}, {Lai}, {Lotz}, {Lucas}, {McGrath}, {Ogaz}, {Rajan},
  {Riess}, {Rodney}, {Strolger}, {Casertano}, {Castellano}, {Dahlen},
  {Dickinson}, {Dolch}, {Fontana}, {Giavalisco}, {Grazian}, {Guo}, {Hathi},
  {Huang}, {van der Wel}, {Yan}, {Acquaviva}, {Alexander}, {Almaini}, {Ashby},
  {Barden}, {Bell}, {Bournaud}, {Brown}, {Caputi}, {Cassata}, {Challis},
  {Chary}, {Cheung}, {Cirasuolo}, {Conselice}, {Roshan Cooray}, {Croton},
  {Daddi}, {Dav{\'e}}, {de Mello}, {de Ravel}, {Dekel}, {Donley}, {Dunlop},
  {Dutton}, {Elbaz}, {Fazio}, {Filippenko}, {Finkelstein}, {Frazer}, {Gardner},
  {Garnavich}, {Gawiser}, {Gruetzbauch}, {Hartley}, {H{\"a}ussler},
  {Herrington}, {Hopkins}, {Huang}, {Jha}, {Johnson}, {Kartaltepe},
  {Khostovan}, {Kirshner}, {Lani}, {Lee}, {Li}, {Madau}, {McCarthy},
  {McIntosh}, {McLure}, {McPartland}, {Mobasher}, {Moreira}, {Mortlock},
  {Moustakas}, {Mozena}, {Nandra}, {Newman}, {Nielsen}, {Niemi}, {Noeske},
  {Papovich}, {Pentericci}, {Pope}, {Primack}, {Ravindranath}, {Reddy},
  {Renzini}, {Rix}, {Robaina}, {Rosario}, {Rosati}, {Salimbeni}, {Scarlata},
  {Siana}, {Simard}, {Smidt}, {Snyder}, {Somerville}, {Spinrad}, {Straughn},
  {Telford}, {Teplitz}, {Trump}, {Vargas}, {Villforth}, {Wagner}, {Wandro},
  {Wechsler}, {Weiner}, {Wiklind}, {Wild}, {Wilson}, {Wuyts}, \&
  {Yun}}]{Koekemoer2011}
{Koekemoer}, A.~M., {Faber}, S.~M., {Ferguson}, H.~C., {et~al.} 2011, \apjs,
  197, 36

\bibitem[{{Kov{\'a}cs} {et~al.}(2006){Kov{\'a}cs}, {Chapman}, {Dowell},
  {Blain}, {Ivison}, {Smail}, \& {Phillips}}]{Kovacs2006}
{Kov{\'a}cs}, A., {Chapman}, S.~C., {Dowell}, C.~D., {et~al.} 2006, \apj, 650,
  592

\bibitem[{{Kriek} {et~al.}(2008){Kriek}, {van Dokkum}, {Franx}, {Illingworth},
  {Marchesini}, {Quadri}, {Rudnick}, {Taylor}, {F{\"o}rster Schreiber},
  {Gawiser}, {Labb{\'e}}, {Lira}, \& {Wuyts}}]{Kriek2008}
{Kriek}, M., {van Dokkum}, P.~G., {Franx}, M., {et~al.} 2008, \apj, 677, 219

\bibitem[{{Kriek} {et~al.}(2009){Kriek}, {van Dokkum}, {Labb{\'e}}, {Franx},
  {Illingworth}, {Marchesini}, \& {Quadri}}]{Kriek2009}
{Kriek}, M., {van Dokkum}, P.~G., {Labb{\'e}}, I., {et~al.} 2009, \apj, 700,
  221

\bibitem[{{Kurk} {et~al.}(2013){Kurk}, {Cimatti}, {Daddi}, {Mignoli},
  {Pozzetti}, {Dickinson}, {Bolzonella}, {Zamorani}, {Cassata}, {Rodighiero},
  {Franceschini}, {Renzini}, {Rosati}, {Halliday}, \& {Berta}}]{Kurk2013}
{Kurk}, J., {Cimatti}, A., {Daddi}, E., {et~al.} 2013, \aap, 549, A63

\bibitem[{{Labb{\'e}} {et~al.}(2015){Labb{\'e}}, {Oesch}, {Illingworth}, {van
  Dokkum}, {Bouwens}, {Franx}, {Carollo}, {Trenti}, {Holden}, {Smit},
  {Gonz{\'a}lez}, {Magee}, {Stiavelli}, \& {Stefanon}}]{Labbe2015}
{Labb{\'e}}, I., {Oesch}, P.~A., {Illingworth}, G.~D., {et~al.} 2015, \apjs,
  221, 23

\bibitem[{{Lamastra} {et~al.}(2013){Lamastra}, {Menci}, {Fiore}, {Santini},
  {Bongiorno}, \& {Piconcelli}}]{Lamastra2013}
{Lamastra}, A., {Menci}, N., {Fiore}, F., {et~al.} 2013, \aap, 559, A56

\bibitem[{{Laporte} {et~al.}(2017){Laporte}, {Bauer}, {Troncoso-Iribarren},
  {Huang}, {Gonz{\'a}lez-L{\'o}pez}, {Kim}, {Anguita}, {Aravena}, {Barrientos},
  {Bouwens}, {Bradley}, {Brammer}, {Carrasco}, {Carvajal}, {Coe}, {Demarco},
  {Ellis}, {Ford}, {Francke}, {Ibar}, {Infante}, {Kneissl}, {Koekemoer},
  {Messias}, {Mu{\~n}oz Arancibia}, {Nagar}, {Padilla}, {Pell{\'o}}, {Postman},
  {Qu{\'e}nard}, {Romero-Ca{\~n}izales}, {Treister}, {Villard}, {Zheng}, \&
  {Zitrin}}]{Laporte2017}
{Laporte}, N., {Bauer}, F.~E., {Troncoso-Iribarren}, P., {et~al.} 2017, \aap,
  604, A132

\bibitem[{{Li} {et~al.}(2016){Li}, {Zheng}, {Gu}, {Wang}, {Wen}, {Guo}, \&
  {An}}]{Li2016}
{Li}, Y., {Zheng}, X.~Z., {Gu}, Q.-S., {et~al.} 2016, \aj, 152, 201

\bibitem[{{Lindner} {et~al.}(2011){Lindner}, {Baker}, {Omont}, {Beelen},
  {Owen}, {Bertoldi}, {Dole}, {Fiolet}, {Harris}, {Ivison}, {Lonsdale}, {Lutz},
  \& {Polletta}}]{Lindner2011}
{Lindner}, R.~R., {Baker}, A.~J., {Omont}, A., {et~al.} 2011, \apj, 737, 83

\bibitem[{{Luo} {et~al.}(2017){Luo}, {Brandt}, {Xue}, {Lehmer}, {Alexander},
  {Bauer}, {Vito}, {Yang}, {Basu-Zych}, {Comastri}, {Gilli}, {Gu},
  {Hornschemeier}, {Koekemoer}, {Liu}, {Mainieri}, {Paolillo}, {Ranalli},
  {Rosati}, {Schneider}, {Shemmer}, {Smail}, {Sun}, {Tozzi}, {Vignali}, \&
  {Wang}}]{Luo2017}
{Luo}, B., {Brandt}, W.~N., {Xue}, Y.~Q., {et~al.} 2017, \apjs, 228, 2

\bibitem[{{Lutz} {et~al.}(2011){Lutz}, {Poglitsch}, {Altieri}, {Andreani},
  {Aussel}, {Berta}, {Bongiovanni}, {Brisbin}, {Cava}, {Cepa}, {Cimatti},
  {Daddi}, {Dominguez-Sanchez}, {Elbaz}, {F{\"o}rster Schreiber}, {Genzel},
  {Grazian}, {Gruppioni}, {Harwit}, {Le Floc'h}, {Magdis}, {Magnelli},
  {Maiolino}, {Nordon}, {P{\'e}rez Garc{\'{\i}}a}, {Popesso}, {Pozzi},
  {Riguccini}, {Rodighiero}, {Saintonge}, {Sanchez Portal}, {Santini}, {Shao},
  {Sturm}, {Tacconi}, {Valtchanov}, {Wetzstein}, \& {Wieprecht}}]{Lutz2011}
{Lutz}, D., {Poglitsch}, A., {Altieri}, B., {et~al.} 2011, \aap, 532, A90

\bibitem[{{Magnelli} {et~al.}(2009){Magnelli}, {Elbaz}, {Chary}, {Dickinson},
  {Le Borgne}, {Frayer}, \& {Willmer}}]{Magnelli2009}
{Magnelli}, B., {Elbaz}, D., {Chary}, R.~R., {et~al.} 2009, \aap, 496, 57

\bibitem[{{Magnelli} {et~al.}(2012){Magnelli}, {Lutz}, {Santini}, {Saintonge},
  {Berta}, {Albrecht}, {Altieri}, {Andreani}, {Aussel}, {Bertoldi},
  {B{\'e}thermin}, {Bongiovanni}, {Capak}, {Chapman}, {Cepa}, {Cimatti},
  {Cooray}, {Daddi}, {Danielson}, {Dannerbauer}, {Dunlop}, {Elbaz}, {Farrah},
  {F{\"o}rster Schreiber}, {Genzel}, {Hwang}, {Ibar}, {Ivison}, {Le Floc'h},
  {Magdis}, {Maiolino}, {Nordon}, {Oliver}, {P{\'e}rez Garc{\'{\i}}a},
  {Poglitsch}, {Popesso}, {Pozzi}, {Riguccini}, {Rodighiero}, {Rosario},
  {Roseboom}, {Salvato}, {Sanchez-Portal}, {Scott}, {Smail}, {Sturm},
  {Swinbank}, {Tacconi}, {Valtchanov}, {Wang}, \& {Wuyts}}]{Magnelli2012}
{Magnelli}, B., {Lutz}, D., {Santini}, P., {et~al.} 2012, \aap, 539, A155

\bibitem[{{Magnelli} {et~al.}(2013){Magnelli}, {Popesso}, {Berta}, {Pozzi},
  {Elbaz}, {Lutz}, {Dickinson}, {Altieri}, {Andreani}, {Aussel},
  {B{\'e}thermin}, {Bongiovanni}, {Cepa}, {Charmandaris}, {Chary}, {Cimatti},
  {Daddi}, {F{\"o}rster Schreiber}, {Genzel}, {Gruppioni}, {Harwit}, {Hwang},
  {Ivison}, {Magdis}, {Maiolino}, {Murphy}, {Nordon}, {Pannella}, {P{\'e}rez
  Garc{\'{\i}}a}, {Poglitsch}, {Rosario}, {Sanchez-Portal}, {Santini}, {Scott},
  {Sturm}, {Tacconi}, \& {Valtchanov}}]{Magnelli2013}
{Magnelli}, B., {Popesso}, P., {Berta}, S., {et~al.} 2013, \aap, 553, A132

\bibitem[{{Maiolino} {et~al.}(2015){Maiolino}, {Carniani}, {Fontana},
  {Vallini}, {Pentericci}, {Ferrara}, {Vanzella}, {Grazian}, {Gallerani},
  {Castellano}, {Cristiani}, {Brammer}, {Santini}, {Wagg}, \&
  {Williams}}]{Maiolino2015}
{Maiolino}, R., {Carniani}, S., {Fontana}, A., {et~al.} 2015, \mnras, 452, 54

\bibitem[{{Mart{\'{\i}}-Vidal} {et~al.}(2012){Mart{\'{\i}}-Vidal},
  {P{\'e}rez-Torres}, \& {Lobanov}}]{Marti-Vidal2012}
{Mart{\'{\i}}-Vidal}, I., {P{\'e}rez-Torres}, M.~A., \& {Lobanov}, A.~P. 2012,
  \aap, 541, A135

\bibitem[{{McMullin} {et~al.}(2007){McMullin}, {Waters}, {Schiebel}, {Young},
  \& {Golap}}]{McMullin2007}
{McMullin}, J.~P., {Waters}, B., {Schiebel}, D., {Young}, W., \& {Golap}, K.
  2007, in Astronomical Society of the Pacific Conference Series, Vol. 376,
  Astronomical Data Analysis Software and Systems XVI, ed. R.~A. {Shaw},
  F.~{Hill}, \& D.~J. {Bell}, 127

\bibitem[{{Micha{\l}owski} {et~al.}(2010){Micha{\l}owski}, {Hjorth}, \&
  {Watson}}]{Michaowski2010}
{Micha{\l}owski}, M., {Hjorth}, J., \& {Watson}, D. 2010, \aap, 514, A67

\bibitem[{{Mohan} \& {Rafferty}(2015)}]{Mohan2015}
{Mohan}, N. \& {Rafferty}, D. 2015, {PyBDSF: Python Blob Detection and Source
  Finder}, Astrophysics Source Code Library

\bibitem[{{Momcheva} {et~al.}(2016){Momcheva}, {Brammer}, {van Dokkum},
  {Skelton}, {Whitaker}, {Nelson}, {Fumagalli}, {Maseda}, {Leja}, {Franx},
  {Rix}, {Bezanson}, {Da Cunha}, {Dickey}, {F{\"o}rster Schreiber},
  {Illingworth}, {Kriek}, {Labb{\'e}}, {Ulf Lange}, {Lundgren}, {Magee},
  {Marchesini}, {Oesch}, {Pacifici}, {Patel}, {Price}, {Tal}, {Wake}, {van der
  Wel}, \& {Wuyts}}]{Momcheva2016}
{Momcheva}, I.~G., {Brammer}, G.~B., {van Dokkum}, P.~G., {et~al.} 2016, \apjs,
  225, 27

\bibitem[{{Moster} {et~al.}(2011){Moster}, {Somerville}, {Newman}, \&
  {Rix}}]{Moster2011}
{Moster}, B.~P., {Somerville}, R.~S., {Newman}, J.~A., \& {Rix}, H.-W. 2011,
  \apj, 731, 113

\bibitem[{{Narayanan} {et~al.}(2010){Narayanan}, {Hayward}, {Cox}, {Hernquist},
  {Jonsson}, {Younger}, \& {Groves}}]{Narayanan2010}
{Narayanan}, D., {Hayward}, C.~C., {Cox}, T.~J., {et~al.} 2010, \mnras, 401,
  1613

\bibitem[{{Noeske} {et~al.}(2007){Noeske}, {Weiner}, {Faber}, {Papovich},
  {Koo}, {Somerville}, {Bundy}, {Conselice}, {Newman}, {Schiminovich}, {Le
  Floc'h}, {Coil}, {Rieke}, {Lotz}, {Primack}, {Barmby}, {Cooper}, {Davis},
  {Ellis}, {Fazio}, {Guhathakurta}, {Huang}, {Kassin}, {Martin}, {Phillips},
  {Rich}, {Small}, {Willmer}, \& {Wilson}}]{Noeske2007}
{Noeske}, K.~G., {Weiner}, B.~J., {Faber}, S.~M., {et~al.} 2007, \apjl, 660,
  L43

\bibitem[{{Nonino} {et~al.}(2009){Nonino}, {Dickinson}, {Rosati}, {Grazian},
  {Reddy}, {Cristiani}, {Giavalisco}, {Kuntschner}, {Vanzella}, {Daddi},
  {Fosbury}, \& {Cesarsky}}]{Nonino2009}
{Nonino}, M., {Dickinson}, M., {Rosati}, P., {et~al.} 2009, \apjs, 183, 244

\bibitem[{{Oke} \& {Gunn}(1983)}]{Oke1983}
{Oke}, J.~B. \& {Gunn}, J.~E. 1983, \apj, 266, 713

\bibitem[{{Ono} {et~al.}(2014){Ono}, {Ouchi}, {Kurono}, \& {Momose}}]{Ono2014}
{Ono}, Y., {Ouchi}, M., {Kurono}, Y., \& {Momose}, R. 2014, \apj, 795, 5

\bibitem[{{Oteo} {et~al.}(2016){Oteo}, {Zwaan}, {Ivison}, {Smail}, \&
  {Biggs}}]{Oteo2015}
{Oteo}, I., {Zwaan}, M.~A., {Ivison}, R.~J., {Smail}, I., \& {Biggs}, A.~D.
  2016, \apj, 822, 36

\bibitem[{{Padilla} \& {Strauss}(2008)}]{Padilla2008}
{Padilla}, N.~D. \& {Strauss}, M.~A. 2008, \mnras, 388, 1321

\bibitem[{{Papovich} {et~al.}(2016){Papovich}, {Labb{\'e}}, {Glazebrook},
  {Quadri}, {Bekiaris}, {Dickinson}, {Finkelstein}, {Fisher}, {Inami},
  {Livermore}, {Spitler}, {Straatman}, \& {Tran}}]{Papovich2016}
{Papovich}, C., {Labb{\'e}}, I., {Glazebrook}, K., {et~al.} 2016, Nature
  Astronomy, 1, 0003

\bibitem[{{Papovich} {et~al.}(2010){Papovich}, {Momcheva}, {Willmer},
  {Finkelstein}, {Finkelstein}, {Tran}, {Brodwin}, {Dunlop}, {Farrah}, {Khan},
  {Lotz}, {McCarthy}, {McLure}, {Rieke}, {Rudnick}, {Sivanandam}, {Pacaud}, \&
  {Pierre}}]{Papovich2010}
{Papovich}, C., {Momcheva}, I., {Willmer}, C.~N.~A., {et~al.} 2010, \apj, 716,
  1503

\bibitem[{{Peng} {et~al.}(2010){Peng}, {Ho}, {Impey}, \& {Rix}}]{Peng2010}
{Peng}, C.~Y., {Ho}, L.~C., {Impey}, C.~D., \& {Rix}, H.-W. 2010, \aj, 139,
  2097

\bibitem[{{Persson} {et~al.}(2013){Persson}, {Murphy}, {Smee}, {Birk},
  {Monson}, {Uomoto}, {Koch}, {Shectman}, {Barkhouser}, {Orndorff}, {Hammond},
  {Harding}, {Scharfstein}, {Kelson}, {Marshall}, \& {McCarthy}}]{Persson2013}
{Persson}, S.~E., {Murphy}, D.~C., {Smee}, S., {et~al.} 2013, \pasp, 125, 654

\bibitem[{{Poglitsch} {et~al.}(2010){Poglitsch}, {Waelkens}, {Geis},
  {Feuchtgruber}, {Vandenbussche}, {Rodriguez}, {Krause}, {Renotte}, {van
  Hoof}, {Saraceno}, {Cepa}, {Kerschbaum}, {Agn{\`e}se}, {Ali}, {Altieri},
  {Andreani}, {Augueres}, {Balog}, {Barl}, {Bauer}, {Belbachir}, {Benedettini},
  {Billot}, {Boulade}, {Bischof}, {Blommaert}, {Callut}, {Cara}, {Cerulli},
  {Cesarsky}, {Contursi}, {Creten}, {De Meester}, {Doublier}, {Doumayrou},
  {Duband}, {Exter}, {Genzel}, {Gillis}, {Gr{\"o}zinger}, {Henning},
  {Herreros}, {Huygen}, {Inguscio}, {Jakob}, {Jamar}, {Jean}, {de Jong},
  {Katterloher}, {Kiss}, {Klaas}, {Lemke}, {Lutz}, {Madden}, {Marquet},
  {Martignac}, {Mazy}, {Merken}, {Montfort}, {Morbidelli}, {M{\"u}ller},
  {Nielbock}, {Okumura}, {Orfei}, {Ottensamer}, {Pezzuto}, {Popesso},
  {Putzeys}, {Regibo}, {Reveret}, {Royer}, {Sauvage}, {Schreiber}, {Stegmaier},
  {Schmitt}, {Schubert}, {Sturm}, {Thiel}, {Tofani}, {Vavrek}, {Wetzstein},
  {Wieprecht}, \& {Wiezorrek}}]{Poglitsch2010}
{Poglitsch}, A., {Waelkens}, C., {Geis}, N., {et~al.} 2010, \aap, 518, L2

\bibitem[{{Pope} {et~al.}(2005){Pope}, {Borys}, {Scott}, {Conselice},
  {Dickinson}, \& {Mobasher}}]{Pope2005}
{Pope}, A., {Borys}, C., {Scott}, D., {et~al.} 2005, \mnras, 358, 149

\bibitem[{{Pope} {et~al.}(2008){Pope}, {Chary}, {Alexander}, {Armus},
  {Dickinson}, {Elbaz}, {Frayer}, {Scott}, \& {Teplitz}}]{Pope2008}
{Pope}, A., {Chary}, R.-R., {Alexander}, D.~M., {et~al.} 2008, \apj, 675, 1171

\bibitem[{{Rafferty} {et~al.}(2011){Rafferty}, {Brandt}, {Alexander}, {Xue},
  {Bauer}, {Lehmer}, {Luo}, \& {Papovich}}]{Rafferty2011}
{Rafferty}, D.~A., {Brandt}, W.~N., {Alexander}, D.~M., {et~al.} 2011, \apj,
  742, 3

\bibitem[{{Ranalli} {et~al.}(2013){Ranalli}, {Comastri}, {Vignali}, {Carrera},
  {Cappelluti}, {Gilli}, {Puccetti}, {Brandt}, {Brunner}, {Brusa},
  {Georgantopoulos}, {Iwasawa}, \& {Mainieri}}]{Ranalli2013}
{Ranalli}, P., {Comastri}, A., {Vignali}, C., {et~al.} 2013, \aap, 555, A42

\bibitem[{{Retzlaff} {et~al.}(2010){Retzlaff}, {Rosati}, {Dickinson},
  {Vandame}, {Rit{\'e}}, {Nonino}, {Cesarsky}, \& {GOODS Team}}]{Retzlaff2010}
{Retzlaff}, J., {Rosati}, P., {Dickinson}, M., {et~al.} 2010, \aap, 511, A50

\bibitem[{{Rieke} {et~al.}(2004){Rieke}, {Young}, {Engelbracht}, {Kelly},
  {Low}, {Haller}, {Beeman}, {Gordon}, {Stansberry}, {Misselt}, {Cadien},
  {Morrison}, {Rivlis}, {Latter}, {Noriega-Crespo}, {Padgett}, {Stapelfeldt},
  {Hines}, {Egami}, {Muzerolle}, {Alonso-Herrero}, {Blaylock}, {Dole}, {Hinz},
  {Le Floc'h}, {Papovich}, {P{\'e}rez-Gonz{\'a}lez}, {Smith}, {Su}, {Bennett},
  {Frayer}, {Henderson}, {Lu}, {Masci}, {Pesenson}, {Rebull}, {Rho}, {Keene},
  {Stolovy}, {Wachter}, {Wheaton}, {Werner}, \& {Richards}}]{Rieke2004}
{Rieke}, G.~H., {Young}, E.~T., {Engelbracht}, C.~W., {et~al.} 2004, \apjs,
  154, 25

\bibitem[{{Rodighiero} {et~al.}(2011){Rodighiero}, {Daddi}, {Baronchelli},
  {Cimatti}, {Renzini}, {Aussel}, {Popesso}, {Lutz}, {Andreani}, {Berta},
  {Cava}, {Elbaz}, {Feltre}, {Fontana}, {F{\"o}rster Schreiber},
  {Franceschini}, {Genzel}, {Grazian}, {Gruppioni}, {Ilbert}, {Le Floch},
  {Magdis}, {Magliocchetti}, {Magnelli}, {Maiolino}, {McCracken}, {Nordon},
  {Poglitsch}, {Santini}, {Pozzi}, {Riguccini}, {Tacconi}, {Wuyts}, \&
  {Zamorani}}]{Rodighiero2011}
{Rodighiero}, G., {Daddi}, E., {Baronchelli}, I., {et~al.} 2011, \apjl, 739,
  L40

\bibitem[{{Rovilos} {et~al.}(2012){Rovilos}, {Comastri}, {Gilli},
  {Georgantopoulos}, {Ranalli}, {Vignali}, {Lusso}, {Cappelluti}, {Zamorani},
  {Elbaz}, {Dickinson}, {Hwang}, {Charmandaris}, {Ivison}, {Merloni}, {Daddi},
  {Carrera}, {Brandt}, {Mullaney}, {Scott}, {Alexander}, {Del Moro},
  {Morrison}, {Murphy}, {Altieri}, {Aussel}, {Dannerbauer}, {Kartaltepe},
  {Leiton}, {Magdis}, {Magnelli}, {Popesso}, \& {Valtchanov}}]{Rovilos2012}
{Rovilos}, E., {Comastri}, A., {Gilli}, R., {et~al.} 2012, \aap, 546, A58

\bibitem[{{Rujopakarn} {et~al.}(2016){Rujopakarn}, {Dunlop}, {Rieke}, {Ivison},
  {Cibinel}, {Nyland}, {Jagannathan}, {Silverman}, {Alexander}, {Biggs},
  {Bhatnagar}, {Ballantyne}, {Dickinson}, {Elbaz}, {Geach}, {Hayward},
  {Kirkpatrick}, {McLure}, {Micha{\l}owski}, {Miller}, {Narayanan}, {Owen},
  {Pannella}, {Papovich}, {Pope}, {Rau}, {Robertson}, {Scott}, {Swinbank}, {van
  der Werf}, {van Kampen}, {Weiner}, \& {Windhorst}}]{Rujopakarn2016}
{Rujopakarn}, W., {Dunlop}, J.~S., {Rieke}, G.~H., {et~al.} 2016, \apj, 833, 12

\bibitem[{{Salpeter}(1955)}]{Salpeter1955}
{Salpeter}, E.~E. 1955, \apj, 121, 161

\bibitem[{{Santini} {et~al.}(2015){Santini}, {Ferguson}, {Fontana}, {Mobasher},
  {Barro}, {Castellano}, {Finkelstein}, {Grazian}, {Hsu}, {Lee}, {Lee},
  {Pforr}, {Salvato}, {Wiklind}, {Wuyts}, {Almaini}, {Cooper}, {Galametz},
  {Weiner}, {Amorin}, {Boutsia}, {Conselice}, {Dahlen}, {Dickinson},
  {Giavalisco}, {Grogin}, {Guo}, {Hathi}, {Kocevski}, {Koekemoer},
  {Kurczynski}, {Merlin}, {Mortlock}, {Newman}, {Paris}, {Pentericci},
  {Simons}, \& {Willner}}]{Santini2015}
{Santini}, P., {Ferguson}, H.~C., {Fontana}, A., {et~al.} 2015, \apj, 801, 97

\bibitem[{{Sargent} {et~al.}(2012){Sargent}, {B{\'e}thermin}, {Daddi}, \&
  {Elbaz}}]{Sargent2012}
{Sargent}, M.~T., {B{\'e}thermin}, M., {Daddi}, E., \& {Elbaz}, D. 2012, \apjl,
  747, L31

\bibitem[{{Schechter}(1976)}]{Schechter1976}
{Schechter}, P. 1976, \apj, 203, 297

\bibitem[{{Schreiber} {et~al.}(2018){Schreiber}, {Elbaz}, {Pannella}, {Ciesla},
  {Wang}, \& {Franco}}]{Schreiber2017}
{Schreiber}, C., {Elbaz}, D., {Pannella}, M., {et~al.} 2018, \aap, 609, A30

\bibitem[{{Schreiber} {et~al.}(2015){Schreiber}, {Pannella}, {Elbaz},
  {B{\'e}thermin}, {Inami}, {Dickinson}, {Magnelli}, {Wang}, {Aussel}, {Daddi},
  {Juneau}, {Shu}, {Sargent}, {Buat}, {Faber}, {Ferguson}, {Giavalisco},
  {Koekemoer}, {Magdis}, {Morrison}, {Papovich}, {Santini}, \&
  {Scott}}]{Schreiber2015}
{Schreiber}, C., {Pannella}, M., {Elbaz}, D., {et~al.} 2015, \aap, 575, A74

\bibitem[{{Schreiber} {et~al.}(2017){Schreiber}, {Pannella}, {Leiton}, {Elbaz},
  {Wang}, {Okumura}, \& {Labb{\'e}}}]{Schreiber2017b}
{Schreiber}, C., {Pannella}, M., {Leiton}, R., {et~al.} 2017, \aap, 599, A134

\bibitem[{{Scott} {et~al.}(2012){Scott}, {Wilson}, {Aretxaga}, {Austermann},
  {Chapin}, {Dunlop}, {Ezawa}, {Halpern}, {Hatsukade}, {Hughes}, {Kawabe},
  {Kim}, {Kohno}, {Lowenthal}, {Monta{\~n}a}, {Nakanishi}, {Oshima}, {Sanders},
  {Scott}, {Scoville}, {Tamura}, {Welch}, {Yun}, \& {Zeballos}}]{Scott2012}
{Scott}, K.~S., {Wilson}, G.~W., {Aretxaga}, I., {et~al.} 2012, \mnras, 423,
  575

\bibitem[{{Scott} {et~al.}(2002){Scott}, {Fox}, {Dunlop}, {Serjeant},
  {Peacock}, {Ivison}, {Oliver}, {Mann}, {Lawrence}, {Efstathiou},
  {Rowan-Robinson}, {Hughes}, {Archibald}, {Blain}, \& {Longair}}]{Scott2002}
{Scott}, S.~E., {Fox}, M.~J., {Dunlop}, J.~S., {et~al.} 2002, \mnras, 331, 817

\bibitem[{{Scudder} {et~al.}(2016){Scudder}, {Oliver}, {Hurley}, {Griffin},
  {Sargent}, {Scott}, {Wang}, \& {Wardlow}}]{Scudder2016}
{Scudder}, J.~M., {Oliver}, S., {Hurley}, P.~D., {et~al.} 2016, \mnras, 460,
  1119

\bibitem[{{Simpson} {et~al.}(2015{\natexlab{a}}){Simpson}, {Smail}, {Swinbank},
  {Almaini}, {Blain}, {Bremer}, {Chapman}, {Chen}, {Conselice}, {Coppin},
  {Danielson}, {Dunlop}, {Edge}, {Farrah}, {Geach}, {Hartley}, {Ivison},
  {Karim}, {Lani}, {Ma}, {Meijerink}, {Micha{\l}owski}, {Mortlock}, {Scott},
  {Simpson}, {Spaans}, {Thomson}, {van Kampen}, \& {van der
  Werf}}]{Simpson2015b}
{Simpson}, J.~M., {Smail}, I., {Swinbank}, A.~M., {et~al.} 2015{\natexlab{a}},
  \apj, 799, 81

\bibitem[{{Simpson} {et~al.}(2015{\natexlab{b}}){Simpson}, {Smail}, {Swinbank},
  {Chapman}, {Geach}, {Ivison}, {Thomson}, {Aretxaga}, {Blain}, {Cowley},
  {Chen}, {Coppin}, {Dunlop}, {Edge}, {Farrah}, {Ibar}, {Karim}, {Knudsen},
  {Meijerink}, {Micha{\l}owski}, {Scott}, {Spaans}, \& {van der
  Werf}}]{Simpson2015}
{Simpson}, J.~M., {Smail}, I., {Swinbank}, A.~M., {et~al.} 2015{\natexlab{b}},
  \apj, 807, 128

\bibitem[{{Simpson} {et~al.}(2014){Simpson}, {Swinbank}, {Smail}, {Alexander},
  {Brandt}, {Bertoldi}, {de Breuck}, {Chapman}, {Coppin}, {da Cunha},
  {Danielson}, {Dannerbauer}, {Greve}, {Hodge}, {Ivison}, {Karim}, {Knudsen},
  {Poggianti}, {Schinnerer}, {Thomson}, {Walter}, {Wardlow}, {Wei{\ss}}, \&
  {van der Werf}}]{Simpson2014}
{Simpson}, J.~M., {Swinbank}, A.~M., {Smail}, I., {et~al.} 2014, \apj, 788, 125

\bibitem[{{Skelton} {et~al.}(2014){Skelton}, {Whitaker}, {Momcheva}, {Brammer},
  {van Dokkum}, {Labb{\'e}}, {Franx}, {van der Wel}, {Bezanson}, {Da Cunha},
  {Fumagalli}, {F{\"o}rster Schreiber}, {Kriek}, {Leja}, {Lundgren}, {Magee},
  {Marchesini}, {Maseda}, {Nelson}, {Oesch}, {Pacifici}, {Patel}, {Price},
  {Rix}, {Tal}, {Wake}, \& {Wuyts}}]{Skelton2014}
{Skelton}, R.~E., {Whitaker}, K.~E., {Momcheva}, I.~G., {et~al.} 2014, \apjs,
  214, 24

\bibitem[{{Smail} {et~al.}(1997){Smail}, {Ivison}, \& {Blain}}]{Smail1997}
{Smail}, I., {Ivison}, R.~J., \& {Blain}, A.~W. 1997, \apjl, 490, L5

\bibitem[{{Straatman} {et~al.}(2016){Straatman}, {Spitler}, {Quadri},
  {Labb{\'e}}, {Glazebrook}, {Persson}, {Papovich}, {Tran}, {Brammer},
  {Cowley}, {Tomczak}, {Nanayakkara}, {Alcorn}, {Allen}, {Broussard}, {van
  Dokkum}, {Forrest}, {van Houdt}, {Kacprzak}, {Kawinwanichakij}, {Kelson},
  {Lee}, {McCarthy}, {Mehrtens}, {Monson}, {Murphy}, {Rees}, {Tilvi}, \&
  {Whitaker}}]{Straatman2016}
{Straatman}, C.~M.~S., {Spitler}, L.~R., {Quadri}, R.~F., {et~al.} 2016, \apj,
  830, 51

\bibitem[{{Swinbank} {et~al.}(2014){Swinbank}, {Simpson}, {Smail}, {Harrison},
  {Hodge}, {Karim}, {Walter}, {Alexander}, {Brandt}, {de Breuck}, {da Cunha},
  {Chapman}, {Coppin}, {Danielson}, {Dannerbauer}, {Decarli}, {Greve},
  {Ivison}, {Knudsen}, {Lagos}, {Schinnerer}, {Thomson}, {Wardlow}, {Wei{\ss}},
  \& {van der Werf}}]{Swinbank2014}
{Swinbank}, A.~M., {Simpson}, J.~M., {Smail}, I., {et~al.} 2014, \mnras, 438,
  1267

\bibitem[{{Tacconi} {et~al.}(2008){Tacconi}, {Genzel}, {Smail}, {Neri},
  {Chapman}, {Ivison}, {Blain}, {Cox}, {Omont}, {Bertoldi}, {Greve},
  {F{\"o}rster Schreiber}, {Genel}, {Lutz}, {Swinbank}, {Shapley}, {Erb},
  {Cimatti}, {Daddi}, \& {Baker}}]{Tacconi2008}
{Tacconi}, L.~J., {Genzel}, R., {Smail}, I., {et~al.} 2008, \apj, 680, 246

\bibitem[{{Tadaki} {et~al.}(2017){Tadaki}, {Genzel}, {Kodama}, {Wuyts},
  {Wisnioski}, {F{\"o}rster Schreiber}, {Burkert}, {Lang}, {Tacconi}, {Lutz},
  {Belli}, {Davies}, {Hatsukade}, {Hayashi}, {Herrera-Camus}, {Ikarashi},
  {Inoue}, {Kohno}, {Koyama}, {Mendel}, {Nakanishi}, {Shimakawa}, {Suzuki},
  {Tamura}, {Tanaka}, {{\"U}bler}, \& {Wilman}}]{Tadaki2017}
{Tadaki}, K.-i., {Genzel}, R., {Kodama}, T., {et~al.} 2017, \apj, 834, 135

\bibitem[{{Talia} {et~al.}(2018){Talia}, {Pozzi}, {Vallini}, {Cimatti},
  {Cassata}, {Fraternali}, {Brusa}, {Daddi}, {Delvecchio}, {Ibar}, {Liuzzo},
  {Vignali}, {Massardi}, {Zamorani}, {Gruppioni}, {Renzini}, {Mignoli},
  {Pozzetti}, \& {Rodighiero}}]{Talia2018}
{Talia}, M., {Pozzi}, F., {Vallini}, L., {et~al.} 2018, \mnras, 476, 3956

\bibitem[{{Tomczak} {et~al.}(2016){Tomczak}, {Quadri}, {Tran}, {Labb{\'e}},
  {Straatman}, {Papovich}, {Glazebrook}, {Allen}, {Brammer}, {Cowley},
  {Dickinson}, {Elbaz}, {Inami}, {Kacprzak}, {Morrison}, {Nanayakkara},
  {Persson}, {Rees}, {Salmon}, {Schreiber}, {Spitler}, \&
  {Whitaker}}]{Tomczak2016}
{Tomczak}, A.~R., {Quadri}, R.~F., {Tran}, K.-V.~H., {et~al.} 2016, \apj, 817,
  118

\bibitem[{{Ueda} {et~al.}(2018){Ueda}, {Hatsukade}, {Kohno}, {Yamaguchi},
  {Tamura}, {Umehata}, {Akiyama}, {Ao}, {Aretxaga}, {Caputi}, {Dunlop},
  {Espada}, {Fujimoto}, {Hayatsu}, {Imanishi}, {Inoue}, {Ivison}, {Kodama},
  {Lee}, {Matsuoka}, {Miyaji}, {Morokuma-Matsui}, {Nagao}, {Nakanishi},
  {Nyland}, {Ohta}, {Ouchi}, {Rujopakarn}, {Saito}, {Tadaki}, {Tanaka},
  {Taniguchi}, {Wang}, {Wang}, {Yoshimura}, \& {Yun}}]{Ueda2018}
{Ueda}, Y., {Hatsukade}, B., {Kohno}, K., {et~al.} 2018, \apj, 853, 24

\bibitem[{{Umehata} {et~al.}(2017){Umehata}, {Tamura}, {Kohno}, {Ivison},
  {Smail}, {Hatsukade}, {Nakanishi}, {Kato}, {Ikarashi}, {Matsuda}, {Fujimoto},
  {Iono}, {Lee}, {Steidel}, {Saito}, {Alexander}, {Yun}, \&
  {Kubo}}]{Umehata2017}
{Umehata}, H., {Tamura}, Y., {Kohno}, K., {et~al.} 2017, \apj, 835, 98

\bibitem[{{van der Wel} {et~al.}(2012){van der Wel}, {Bell}, {H{\"a}ussler},
  {McGrath}, {Chang}, {Guo}, {McIntosh}, {Rix}, {Barden}, {Cheung}, {Faber},
  {Ferguson}, {Galametz}, {Grogin}, {Hartley}, {Kartaltepe}, {Kocevski},
  {Koekemoer}, {Lotz}, {Mozena}, {Peth}, \& {Peng}}]{van_der_Wel2012}
{van der Wel}, A., {Bell}, E.~F., {H{\"a}ussler}, B., {et~al.} 2012, \apjs,
  203, 24

\bibitem[{{van der Wel} {et~al.}(2014){van der Wel}, {Franx}, {van Dokkum},
  {Skelton}, {Momcheva}, {Whitaker}, {Brammer}, {Bell}, {Rix}, {Wuyts},
  {Ferguson}, {Holden}, {Barro}, {Koekemoer}, {Chang}, {McGrath},
  {H{\"a}ussler}, {Dekel}, {Behroozi}, {Fumagalli}, {Leja}, {Lundgren},
  {Maseda}, {Nelson}, {Wake}, {Patel}, {Labb{\'e}}, {Faber}, {Grogin}, \&
  {Kocevski}}]{VanderWel2014}
{van der Wel}, A., {Franx}, M., {van Dokkum}, P.~G., {et~al.} 2014, \apj, 788,
  28

\bibitem[{{van Dokkum} {et~al.}(2015){van Dokkum}, {Nelson}, {Franx}, {Oesch},
  {Momcheva}, {Brammer}, {F{\"o}rster Schreiber}, {Skelton}, {Whitaker}, {van
  der Wel}, {Bezanson}, {Fumagalli}, {Illingworth}, {Kriek}, {Leja}, \&
  {Wuyts}}]{van_Dokkum2015}
{van Dokkum}, P.~G., {Nelson}, E.~J., {Franx}, M., {et~al.} 2015, \apj, 813, 23

\bibitem[{{van Dokkum} {et~al.}(2010){van Dokkum}, {Whitaker}, {Brammer},
  {Franx}, {Kriek}, {Labb{\'e}}, {Marchesini}, {Quadri}, {Bezanson},
  {Illingworth}, {Muzzin}, {Rudnick}, {Tal}, \& {Wake}}]{VanDokkum2010}
{van Dokkum}, P.~G., {Whitaker}, K.~E., {Brammer}, G., {et~al.} 2010, \apj,
  709, 1018

\bibitem[{{Vanzella} {et~al.}(2008){Vanzella}, {Cristiani}, {Dickinson},
  {Giavalisco}, {Kuntschner}, {Haase}, {Nonino}, {Rosati}, {Cesarsky},
  {Ferguson}, {Fosbury}, {Grazian}, {Moustakas}, {Rettura}, {Popesso},
  {Renzini}, {Stern}, \& {GOODS Team}}]{Vanzella2008}
{Vanzella}, E., {Cristiani}, S., {Dickinson}, M., {et~al.} 2008, \aap, 478, 83

\bibitem[{{Wang} {et~al.}(2013){Wang}, {Brandt}, {Luo}, {Smail}, {Alexander},
  {Danielson}, {Hodge}, {Karim}, {Lehmer}, {Simpson}, {Swinbank}, {Walter},
  {Wardlow}, {Xue}, {Chapman}, {Coppin}, {Dannerbauer}, {De Breuck}, {Menten},
  \& {van der Werf}}]{Wang2013}
{Wang}, S.~X., {Brandt}, W.~N., {Luo}, B., {et~al.} 2013, \apj, 778, 179

\bibitem[{{Wang} {et~al.}(2016){Wang}, {Elbaz}, {Schreiber}, {Pannella}, {Shu},
  {Willner}, {Ashby}, {Huang}, {Fontana}, {Dekel}, {Daddi}, {Ferguson},
  {Dunlop}, {Ciesla}, {Koekemoer}, {Giavalisco}, {Boutsia}, {Finkelstein},
  {Juneau}, {Barro}, {Koo}, {Micha{\l}owski}, {Orellana}, {Lu}, {Castellano},
  {Bourne}, {Buitrago}, {Santini}, {Faber}, {Hathi}, {Lucas}, \&
  {P{\'e}rez-Gonz{\'a}lez}}]{Wang2016}
{Wang}, T., {Elbaz}, D., {Schreiber}, C., {et~al.} 2016, \apj, 816, 84

\bibitem[{{Wang} {et~al.}(2009){Wang}, {Barger}, \& {Cowie}}]{Wang2009}
{Wang}, W.-H., {Barger}, A.~J., \& {Cowie}, L.~L. 2009, \apj, 690, 319

\bibitem[{{Wardlow} {et~al.}(2011){Wardlow}, {Smail}, {Coppin}, {Alexander},
  {Brandt}, {Danielson}, {Luo}, {Swinbank}, {Walter}, {Wei{\ss}}, {Xue},
  {Zibetti}, {Bertoldi}, {Biggs}, {Chapman}, {Dannerbauer}, {Dunlop},
  {Gawiser}, {Ivison}, {Knudsen}, {Kov{\'a}cs}, {Lacey}, {Menten}, {Padilla},
  {Rix}, \& {van der Werf}}]{Wardlow2011}
{Wardlow}, J.~L., {Smail}, I., {Coppin}, K.~E.~K., {et~al.} 2011, \mnras, 415,
  1479

\bibitem[{{Wei{\ss}} {et~al.}(2009){Wei{\ss}}, {Kov{\'a}cs}, {Coppin}, {Greve},
  {Walter}, {Smail}, {Dunlop}, {Knudsen}, {Alexander}, {Bertoldi}, {Brandt},
  {Chapman}, {Cox}, {Dannerbauer}, {De Breuck}, {Gawiser}, {Ivison}, {Lutz},
  {Menten}, {Koekemoer}, {Kreysa}, {Kurczynski}, {Rix}, {Schinnerer}, \& {van
  der Werf}}]{Weiss2009}
{Wei{\ss}}, A., {Kov{\'a}cs}, A., {Coppin}, K., {et~al.} 2009, \apj, 707, 1201

\bibitem[{{Wiklind} {et~al.}(2014){Wiklind}, {Conselice}, {Dahlen},
  {Dickinson}, {Ferguson}, {Grogin}, {Guo}, {Koekemoer}, {Mobasher},
  {Mortlock}, {Fontana}, {Dav{\'e}}, {Yan}, {Acquaviva}, {Ashby}, {Barro},
  {Caputi}, {Castellano}, {Dekel}, {Donley}, {Fazio}, {Giavalisco}, {Grazian},
  {Hathi}, {Kurczynski}, {Lu}, {McGrath}, {de Mello}, {Peth}, {Safarzadeh},
  {Stefanon}, \& {Targett}}]{Wiklind2014}
{Wiklind}, T., {Conselice}, C.~J., {Dahlen}, T., {et~al.} 2014, \apj, 785, 111

\bibitem[{{Williams} {et~al.}(2014){Williams}, {Giavalisco}, {Cassata},
  {Tundo}, {Wiklind}, {Guo}, {Lee}, {Barro}, {Wuyts}, {Bell}, {Conselice},
  {Dekel}, {Faber}, {Ferguson}, {Grogin}, {Hathi}, {Huang}, {Kocevski},
  {Koekemoer}, {Koo}, {Ravindranath}, \& {Salimbeni}}]{Williams2014}
{Williams}, C.~C., {Giavalisco}, M., {Cassata}, P., {et~al.} 2014, \apj, 780, 1

\bibitem[{{Williams} {et~al.}(2009){Williams}, {Quadri}, {Franx}, {van Dokkum},
  \& {Labb{\'e}}}]{Williams2009}
{Williams}, R.~J., {Quadri}, R.~F., {Franx}, M., {van Dokkum}, P., \&
  {Labb{\'e}}, I. 2009, \apj, 691, 1879

\bibitem[{{Wuyts} {et~al.}(2008){Wuyts}, {Labb{\'e}}, {F{\"o}rster Schreiber},
  {Franx}, {Rudnick}, {Brammer}, \& {van Dokkum}}]{Wuyts2008}
{Wuyts}, S., {Labb{\'e}}, I., {F{\"o}rster Schreiber}, N.~M., {et~al.} 2008,
  \apj, 682, 985

\bibitem[{{Xue} {et~al.}(2011){Xue}, {Luo}, {Brandt}, {Bauer}, {Lehmer},
  {Broos}, {Schneider}, {Alexander}, {Brusa}, {Comastri}, {Fabian}, {Gilli},
  {Hasinger}, {Hornschemeier}, {Koekemoer}, {Liu}, {Mainieri}, {Paolillo},
  {Rafferty}, {Rosati}, {Shemmer}, {Silverman}, {Smail}, {Tozzi}, \&
  {Vignali}}]{Xue2011}
{Xue}, Y.~Q., {Luo}, B., {Brandt}, W.~N., {et~al.} 2011, \apjs, 195, 10

\bibitem[{{Yamaguchi} {et~al.}(2016){Yamaguchi}, {Tamura}, {Kohno}, {Aretxaga},
  {Dunlop}, {Hatsukade}, {Hughes}, {Ikarashi}, {Ishii}, {Ivison}, {Izumi},
  {Kawabe}, {Kodama}, {Lee}, {Makiya}, {Matsuda}, {Nakanishi}, {Ohta},
  {Rujopakarn}, {Tadaki}, {Umehata}, {Wang}, {Wilson}, {Yabe}, \&
  {Yun}}]{Yamaguchi2016}
{Yamaguchi}, Y., {Tamura}, Y., {Kohno}, K., {et~al.} 2016, \pasj, 68, 82

\bibitem[{{Yun} {et~al.}(2012){Yun}, {Scott}, {Guo}, {Aretxaga}, {Giavalisco},
  {Austermann}, {Capak}, {Chen}, {Ezawa}, {Hatsukade}, {Hughes}, {Iono},
  {Johnson}, {Kawabe}, {Kohno}, {Lowenthal}, {Miller}, {Morrison}, {Oshima},
  {Perera}, {Salvato}, {Silverman}, {Tamura}, {Williams}, \&
  {Wilson}}]{Yun2012}
{Yun}, M.~S., {Scott}, K.~S., {Guo}, Y., {et~al.} 2012, \mnras, 420, 957

\end{thebibliography}

\end{document}